\documentclass[a4paper,openright,twoside,11pt]{report}
\usepackage{graphicx}
\usepackage{amsmath}
\usepackage{amsfonts}
\usepackage{amssymb}
\usepackage{slashed}
\usepackage{feynmp}
\usepackage{url}
\usepackage[numbers,sort&compress]{natbib}
\usepackage{bm}
\usepackage{caption}
\def \sd {\mathbf{D}}
\def \bs {\boldsymbol{\sigma}}
\def \bb {\mathbf{B}}
\def \be {\mathbf{E}}
\def \ba {\mathbf{A}}
\newcommand{\Tr}[1]{\textrm{Tr}\left[#1\right]}

\def \br {\mathbf{r}}
\def \brg {\mathbf{R}}
\def \bx {\mathbf{x}}
\def \by {\mathbf{y}}
\def \bq {\mathbf{q}}

\def \cf {C_F}
\def \nc {N_c}
\def \ca {C_A}
\def \nf {n_f}
\def \tf {T_F}

\def \id {\mathbf{1}}

\def \bk {\mathbf{k}}
\def \mbk {\vert\bk\vert}
\def \bp {\mathbf{p}}
\def \bpg {\mathbf{P}}
\def \bsg {\mathbf{S}}
\newcommand{\Tint}[1]{{\hbox{$\sum$}\!\!\!\!\!\!\!\int\,}_{\!\!\!\!\raise-0.9ex\hbox{$\scriptstyle{#1}$}}}

\def \mbq {\vert\bq\vert}
\def \lag {\mathcal{L}}
\def \qbar {\overline{q}}

\def\siml{{\ \lower-1.2pt\vbox{\hbox{\rlap{$<$}\lower6pt\vbox{\hbox{$\sim$}}}}\ }}
\def\simg{{\ \lower-1.2pt\vbox{\hbox{\rlap{$>$}\lower6pt\vbox{\hbox{$\sim$}}}}\ }}
\def \sd {\mathbf{D}}
\def \bs {\boldsymbol{\sigma}}
\def \bb {\mathbf{B}}
\def \be {\mathbf{E}}
\def \ba {\mathbf{A}}

\def \br {\mathbf{r}}
\def \brg {\mathbf{R}}

\def \bx {\mathbf{x}}
\def \by {\mathbf{y}}
\def \bq {\mathbf{q}}

\def \crr {C_R}

\def \bfnabla {\boldsymbol{\nabla}}
\def \bk {\mathbf{k}}
\def \mbk {\vert\bk\vert}
\def \bp {\mathbf{p}}
\def \bpg {\mathbf{P}}
\def \bsg {\mathbf{S}}

\def \mbq {\vert\bq\vert}
\def \mbp {\vert\bp\vert}
\def \lag {\mathcal{L}}
\def \qbar {\overline{q}}

\def \trt {\tilde{\mathrm{Tr}}}
\def \als {\alpha_{\mathrm{s}}}

\def \m2   {\mu^{2 \epsilon}}

\newcommand{\order}[1]{\mathcal{O}\left(#1\right)}
\newcommand{\MS}{{\overline{\rm MS}}}
\def\alVs{\alpha_{V_s}}
\def\alVo{\alpha_{V_o}}
\def\siml{{\ \lower-1.2pt\vbox{\hbox{\rlap{$<$}\lower6pt\vbox{\hbox{$\sim$}}}}\ }}
\def\simg{{\ \lower-1.2pt\vbox{\hbox{\rlap{$>$}\lower6pt\vbox{\hbox{$\sim$}}}}\ }}
\def\lqcd{\Lambda_\mathrm{QCD}}
\def\lQ{\Lambda_\mathrm{QCD}}
\def\nn {\nonumber}
\def\bfsigma{\mbox{\boldmath $\sigma$}}
\def\sgn{\mathrm{sgn}}

\author{Jacopo Ghiglieri}
\title{PhD Thesis}
\begin{document}

\begin{titlepage}
\begin{center}


\parbox{.38\textwidth}{\begin{flushleft}\quad\phantom{a}\includegraphics[height=1.5cm]{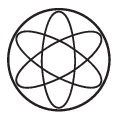}\end{flushleft}}\hfill
\parbox{.58\textwidth}{\begin{flushright} \includegraphics[height=1.5cm]{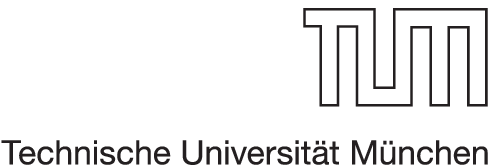}\quad\phantom{a}\end{flushright}}

\vspace{1cm}
TECHNISCHE UNIVERSIT\"AT M\"UNCHEN\\
Institut f{\"u}r Theoretische Physik T30f\\

\vspace{2.5cm}


{\Huge \bf Effective Field Theories of QCD}\\
\vspace{2mm}
{\Huge \bf for Heavy Quarkonia}\\
\vspace{2mm}
{\Huge \bf at Finite Temperature}
%

\vspace{1.5cm}
 
{Jacopo Ghiglieri
}
\end{center}
\vfill

{\noindent Vollst{\"a}ndiger Abdruck der von der Fakult{\"a}t f{\"u}r Physik der Technischen Universit{\"a}t M{\"u}nchen zur Erlangung des akademischen Grades eines
\vspace{.4cm}
\begin{center} 
\emph{Doktors der Naturwissenschaften (Dr.~rer.~nat.)}
\end{center}
\vspace{.4cm}
genehmigten Dissertation.

\vspace{1cm}
\begin{tabular}{lll}
Vorsitzender:&&Univ.-Prof. Dr. Laura Fabbietti\\\\
Pr{\"u}fer der Dissertation:&\qquad1.&Univ.-Prof. Dr. Nora Brambilla\\
&\qquad 2.&Univ.-Prof. Dr. Wolfram Weise
\end{tabular}
\vspace{1cm}

\noindent Die Dissertation wurde am 07.07.2011 bei der Technischen Universit{\"a}t M{\"u}nchen eingereicht und durch die Fakult{\"a}t f{\"u}r Physik am 27.07.2011 angenommen.
}

\end{titlepage}

	 \cleardoublepage
	\newpage
	 \begin{flushright}
	 \vspace*{2in}
	 {\it To my grandfather Eugenio}
	 \end{flushright}
	 \cleardoublepage
	
\section*{Zusammenfassung}
Quarkonia, das hei\ss t gebundene Zust\"ande aus einem Quark und dem dazugeh\"origen Antiquark, repr\"asentieren eine der wichtigsten Sonden in der experimentellen Erforschung des Bereiches hoher Temperaturen des QCD-Phasendiagramms mittels der Kollision von Schwerionen. In diesem Regime wird eine \"Ubergang zum Quark-Gluon-Plasma, einem Medium in dem kein ``Confinement'' mehr vorliegt, erwartet.
Es wurde angenommen, dass gebundene Zust\"ande aufgrund der Abschirmung der Farbladungen in diesem Plasma aufbrechen. 
Experimentelle Daten von SPS, RHIC und vor kurzem auch vom LHC best\"atigen in der Tat diese Hypothese. In der vorliegenden Doktorarbeit dehnen wir  den etablierten, erfolgreichen Rahmen nicht-relativistischer (NR) effektiver Feldtheorien (EFTs) (NRQCD, pNRQCD) zur Untersuchung schwerer Quarkonia (Erzeugung, Spektroskopie, Zer¨f\"alle, ...)  bei verschwindender Temperatur auf nicht verschwindende Temperaturen aus. Das wird durch die sequentielle Ausintegration der Energieskalen, die einen nicht-relativistischen Zustand kennzeichnen, und jenen, die ein thermisches Medium beschreiben, in allen m\"oglichen Hiearchien, die f\"ur Quarkonia im Quark-Gluon-Plasma  von Bedeutung sind, ausgef\"uhrt. 
In diesem Rahmen zeigen wir, wie das Potential, das die zeitliche Entiwicklung des Quark-Antiquark Paares steuert, 
in einer modernen und rigorosen Weise aus der QCD hergeleitet wird, und \"uberbr\"ucken damit die L\"ucke zwischen Potential-Modellen und QCD. Wir zeigen, wie die effektive Feldtheorien systematisch verbessert werden k\"onnen und wie Effekte, die nich mittels eines Potentials beschrieben werden k\"onnen, in dem effektiven Feldtheorie in nat\"urlicher Weise auftreten und neue M\"oglichkeiten f\"ur den Dissoziationsprozess er\"offnen. Wir nutzen diesen EFT-Rahmen, um das Spektrum und die Zerfallsbreite von Quarkonia in einer besonderen Konfiguration der Energieskalen, die f\"ur die Ph\"anomenologie der Gr\"undzust\"ande des Bottomoniums am LHC von Bedeutung ist, zu berechnen. Ferner untersuchen wir in diesem Rahmen den Korrelator von Polyakov-Loops, der  mit den thermodynamischen freien Energien schwerer Quark-Antiquark Paare im Medium in Beziehung steht. Als Input f\"ur ph\"anomenologische Potential-Modell wurde dieser daher oft mittels Berechnungen im Rahmen der Gitter-QCD bestimmt. Unserer Methode erlaubt es uns, die Beziehung zwischen diesen freien Energien und dem Echtzeit Potential, das die Dynamik von Quarkonia beschreibt,  aufzukl\"aren; wir stellen fest, dass diese beiden Gr\"o\ss en nicht nur in ihrem wichtigen Imagin\"arteil, den erstere Gr\"o\ss en \"uberhaupt nicht aufweisen, sondern auch in ihrem Realteil voneinander abweichen.

\section*{Abstract}
Quarkonia, i.e. heavy quark-antiquark bound states, represent one of the most important probes in the experimental investigation, through heavy-ion collisions, of the high-temperature region of the phase diagram of QCD, where the onset of a deconfined medium, the quark-gluon plasma, is expected. Such bound states were hypothesized to dissociate in this plasma due to the screening of the colour charges and experimental data from SPS, RHIC and very recently also LHC indeed show a suppression pattern. In this thesis we extend the well-established and successful zero temperature framework of Non-Relativistic (NR) Effective Field Theories (EFTs) (NRQCD, pNRQCD) for the study of heavy quarkonia (production, spectroscopy, decays, ...)  to finite temperatures. This is achieved by integrating out in sequence the scales that characterize a NR bound state and those that are typical of a thermal medium, in the possible hierarchies that are relevant for quarkonia in the quark-gluon plasma. Within this framework we show how the potential that governs the evolution of the quark-antiquark pair is derived from QCD in a modern and rigorous way, thereby bridging the gap between phenomenological potential models and QCD. We show how the EFTs can be systematically improved and how effects that cannot be encoded in a potential arise naturally in the EFT, giving rise to new mechanisms of dissociation. We use this EFT framework to compute  the spectrum and width of quarkonia in a particular setting that is relevant for the phenomenology of the ground states of bottomonium at the LHC. We also analyze within this framework the correlator of Polyakov loops, which is related to the thermodynamical free energy of heavy quark-antiquark pairs in the medium. As such, lattice computations thereof were frequently used as input for potential models. With our approach we are able to clarify the relation between these free energies and the real-time potential describing the dynamics of quarkonia, finding that the two are different not only in the important imaginary parts, that the former completely lack, but also in the real parts.
	\tableofcontents
	\chapter*{Introduction}
		\addcontentsline{toc}{chapter}{{}Introduction}
	 	Ever since the discovery of the $J/\psi$ meson \cite{Augustin:1974xw,Aubert:1974js}, quarkonia, i.e. heavy quark-antiquark ($Q\overline{Q}$) bound states, have represented an extremely valuable tool in the understanding of the strong interactions. As an example, the analysis of the decay width of the $J/\psi$ \cite{Appelquist:1974zd}, performed within months of its discovery, helped the identification of asymptotically-free Quantum Chromodynamics (QCD) \cite{Fritzsch:1972jv,Fritzsch:1973pi}, a non-Abelian $SU(3)$ gauge theory minimally coupled to quarks, as the theory of the strong interactions. Non-Abelian gauge theories had been proved to be asymptotically free just the year before \cite{Politzer:1973fx,Gross:1973id}.

In this thesis we consider another aspect of the study of strongly-interacting matter where quarkonia play an important role. While at sufficiently high energies QCD is asymptotically free, it instead exhibits confinement at low energies, causing, under ordinary conditions, its asymptotic states to be colourless hadrons composed of the elementary, coloured degrees of freedom, quarks and gluons. At sufficiently high temperatures, however, the phase diagram of QCD is theorized to exhibit, for low chemical potential, a crossover to a deconfined phase called \emph{quark-gluon plasma}, where quarks and gluons are no longed confined into hadrons. This phase has been and is actively investigated in past and present heavy-ion collision experiments at the Super Proton Synchrotron and Large Hadron Collider at CERN and at the Relativistic Heavy Ion Collider at Brookhaven National Laboratory. In these experiments, large nuclei ($A\approx 200$) are collided at energies up to $\sqrt{s_\mathrm{NN}}\approx 5$ TeV per nucleon pair at the LHC, resulting in extremely high particle multiplicities in the final state, which in turn demand reliable, easily identifiable probes of the produced medium.\\
In 1986 Matsui and Satz \cite{Matsui:1986dk} hypothesized that the suppression of the $J/\psi$ in heavy-ion collisions would have represented a striking signature of the formation of a deconfined medium. In particular, their qualitative argument for the suppression relied on the change of the potential governing the evolution of the non-relativistic bound state from a Coulomb+linear to a screened Yukawa potential, motivated by the colour screening induced by the medium. Since then this hypothesis has been intensely investigated, both theoretically and experimentally. On the latter side, a suppressed $J/\psi$ yield in the dilepton channel, with respect to the scaled $pp$ yield, was observed at the SPS and RHIC experiments. The first, very recent data from the heavy-ion collisions at the LHC at the end of 2010 also points to a substantial $J/\psi$ suppression and represents the first quality data on the $\Upsilon$ family of bottomonium ($b\overline{b}$) bound states in heavy-ion collisions.\\
On the theory side, a great deal of the studies of the in-medium dynamics of the $Q\overline{Q}$ bound states has been carried out with phenomenological potential models, first introduced in \cite{Karsch:1987pv}, where all medium effects are encoded in a $T$-dependent potential plugged in a Schr\"odinger equation.  We refer to \cite{Rapp:2008tf,Kluberg:2009wc,Bazavov:2009us} for recent reviews. The derivation of such models from QCD was however not established. Moreover, lattice calculations of free energies and other quantities \cite{McLerran:1981pb,Nadkarni:1986as} obtained from correlation functions of Polyakov loops are often taken as input for the $T$-dependent potential. Although these quantities have been thought to be related to the colour-singlet and colour-octet heavy quark potentials at finite temperature \cite{Nadkarni:1986cz,Nadkarni:1986as}, a precise relation  was still lacking in the literature, as pointed out in \cite{Philipsen:2008qx}.

On the other hand, at zero temperature a framework of Non-Relativistic (NR) Effective Field Theories (EFTs) has been developed in the past decades. These EFTs exploit the hierarchy $m\gg \lqcd$ between the mass $m$ of the heavy quarks $c$, $b$ and $t$ and the QCD scale $\Lambda_\mathrm{QCD}\approx 200$ MeV, as well as, for what concerns their bound states,\footnote{The top quark has a  width so large that it decays before being able to form a bound state. Bound-state effects are however important in the study of $t\overline{t}$ production at threshold.} the non-relativistic hierarchy $m\gg mv\gg mv^2$ between the mass, the typical momentum transfer $mv$, $v$ being the relative velocity, and the kinetic/binding energy $mv^2$. In this latter case, the EFTs that follow from integrating out in sequence the scales $m$ and $mv$ from QCD are respectively Non-Relativistic QCD (NRQCD) \cite{Caswell:1985ui,Bodwin:1994jh} and Potential Non-Relativistic QCD (pNRQCD) \cite{Pineda:1997bj,Brambilla:1999xf,Brambilla:2004jw}. In this second theory the heavy quark potential is given a modern, rigorous definition as a \emph{matching coefficient} of the EFT, in a way bridging the gap between the $T=0$ non-relativistic potential models and QCD. Furthermore in pRNQCD non-potential effects, i.e. effects that cannot be encoded in a potential appearing in a Schr\"odinger equation, such as the retardation effects that give rise to the Lamb shift in QED, are consistently taken into account, as well as relativistic corrections. Finally the theory can be systematically improved by including more operators in the expansion of the small energy scales over the large energy scales, as in any EFT.

In this thesis we then aim at extending this well-established $T=0$ framework to finite temperatures, with the ultimate goal of creating a theory that can describe the dynamics of quarkonium in a deconfined medium, allowing for predictions on its suppression. We will then construct a set of EFTs that generalize NRQCD and pNRQCD with the inclusion of the thermodynamical scales, in the different possible hierarchies that can arise between these and the non-relativistic bound-state hierarchy. We will also take advantage of existing finite-temperature EFTs for the description of the light degrees of freedom, gluons and light quarks. These EFTs will bring to a modern, rigorous QCD derivation of the real-time potential governing the evolution of the heavy quark-antiquark pair, as well as to the systematic consideration of new medium effects, such as the thermal widths induced by the large imaginary parts of the potentials, first observed in \cite{Laine:2006ns}, that are generally not accounted for by phenomenological potential models. We will also analyze the correlation function of Polyakov loops in this EFT framework, in this way obtaining a better understanding of its relation with the real-time potentials that appear in the EFT.

The thesis is organized as follows. Part~\ref{part_intro} is dedicated to introducing in more detail the physical motivation and the theoretical tools that have been mentioned here. In particular, Chap.~\ref{chap_EFT} will be devoted to an overview of QCD and  Effective Field Theories, with particular emphasis on the $T=0$ non-relativistic EFT framework for heavy quarkonium. In Chap.~\ref{chap_thermal} we will concentrate on QCD at finite temperature, introducing its phase diagram and the experimental investigation through heavy-ion collision. We will explain in detail the relevance of quarkonium in these experiments and introduce the most common theoretical tools for its study. Finally, we will give an overview of Thermal Field Theory and of finite-temperature EFTs of QCD.\\
In Part~\ref{part_realtime} we will generalize the NR EFT framework of NRQCD and pNRQCD to finite temperatures. In Chap.~\ref{chap_realtime} we will give an overview of the subject and introduce the hierarchies that can exist between the temperature (and other thermodynamical scales) and the bound-state scales that are relevant for the phenomenology of quarkonium in heavy-ion collisions. In the subsequent Chapters~\ref{chap_Tggr} and \ref{chap_rggT} we consider in detail the two complementary cases $T\gg mv$ and $mv\gg T$, with $m\gg T$ in both cases. In the former we reobtain the potential of  \cite{Laine:2006ns}, with its large imaginary parts, in a rigorous EFT derivation, and we show how a dissociation temperature can be estimated from these results. In the latter we proceed to a computation of the shift of the energy levels of quarkonia and of the width induced by the thermal medium, to a fixed accuracy in the power counting of the EFT. We will also see how new in-medium decay channels arise in this situation and are systematically accounted for by our framework. In both cases we discuss the phenomenological implications of our results, which have been published in \cite{Brambilla:2008cx} and \cite{Brambilla:2010vq}. This Part is concluded by Chap.~\ref{chap_poincare}, where the effects of the explicit breaking of Lorentz invariance caused by the thermal bath are analyzed. We focus on the spin-orbit part of the potential and calculate contributions thereto induced by the medium, showing how they break the realization of Poincar\'e invariance in the NR EFT. The consequences of these results, recently published in \cite{poincare}, for quarkonia moving with different velocities in the preferred reference frame introduced by the bath are analyzed.

Part~\ref{part_imtime} is instead focussed on the thermodynamical heavy-quark free energies extracted from correlation functions of Polyakov loop. After an introduction to the subject in Chap.~\ref{chap_imtimeintro}, we first perform in Chap.~\ref{chap_imtimepert} a next-to-next-to-leading order perturbative calculation of the Polyakov loop and of the Polyakov-loop correlator. In the first case, we find a result that differs from the long-time accepted result of \cite{Gava:1981qd} and we investigate the discrepancy, while in the second case our results are new. In order to have a more transparent physical interpretation of the perturbative result for the correlator, in Chap.~\ref{chap_imtimeEFT} we construct an EFT framework that can be seen as the Euclidean counterpart of the one introduced in Part~\ref{part_realtime}. As we shall show, the Polyakov loop is an observable that is naturally defined in Euclidean space-time and relies on the periodic boundary conditions on imaginary time introduced by the definition of thermal average to be gauge invariant. Within this EFT framework, we will be able to show that, up to a certain accuracy, the correlator can indeed be written as the sum of a colour-singlet and a colour-octet correlator, thus giving a rigorous footing and validity region to the previous statements in the literature. By computing the thermal contributions to these correlator we will be able to recover our previous perturbative result and define gauge-invariant singlet and octet free energies. The results of this Part have been published in \cite{Brambilla:2010xn}.

Finally, Part~\ref{part_concl} contains our conclusions, whereas technical details on the calculations can be found in the appendices.	
\part{Building blocks}
\label{part_intro}
	\chapter{QCD and Non-Relativistic Effective Field Theories at zero temperature}
		\label{chap_EFT}
		In this first Chapter  we shall introduce Quantum Chromodynamics and Non-Relativistic Effective Field Theories thereof. In Sec.~\ref{sec:quantum_chromodynamics} we will lay down the basics of QCD, thereby establishing the notation used throughout this thesis. Subsequently we will give a brief primer to low-energy Effective Field Theories in Sec.~\ref{sec_princ_eft}, showing how an EFT is constructed. In Secs.~\ref{sec_nrqcd} and \ref{sec_pnrqcd} we shall employ this method to construct Non-Relativistic QCD and Potential Non-Relativistic QCD, which are the two Non-Relativistic EFTs of QCD that constitute the $T=0$ EFT framework for heavy quarkonium. 
\section{Quantum Chromodynamics} 
\label{sec:quantum_chromodynamics}
We start by considering Quantum Chromodynamics (QCD) and by writing down its Lagrangian, which will also be useful in clarifying some of the conventions used throughout this thesis.\\
Quantum Chromodynamics \cite{Fritzsch:1972jv,Fritzsch:1973pi} is the accepted theory of strong interactions. It is a non-Abelian gauge theory, often called Yang-Mills theory after its discoverers \cite{Yang:1954ek}, minimally coupled to fermions, the quarks. The non-Abelian gauge group of QCD is $SU(3)$, whose associated charge is called colour. The six flavours of quarks belong to the fundamental representation of this group, which for a gauge group $SU(\nc)$ has dimension $\nc$; therefore $\nc$ is called the number of colours and quarks are then said to have 3 different colours. The gauge bosons mediating the colour interaction are called gluons, are massless and transform like connections in the adjoint representation. This representation having dimension $\nc^2-1$, QCD possesses 8 gluons.\\
The requests of local gauge invariance, Poincar\'e invariance, renormalizability and Parity, Time reversal and Charge conjugation invariance dictate this shape for the Lagrangian\footnote{There exists another gauge invariant operator of mass dimension four, i.e. not spoiling renormalizability, which can be added to the QCD Lagrangian. Such a term is called the $\theta$-term and reads
\begin{equation}
	\lag_\theta=\frac{\theta g^2}{16\pi^2}F^a_{\mu\nu}\tilde F^{\mu\nu\,a},
	\label{thetaterm}
\end{equation}
where $\tilde F$ is the dual tensor of $F$, defined in general as $\tilde T_{\mu\nu}=\frac{1}{2}\varepsilon_{\mu\nu\rho\sigma}T^{\rho\sigma}$.\\
The $\theta$-term can be rewritten as a total divergence, which thus contributes only to a surface term in the action and is irrelevant in the context of perturbation theory, and therefore never enters the Feynman rules. The QCD vacuum can however have a non-trivial topological structure and in this case the surface term above cannot be neglected, as it gives rise to CP-violating effects such as an Electric Dipole Moment (EDM) for the neutron. Experimental searches of the latter allow to put an upper limit on the magnitude of $\theta$ as $\theta<10^{-10}$. For its smallness and its irrelevance in perturbative calculations, we will not consider the $\theta$-term in the rest of this work, thus considering C, P and T as exact symmetries of QCD. We refer the reader to \cite{Dine:2000cj} for a review on the so-called strong CP problem.}
\begin{equation}
	\label{lagrqcd}
	\mathcal{L}_\mathrm{QCD}=-\frac{1}{4}F^{\mu\nu\,a}F^a_{\mu\nu}+\sum_f\overline{\psi}_f(i\gamma^\mu D_\mu -m_f)\psi_f,
\end{equation}
where $f$ is the flavour index ($f=u,\,d,\, c,\, s,\, t,\, b$), $F_{\mu\nu}^a$ is the Yang-Mills field strength tensor
\begin{equation}
	\label{fmunu}
	F_{\mu\nu}^a=\partial_\mu A_\nu^a -\partial_\nu A_\mu^a - gf^{abc}A_\mu^bA_\nu^c,
\end{equation}
$f^{abc}$ being the \emph{structure constants} of the gauge group, defined by the commutator of its generators $T^a$ in the fundamental representation as 
\begin{equation}
	\label{commutator}
	\left[T^a,T^b\right]=if^{abc}T^c.
\end{equation}
$D_\mu$ is the gauge-covariant derivative
\begin{equation}
	\label{derivcov}
	D_\mu=\partial_\mu+igA_\mu,
\end{equation}
where the gauge field has been written in compact form as $A_\mu=A_\mu^aT^a$, with $a=1\ldots8$ and the repeated colour index is understood to be implicitly summed. In the fundamental representation of SU(3) the generators $T^a_{ij}$, with $i,j=1\ldots3$, are 8 Hermitean and traceless $3\times3$ matrices. An explicit representation is given through the Gell-Mann matrices $\lambda^a$ as $T^a=\frac12\lambda^a$. In the adjoint representation $(T^a)_{bc}=-if^{abc}$.\\
Our convention for the normalization of the trace follows from our identification $T^a=\frac12\lambda^a$, but holds for a general $SU(\nc)$ as well. Define $T_R$ as
\begin{equation}
	\Tr{T^aT^b}\equiv T_R\delta^{ab}\,
	\label{deftf}
\end{equation}
where $R=F,A$ labels the representation, either fundamental or adjoint. The properties of the Gell-Mann matrices then yield $T_F=\frac12$. With this condition one can then prove the following identities for the quadratic Casimir
operators of the fundamental and adjoint representations
\begin{equation}
	\label{defcd}
	T^a_{ij}T^a_{jk}\equiv C_F\delta_{ik},\quad C_F=\frac{\nc^2-1}{2\nc},
\end{equation}
\begin{equation}
	\label{defca}
	(T^a)_{cd}(T^b)_{dc}\equiv T_A\delta^{ab}=f^{acd}f^{bcd}=(T^c)_{ad}(T^c)_{db}\equiv C_A\delta^{ab},\,\Rightarrow T_A=C_A=\nc\,.
\end{equation}
For the anticommutator of the generators in the fundamental representation we have
\begin{equation}
\{T^a,T^b\}=\frac{\delta^{ab}}{\nc}\id+d^{abc}T^c\,,\quad d^{abc}d^{abd}=\frac{\nc^2-4}{\nc}\delta^{cd},\quad d^{aab}\equiv 0\,.
	\label{defdabc}
\end{equation}
\subsection{Quantization, renormalization and running coupling}
In this subsection we deal with the quantization of QCD within perturbation theory, which is the approach we will use throughout most of this thesis. Nonperturbative approaches, such as lattice gauge theory, will not be analyzed here.\\
The quantization of gauge theories, and in particular of non-Abelian ones such as QCD, is best performed in the functional formalism, where the procedure of gauge-fixing is carried out through the Faddeev-Popov method. We refer to textbooks such as \cite{Ellis:1991qj,Muta:1998vi,Peskin} for an illustration of the quantization procedure and a derivation of the Feynman rules, which we summarize in App.~\ref{sub_feyn_qcd_realtime}.\\
The quantization of the theory introduces divergences in loop integrals; QCD being a renormalizable theory, the number of superficially divergent amplitudes is finite and these divergences can be absorbed by replacing the bare parameters of the QCD Lagrangian~\eqref{lagrqcd}, i.e. the bare coupling $g_\mathrm{B}$ and the fermion masses $m_{f\,\mathrm{B}}$ with the renormalized parameters measured at an arbitrary scale $\mu$, the \emph{renormalization scale}.\\
The requirement that physical observables need to be independent of the renormalization scale is at the base of the concept of the  \emph{renormalization group} and gives predictive power to the procedure of renormalization. This is achieved through a set of \emph{renormalization group equations} that specify the running of the renormalized parameters as a function of the energy. For what concerns the strong coupling constant $\als\equiv\frac{g^2}{4\pi}$, assuming it is known at an energy scale $Q^2$, the relevant renormalization group equation (RGE) yielding the energy dependence can be derived from the Callan-Symanzik equation. The RGE for $\als$ then reads, considering only massless fermions 
\begin{equation}
	\label{betafunc}
	Q^2\frac{\partial}{\partial Q^2}\als\left(Q^2\right)=\beta\left(\als\left(Q^2\right)\right),
\end{equation}
where we have implicitly defined the so-called $\beta$-function of QCD, which can be computed perturbatively order-by-order by calculating the relevant Green functions and can in general be expressed as an expansion in the coupling. For further convenience and in order to fix our notation we write this expansion as
\begin{equation}
	\label{betaexpand}
	\beta=-2\alpha_s\left(\frac{\alpha_s}{4\pi}\beta_0+\frac{\alpha_s^2}{(4\pi)^2}\beta_1+\ldots\right).
\end{equation}
It is worth mentioning that the QCD $\beta$-function is neither gauge- nor renormalization scheme-independent. However it can be shown that in a mass-independent regularization scheme, such as the MS and $\overline{\mathrm{MS}}$,  the $\beta$-function is gauge independent \cite{Caswell:1974cj,Gross:1975vu} and its first two coefficients $\beta_0$ and $\beta_1$ turn out not to depend on the particular mass-independent scheme adopted.\\
The well-known one-loop result yields \cite{Politzer:1973fx,Gross:1973id}
\begin{equation}
	\label{beta0}
	\beta_0=\left(\frac{11}{3}C_A-\frac{2}{3}\nf\right),		
\end{equation}
where $n_f$ is the number of active quark flavours, i.e. those that can be considered massless at the energy scale considered\footnote{Mass-independent regularization schemes require special care in handling the contribution of massive quarks to the running coupling. When the momentum scale $Q$ falls below the mass of a quark, its contribution does not decouple automatically, as it happens instead in mass-dependent schemes. Decoupling needs then to be performed explicitly, in fact matching the theory to an Effective Field Theory where the quark has been integrated out. We refer to \cite{Rodrigo:1993hc} for a review on the subject of quark mass thresholds, and to \cite{Chetyrkin:1997sg} for the most updated results for this matching.}.\\
Truncating Eq.~\eqref{betaexpand} to its leading order term $\beta_0$ leads to a simple approximate solution of the renormalization group equation~\eqref{betafunc}. It reads
\begin{equation}
	\label{runningalphaQ}
	\als(Q^2)=\frac{\als(\mu^2)}{1+\beta_0\frac{\als(\mu^2)}{\pi}\log\frac{Q^2}{\mu^2}}.
\end{equation}
It is convenient to define the QCD scale  $\lqcd$ as the scale which causes the coupling $\als(\lqcd)$ to diverge, thus bringing to a breakdown of perturbation theory. At the leading order one then has 
\begin{equation}
	\label{lambdaqcd}
	1= \beta_0\frac{\als(\mu)}{2\pi}\log\frac{\mu}{\Lambda_\mathrm{QCD}}.
\end{equation}
A recent world average \cite{Bethke:2009jm} of various different measurements of $\als$ yields $\als \left(M_{Z_0} \right) = 0.1184 \pm 0.0007$, whose corresponding value for the QCD scale is $\Lambda^{(5)}_{\mathrm{QCD}\,\overline{MS}} =(213\pm9)\,\mathrm{MeV}$, where the $(5)$ apex signifies that five flavours of quarks (the up, down, strange, charm and bottom) have been considered massless at the mass $M_{Z_0}=91.2\,\mathrm{GeV}$ of the neutral $Z_0$ boson. This values of $\als$ and $\lqcd$ were obtained using the QCD $\beta$-function and its corresponding running coupling up to 4 loops \cite{vanRitbergen:1997va,Czakon:2004bu,Chetyrkin:1997sg}.\\
We remark again that the perturbative expansion of QCD can only be trusted when $Q\gg\lqcd$; indeed the evolution of $\als\left(Q^2\right)$ to lower values of $Q^2$, according to Eq.~\eqref{betafunc}, yields $\als\sim1$ in the range $Q\sim0.5-1\,\mathrm{GeV}$.

	 	\section{Principles and construction of Effective Field Theories}
\label{sec_princ_eft}
A recurring aspect in the description of many physical phenomena is the presence of different, well-separated energy and/or momentum scales. As a first example, in the study of the hydrogen atom one encounters in descending order the masses of the proton and of the electron, the inverse Bohr radius and the binding energy, with large separations (at least roughly two orders of magnitude) between each step. A second example is represented by the large hierarchical separation between the mass of a muon, when considering its weak decay, and that of the W boson mediating the interaction.\\
The idea at the base of the construction of Effective Field Theories (EFTs) is that,
when a problem is characterized by two or more separated energy scales, the physics at
one energy scale should not be sensitive to the details of the physics at the other. Qualitatively we can then think that, in a first approximation, the scales that are very large (or very small) with respect to the relevant one can be sent to infinity (or zero), and the approximation can be systematically improved by considering the subsequent terms in an expansion in the ratio of the small scale over the large one. In our example,
when studying the weak decay of a muon, a theory describing the full dynamics
of the W boson is not necessary for the problem at hand. In this sense the Fermi theory of weak decays can be considered an Effective Field Theory ``ante litteram'', corresponding to a leading-order term in the expansion in the ratio of the light scale over the W boson mass.\\
The concept of an Effective Field Theory was first introduced by Weinberg in the context of chiral dynamics under the name of ``Effective Lagrangians'' in \cite{Weinberg:1978kz} (see also \cite{Weinberg:2009bg} for an historical overview), from which we quote this excerpt summarizing the main properties of EFTs:
\begin{quote}
For a given set of asymptotic states, perturbation theory with the most
general Lagrangian containing all terms allowed by the assumed symmetries
will yield the most general S-matrix elements consistent with
analyticity, perturbative unitarity, cluster decomposition and the assumed symmetries.
\end{quote}
In the remainder of this Section we will give an overview of the realization of an EFTs in the framework of Quantum Field Theory. For an in-depth review we refer to \cite{Pich:1998xt}. 
\subsection{Construction of an EFT}
\label{sub_eft_basics}
We consider a generic Quantum Field Theory with Lagrangian $\lag_\mathrm{F}$, which we call for simplicity our \emph{fundamental theory}, even though it need not be a fundamental theory but can itself be an EFT. We assume that this theory describes a system with (at least) two well-separated energy/momentum scales, which we call $M\gg\Lambda$. It is worth remarking that these scales need not be necessarily parameters of the Lagrangian $\lag_\mathrm{F}$, such as the masses of the fields, but can also be generated at the quantum level; $\lqcd$ plays indeed very often the role of the low-energy scale in EFTs of QCD.\\
We now introduce a \emph{cutoff scale} $\nu$, with $M\gg\nu\gg\Lambda$ and set out to construct an Effective Field Theory of our original theory, which we require to describe the low energy degrees of freedom (d.o.f.) of the original theory, i.e. those whose energy $E$ is smaller than $\nu$. The high energy degrees of freedom are said to be \emph{integrated out}; they are omitted as explicit degrees of freedom in the EFT, but their physical effect on the low energy physics is taken into account through parameters of the EFT. In this way the effective theory has the same infrared behaviour (but a different UV one) of the underlying fundamental theory.\\
The construction procedure of the EFT can be summarized in the following six points \cite{BraatenBenasque,Pich:1998xt}.
\begin{enumerate}
	\item \emph{Identify the hierarchy of scales and the relevant low-energy degrees of freedom.}\\
		In the simple example the hierarchy would then be $M\gg\Lambda$ and the low-energy d.o.f.s those with energy $E\sim\Lambda$.
	\item \emph{Identify the symmetries to be preserved.}\\
		One has to identify the subset of the symmetries of the fundamental theory that are preserved in the low-energy sector of interest.  These symmetries must then clearly be symmetries of the EFT as well.\\
		For instance, wherever a symmetry group is spontaneously broken at low energies, only the unbroken subgroup (if any) need to be preserved in the EFT.
	\item \emph{Construct the most general theory consistent with these symmetries.}\\
		Once the degrees of freedom and the symmetries have been identified, one can construct the most general theory consistent with the symmetries. With most general it is meant that all operators allowed by the symmetries are to be included, irrespectively of their mass dimension. The Lagrangian $\lag_\mathrm{EFT}$ of the EFT can then be conveniently organized as an expansion in the inverse of the large scale, in our case $M$. We write it as
\begin{equation}
\label{towerop}
\mathcal{L}_\mathrm{EFT}=\sum_i c_i\left(\frac{\mu}{M}\right)\frac{O_i}{M^{d_i-4}},
\end{equation}
where $d_i=[O_i]$ is the mass dimension of the operator $O_i$ and for simplicity we have set the number of space-time dimensions to four. The coefficients $c_i(\mu/M)$, $\mu$ being the renormalization scale, are the \emph{Wilson coefficients}, or \emph{matching coefficients}, of the EFT, and will be determined in a subsequent step of this procedure.\\
If our high-energy degree of freedom at the scale $M$ is actually a heavy particle, it is easy to see that the terms in Eq.~\eqref{towerop} with $d_i>4$ correspond to having replaced the non-local exchange of the heavy particle among the ligher d.o.f.s with a tower of local interactions.
\item \emph{``Power counting'': determination of the relative importance of the terms.}\\
	Eq.~\eqref{towerop} does not seem particularly useful or predictive: it defines a non-renor\-malizable theory with an infinite number of operators and of parameters to be determined, the Wilson coefficients. However the hierarchy of scales easily allows to establish a power counting, enabling us to gauge the relative importance of the terms in $\lag_\mathrm{EFT}$. For instance the light fields appearing in $O_i$ will scale like $\Lambda^d$, where $d$ is the mass dimension of the field, derivatives scale like the momenta of the light d.o.f.s, and thus like $\Lambda$ in our simple example. It is then possible to estimate the relative size of the various operators $O_i/M^{d_i-4}$ in terms of powers of the ratio $\Lambda/M$.  
\item \emph{Choose the desired accuracy.}\\
	One can then decide to fix the accuracy to a certain order $\left(\frac\Lambda M\right)^N$, $N$ being some positive power. The choice of $N$ reflects in general the accuracy one intends to achieve in the calculation of some physical observable through the EFT. It is then clear that, for a fixed $N$, the number of terms in $\lag_\mathrm{EFT}$ is finite, corresponding to a finite number of divergent amplitudes, making the theory renormalizable order by order.
\item \emph{``Matching'': determination of the parameters.}\\
	Once the accuracy has been fixed, the (finite) set of Wilson coefficients needs to be determined. The basic requirement of the EFT is that it describe the same physics of the fundamental theory below the cutoff $\nu$. This is achieved by imposing a \emph{matching condition} between Green functions in the fundamental theory and in the effective theory, i.e. requiring that, at an energy scale $E_m\simeq\nu$ where both theories are valid,  the two sets of Green functions match (hence the name).\\
	This matching is in general performed order by order in the expansion parameter, in our case $\Lambda/M$, and, if the theory allows for a perturbative expansion, in its couplings. The matching furthermore usually relies on Dimensional Regularization (DR) for the treatment of divergent amplitudes. In DR scaleless integrals vanish by definition; this turns out to be extremely useful, since in the matching the lower scales are put to zero. On the EFT side of the matching, the matching scale $E_m\simeq\nu$ is then much larger than $\Lambda$. Therefore the latter, which is the natural scale of loop integrals in the EFT, can be put to zero, effectively causing all loop diagrams to vanish in the matching condition.\\
	For what concerns the fundamental theory side, the external momenta $q_i$ at the matching scale  are fixed to be much smaller than the large scale, i.e. $q_i\simeq\nu\ll M$. One  can then expand in powers of  $q_i/M$. In this way non-analytic terms in the momenta, most typically logarithms thereof, which are not needed in the matching since the EFT is analytic in the external momenta, are explicitly excluded.\\
	UV divergences are cancelled by renormalization counterterms of the fundamental theory itself, thus causing the Wilson coefficients to depend on the renormalization scale $\mu$ and to have a possible non-analiticity in the large scale $M$, most typically through logarithms of $\mu/M$.\\ 
IR divergences instead are not cancelled by the renormalization of the fundamental theory, but are required to vanish when a physical observable is computed. What usually happens is that infrared poles in the matching coefficients, coming from IR divergences in the fundamental theory, are cancelled by corresponding opposite UV poles from divergent amplitudes in the EFT, eventually yielding a finite result. In the following chapters we will encounter many such examples.
\end{enumerate}
We remark that the procedure we have just sketched is perfectly iterable: suppose that there are more scales in the problem, such as $M\gg\Lambda\gg\lambda$. Then the scale $\Lambda$ can be integrated out as well, using the EFT we have just obtained as our ``fundamental'' theory and proceeding to obtain a new EFT that is valid only for $E\ll\Lambda$. We also point out that, when integrating out the largest scale $M$ in a multiscale situation, the EFT one obtains does not depend on the hierarchy (if any) of the other scales, provided of course they are all much smaller than $M$.\\
Due to the many advantages we have briefly illustrated, EFTs have in the past decades been developed for many problems in high energy physics, nuclear physics, gravitation and many other fields. In particular, for what concerns the strong interactions, which represent the birthplace of EFTs,  Weinberg's original idea for chiral dynamics \cite{Weinberg:1978kz} was further developed by Gasser and Leutwyler into \emph{Chiral Perturbation Theory} ($\chi$PT) \cite{Gasser:1983yg,Leutwyler:1993iq,Ecker:1998ai,Pich:1998xt}, which describes the physics of the pseudoscalar Goldstone bosons of QCD at low energies, the expansion parameter being the momentum of the light bosons over the mass of the first vector resonances. Over the years this theory has been extended in various directions, including baryons as well.\\
Other relevant EFTs of QCD for particle physics include Soft Collinear Effective Theory (SCET) \cite{Bauer:2000ew,Bauer:2000yr,Bauer:2001ct,Bauer:2001yt,Bauer:2002nz,Beneke:2002ph}, which exploits the hierarchy between hard, collinear and soft scales in jet physics, and non-relativistic (NR) EFTs of QCD, which are suited for bound states made by at least one heavy quark $Q$. The heavy quark mass plays the role of the large scale, and by integrating it out non-relativistic EFTs are obtained. These theories include Heavy Quark Effective Theory (HQET) \cite{Isgur:1989vq,Isgur:1989ed} for heavy-light mesons ($Q\overline{q}$ states, $D$ and $B$ mesons) and Non-Relativistic QCD (NRQCD) \cite{Caswell:1985ui,Bodwin:1994jh} for heavy quarkonia ($Q\overline{Q}$). In the former case the hierarchy is simply $m\gg\lqcd$, where $m$ is the heavy quark mass and $\lqcd$ is associated with the nonperturbative
dynamics of the light degrees of freedom. In the latter case the hierarchy is instead more complicated, with the appearance of the non-relativistic bound state scales. They are the typical momentum transfer $mv$, $v$ being the relative velocity, and the typical binding energy
$mv^2$. In a non-relativistic bound state $v\ll1$ and hierarchy becomes $m\gg mv\gg mv^2$, with the three scales being called the \emph{hard}, \emph{soft} and \emph{ultrasoft} (US) scale. If one is interested only in the physics at the US scale, it is possible to integrate out the soft scale from NRQCD, obtaining a further EFT called potential Non-Relativistic QCD (pNRQCD) \cite{Pineda:1997bj,Brambilla:1999xf,Brambilla:2004jw}.\\
We dedicate the final Sections of this Chapter to an introduction to these non-relativistic EFTs. In particular Sec.~\ref{sec_nrqcd} will be dedicated to NRQCD, with some mentions of the similarities and differences with HQET, and the subsequent Sec.~\ref{sec_pnrqcd} shall be devoted to an overview of pNRQCD.
\section{Non-Relativistic QCD}
\label{sec_nrqcd}
As we just mentioned, NRQCD is a NR EFT for bound states of heavy quarks and antiquarks, where the hierarchy is that of a NR bound state, $m\gg mv\gg mv^2$. Estimates for the lowest-lying resonances of charmonium and bottomonium give $v^2\approx0.3$ for the former and $v^2\approx0.1$ for the latter: therefore a non-relativistic treatment is viable, but relativistic corrections need to be considered, especially for charmonium. 

Non-Relativistic QCD is then obtained by integrating out the hard scale $m$ from the QCD Lagrangian~\eqref{lagrqcd}, along the method of the previous section. In QCD there is of course another intrinsic scale, $\lqcd$: the position of this scale with respect to the others will play an important role in the following section, for now we just assume that $m\gg\lqcd$, which is in a way the defining aspect of a heavy quark. In a non-relativistic system energy and three-momentum scale differently; however for NRQCD we define a single UV cut-off $\nu_\mathrm{NR}=\{\nu_p,\nu_s\}$ satisfying $m\gg\nu_\mathrm{NR}\gg\lqcd,E,\vert\bp\vert$. $\nu_p$  is the cut-off of the relative three-momentum $\vert\bp\vert$ of the heavy quarks, $\nu_s$ is the cut-off of the energy $E$ of the heavy quarks and of the four-momenta of gluons and light quarks. Moreover the relation $\nu_\mathrm{NR}\gg\lqcd$ implies that the integration of the hard scale can be done perturbatively. The degrees of freedom of the EFT will then be the heavy quark and antiquark and light quarks and gluons with energies and momenta smaller than these cutoffs.\\
For what concerns the power counting of the EFT, below the cut-off $\nu_\mathrm{NR}$ several scales ($E$, $\vert\bp\vert$, $\lqcd$) remain dynamical and one cannot assign a size to each operator without further assumptions. In the original power counting by Bodwin, Braaten and Lepage \cite{Bodwin:1994jh}, it was assumed that $E\sim\lqcd\sim mv^2$, $\mbp\sim mv$ and $v\sim\als(mv)\ll1$. As we shall see in the next section, this implies a Coulombic bound state and it is not the only possibility; a further EFT, pNRQCD, allows an easier disentangling of these scales. The power counting of NRQCD has been further analyzed in \cite{Beneke:1997av,Fleming:2000ib,Brambilla:2002nu,Brambilla:2006ph,Brambilla:2008zg}.\\

\begin{figure}[ht]
	\begin{center}
		\includegraphics{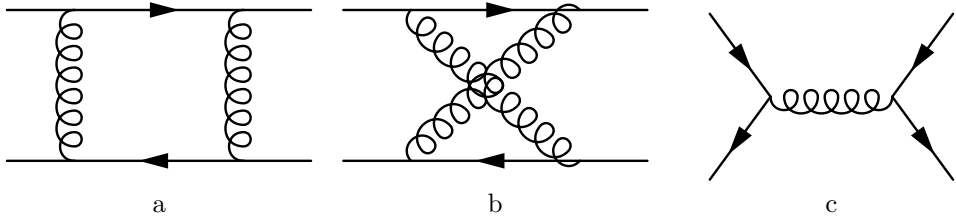}
	\end{center}
	\caption{Leading order QCD diagrams giving rise to the operators in $\lag_{\psi\chi}$. In the unequal mass case $m_1\ne m_2$ only diagrams a and b contribute, whereas for $m_1=m_2$ annihilation diagrams such as c have to be considered as well. Loop corrections to diagram $c$ give rise to the imaginary parts of the matching coefficients.\\
	In QCD the diagrams shown here include the momentum region where the loop momentum (in diagrams a and b) or the gluon momentum (in diagram c) is of the order of the heavy quark mass. In the EFT this momentum region is no longer present and its effects are replaced by the terms in $\lag_{\psi\chi}$.	}
	\label{fig_nrqcd4}
\end{figure}

Coming to the Lagrangian, in this EFT heavy quark-antiquark pairs cannot be created anymore, so it is convenient to use non-relativistic Pauli spinors instead of Dirac spinors. Let then $\psi(x)$ be the Pauli spinor field annihilating a heavy quark and $\chi(x)$ the one creating a heavy antiquark. Furthermore if the quark-antiquark pair is of the same flavour it can annihilate to hard gluons, which have been integrated out: in order to preserve this physical aspects the NRQCD Lagrangian contains imaginary Wilson coefficients.\\
The NRQCD Lagrangian will thus be expressed as a power expansion in $\frac{1}{m_1}$ and $\frac{1}{m_2}$, $m_1$ and $m_2$ being the masses of the heavy quark and antiquark. 
The symmetries to be preserved are those of QCD, although full Lorentz invariance will not be linearly realized in the heavy quark sector of the EFT, as we will discuss in more detail later on. The Lagrangian is then, with two different masses $m_1$ and $m_2$ and at order $\frac{1}{m_1^a m_2^b}$ with $a+b=2$  \cite{Caswell:1985ui,Bodwin:1994jh,Manohar:1997qy,Bauer:1997gs,Brambilla:2004jw}
\begin{equation}
	\label{lagrnrqcd}
	\mathcal{L}_\mathrm{NRQCD}=\lag_\psi+\lag_\chi+\lag_{\psi\chi}+\lag_{g}+\lag_l\,,
\end{equation}
where  $\lag_\psi$ and $\lag_\chi$ are the heavy quark and antiquark terms, $\lag_{\psi\chi}$ contains the heavy quark-antiquark interaction terms expressed as four-fermion terms at this order, $\lag_{g}$ is the gluon Lagrangian and $\lag_l$ is the light quark one.\\
The heavy quark sector Lagrangian is
 \begin{eqnarray}
 	\nonumber \lag_\psi &=& \psi^\dagger \left\{ iD^0 + c_k^{(1)} \frac{\sd^2}{2m_1} + c_4^{(1)} \frac{\sd^4}{8m_1^3} + c_F^{(1)} g \frac{\bs\cdot\bb}{2m_1} + c_D^{(1)}  g\frac{\sd\cdot\be-\be\cdot\sd}{8m_1^2}\right.\\ 
 \nonumber&&+\left.ic_S^{(1)}g\frac{\bs\cdot(\sd\times\be-\be\times\sd)}{8m_1^2}\right\} \psi\\
 \nonumber&&+\frac{c_1^{hl(1)}}{8m_1^2}g^2\sum_{i=1}^{\nf}\psi^\dagger T^a\psi\;\qbar_i\gamma_0T^aq_i+\frac{c_2^{hl(1)}}{8m_1^2}g^2\sum_{i=1}^{\nf}\psi^\dagger\gamma^\mu\gamma_5 T^a\psi\;\qbar_i\gamma^\mu\gamma_5T^aq_i\\
&&+\frac{c_3^{hl(1)}}{8m_1^2}g^2\sum_{i=1}^{\nf}\psi^\dagger \psi\;\qbar_i\gamma_0q_i+\frac{c_4^{hl(1)}}{8m_1^2}g^2\sum_{i=1}^{\nf}\psi^\dagger\gamma^\mu\gamma_5\psi\;\qbar_i\gamma^\mu\gamma_5q_i,
\label{lagpsi}
 \end{eqnarray}
 where the part in brackets is the free field part plus the coupling to soft gluons, whereas the remaining terms describe the local interaction with light quarks, again mediated by heavy quark loops. 
 Here $\bs$ are the Pauli matrices, $\sd=\nabla-ig\mathbf{A}$, $\be^i=F^{i0}$ is the chromoelectric field and $\bb^i=-\varepsilon_{ijk}F^{jk}/2$ is the chromomagnetic field. The term $c_4^{(1)} \frac{\sd^4}{8m_1^3}$ is of order $\frac{1}{m^3}$, but it is nonetheless included: in the power counting of Bodwin, Braaten and Lepage \cite{Bodwin:1994jh} $g\be\sim m^2v^3$, $g\bb\sim m^2v^4$ and of course $\sd\sim mv$, so this term is of the same size $mv^4$ as the other three after it in the brackets.\\
We furthermore remark that the mass term, i.e. the operator  $-\psi^\dagger m_1\psi$ in the bilinear part of $\lag_\psi$, has been removed via a \emph{field redefinition}. Field redefinitions are a very important tool in EFTs and are widely used to reduce the number of operators, as we shall see also in the other sectors of the NRQCD Lagrangian. In this particular case, the field redefinition we have employed is
\begin{equation}
	\psi(\bx,t)\to\exp\left(-im_1t\right)\psi(\bx,t)\,.
	\label{fieldredef}
\end{equation}
The heavy antiquark sector can be obtained by applying charge conjugation to this Lagrangian, i.e. $\psi^c=-i\sigma^2\chi^*$, $A_\mu^c=-A_\mu^T$. The heavy quark-heavy antiquark interaction is described by these terms 
\begin{eqnarray}
	\nonumber \lag_{\psi\chi}&=&\frac{f_1({}^1S_0)}{m_1m_2}\psi^\dagger\chi\chi^\dagger\psi+\frac{f_1({}^3S_1)}{m_1m_2}\psi^\dagger\bs\chi\chi^\dagger\bs\psi\\
	&&+\frac{f_8({}^1S_0)}{m_1m_2}\psi^\dagger T^a\chi\chi^\dagger T^a\psi+\frac{f_8({}^3S_1)}{m_1m_2}\psi^\dagger T^a\bs\chi\chi^\dagger T^a\bs\psi,
\end{eqnarray}
where $f_{1,8}$ refers to colour singlet or octet states, while the spectroscopic notation labels angular momentum states. In the case $m_1=m_2$ these coefficients have also an imaginary part that describes the decay into light particles. The QCD processes giving rise to these terms are sketched in Fig.~\ref{fig_nrqcd4}.
\\
The gluon Lagrangian is 
\begin{equation}
	\label{lagg}
	\lag_g=-\frac{1}{4}F^{\mu\nu\, a}F_{\mu\nu}^a+\frac{1}{4}\left(\frac{c_1^{g(1)}}{m_1^2}+\frac{c_1^{g(2)}}{m_2^2}\right)gf_{abc}F_{\mu\nu}^aF^{b\,\mu}_\alpha F^{\nu\alpha\,c}\,,
\end{equation}
where the second term on the r.h.s. encodes the dependence on the integrated-out hard degrees of freedom. The index $(1)$ labels the contribution from the heavy mass $m_1$, while the index $(2)$ labels the analogous contribution from the mass $m_2$; the apex $g$ stands for gluon. This term comes from the radiative corrections to the three- and four-gluon vertices mediated by a heavy quark loop. A field redefinition has been used to have the usual value of $-\frac{1}{4}$ in front of the first term: this corresponds to redefining the coupling constant in such a way that it runs with $\nf-2$ flavours ($\nf-1$ for $m_1=m_2$)\footnote{Further identities and field redefinitions have been used to reduce the number of terms and coefficients. For details see \cite{Manohar:1997qy,Pineda:1998kj,Pineda:2000sz}}. Finally, the light quark sector is described by this Lagrangian
\begin{eqnarray}
	\nonumber \lag_l&=&\sum_{i=1}^{\nf}\qbar_i i\gamma^\mu D_\mu q_i+\frac{1}{8}\left(\frac{c_1^{ll(1)}}{m_1^2}+\frac{c_1^{ll(2)}}{m_2^2}\right)g^2\sum_{ij=1}^{\nf}
\qbar_iT^a\gamma^\mu q_i\;\qbar_iT^a\gamma_\mu q_i\\
\nonumber &&+\frac{1}{8}\left(\frac{c_2^{ll(1)}}{m_1^2}+\frac{c_2^{ll(2)}}{m_2^2}\right)g^2\sum_{ij=1}^{\nf}
\qbar_iT^a\gamma^\mu\gamma_5q_i\;\qbar_iT^a\gamma_\mu\gamma_5 q_i\\
\nonumber &&+\frac{1}{8}\left(\frac{c_3^{ll(1)}}{m_1^2}+\frac{c_3^{ll(2)}}{m_2^2}\right)g^2\sum_{ij=1}^{\nf}
\qbar_i\gamma^\mu q_i\;\qbar_i\gamma_\mu q_i\\
 &&+\frac{1}{8}\left(\frac{c_4^{ll(1)}}{m_1^2}+\frac{c_4^{ll(2)}}{m_2^2}\right)g^2\sum_{ij=1}^{\nf}
\qbar_i\gamma^\mu\gamma_5 q_i\;\qbar_i\gamma_\mu \gamma_5q_i,
\end{eqnarray}
where the Fermi-like four-fermion operators describe the interaction between light quarks at the hard scale, mediated by heavy quarks in the loops. $D_\mu$ is the usual covariant derivative.\\
We remark that, when only a single heavy quark is considered, which is tantamount to omitting $\lag_\chi$ and $\lag_{\psi\chi}$, the theory simplifies and the resulting Lagrangian can be brought with a field redefinition of the heavy field $\psi$ (see \cite{Manohar:1997qy}) to that of Heavy Quark Effective Theory (HQET), whose power counting is however different, since there one has $E\sim\lqcd$ and $\mbp\sim\lqcd$.\\

For what concerns the determination of the Wilson coefficients appearing in the NRQCD Lagrangian~\eqref{lagrnrqcd}, they have to be computed by matching on-shell scattering amplitudes\footnote{The matching can be done also with off-shell amplitudes, but it turns out to be more intricate, since the loop diagrams in the NRQCD side are not automatically scaleless \cite{Hoang:2002ae}.} in QCD and NRQCD, imposing that they be equal for scales below the cut-off $\nu_\mathrm{NR}$ at the desired order in the expansion in $\als$ and $1/m_i$. The matching coefficients at order $\als/m^3$ in the heavy quark and gauge sector can be found in \cite{Manohar:1997qy}, whereas the heavy quark-heavy antiquark coefficients have been obtained at one loop up to order $1/m^2$ in \cite{Pineda:1998kj}. Higher order expression in $\als$ and $1/m$ have been obtained in the literature; a review can be found in \cite{Vairo:2003gh}. Here we just remark that a limited subset of the matching coefficients is nonzero at order $\als^0$. They are $c_k^{(i)}=c_4^{(i)}=c_F^{(i)}=c_D^{(i)}=c_S^{(i)}=1+\order{\als}$, where we stress that the equality holds only at order $\als^0$, and both for the heavy quark and the heavy antiquark. All other matching coefficients are at least of order $\als$.\\

As we mentioned above, Poincar\'e invariance is not linearly realized in the heavy quark sector of NRQCD. However the EFT is by construction equivalent order by order in the $1/m$ and $\als$ expansions to QCD, which is fully Poincar\'e invariant; it is then clear that this invariance should reflect itself in the matching coefficients of the EFT. A first analysis in the context of HQET, corresponding to the one-quark or one-antiquark sectors of NRQCD, was carried out under the name of \emph{reparametrization invariance} in \cite{Luke:1992cs,Chen:1993sx,Manohar:1997qy,Sundrum:1997ut}, where the following exact relations were found
\begin{equation}
	c_k^{(i)}=c_4^{(i)}=1,\qquad 2c_F^{(i)}-c_S^{(i)}-1=0\,.
	\label{poincnrqcd}
\end{equation}
These results are confirmed by an alternative, more general derivation \cite{Brambilla:2001xk,Brambilla:2003nt} which imposes the Poincar\'e algebra on the generators of Poincar\'e transformations of the EFT. Such a framework, contrary to reparametrization invariance, can be applied to any non-relativistic EFT. It has furthermore been employed to obtain constraints on the $\lag_{\psi\chi}$ part of the Lagrangian: in \cite{Brambilla:2008zg} a set of relations between the imaginary parts of the matching coefficients of the four-fermion operators at order $1/m^2$ with those at order $1/m^4$ (in the equal mass case) was obtained.\\

We conclude this overview of NRQCD by mentioning some of its successful applications in the study of the production and decay of heavy quarkonium, where the EFT allows to factorize the hard dynamics at the scale $m$ from the soft dynamics at the lower scales. The first can then be computed in perturbation theory, whereas the latter are encoded into long distance matrix elements between operators of specific angular momentum and colour and the specific quarkonium state being considered. These matrix elements can then be extracted from experimental data, or evaluated on the lattice. For reviews on the subject we refer to \cite{Brambilla:2004wf,Brambilla:2010cs}.\\
It is worth mentioning that NRQCD is also employed on the lattice, especially for the study of bottomonium, whose large mass, when studied in QCD, would require a very fine lattice spacing, which in turn requires large computational resources. We refer to \cite{Kronfeld:2003sd,Lepage:2005eg} for reviews on NRQCD on the lattice. Recent results are summarized in \cite{Brambilla:2010cs}.

\section{Introduction to pNRQCD}
\label{sec_pnrqcd}
As we mentioned in the previous section, the scales $E\sim mv^2$, $\mbp\sim mv$ and $\lqcd$ are still dynamical in NRQCD, with $mv\gg mv^2$.
Our aim is now to obtain a further EFT of NRQCD, potential Non-Relativistic QCD (pNRQCD) \cite{Pineda:1997bj,Brambilla:1999xf,Brambilla:2004jw}, that focuses on the physics at the ultrasoft scale E, relevant for the study of the binding of quarkonium, and therefore for its spectroscopy.\\
To this end it is necessary to integrate out the soft scale $\mbp$, under a further hierarchical assumption on the relative magnitude of $\lqcd$ with respect to $\mbp$ and $E$. According to the employed assumption the resulting EFT and its degrees of freedom will be different: if one assumes that $\mbp\gg\lqcd$ than the integration of the soft scale can be carried out in perturbation theory. If furthermore $\mbp\gg E\simg\lqcd$, than the EFT one obtains after integrating out $\mbp$ is the desired one, and it is normally termed \emph{weakly coupled} pNRQCD. If instead $\mbp\gg\lqcd\gg E$ one needs to integrate out with nonperturbative techniques $\lqcd$ after $\mbp$ to obtain pNRQCD. In the last case, when $\mbp\simg\lqcd$, the integration of the soft scale has to be done nonperturbatively from the outset. These last two scenarios go under the name of \emph{strongly coupled} pNRQCD. 
We discuss here the weak-coupling scenario only. Our exposition follows the one in \cite{Brambilla:2004jw}, to which we refer for further details and for an introduction to the strong-coupling regime.
\subsection{pNRQCD in the weak coupling regime}
\label{sub_weakpnrqcd}
In this scenario the scales are assumed to fulfill the hierarchy $\mbp\gg E\simg\lqcd$; furthermore, one has to assume that the considered states are far from any open heavy flavour threshold. These assumptions are expected to hold for the lowest-lying resonances of bottomonium ($\Upsilon(1S)$, $\eta_b$) and, to a lesser extent, charmonium ($J/\psi$, $\eta_c$).\\
The integration of the degrees of freedom at the energy scale $\vert\bp\vert$ is then perturbative and results in a lowering of the UV cut-off without a qualitative change in the degrees of freedom with respect to NRQCD. We thus have $\nu_\mathrm{pNR}=\{\nu_p,\nu_\mathrm{US}\}$, where $\nu_p$ is the cut-off in the three-momenta of the heavy quarks and $\nu_\mathrm{US}$ is the ultrasoft cut-off in the energy of the heavy quarks and in the four-momenta of the light degrees of freedom. The scales obey $\vert\bp\vert\ll\nu_p\ll m$ and $E\sim\bp^2/m\ll\nu_\mathrm{US}\ll\vert\bp\vert$. The Wilson coefficients will then depend on the three-momenta of the heavy quark-antiquark pair, usually through the momentum transfer $\bp$, giving rise to non-local potential terms in position space. It is thus convenient to take advantage of the fact that the relative three-momentum of the heavy quarks is always larger than the four-momentum of the light degrees of freedom. To translate this fact in position space let $\brg=(\bx_1+\bx_2)/2$ be the center-of-mass coordinate of the $Q\overline{Q}$ system\footnote{In this section we restrict ourselves to the equal mass case $m_1=m_2=m$.} and let $\br=\bx_1-\bx_2$ be the relative coordinate: then the gauge fields can be evaluated in $\brg$ and $t$, i.e. $A^\mu=A^\mu(\brg,t)$ dropping the explicit $\br$-dependence and multipole-expanding with respect to this variable. This is possible since the typical size of $\br$ is the inverse of the soft scale, i.e. $1/r\sim\mbp$.\\
It is furthermore convenient, as we shall see, to decompose the $Q\overline{Q}$ system according to its colour state in a singlet state $\mathrm{S}(\br,\brg,t)$ and in the octet states $\mathrm{O}(\br,\brg,t)$. Since the hadronic scale $\lqcd$ has not been integrated out and $E\simg\lqcd$, colour-octet $Q\overline{Q}$ states are then an explicit degree of freedom in the EFT.\\

For what concerns the power counting of the EFT, we have that  $m$ and $\als(m)$, inherited from the
hard matching coefficients of NRQCD, have well-known values. Derivatives with respect to the relative
coordinate $i\nabla_\br$ and $1/r\sim k$ (the momentum transfer) must be assigned the soft scale $\sim \mbp$.
On the other hand, time derivatives $i\partial_0$ , centre-of-mass derivatives $i\nabla_\brg$ , and the fields of the light degrees of
freedom must be assigned the US scale $E \sim \bp^2 /m$. For what concerns the coupling, the $\als$ arising in the matching calculation from NRQCD, namely those in the potentials, are naturally assigned the size $\als(1/r)$ and
those associated with the light d.o.f.s the size $\als(E)$. The identification $v\sim\als(1/r)\ll1$, which corresponds to a Coulombic bound state, will be justified a posteriori by computing the spectrum of the bound state in this EFT.\\
We remark that, if $\lqcd$ did not exist (like in QED), this counting would be homogeneous, in that
each term has a well-defined size. If $E\sim\lqcd$ this is also true, but clearly the scale $E$ is not perturbative any longer.
If on the other hand $E\gg\lqcd$, the US scale is perturbative but the counting is inhomogeneous (i.e. it is not possible to assign a priori
a unique size to each term) since the light degrees of freedom may have contributions both
at the scales $E$ and $\lqcd$. It is however still possible to identify the largest term as before.\\

In order to write the Lagrangian it is convenient to represent the quark-antiquark pair by a wavefunction field $\Psi(\bx_1,\bx_2,t)$, where $\bx_1$ is the spatial position of the quark, $\bx_2$ is the position of the antiquark and both  are evaluated at the same time $t$: since the system is non-relativistic time is universal and we can restrict ourselves to calculating equal-time correlators. $\Psi(\bx_1,\bx_2,t)$  thus spans a subspace of the Fock space:
\begin{equation}
	\label{psidef}
	\int d^3\bx_1 d^3\bx_2\Psi(\bx_1,\bx_2,t)\psi^\dagger(\bx_1,t)\chi(\bx_2,t)\left\vert\mathrm{US\;gluons}\right\rangle,
\end{equation}
where $\left\vert\mathrm{US\;gluons}\right\rangle$ is a state composed by an arbitrary number of ultrasoft gluons (the light degrees of freedom) and no heavy quarks or antiquarks. This subspace of the Fock space describes the heavy quark-antiquark sector, which is our sector of interest. Since in the NR EFT heavy quark and heavy antiquark numbers are separately conserved we can thus project our theory to this subspace. It is furthermore convenient, as we mentioned before, to decompose this field into its colour singlet and colour octet components with homogeneous ultrasoft gauge transformations with respect to the center-of-mass coordinate. Quantitatively this translates to
\begin{equation}
	\label{decomposesingoct}
	\Psi(\bx_1,\bx_2,t)=P\left[e^{ig\int_{\bx_2}^{\bx_1}\mathbf{A}\cdot d\bx}\right]\mathrm{S}(\br,\brg,t)+P\left[e^{ig\int_{\brg}^{\bx_1}\mathbf{A}\cdot d\bx}\right]\mathrm{O}(\br,\brg,t)P\left[e^{ig\int_{\bx_2}^{\brg}\mathbf{A}\cdot d\bx}\right],
\end{equation}
where $P$ is the path-ordering operator and $\mathbf{A}$ is the spatial part of the gauge field. Under ultrasoft gauge transformations $g(\brg,t)$ we have
\begin{equation}
	\label{gaugesingoct}
	\mathrm{S}(\br,\brg,t)\to \mathrm{S}(\br,\brg,t),\qquad \mathrm{O}(\br,\brg,t)\to g(\brg,t)\mathrm{O}(\br,\brg,t)g^{-1}(\brg,t).
\end{equation}
Using the singlet and octet fields makes the relative and center-of-mass coordinates $\br$ and $\brg$ explicit, because of this transformation property, and allows us to exploit the fact that $\br$ is much smaller than the typical length scale of the light d.o.f.s by a multipole expansion of the gauge field in this variable. Consider for example the time component of the covariant derivative
\begin{equation}
	\label{der0}
	iD_0\Psi(\bx_1,\bx_2,t)=i\partial_0\Psi(\bx_1,\bx_2,t)-gA_0(t,\bx_1)\Psi(\bx_1,\bx_2,t)+\Psi(\bx_1,\bx_2,t)gA_0(t,\bx2),
\end{equation}
where the relative sign of the last two terms is a consequence of charge conjugation. Then multipole expanding this equation in $\br$ yields
\begin{eqnarray}
	\nonumber iD_0\Psi(\bx_1,\bx_2,t)=i\partial_0\Psi(\bx_1,\bx_2,t)-[gA_0(t,\brg),\Psi(\bx_1,\bx_2,t)]\\
	\label{multipole0}
	-\frac{1}{2}\br^i(\partial_igA_0(t,\brg))\Psi(\bx_1,\bx_2,t)-\frac{1}{2}\br^i\Psi(\bx_1,\bx_2,t)(\partial_igA_0(t,\brg))+\order{r^2}\,,
\end{eqnarray}
and analogously for the spatial terms
\begin{eqnarray}
	\nonumber i\sd_{\bx_{1(2)}}\Psi(t,\bx_1,\bx_2)&=&i\bfnabla_{\bx_1(2)}\Psi(t,\bx_1,\bx_2)+g\mathbf{A}(t,\bx_{1(2)})\Psi(t,\bx_1,\bx_2)\\
\nn&=&\left(+(-)i\nabla_{\br}+\frac{i}{2}\nabla_\brg+g\mathbf{A}(t,\brg)+(-)\frac{\br^i}{2}(\partial_ig\mathbf{A}(t,\brg))\right)\\
	\label{multipole1}
	&&\times\Psi(t,\bx_1,\bx_2)	+\order{r^2}\,.
\end{eqnarray}
The Lagrangian will have this general form
\begin{equation}
	\label{lagrpnrgeneral}
	L_\mathrm{pNRQCD}=L_\mathrm{NRQCD}^\mathrm{US}+L_\mathrm{pot},
\end{equation}
where $L_\mathrm{NRQCD}^\mathrm{US}$ is the NRQCD Lagrangian where all light degrees of freedom are intended as ultrasoft and $L_\mathrm{pot}$ are the new non-local potential terms. We thus have, after projection on the subspace spanned by~\eqref{psidef}, best carried out in the Hamiltonian formalism, but before multipole-expanding
\begin{eqnarray}
	\nn L_\mathrm{pNRQCD}&=&\int d^3\bx_1d^3\bx_2\Tr{\Psi^\dagger(\bx_1,\bx_2,t)\left( iD_0+\frac{\sd^2_{\bx_1}}{2m_1}+\frac{\sd^2_{\bx_2}}{2m_2}+\ldots\right)\Psi(\bx_1,\bx_2,t)}\\
	\nn &&-\int d^3x\frac{1}{4}F^{\mu\nu\,a}(x)F_{\mu\nu}^a(x)+\int d^3x\sum_{i=1}^{\nf}\qbar_i(x)i\gamma^\mu D_\mu q_i(x)+\ldots\\
	\nn&&+\int d^3\bx_1 d^3\bx_2\mathrm{Tr}\left[\Psi^\dagger(\bx_1,\bx_2,t)V(\br,\bp_1,\bp_2,\bsg_1,\bsg_2)\right.\\
	&&\hspace{5cm}\left.\times(\text{US gluon fields})\Psi(\bx_1,\bx_2,t)\right],
	\label{lagprnqcdnomulti}
\end{eqnarray}
where in the last term $V(\br,\bp_1,\bp_2,\bsg_1,\bsg_2)$ is the non-local potential, depending on the relative coordinate $\br$, on the momenta $\bp$ and on the spins $\bsg^{(i)}\equiv\bfsigma^{(i)}/2$ of the heavy quark and antiquark. Ultrasoft gluons typically appear at higher orders: at the lowest order the potential term is a four-fermion non-local operator. At the leading order in $\als$, $1/m$ and in the multipole expansion the trace appearing in the last line of Eq.~\eqref{lagprnqcdnomulti} is simply a Coulomb term given by one-gluon exchange, reading
\begin{equation}
	\frac{\als}{\vert \bx_1-\bx_2\vert}\Tr{T^a\Psi^\dagger(t,\bx_1,\bx_2)T^a\Psi(t,\bx_1,\bx_2)},
	\label{coulombpot}
\end{equation}
and it is easy to understand that its non-local nature comes from having integrated out the gluons exchanged between the heavy quark-antiquark pair with momenta of the order of the soft scale $\mbp\sim1/r$.\\
We can exploit Eqs.~\eqref{decomposesingoct},~\eqref{multipole0} and~\eqref{multipole1} to introduce the colour singlet and octet fields and perform the multipole expansion. We choose the following normalizations in colour space
\begin{equation}
	\label{normsingoct}
	\mathrm{S}=S\frac{\id_c}{\sqrt{\nc}},\qquad \mathrm{O}=O^a\frac{T^a}{\sqrt{\tf}},
\end{equation}
where $\id_c$ is the $SU(\nc)$ identity. After the multipole expansion the Lagrangian can be organized as an expansion in powers of $1/m$, $r$ and, in the perturbative regime we are considering, $\als(1/r)$. Up to order $p^3/m^2$ we then have this Lagrangian density \cite{Pineda:1997bj,Brambilla:1999xf,Brambilla:2004jw}
\begin{align}
	 \nonumber \mathcal{L}_\mathrm{pNRQCD}=&\int d^3\br\,\mathrm{Tr}\left[\mathrm{S}^\dagger(i\partial_0-h_s(\br,\bp,\bpg,\bsg_1,\bsg_2))\mathrm{S}+\mathrm{O}^\dagger(iD_0-h_o(\br,\bp,\bpg,\bsg_1,\bsg_2))\mathrm{O}\right]\\
	 \nonumber &+gV_A(r)\Tr{\mathrm{O}^\dagger\br\cdot\be \mathrm{S}+\mathrm{S}^\dagger\br\cdot\be \mathrm{O}}+g\frac{V_B(r)}{2}\Tr{\mathrm{O}^\dagger\br\cdot\be \mathrm{O}+\mathrm{O}^\dagger \mathrm{O}\br\cdot\be}\\
\label{lagrpnrqcd}
&-\frac{1}{4}F^{\mu\nu\,a}F_{\mu\nu}^a+\sum_{i=1}^{\nf}\qbar_ii\gamma^\mu D_\mu q_i,
 \end{align}
 where $iD_0\mathrm{O}=i\partial_0\mathrm{O}-g[A_0(\brg,t),\mathrm{O}]$. The operators $h_s$ and $h_o$ appearing in the first line contain the kinetic terms and the potentials. They are
\begin{equation}
	h_{s,o}=\frac{\bp^2}{m}+\frac{\bpg^2}{4m}
+V^{(0)}_{s,o}
+\frac{V^{(1)}_{s,o}}{m}+\frac{V^{(2)}_{s,o}}{m^2}+\ldots,
\label{sinoctham}
\end{equation}
where  $\bpg=-i\sd_\brg$, $\bp=-i\nabla_\br$ and the potentials $V_s$ and $V_o$ have been arranged as an expansion in $1/m$. $V^{(0)}$ is the \emph{static potential}, $V^{(1)}$ represents its first non-static correction and  $V^{(2)}$ consists of a sum of many terms, such as a $\bp$-dependent term, terms
depending on the angular momentum, on the heavy quark-antiquark spins
and a spin-orbit term. They have also an imaginary, local part proportional to $\delta^3(\br)$ that governs the decays. We refer to Appendix~\ref{app_pnrqcd} for a listing of $V^{(2)}$.\\
The operators shown on the second line of the pNRQCD Lagrangian~\eqref{lagrpnrqcd} represent in fact the terms appearing in the multipole expansion at order $r^1/m^0$. They are chromoelectric dipole terms that can cause singlet-octet and octet-octet transitions. If one is interested in obtaining the quarkonium spectrum up to order $m\als^5$, as we shall see, as well as for most of the applications discussed in this thesis, the displayed Lagrangian is sufficient. We refer again to  App.~\ref{app_multipole} for further terms in the multipole expansion.\\
We furthermore remark  that, from the decomposition~\eqref{decomposesingoct} and its consequence~\eqref{gaugesingoct} it follows that the Lagrangian is written in an explicitly gauge invariant form. Moreover at the zeroth order in $r$, the Lagrangian reduces for the singlet field to a Schr\"odinger Lagrangian, where the potential takes a modern and rigorous definition as a Wilson coefficient of the theory.\\

The functions $V_s$, $V_o$, $V_A$, $V_B$ are the Wilson coefficients of the theory and have to be matched to NRQCD. For what concerns the multipole coefficients we have $V_A=1+\order{\als^2}$ and $V_B=1+\order{\als^2}$ \cite{Brambilla:2006wp}. Moving to the potentials, the matching can be done either through a diagrammatic matching, order by order in $\als$ and $1/m$ \cite{Pineda:1997bj,Pineda:1997ie,Pineda:1998kn}, or through a Wilson loop matching \cite{Brambilla:1999xf,Brambilla:2000gk,Pineda:2000sz}, where the Wilson loop is expressed in NRQCD and pNRQCD as a function of the fields appearing in each theory at each order in the expansions. This second approach is valid in the non-perturbative regime as well.\\
For the static potentials one obtains 
\begin{equation}
V^{(0)}_s=-C_F\frac{\alVs}{r}\,,\qquad V^{(0)}_o=\frac{1}{2N_c}\frac{\alVo}{r}\,,
\label{staticpot}
\end{equation}
where $\alVs$ and $\alVo$ are series in
$\als$ and at the leading order $\alVs=\alVo=\als$. $\alVs$ is known
up to three loops \cite{Smirnov:2008pn,Smirnov:2009fh,Anzai:2009tm}, whereas $\alVo$
to two loops \cite{Kniehl:2004rk}. Starting from order $\als^4$, corresponding to three loops,
$\alVs$ is infrared divergent. This divergence was first identified in the Wilson loop in
\cite{Appelquist:1977es}, and later analyzed in the context of this EFT as an IR-divergent contribution to $\alVs$ in \cite{Brambilla:1999qa}. As we shall see later in this section, this divergence cancels when computing the related physical observable, the spectrum or, when considering the static limit only, the static energy.  We would like to stress again that in pNRQCD the potential is a Wilson coefficient and is not an observable itself; therefore the fact that it is divergent does not represent, in this EFT context, an issue.\footnote{$\alVo$ is IR divergent at order $\als^4$ as well. A pNRQCD treatment of this divergence, as well as a calculation of the renormalization-group logarithms for the octet potential at order $\als^4$, has been performed recently in \cite{Pineda:2011db}.}\\
The non-static potentials are affected by similar IR divergences; they manifest themselves at order $\als^3$ (two loops) in $V^{(1)}$, and at order $\als^2$ (one loop) in $V^{(2)}$. As we shall see, the divergent terms in $V_s^{(0)}$, $V_s^{(1)}$ and $V_s^{(2)}$ all contribute to the spectrum at the same order, $m\als^5$. We refer to Appendix~\ref{app_pnrqcd} for the detailed expressions of $\alVs$ and $\alVo$, as well as for the matching of $V_s^{(1)}$ and $V_s^{(2)}$. In the octet sector, the non-static potentials are not known beyond tree level.\\

In the previous section we have mentioned how the constraints of Poincar\'e invariance can be imposed on NRQCD, by demanding
that the generators of Poincar\'e transformations in the EFT obey the Poincar\'e algebra. The same analysis can be carried 
out in pNRQCD \cite{Brambilla:2001xk,Brambilla:2003nt}. This enables to obtain a set of exact relations between the matching coefficients of
the theory; for instance, the matching coefficients that should in principle accompany the kinetic terms in Eq.~\eqref{sinoctham} 
are found to be exactly equal to unity, and for this reason they have been omitted there in order to keep a lighter notation. Other relations
include the so-called \emph{Gromes relation} \cite{Gromes:1984ma}, linking the derivative of the static potential to the spin-orbit potential, which is part of $V^{(2)}$. Such a relation had originally been introduced in the context of the transformation properties under Lorentz boosts of Wilson loops. For more details on this relation we refer to App.~\ref{app_pnrqcd} and to Chap.~\ref{chap_poincare}, where we will consider this relation in the context of an explicitly Lorentz-breaking thermal medium.\\

We now move to the evaluation of the quarkonium spectrum. We then consider the spectrum of singlet states: as we already mentioned, the equation of motion for the singlet field in the Lagrangian~\eqref{lagrpnrqcd} is at the zeroth-order in the multipole expansion a Schr\"odinger equation. In order to ensure coherence with the assumed power counting, i.e.  $v\sim\als(1/r)\ll1$, it is customary to solve this Schr\"odinger equation with the leading terms in the Hamiltonian only, which turn out to be the kinetic term and the leading-order static potential
\begin{equation}
	h^{(0)}_s=\frac{\bp^2}{m}-\cf\frac{\als}{r}\,.
\label{leadingham}
\end{equation}
In the octet sector the same expansion yields
\begin{equation}
h^{(0)}_o=\frac{\bp^2}{m}+\frac{1}{2\nc}\frac{\als}{r}\,.
\label{leadinghamoctet}
\end{equation}
It is easy to see that the size of $h^{(0)}$ is $m\als^2$. The spectrum of the leading-order singlet Hamiltonian~\eqref{leadingham} is given by the Coulomb levels
\begin{equation}
E_n=-\frac{mC_F^2\als^2}{4n^2}=-\frac{1}{ma_0^2n^2}\,,\qquad a_0\equiv\frac{2}{m\cf\als}\,,
\label{coulomblevels}
\end{equation}
and justifies a posteriori the assumption $v\sim\als$, since now one has $\left\langle\frac{1}{r}\right\rangle\sim m\als$ and $E\sim m\als^2$.\\
The subleading terms in $h_s$, such as the loop corrections to the static potential and the non-static potentials, are then treated in quantum-mechanical perturbation theory. For a somewhat different approach to the power counting, in which all known loop corrections to the static potential are included in $h^{(0)}_s$, we refer to \cite{Kiyo:2010jm}.\begin{figure}
	\begin{center}
		\includegraphics{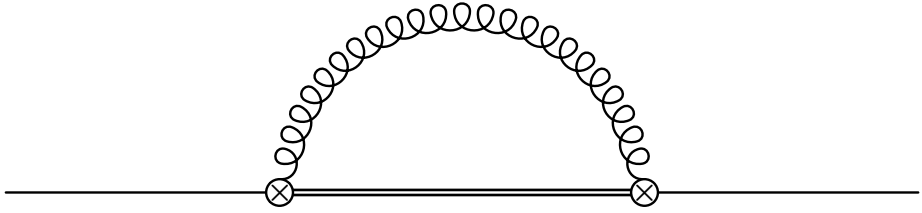}
		\put(-109.5,21){$\boldsymbol\times$}
		\put(-246,21){$\boldsymbol\times$}
	\end{center}
	\caption{The heavy quarkonium self-energy diagram. The single line is a singlet propagator, the double line is an octet propagator, the curly line is a gluon and the vertices are chromoelectric dipoles vertices.}
	\label{fig:hqself}
\end{figure}\\
As we stated before, the IR-divergent terms in the potential start to contribute to the spectrum at order $m\als^5$. At the same order we also encounter the first loop correction in the EFT, depicted in Fig.~\ref{fig:hqself}. In this diagram a colour-singlet state emits and then reabsorbs a US chromoelectric gluon through the dipole vertex in the second line of Eq.~\eqref{lagrpnrqcd}, propagating as an intermediate colour-octet state. This diagram  can be evaluated with the Feynman rules of pNRQCD, which are shown in App.~\ref{sub_feyn_pnrqcd}; there, the same expansion around the leading Hamiltonians~\eqref{leadingham} and~\eqref{leadinghamoctet} has been performed in the singlet and octet propagators. The diagram then turns out to be UV divergent, and the divergence exactly cancels the one in the potentials, thereby yielding a finite spectrum. This cancellation was first discussed at the static level in \cite{Brambilla:1999qa} (see also \cite{Brambilla:1999xf} for more detail), and at the non-static level in \cite{Brambilla:1999xj,Kniehl:1999ud,Kniehl:2002br,Penin:2002zv}, which also contain the calculation of the spectrum up to order $m\als^5$.\\
Applications of weakly-coupled pNRQCD at $T=0$ include the analysis of inclusive decays of quarkonium, through the imaginary part of the matching coefficients, the determination of the heavy quark masses $m_c$ and $m_b$ from the calculation of the spectrum and the comparison with the lattice data for the static energy, allowing the determination of lattice parameters \cite{Brambilla:2006wp,Brambilla:2009bi,Brambilla:2010pp}. These last applications furthermore require the subtraction of renormalons; we refer to \cite{Brambilla:2004jw} for a review on the applications and on renormalons, and to \cite{Brambilla:2010cs} for more updated results.

	\chapter{QCD at finite temperature and heavy-ion collision experiments}
		\label{chap_thermal}
		In this Chapter we will give a brief overview to the theoretical and experimental status of Quantum Chromodynamics at finite temperatures and densities. We shall start by introducing the current understanding of the QCD phase diagram in Sec.~\ref{sec_pd_qcd}, with particular emphasis on the deconfined quark-gluon plasma phase. In the following Sec.~\ref{sec_exp_quarkonium} we will give an overview of the experimental exploration of this phase through heavy-ion collisions, stressing the importance of quarkonium as a probe of the medium produced in such experiments. Sec.~\ref{sec_tft} will be dedicated instead to the introduction to the chief theoretical tools of Thermal Field Theory, concentrating on the imaginary-time and real-time formalisms. Finally, Sec.~\ref{sec_htl} shall be dedicated to some subtleties that arise in Thermal Field Theory, and in QCD in particular, in the infrared sector. Their cure through Effective Field Theories will be introduced and their physical meaning and relevance will be explained.
\section{The phase diagram of QCD}
\label{sec_pd_qcd}
\begin{figure}
	\begin{center}
		\includegraphics[scale=0.7]{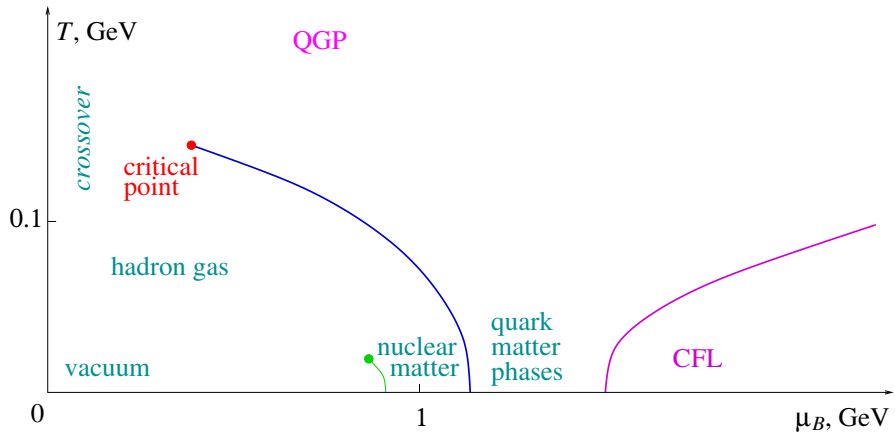}
	\end{center}
	\caption{A sketch of the QCD phase diagram. Figure taken from \cite{Stephanov:2007fk}.}
	\label{fig_qcdpd}
\end{figure}

\begin{figure}
	\begin{center}
		\includegraphics[scale=0.62]{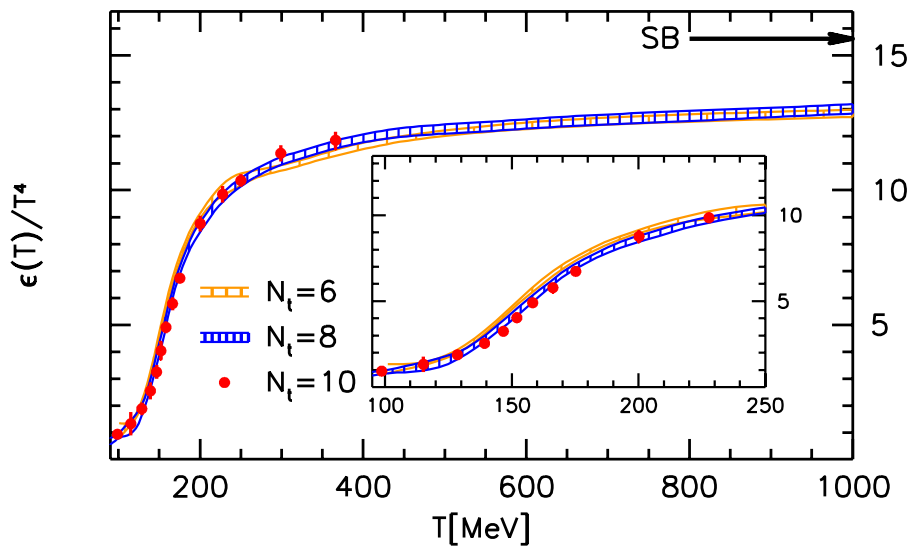}
	\end{center}
	\caption{Recent lattice results \cite{Borsanyi:2010cj} for the energy density $\epsilon(T)$. The calculations were performed  with 2+1 flavours at the physical quark mass with the stout action, a particular implementation of the staggered formulation for fermions. $N_t$ is the number of lattice sites on the imaginary time axis.  In the top right corner the Stefan-Boltzmann limit of an ideal quantum gas is shown. Figure taken from \cite{Borsanyi:2010cj}.}
	\label{fig_e_lattice}
\end{figure}

The phase diagram of hadronic matter was, to the best of our knowledge, first analyzed by Cabibbo and Parisi in 1975: in \cite{Cabibbo:1975ig} the authors hypothesized a deconfined phase at sufficiently high temperatures and/or densities. In this deconfined phase quarks and gluons were supposed to be no longer bound into colourless hadrons. In the subsequent three decades, intense research, both theoretical and experimental, has brought to a deeper understanding of the phase diagram of QCD. We refer to \cite{McLerran:2002jb} for a historical summary of its ``time evolution''.  In Fig.~\ref{fig_qcdpd} a sketch of the current understanding of the phase diagram is shown. Quoting its author \cite{Stephanov:2007fk}, the plot ``is a compilation of a body of results from model calculations, empirical nuclear physics, as well as first principle lattice QCD calculations and perturbative calculations in asymptotic regimes''.\\
The phase diagram is plotted in the $(T,\mu_B)$ plane, where $\mu_B$ is the baryon chemical potential and $T$ is the temperature\footnote{We adopt a system of units where the Boltzmann constant $k_\mathrm{B}$ is equal to unity; therefore a temperature of 1 GeV corresponds in SI units to approximately $1.16\times10^{13}$ K.}. In the bottom left corner, for low temperatures and chemical potentials, there is the \emph{hadronic matter} phase, where quarks and gluons are confined into hadrons and the approximate chiral symmetry of QCD is spontaneously broken. These hadrons form a gas which, at sufficiently high chemical potential and low temperatures undergoes a phase transition to a liquid phase. The critical line and its endpoint are shown in green in the diagram and are of great interest for nuclear physics, since they are in the same $(T,\mu_B)$ region of nuclear matter.\\
Moving further to the right at low temperatures and increasing chemical potentials, one encounters the quark matter phases, which can be described by a degenerate Fermi liquid and might be of relevance for the description of the cores of compact/neutron stars. At asymptotically large chemical potentials there is a growing consensus for the existence of a Colour SuperConductor (CSC) phase, possibly in its particular Colour-Flavour Locked (CFL) flavour \cite{Alford:1998mk}. We refer to \cite{Alford:2007xm} for a review on colour superconductivity. It is also worth mentioning that, for $SU(\nc)$ gauge theories in the large-$\nc$ limit, the existence of a confined but chirally symmetric phase, called \emph{quarkyonic matter} has recently been proposed \cite{McLerran:2007qj}. This phase would occur in the region of the phase diagram of the large-$\nc$ theory corresponding to the quark matter region of the QCD phase diagram.\\
Our sector of interest is instead the upper-left part of the diagram, which is occupied by the \emph{quark-gluon plasma} (QGP) phase. In this phase, whose name is due to Shuryak \cite{Shuryak:1978ij}, quarks and gluons are no longer confined into hadrons, but rather unbound in a gas of coloured particles and the approximate chiral symmetry is restored. This phase has been actively searched for in heavy ion collision experiments in the past decades, from the pioneering experiments at the Alternating Gradient Synchrotron (AGS) at Brookhaven National Laboratory (BNL) and at the Super Proton Synchrotron (SPS) at CERN in the 1980s and 1990s, to the ongoing experiments at the Relativistic Heavy Ion Collider (RHIC) at BNL and the recent first heavy ion run at the Large Hadron Collider (LHC) at CERN. In these experiments the chemical potential is fixed by the conserved baryon number of the colliding nuclei, whereas the temperature increases with the collision energy $\sqrt{s}$ as $\sqrt{s}\propto T^4$. Therefore, the higher the collision energy, the closer to the vertical axis is the region  of the phase diagram probed by the experiment. In particular, for what concerns the RHIC and LHC experiments, where in the former gold ions with energy per nucleon pair $\sqrt{s_\mathrm{NN}}=200$ GeV are collided and in the latter lead ions with $\sqrt{s_\mathrm{NN}}=5.5$ TeV \footnote{$\sqrt{s_\mathrm{NN}}=5.5$ TeV represents the design energy, corresponding to $pp$ collisions at 14 TeV. The first heavy ion run in November 2010 was limited to $\sqrt{s_\mathrm{NN}}=2.76$ TeV, and this will also be the energy of the 2011 run.}, one has $T\gg\mu_B$ and finite density effect can be neglected. On the theory side, lattice QCD calculations, which can only probe the $\mu_B\simeq0$ region because of the sign problem introduced by a finite chemical potential, show that, close to the vertical axis,  one has a crossover from a hadron gas to the QGP, rather than a phase transition\footnote{The existence of a phase transition and its order at $\mu_B=0$ depend strongly on the masses of the light quarks, as well as on their number. For infinite masses (pure gauge theory) one has a first-order phase transition, whereas at the physical point one encounters a crossover, as we mentioned. Other scenarios are summarized in the so-called Columbia plot \cite{Brown:1990ev}.}. In Fig.~\ref{fig_e_lattice}, the energy density $\epsilon(T,\mu_B=0)$ is plotted from a recent lattice calculation \cite{Borsanyi:2010cj} of the equation of state. The plot shows clearly how in the region between $T\sim 150$ MeV and $T\sim 350$ MeV the energy density increases by roughly an order of magnitude, signalling a qualitative change in the degrees of freedom of the system, from the hadronic ones to the partonic ones. At higher temperatures the system should approach asymptotically the Stefan-Boltzmann limit (shown  in the top right corner) of an ideal quantum gas.\\
The transition being a crossover, the concepts of a \emph{critical temperature} and of an \emph{order parameter} are in themselves not wholly defined. Indeed, on the lattice one considers different observables, and for  each of them calculates an observable-dependent pseudocritical temperature $T_c$, defined as the inflection point or peak position of the considered observable. In the most recent results of the Budapest-Wuppertal collaboration \cite{Borsanyi:2010bp,Borsanyi:2010cj} these pseudocritical temperatures are found to be in the range $147-165$ MeV for observables such as the Polyakov loop, the chiral condensate and the trace anomaly. The results from the competing hotQCD collaboration, which seemed to point to a higher pseudocritical temperature $T_c\sim190$ MeV \cite{Cheng:2009zi}, are now converging to smaller values \cite{Soldner:2010xk,Bazavov:2010pg}, compatible with those of \cite{Borsanyi:2010bp,Borsanyi:2010cj}, and the past discrepancy has been attributed to larger discretization errors at low $T$ in the particular implementations of the staggered fermion action employed by the hotQCD collaboration.\\ 
For higher values of the baryon chemical potential a real phase transition is expected to occur. Searches for it, as well as for the critical point, are ongoing in beam energy scans at the RHIC, and in the planned Compressed Baryonic Matter experiment at the Facility for Antiprotons and Ion Research (FAIR) at the Gesellschaft f\"ur SchwerIonenforschung (GSI) in Germany. On the theory side some of the most common tools are effective models, such as the Polyakov-extended Nambu-Jona-Lasino model \cite{Fukushima:2003fw,Ratti:2005jh} and extensions of lattice QCD to finite, small $\mu_B$, for instance through Taylor expansions or imaginary chemical potentials. We refer to rewiews such as \cite{Fukushima:2010bq} for a summary on finite-$\mu_B$ aspects of the transition to the deconfined phase. \\

We furthermore remark that the study of the phase diagram of QCD is also relevant for cosmological and astrophysical reasons. As we mentioned before, the high $\mu_B$, low $T$ region is of interest for the cores of compact stars, while, for what concerns cosmology, the accepted theories of the early universe indicate that it went through a quark-gluon plasma phase until about $10^{-6}$ s after the Big Bang, before cooling down below the (pseudo)critical temperature. 

\section{Heavy-ion collision experiments and quarkonium suppression}
\label{sec_exp_quarkonium}
As we mentioned before, the ongoing experiments for the study of the QGP are being carried out with heavy-ion collisions at RHIC and LHC. There, a collision of two heavy nuclei, and especially a ``central'' one, is thought to have enough energy density for the formation of the deconfined medium. By central collisions it is meant those with the smallest impact parameter, corresponding to the highest number of participating nucleons. The sketchy timeline of such a collision starts with the proper time $\tau=0$ at the collision; after that it takes a (proper) time $\tau_0\sim 1$ fm/c, the typical strong interaction timescale, for the medium to be formed and reach a near-equilibrium state. This medium, sometimes called \emph{fireball}, will then rapidly expand and cool down, eventually going below the pseudocritical temperature and producing a shower of hadrons. The lifetime of the medium is estimated to be of a few fm/c, up to $\sim10$ at the LHC. The temperatures reached in these collisions are estimated to be in the range $1-2$ $T_c$ at RHIC and up to $3-4$ $T_c$ at the LHC for central collisions.\\
In the case of less central collisions, the overlap region of the two colliding, Lorentz-contracted nuclei is roughly elliptical. This brings to further pressure gradients, besides those coming from the Lorentz-contracted geometry of the collision volume, and to different expansion velocities along the azimuthal angle. The study of \emph{bulk properties}, such as the particle distribution along the azimuthal angle, called \emph{elliptic flow}, gave a first indication \cite{Ackermann:2000tr} that the medium produced at RHIC was well described by an almost perfect liquid, with a shear viscosity to entropy ration $\eta/s$ very close to the conjectured lower bound of $1/{4\pi}$ saturated from strongly-coupled supersymmetric Yang-Mills theories \cite{Kovtun:2004de}, and much smaller than that predicted by perturbative QCD \cite{Arnold:2000dr,Arnold:2003zc}. Early LHC results \cite{Aamodt:2010pa} also show a substantial elliptic flow. For a review on the hydrodynamical description of heavy-ion collisions we refer to \cite{Romatschke:2009im}.\\
Another feature of these experiments is that very high particle multiplicities are recorded; for instance, the ALICE collaboration at the LHC recently reported \cite{Aamodt:2010pb} the highest multiplicities ever recorded, during the 2010 Pb+Pb run at $\sqrt{s_\mathrm{NN}}=2.76$ TeV. The collaboration measured a pseudo-rapidity density of primary charged particles at mid-rapidity $dN_\mathrm{ch}/d\eta=$ 1584 $\pm$ 4 (\emph{stat.}) $\pm$ 76 (\emph{sys.}) for the 5\% most central collisions. \\

Answering the most basic questions, such as whether a deconfined medium is actually produced, and in case measuring its properties such as the temperature, is then made difficult by these high multiplicities. To this end, it is common to resort to the so-called \emph{hard probes}. The term \emph{probes} indicates particles that are not in thermal equilibrium, whereas \emph{hard} labels a large energy and momentum, much larger than the temperature.\\
The most studied hard probes include the electromagnetic probes, i.e. photons and leptons, which have the advantage of not interacting strongly with the medium. High-$p_\mathrm{T}$ quarks and gluons are also of great importance; as in the vacuum, they generate jets. However, in a heavy ion collision, these particles have to cross the medium, which yields to the phenomenon of \emph{jet quenching}, that is a substantial energy loss and transverse momentum broadening of the jet. Finally, quarkonia are also widely studied in the context of heavy ion collision, since the initial proposal by Matsui and Satz \cite{Matsui:1986dk}. In this reference, the authors pointed out that the suppression of the $J/\psi$ in heavy ion collision experiments would have provided a striking signature of the formation of the deconfined medium.\\
The basic qualitative reasoning in \cite{Matsui:1986dk} is that the $c\overline{c}$ pair is produced in a hard process, with a time of the order of the inverse charm mass, much shorter than the formation time $\tau_0$ of the medium. The formation of a $J/\psi$ from the $c\overline{c}$ pair would then require, according to the authors, a time comparable with $\tau_0$. The mesonic bound state would at that point find itself in the deconfined medium, where the authors argued that the leading dissociation mechanism was the colour screening of the strong interaction binding its constituent $Q\overline{Q}$ pair. This phenomenon, which is typical of a plasma, when applied to a rough non-relativistic potential model for the $J/\psi$, would change the potential into a Debye potential, a Yukawa potential where the screening mass, or Debye mass, as we shall see later on, depends linearly on the temperature. Therefore at some dissociation temperature $T_d$ the bound state would cease to exist, as the Yukawa potential supports only a finite number of bound states.\\
The authors then argued that no other non-plasma suppression mechanism could compete with colour screening, and that the suppression could not realistically be compensated by a recombination of unbound $c\overline{c}$ pairs in the hadronization phase, when the plasma is cooling down, a few fm/c after $\tau_0$. Finally, any surviving $J/\psi$ meson would decay according to the usual $T=0$ branching ratios, since the lifetime of the $J/\psi$ is orders of magnitude larger than that of the plasma. Of particular interest is the dilepton decay channel, which has a significant branching ratio (BR($J/\psi\to e^+e^-=(5.94\pm0.06)\times10^{-2}$), BR($J/\psi\to \mu^+\mu^-=(5.93\pm0.06)\times10^{-2}$) \cite{Nakamura:2010zzi}). Dilepton pairs are more easy to identify amidst the wealth of produced particles, and the background was argued by the authors to be small.\\

\begin{figure}[t]
	\begin{center}
		\includegraphics[scale=0.32]{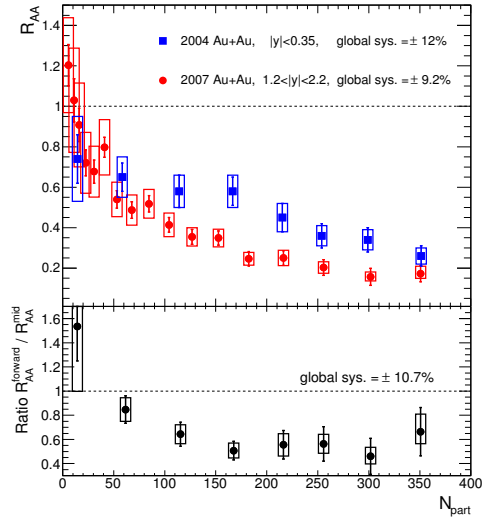}
	\end{center}
	\caption{In the top pane the PHENIX Au+Au $J/\psi$ $R_{AA}$ \cite{Adare:2011yf} for $\sqrt{s_\mathrm{NN}}=$ 200 GeV is plotted as a function of the number of participants at mid-rapidity ($\vert y\vert < 0.35$, 2004 run) and at forward rapidity ($1.2<\vert y\vert<2.2$, 2007 run). The bottom pane shows a plot of the ratio of $R_{AA}$ in the two different rapidity ranges, showing a larger suppression at forward rapidity.}
	\label{fig_phenix}
\end{figure}

\begin{figure}
	\begin{center}
		\includegraphics[scale=0.3]{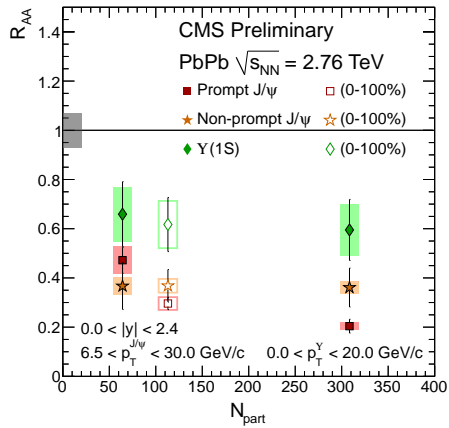}
	\end{center}
	\caption{$J/\psi$ and $\Upsilon(1S)$ data from the CMS experiment \cite{CMS-PAS-HIN-10-006}. The nuclear modification factor $R_{AA}$ at mid-rapidity is plotted as a function of the number of participating nucleons for Pb-Pb collisions at $\sqrt{s_\mathrm{NN}}=2.76$ TeV. Non-prompt $J/\psi$ come from decays of $B$ hadrons, whereas prompt ones do not. They might however come from feed-down from higher charmonium resonances. At the highest number of participants, the suppression of prompt $J/\psi$ at mid-rapidity is comparable to that reported by PHENIX (without non-prompt separation) at forward rapidity, as shown in Fig.~\ref{fig_phenix}. Picture taken from \cite{CMS-PAS-HIN-10-006}.}
	\label{fig_cms}
\end{figure}

The research, both theoretical and experimental, that has been devoted to this hypothesis in the past 25 years has brought to a deeper understanding of the problem. On the experimental side the measured observable at hand is the so-called \emph{nuclear modification factor} $R_{AA}$, which for a generic process is defined as the yield observed in a heavy-ion ($AA$) collision, divided by the yield in a $pp$ collision multiplied by the (measured) number of participants in the $AA$ collision. Any deviation from unity is a signal of some form of nuclear modification to the process.\\
It was however quickly understood that the so-called \emph {Cold Nuclear Matter} (CNM) effects play an important role in studies of quarkonium suppression: these effects are those caused by a nuclear, non-deconfined medium, as opposed to the \emph{Hot Nuclear Matter} effects caused instead by a deconfined medium. CNM effects include initial-state effects, such as modification of the parton distribution functions of the nucleon inside the nucleus (\emph{shadowing}),  and energy loss of the parton traversing the nucleus before the hard scattering, which both affect the production mechanism. Final-state effects include the absorption, i.e. destruction, of the produced quarkonium state as it crosses the nucleus. As a consequence, in order to disentangle hot and cold nuclear matter effects, heavy-ion experiments must establish the so-called \emph{CNM baseline}. This is achieved by performing proton-nucleus ($pA$) or deuteron-nucleus ($dA$) collision experiments and measuring $R_{pA}$ or $R_{dA}$.\\
In Fig.~\ref{fig_phenix} the $J/\psi$ $R_{AA}$  measured from the PHENIX collaboration \cite{Adare:2011yf} at RHIC is plotted. The data show a clear suppression pattern with increasing centrality and rapidity. The authors remark that, while the quantitative disentangling of the hot matter suppression from the CNM effects is at the moment not possible, the data shows a larger suppression than that predicted by CNM models and parametrizations alone, thus pointing to significant quark-gluon plasma effects. Other issues in the interpretation of the data stem from the estimation of the loss of feed-down \cite{Karsch:2005nk}: a significant percentage of the $J/\psi$ yield in $pp$ collision comes from decays of higher excitations of charmonium or from beauty hadrons to the $J/\psi$. In the thermal medium the excited states are instead expected to be much more suppressed than the $J/\psi$, due to their looser binding.
\\
The $J/\psi$ is of course not the only quarkonium probe available, even though it certainly is the most studied. $b\overline{b}$ bound states are equally interesting, and the vector resonances $\Upsilon(nS)$ share the appealing features of the $J/\psi$, among which the clean dileptonic decays. The $\Upsilon(1S)$, which is the bottomonium analogue of the $J/\psi$, is, due to the larger mass of the $b$ quark, more tightly bound than the $J/\psi$ and hence, by the qualitative approach of Matsui and Satz, expected to dissociate at higher temperatures. The \emph{dissociation temperature} which can be estimated from our results is actually above the temperature ranges of the RHIC and LHC for the $\Upsilon(1S)$, as we shall see in Part~\ref{part_realtime}.\\
On the experimental side, at RHIC energies the cross section for $b\overline{b}$ is quite small and the detectors do not have the resolution needed to resolve each of the first three $\Upsilon(nS)$ resonances, which are then measured together. Recent and current detector upgrades should solve this issue and a future luminosity increase should give more statistics, compensating the low cross section. As of now, the combined data for the three vector resonances  hints at a large suppression, with PHENIX \cite{Atomssa:2009ek} reporting an upper limit for the nuclear modification factor $R_{AA}<0.64$, independent of centrality. Preliminary STAR data \cite{Masui:2011qi} confirm this result, with $R_{AA} (0-10\%)=0.34\pm0.17\;(\text{stat})\,+0.06/-0.07\;(\text{syst})$ for the 10\% most central collisions.\\
At the LHC the $b\overline{b}$ cross section increases by roughly two orders of magnitude \cite{Giubellino:2008yy} and the detectors have a very good resolution in the bottomonium mass region. At the time of writing the first data from the CMS experiment  \cite{Collaboration:2011pe,CMS-PAS-HIN-10-006} point to a substantial suppression of the radial excitations $\Upsilon(2S)$ and $\Upsilon(3S)$: the double ratio of $(\Upsilon(2S)+\Upsilon(3S))/\Upsilon(1S)$ events in Pb-Pb collisions over $(\Upsilon(2S)+\Upsilon(3S))/\Upsilon(1S)$ events in $pp$ collisions is found to be $0.31 - 0.15 + 0.19\;(\text{stat})\, \pm 0.03\;(\text{syst})$. For the $\Upsilon(1S)$ the measured nuclear modification factor for the 20\% most central collisions is $R_{AA} =0.60\pm 0.12\;(\text{stat})\,\pm 0.10\;(\text{syst})$. According to the collaboration, this value can be explained qualitatively by the loss of feed-down resulting from the suppression of the excited states, thus hinting at modest direct suppression of the $\Upsilon(1S)$ in the medium created at the LHC. The nuclear modification factor measured by CMS for the $J/\psi$ and $\Upsilon(1S)$ is shown in Fig.~\ref{fig_cms}.
\\

On the theory side, the relevant points for a understanding of quarkonium in heavy-ion collisions can be summarized in
\begin{enumerate}
	\item Understanding the intricacies of quarkonium production.
	\item Cold Nuclear Matter effects, which are intertwined with the study of production.
	\item Hot Nuclear Matter effects on the bound state.
	\item Hadronization and recombination effects of $Q\overline{Q}$ pairs.
\end{enumerate}
In the rest of this thesis we will concentrate on point 3, we refer to \cite{Kluberg:2009wc} for a more general review on quarkonium in media and to \cite{Brambilla:2010cs} for recent results.\\
$Q\overline{Q}$ bound states in a deconfined medium have been studied and are being studied theoretically through a variety of approaches. Based on the original idea of Matsui and Satz and  on the assumption that medium effects can be entirely described by a Schr\"odinger equation with a temperature-dependent potential, the problem of quarkonium dissociation has been addressed in terms of \emph{potential models} with
screened temperature-dependent potentials over the past 20 years (see e.g. 
Refs.~\cite{Karsch:1987pv,Digal:2001iu,Mocsy:2005qw,Alberico:2005xw,Riek:2010fk} for some representative works), where the potential was often derived from lattice QCD calculation of thermodynamical quantities.  Another approach relies on calculations of Euclidean correlation functions 
in lattice QCD and the reconstruction of the corresponding spectral functions  
using the Maximum Entropy Method \cite{Asakawa:2000tr}. At present, however, a reliable determination of the
quarkonium spectral functions from the lattice data appears very difficult 
due to statistical errors and lattice discretization effects (see discussion in Ref.~\cite{Jakovac:2006sf} 
and references therein). More information on potential models and lattice QCD spectral functions can be found in these recent reviews \cite{Rapp:2008tf,Kluberg:2009wc,Bazavov:2009us}.\\
Other approaches include the application of QCD sum rules to quarkonium states (see for instance \cite{Morita:2009qk}) and the usage of the AdS/CFT correspondence \cite{Maldacena:1997re,Witten:1998qj,Gubser:1998bc} to model in-medium quarkonium properties, such as the potential or the width (see \cite{Albacete:2008dz,Noronha:2009da,Mia:2010tc,Mia:2010zu} for some recent works).

As we mentioned in the Introduction, In Part~\ref{part_realtime} we will extend the NR EFT framework of NRQCD and pNRQCD, as exposed in Chapter~\ref{chap_EFT}, to finite temperature. Within this context it will be possible to give a modern, rigorous QCD definition of the potential, as well as to take systematically into account relativistic corrections. In Part~\ref{part_imtime} we will study the Polyakov loop and the correlator of Polyakov loops, two observables that are linked to the free energies of a single static quark and of a static quark-antiquark pair in the medium; as such, they are extensively measured on the lattice and used as input for the potential models. We will perform a perturbative computation and then use our EFT framework to investigate the relation between these thermodynamical quantities and the potentials that appear in the EFT.\\
The calculations of these chapters require the generalization of relativistic Quantum Field Theory to finite temperatures; this branch of QFT goes under the name of Thermal Field Theory (TFT). The remainder of this chapter is dedicated to the exposition of its basic principles, with particular attention to their application to QCD. Section~\ref{sec_htl} is in fact devoted to the exposition of some finite-temperature EFTs of QCD which will be used in the following chapters.

\section{Introduction to Thermal Field Theory}
\label{sec_tft}
The two formulations of Thermal Field Theory we are going to illustrate here and use in the thesis are the Imaginary Time Formalism (ITF), also known as Matsubara formalism after its author \cite{Matsubara:1955ws}, and Real Time Formalism (RTF). They both stem from the need to evaluate thermal expectation values in a QFT context. \\
Let us consider first a quantum-mechanical system described by a Hamiltonian $H$, at thermal equilibrium at a temperature $T$ in a thermal bath. The thermal average for an operator $O$ is defined as
\begin{equation}
	\label{defthermalaverage}
	\langle O\rangle_T\equiv\frac{1}{Z}\mathrm{Tr}\left\{{O}e^{-\frac{{H}}{T}}\right\},\qquad Z\equiv\mathrm{Tr}\left\{e^{-\frac{{H}}{T}}\right\}.
\end{equation}
For simplicity we are restricting ourselves to $\mu=0$. The exponential $e^{-H/T}$ is called the \emph{Boltzmann factor}, $Z$ is named the \emph{partition function} of the system, and the traces are intended over the Hilbert space of the theory. They are thus to be evaluated using a complete set of states, such as the eigenstates $\vert n\rangle$ of the Hamiltonian. In this case one has explicitly
\begin{equation}
	\label{traceexplicit}
	Z=\mathrm{Tr}\left\{e^{-\frac{{H}}{T}}\right\}=\sum_n e^{- \frac{E_n}{T}}\,,
\end{equation}
where $E_n$ are the eigenvalues of the Hamiltonian, i.e. $H\vert n\rangle=E_n\vert n\rangle$.\\
In the next subsections we will illustrate how these relations can be brought to the context of a Quantum Field Theory, introducing the ITF and the RTF.
\subsection{The Imaginary Time Formalism}
\label{sec_imtime}
Even though our aim is to describe QCD and Effective Field Theories thereof, let us consider here as an example the case of a scalar field $\phi$. For reasons which will become immediately clear, we work in Euclidean space-time. Only at the end we will be allowed to perform an analytical continuation back to real times. We label the imaginary time axis as 
\begin{equation}
	\label{wicktime}
	t\to-i\tau\,,
\end{equation}
In Eq.~\eqref{traceexplicit} any complete set of states can be employed. One can then choose to use the eigenstates of the field operator $\phi$ and, using the standard techniques introduced by Feynman \cite{Feynmanbook}, one can derive a functional integral formulation of Eqs.~\eqref{defthermalaverage} and~\eqref{traceexplicit} \footnote{It is also possible to obtain the results presented here in the canonical operator approach, see for instance Appendix B of \cite{LeBellac:1996}.}. The Boltzmann factor $e^{-\frac{{H}}{T}}$ can be interpreted as a time evolution operator for an imaginary time $\tau=\frac{1}{T}\equiv\beta$. The partition function can be written as a path integral
\begin{equation}
	\label{pathz}
	Z(T)=\int \mathcal{D}\phi\exp\left(-\int_0^\beta d^4x \lag\right),
\end{equation}
where $\lag$ is the Lagrangian density of the Euclidean theory and the integration is defined as
\begin{equation}
	\label{defineint}
	\int_0^\beta d^4x\equiv\int_0^\beta d\tau\int d^3\bx\,.
\end{equation}
Since the original definition of $Z$ in Eq.~\eqref{traceexplicit} contains a Hilbert space trace, the functional integration is furthermore subject to the boundary condition $\phi(\bx,0)=\phi(\bx,\beta)$.\\
Similarly the expectation value of any operator can be written as a path integral and Eq.~\eqref{defthermalaverage} becomes
\begin{equation}
	\langle O\rangle=\frac{\int \mathcal{D}\phi\,O\,\exp\left(-\int_0^\beta d^4x \lag\right)}{\int \mathcal{D}\phi\exp\left(-\int_0^\beta d^4x \lag\right)},
	\label{thermalpathintegral}
\end{equation}
where we have omitted the ${}_T$ pedix in $\langle O\rangle$; from now on, all expectation values are understood to be thermal ones according to the definition~\eqref{defthermalaverage}.\\
As usual in Quantum Field Theory we can add a source term to the partition function~\eqref{pathz}, thus defining a generating functional $Z(T;j)$ as 
\begin{equation}
	\label{pathzj}
	Z(T;j)=\int \mathcal{D}\phi\exp\left(-\int_0^\beta d^4x \lag+\int_0^\beta d^4x j(x)\phi(x)\right).
\end{equation}
Any $n$-pointh function of the field $\phi$ can then be obtained by differentiation of the generating functional with respect to $j$, thus allowing us to obtain the Feynman rules of the thermal theory.\\ 
We start by defining the propagator of the thermal theory for the scalar field $\phi$ as
\begin{equation}
	\label{propagator}
	\Delta(\bx,\tau)=\langle T\phi(\bx,\tau)\phi(\boldsymbol{0},0)\rangle\,,
\end{equation}
where $T$ stands for the time ordering in imaginary time and we note that the periodicity condition on the path integral~\eqref{pathz} implies that $\tau$ is constrained in the interval $[0,\beta]$ and has a periodicity in $\tau\to \tau-\beta$:
\begin{equation}
	\label{periodicity}
	\langle T\phi(\bx,\tau-\beta)\phi(\boldsymbol{0},0)\rangle=\langle T\phi(\bx,\tau)\phi(\boldsymbol{0},0)\rangle
\end{equation}
Applying the usual methods of quantum field theory the generating functional for the free field can be written as
\begin{equation}
	\label{generatingf2}
	Z_F(T;j)=Z_F(T)\exp\left(\frac{1}{2}\int_0^\beta d^4xd^4yj(x)\Delta_F(\bx-\by,\tau_x-\tau_y)j(y)\right),
\end{equation}
where $F$ stands for free and the propagator is the solution of the differential equation
\begin{equation}
	\label{greenpropscalar}
	\left(-\frac{\partial^2}{\partial\tau^2}-\nabla^2+m^2\right)\Delta_F(x-y)=\delta(\tau_x-\tau_y)\delta^3(\bx-\by),
\end{equation}
where $\delta(\tau_x-\tau_y)=\delta(\tau_x-\tau_y-p\beta)$, with $p\in\mathbb{Z}$. This equation is best solved in momentum space; the periodicity condition over time then implies the following solution
\begin{equation}
	\label{matsubaramom}
	\Delta_F(\omega_n,\bk)=\frac{1}{\omega_n^2+\bk^2+m^2},
\end{equation}
where the frequencies $\omega_n$ are constrained to be discrete, i.e. 
\begin{equation}
	\label{matsubarafreq}
	\omega_n=2\pi n T\qquad n\in\mathbb{Z}.
\end{equation}
These frequencies are called \emph{Matsubara frequencies} and the propagator is called the \emph{Matsubara propagator} \cite{Matsubara:1955ws}. It is often convenient to express this propagator in the mixed representation, that is performing the discrete Fourier transform over the frequencies. One then has
\begin{equation}
	\label{matsumixed}
	\Delta_F(\tau,\bk)=T\sum_{n=-\infty}^{+\infty} e^{i\omega_n\tau}\Delta_F(\omega_n,\bk)=\frac{1}{2\omega_k}\left[(1+n_\mathrm{B}(\omega_k))e^{-\omega_k\tau}+n_\mathrm{B}(\omega_k)e^{\omega_k\tau}\right],
\end{equation}
where $\omega_k=\sqrt{\bk^2+m^2}$ and $n_\mathrm{B}(\omega)$ is the \emph{Bose--Einstein distribution}, defined as
\begin{equation}
	\label{bosedistr}
	n_\mathrm{B}(\omega)\equiv\frac{1}{e^{\frac{\omega}{T}}-1}.
\end{equation}
In the r.h.s of Eq.~\eqref{matsumixed} we call the \emph{vacuum part} the term $e^{-\omega_k\tau}/(2\omega_k)$, since it is the only term that would survive if the discrete Fourier transform were replaced by the standard $T=0$ integral, i.e. $\int\,d\omega/(2\pi)$. Conversely the other two terms, i.e. those proportional to the Bose--Einstein distribution, are often called the \emph{matter part} or \emph{thermal part}, as they only appear for $T>0$ and describe the contribution of the thermal bath to the free propagator.\\
Since in general one is ultimately interested in computing correlation functions in real time it is convenient to define the spectral representation of the propagator as
\begin{equation}
	\label{spectralscalar}
	\frac{1}{\omega_n^2+\bk^2+m^2}=\int_{-\infty}^{\infty}\frac{dk_0}{2\pi}\frac{\rho(k_0)}{k_0-i\omega_n},
\end{equation}
where $\rho(k_0)$ is the spectral density and the above expression can be analytically continued  to arbitrary (non-discrete) values of the frequencies: $i\omega_n\to z$. This continuation is unique provided that  $\vert\Delta(z,\bk)\vert\to0$ if $\vert z \vert\to0$ and $\Delta(z,\bk)$ is analytic outside the real axis. Then the spectral density is
\begin{equation}
	\label{defrho}
	\rho(k_0)=\frac{1}{i}\left[\Delta(z\to k_0+i\eta)-\Delta(z\to k_0-i\eta)\right]\qquad\eta\to0^+\,.
\end{equation}
The computation of any $n$-point function is then carried out with the Feynman rules of the Euclidean $T=0$ theory, with the replacement of the Feynman propagator with the Matsubara propagator. Any integral over the four-momentum $d^4k$ becomes
\begin{equation}
	\label{intmom}
	\int \frac{d^4k}{(2\pi)^4}\to \Tint{k}\equiv T\sum_{n=-\infty}^{+\infty}\int\frac{d^3\bk}{(2\pi)^3}.
\end{equation}
In the computation of a time-dependent physical observable one is allowed to analytically continue the result to Minkowskian space-time only after the summation of the Matsubara modes has been performed. This can turn out to be cumbersome, and the real-time formalism is very likely to be better suited for such a computation. On the other hand, the Matsubara formalism proves to be very convenient if one is interested in a time-independent observable, such as a thermodynamical quantity like the pressure, for instance.\\

The application of the ITF to a gauge theory, such as QCD, is more complicated. We write the QCD Lagrangian in Euclidean space-time as
\begin{equation}
{\cal L}_{\rm QCD} = \frac{1}{4}F^a_{\mu\nu}F^a_{\mu\nu} + \sum_{f=1}^{n_f} \bar{q}_fD\!\!\!\!/
\, q_f \,,
\label{QCDimtime}
\end{equation}
where we maintain our convention $D_\mu = \partial_\mu + igA_\mu$. Now, if we were to use this Lagrangian in the partition function~\eqref{pathz}, we would have to require periodic boundary conditions for the gauge fields and antiperiodic boundary conditions for the quark fields, as a consequence of their fermionic statistics. The resulting functional integral would however run over all the unphysical gauge configurations as well as the physical ones, and the corresponding partition function would be unphysical. This can be cured by the usual Faddeev-Popov method; we refer to textbooks such as \cite{LeBellac:1996} and \cite{Kapusta:2006pm} for the detailed implementation of the gauge-fixing procedure at finite temperature, while in App.~\ref{sec_im_time} we list the Feynman rules for QCD in imaginary time. We wish however to make two remarks at this point: first we observe that the antiperiodic boundary conditions cause the fermionic Matsubara frequencies to be odd, i.e. $\tilde{\omega}_n\equiv(2n+1)\pi T$, $n\in\mathbb{Z}$. In the corresponding mixed representation (see Eq.~\eqref{matsumixed}) the Bose--Einstein distribution is replaced by the \emph{Fermi--Dirac distribution}, defined as
\begin{equation}
	\label{fermidistr}
	n_\mathrm{F}(\omega_k)\equiv\frac{1}{e^{\frac{\omega_k}{T}}+1},
\end{equation}
and, as required by fermionic statistics, $n_\mathrm{F}(\omega_k)\le1$ $\forall\,\omega_k\ge0$.\\
Secondly we remark that, if we were to quantize the theory in the $A_0=0$ gauge, alternatively called temporal gauge, which is sometimes convenient at zero temperature, especially when quantizing the theory in the canonical formalism, we would have to impose the Gau\ss{} condition on the physical states. This can in turn be achieved by adding to the path integral a projector in the form of a functional integration over an auxiliary field. It can however be shown \cite{Gross:1980br} that, once the periodic boundary conditions necessary at finite temperature are imposed, the functional integration over this auxiliary field becomes identical to that over the $A_0$ field which was originally removed, in fact reintroducing it. Therefore the choice $A_0=0$ is not a sensible one in this formalism, while it can be used in the real-time one \cite{James:1990fd}.\\
The issue with the $A_0=0$ gauge in the ITF is also confirmed by the existence at finite temperature of operators that exploit the periodic boundary conditions to be gauge-invariant. The most important among them is the Polyakov loop \cite{Polyakov:1978vu,McLerran:1981pb}, defined as the trace of a Wilson line spanning around the periodic time axis from 0 to $1/T$, i.e.
\begin{equation}
\label{defploop}
L(\bx)\equiv  \mathrm{Tr}\,{\rm P} \exp\left(-ig\int_0^{1/T} d\tau A_0(\bx,\tau)\right)\,.
\end{equation}
Part~\ref{part_imtime} will be devoted to the study of the thermal expectation value of this operator, which, as we shall see, is different from unity, the value it would have in the $A_0=0$ gauge. As we just observed, this operator is gauge invariant, so this mismatch highlights the issues one would incur into when naively adopting the temporal gauge. As we shall see in Part~\ref{part_imtime}, there exists a class of gauges which is advantageous for the computation of this operator. This class is called the \emph{static gauge} \cite{D'Hoker:1981us} and is defined by the condition $\partial_0A_0=0$.
\subsection{The Real Time Formalism}
\label{sec_realtime}
\begin{figure}
	\begin{center}
		\includegraphics[scale=0.61]{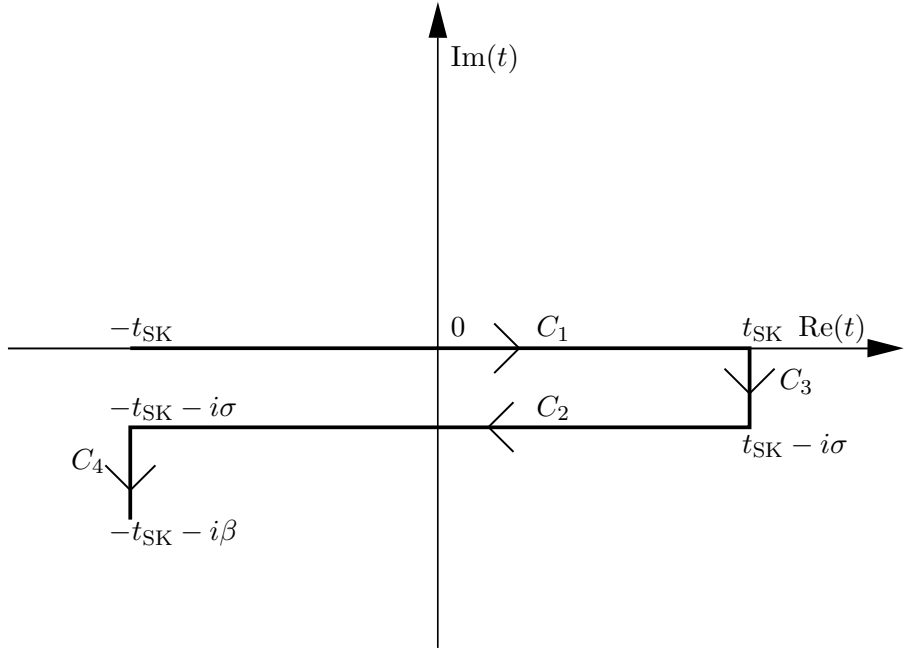}
		\put(-170,118){$0$}
		\put(-298,118){$-t_\mathrm{SK}$}
		\put(-298,88){$-t_\mathrm{SK}-i\sigma$}
		\put(-138,88){$C_2$}
		\put(-138,118){$C_1$}
		\put(-47,98){$C_3$}
		\put(-312,68){$C_4$}
		\put(-298,41){$-t_\mathrm{SK}-i\beta$}
		\put(-61,118){$t_\mathrm{SK}$}
		\put(-61,74){$t_\mathrm{SK}-i\sigma$}
		\put(-40,118){$\mathrm{Re}(t)$}
		\put(-170,220){$\mathrm{Im}(t)$}
	\end{center}
	\caption{The Schwinger-Keldysh contour \cite{Schwinger:1960qe,Keldysh:1964ud} for an arbitrary choice of the parameter $\sigma$.}
	\label{fig_sk}
\end{figure}

In the previous subsection we have discussed how the ITF requires an analytic continuation to real values of time for the computation of time-dependent observables. We will now introduce the Real Time Formalism, that overcomes this difficulty at the price of, as we shall see, a doubling of the field degrees of freedom.\\
Before we discuss how to generalize Eq.~\eqref{pathz} to real values of time, we review the different time orderings for the propagators. As before we use the scalar theory for the exposition. We then have the time-ordered, or Feynman, propagator 
\begin{equation}
	D(x)\equiv\langle T \phi(x)\phi(0)\rangle\,,\qquad D_F(k)=\frac{i}{k_0^2-\bk^2-m^2+i\eta}\,,
	\label{propfeynman}
\end{equation}
where we are again working in Minkowskian space-time, the $T$ in the definition stands again for the time-ordering operator and the $F$ subscript in the second equation stands for the free propagator. This second equation holds at $T=0$. Our conventions for the momenta is that the italic letter $k$ labels the four-momentum, i.e. $k=(k_0,\bk)$ and $k^2=k_0^2-\bk^2$. We next introduce the forward and backward, or Wightman, propagators. They read
\begin{eqnarray}
\nn	D^>(x)&\equiv&\langle  \phi(x)\phi(0)\rangle\,,\qquad D^>_F(k)=\theta(k_0)\delta(k_0^2-\bk^2-m^2)\,,\\
	D^<(x)&\equiv&\langle  \phi(0)\phi(x)\rangle=D^>(-x)\,,\quad D^<_F(k)=\theta(-k_0)\delta(k_0^2-\bk^2-m^2)\,.
	\label{propwightman}
\end{eqnarray}
For further convenience we also define the retarded and advanced propagators as
\begin{equation}
	D_{\mathrm{R},\mathrm{A}}(x)\equiv\pm\theta(\pm x_0)\left\langle\left[\phi(x),\phi(0)\right]\right\rangle\,,\quad D_{\mathrm{R},\mathrm{A}\,F}(k)=\frac{i}{k_0^2-\bk^2-m^2\pm i\sgn(k_0)\eta}\,,
	\label{propretadv}
\end{equation}
where the upper sign refers to the retarded (R) propagator and the lower one to the advanced (A) propagator. The Equations on the right in~\eqref{propwightman} and~\eqref{propretadv} are again valid only at $T=0$.\\
In \cite{Dolan:1973qd} Dolan and Jackiw generalized the free time-ordered propagator~\eqref{propfeynman} to finite temperature, through an analytic continuation of the Matsubara propagator, along the lines sketched in Eqs.~\eqref{spectralscalar} and~\eqref{defrho}. They obtained
\begin{equation}
	D_F(k)=\frac{i}{k_0^2-\bk^2-m^2+i\eta}+2\pi\delta(k_0^2-\bk^2-m^2)n_\mathrm{B}(\vert k_0\vert)\,,
	\label{propfeynmanfiniteT}
\end{equation}
where $n_\mathrm{B}$ is again the Bose--Einstein distribution, defined in Eq.~\eqref{bosedistr}. This propagator looks very convenient with respect to the Matsubara one in Eq.~\eqref{matsumixed}: it consists of a sum of the standard $T=0$ Feynman propagator, the \emph{vacuum part}, and of a temperature-dependent part, the \emph{thermal part} or \emph{matter part}. However, with respect to the ITF case, here the separation between the former and the latter is immediately apparent, without the need to perform summations over the Matsubara frequencies, which can be very cumbersome, especially in the case of loop integrals. We furthermore observe that, in the thermal part, the Dirac $\delta$-function enforces that the particle be on shell, and the Bose-Einstein distribution specifies the thermal distribution of the momenta. One then sees clearly how, in the free approximation, i.e. neglecting interactions, the medium is composed of thermalized, on-shell particles.\\
One would then naively expect that it suffice to use this propagator with the standard Feynman rules of the Minkowskian theory to perform finite-temperature calculations in real time; however, as the authors already proved \cite{Dolan:1973qd}, this turns out to be false, since one encounters problems such as pinch singularities when products of $\delta$-functions of the same argument appear in loop integrals. We refer to \cite{Landsman:1986uw,Brambilla:2004jw} for more details on pinch singularities.\\
The cure for these issues requires a rethinking of the functional approach. Let us go back to the partition function~\eqref{pathz}; there, the time axis is taken to go straight along the imaginary direction from the initial real time $t_i$ to $t_i-i\beta$. It is however possible to deform this contour in the complex time plane in order to include the real axis. This deformed contour is called the Schwinger-Keldysh contour after its authors \cite{Schwinger:1960qe,Keldysh:1964ud} and is depicted in Fig.~\ref{fig_sk}. One starts from an initial time $t_i=-t_\mathrm{SK}$ along the negative real axis ($t_\mathrm{SK}>0$), then evolves along the real axis to the positive real time $t_\mathrm{SK}$. There one starts to move down the imaginary time axis to $t_\mathrm{SK}-i\sigma$, where $0\le\sigma\le\beta$ and then horizontally to $-t_\mathrm{SK}-i\sigma$. From there the last step brings to $-t_\mathrm{SK}-i\beta$.
\\
Going back to the functional integral, one can then obtain the action from the Lagrangian by performing the time integration along this contour. The generating functional becomes 
\begin{equation}
	\label{pathzjreal}
	Z_C(T;j)=\int \mathcal{D}\phi\exp\left(i\int_C d^4x (\lag(x)+ j(x)\phi(x))\right),
\end{equation}
where the boundary condition is now $\phi(x^0,\bx)=\phi(x^0-i\beta,\bx)$ and the letter $C$ stands for the Schwinger-Keldysh contour. If we now take the limit $t_\mathrm{SK}\to\infty$ in this contour, the entire real axis is spanned and the contribution from the two vertical legs $C_3$ and $C_4$ factorizes, giving only a multiplicative constant to the free field generating functional, as was shown by Niemi and Semenoff in \cite{Niemi:1983nf,Niemi:1983ea}. The critical point made by these authors is then to associate two different fields, labeled $\phi_1$ and $\phi_2$, with the two horizontal legs $C_1$ and $C_2$. The fields of kind ``1'' are associated with physical particles, while fields of kind ``2'' act like virtual particles or ghosts, i.e. they do not appear as external lines in the computation of a physical amplitude but are to be included in internal lines. It is exactly the contribution of this additional field that cures the issues mentioned before.\\
In detail, for a real scalar Lagrangian with an interaction term $\lag_{int}=-\frac{\lambda}{4!}\phi^4$ the generating functional then becomes
\begin{eqnarray}
 	\nonumber Z^F_C(T;j)&=&\int \mathcal{D}\phi_1  \mathcal{D}\phi_2  \exp\left(-\frac{1}{2}\int_{-\infty}^{\infty}d^4xd^4x'\phi_a(x)(D_F^{-1})_{ab}(x-x') \phi_b(x')\right.\\
 	\label{pathzjrealf}
&&\left.-i\int_{-\infty}^{\infty}d^4x\left( \frac{\lambda}{4!}\phi_1^4(x)-\frac{\lambda}{4!}\phi_2^4(x)\right)+i\int_{-\infty}^\infty d^4xj_a(x)\phi_a(x)\right),
 \end{eqnarray}
 where $a,b=1,2$ and the sources $j_a(x)$ are defined as
\begin{equation}
 	\label{defsources}
 	j_1(x)=j(t,\bx)\qquad j_2(x)=j(t-i\sigma,\bx)
 \end{equation}
 and the functional differentiation is intended as
 \begin{equation}
 	\label{fundiffreal}
 	\frac{\delta j_a(x)}{\delta j_b(x')}=\delta_{ab}\delta^4(x-x').
 \end{equation}
 The propagator thus becomes a $2\times2$ matrix
\begin{equation}
	\label{genericrealprop}
	\mathbf{D}=\left(\begin{array}{cc}D_{11}&D_{12}\\D_{21}&D_{22}\end{array}\right)
\end{equation}
where the off-diagonal elements transform fields of type ``1'' into fields of type ``2'' and viceversa. The off-diagonal matrix elements have a dependence on $\sigma$ that however enters no physical observable \cite{Matsumoto:1982ry,Matsumoto:1984au,Landsman:1986uw}. In this thesis we adopt the popular choice $\sigma=0^+$, corresponding to the original Schwinger-Keldysh contour. Since fields of type ``2'' always come at later times than fields of type ``1'', the time ordering being given by the direction of the contour, the scalar propagator can then be shown to be
\begin{equation}
	\mathbf{D}(x)=\left(\begin{matrix}\langle T\phi(x)\phi(0)\rangle&\langle \phi(0)\phi(x)\rangle\\\langle \phi(x)\phi(0)\rangle&\langle T^*\phi(x)\phi(0)\rangle\end{matrix}\right),
	\label{rtmatrixpropposspace}
\end{equation}
where $T^*$ stands for the anti-time ordering. In momentum space this becomes
\begin{equation}
	\mathbf{D}(k)=\left(\begin{matrix}D(k)&D^<(k)\\D^>(k)&D^*(k)\end{matrix}\right).
	\label{rtmatrixprop}
\end{equation}
In the free case, $D(k)$ is given by Eq.~\eqref{propfeynmanfiniteT}, $D^*$ is its complex conjugate and the other elements of the matrix can be obtained by introducing the contour-ordered propagator $D_C(x)\equiv\langle T_C\phi(x)\phi(0)\rangle$, where the pedix $C$ indicates that the time ordering is to be intended along the contour, in the sense indicated by the arrows. This propagator can then be shown to be \cite{LeBellac:1996}
\begin{equation}
	D_C(x)=\int\frac{d^4k}{(2\pi)^4}e^{-ik\cdot x}\left[\theta_C(x)+n_\mathrm{B}(k_0)\right]\rho(k_0)\,,
	\label{contourprop}
\end{equation}
where $\theta_C$ is the $\theta$-function along the contour and $\rho(k_0)$ is the spectral density, as in Eq.~\eqref{defrho}. One then has $D^>(t,\bx)=D_{C}(t-i0^+,\bx)$ and $D^<(t,\bx)=D_{C}(t+i0^+,\bx)$. For a free scalar field one has $\rho_F(k_0)=2\pi\sgn(k_0)\delta(k_0^2-\bk^2-m^2)$ and the matrix~\eqref{rtmatrixprop} becomes
\begin{eqnarray}
	\nonumber \mathbf{D}_F(k)&=&\left[\left(\begin{matrix}\displaystyle\frac{i}{k_0^2-\bk^2-m^2+i\eta}&\theta(-k_0)2\pi\delta(k_0^2-\bk^2-m^2)\\\theta(k_0)2\pi\delta(k_0^2-\bk^2-m^2)&\displaystyle-\frac{i}{k_0^2-\bk^2-m^2-i\eta}\end{matrix}\right)\right.\\
	\label{propreal}
	&&\hspace{3.6cm}+2\pi\delta(k_0^2-\bk^2-m^2)\,n_\mathrm{B}(\vert k_0\vert)\left(\begin{matrix}1&1\\1&1\end{matrix}\right)\Bigg].
\end{eqnarray}
The so-called Kubo-Martin-Schwinger (KMS) relation \cite{Kubo:1957mj,Martin:1959jp}, that links the various time orderings through the periodicity of the fields in the imaginary component of the complex time, imposes some constraints on the elements of the matrix in Eq.~\eqref{rtmatrixprop}, such as
\begin{equation}
	D^<(k)=e^{-\frac{k_0}{T}}D^>(k)\,,\qquad D_{11}+D_{22}=D_{21}+D_{12}\,.
	\label{kms}
\end{equation}
These constraints, together with the definition of the retarded and advanced propagators in Eq.~\eqref{propretadv}, allow one to obtain a relation between those propagators  and the time-ordered, or ``11'', propagator. It reads, for further use
\begin{equation}
D_{11}(k)= \frac{ D_{\rm R}(k) + D_{\rm A}(k)}{2}
+ \left(\frac{1}{2} + n_{\rm B}(k_0)\right)\left(D_{\rm R}(k) - D_{\rm A}(k)\right),
\label{11component}
\end{equation}
which holds for the tree level propagator as well as for the full one. 
The second term on the right-hand side, proportional to the difference 
between the retarded and advanced propagators, 
is often termed the \emph{symmetric propagator}.\\
For what concerns the interactions, we notice from Eq.~\eqref{pathzjrealf} that the vertices are diagonal, i.e. all fields entering a vertex are of the same kind, and the backward time propagation of $C_2$ causes the vertices of type ``2'' to have an opposite sign.\\

The extension of this methodology to gauge theories and to QCD in particular is rather straightforward. The resulting Feynman rules are summarized in App.~\ref{app_feyn_rt}. For what concerns the gauge-fixing procedure, we remark that, in the free case, the thermal part is strictly related to physical particles. Hence, for what concerns the ghosts and the unphysical components of the gauge fields, it has been shown that their thermal parts always cancel each other and can consistently be omitted from the outset \cite{Landshoff:1992ne,Landshoff:1993ag}. This is particularly convenient in the covariant gauges, while in Coulomb gauge, which is our choice for all real-time calculations, the spectral densities of ghosts and longitudinal gluons vanish, effectively freeing them of their thermal parts, as can be seen from Eq.~\eqref{transgluonpropreal}.
\section{Infrared problems and Effective Field Theories of QCD at finite temperature}
\label{sec_htl}
Let us consider the ultraviolet and infrared behaviour of Thermal Field Theory, and of QCD in particular, at finite temperature. As we have shown, the propagators can be separated into a vacuum part and a thermal part in both formalisms. If we then examine the UV behaviour of the theory, the only possible divergences are those of the vacuum part, since in the matter part the thermal distributions act as an exponential UV cut-off. So, for QCD, no new UV divergences appear besides the $T=0$ ones, which are cured by the standard renormalization procedure.\\
The infrared behaviour is, on the other hand, quite different. Let us examine the low-momentum expansion of the Bose--Einstein distribution. When $k\ll T$ it reads 
\begin{equation}
\label{boseexp}
\frac{1}{e^{k/T}-1}=\frac{T}{k}-\frac{1}{2}+\frac{k}{12 \, T}+\order{\frac{k^3}{T^3}}.
\end{equation}
It then appears clear that the infrared behaviour changes drastically: the first term in this expansion, sometimes called the Bose enhancement, renders any loop diagram more sensitive to the infrared region, whereas the second one cancels exactly the leading IR term of the vacuum part, as can be seen easily by applying this expansion to the propagator in the mixed representation, as in Eq.~\eqref{matsumixed}, for $\Delta_F(\tau=0,\mbk\ll T)$. In detail, for a massless boson
\begin{equation} 
	\Delta_F(\tau=0,\mbk\ll T)=\frac{T}{\mbk^2}+\frac{1}{12T}+\order{\frac{\mbk^2}{T^3}},
	\label{propexpandlowk}
\end{equation}
and we notice that the $1/(2\mbk)$ term, i.e. the vacuum part, has vanished. We remark that the analogue for fermions is different, since there is no $T/k$ term in the expansion of the Fermi--Dirac distribution. The first term is $+1/2$, which again cancels with the vacuum part.\\
In a gauge theory, if we were to perform the IR expansion for the thermal part of the massless gauge boson, the photon or the gluon, we would then indeed get one more power of the momentum at the denominator, which can clearly render a IR finite diagram at zero temperature IR divergent at finite temperature.\\
The cure for these divergences relies on the resummation of a particular class of amplitudes, which are called \emph{Hard Thermal Loops} after Braaten and Pisarski \cite{Braaten:1989kk}. A Hard Thermal Loop (HTL) is a particular momentum region of  a loop diagram, the one where the internal momentum $p$ is assumed to be of the order of the temperature and the $n$ external legs have momenta $k_n$ that are assumed to be much smaller than the temperature, i.e. $k_n\ll p\sim T$. In the case of a gauge theory, for what concerns the gauge boson propagator, it can be shown that in this approximation its longitudinal and transverse component develop different self-energies, which by dimensional analysis behave like $g^2T^2$ or $e^2T^2$. Such self-energies, and also the other Hard Thermal Loop amplitudes, can be shown to be gauge-invariant \cite{Braaten:1989mz,Braaten:1990az,Frenkel:1989br}.\\
The longitudinal HTL vacuum polarization then reads, in imaginary time and at the leading order in perturbation theory \cite{Kalashnikov:1979cy,Weldon:1982aq}
\begin{equation}
	\label{pie}
	\Pi_L^{\mathrm{HTL}}(\omega_n,\bk)=m_D^2\left[1-\frac{(i\omega_n)^2}{\bk^2}\right]\left[1-\frac{i\omega_n}{2\vert\bk\vert}\log\frac{i\omega_n+\vert\bk\vert}{i\omega_n-\vert\bk\vert}\right],
\end{equation}
where 
\begin{equation}
m_D^2 = \frac{g^2T^2}{3}\left(N_c + T_F\,n_f \right),
\label{mD}
\end{equation}
is the \emph{Debye mass}, since one has $\Pi_L^{\mathrm{HTL}}(0,\bk)=m_D^2$, i.e. static\footnote{\label{foot_static}In the language of Thermal Field Theory the term ``static'' is used to identify the Matsubara zero-mode.} longitudinal gluons acquire a mass thanks to the interactions with the medium.\\
The transverse HTL vacuum polarization reads instead
\begin{equation}
	\label{pit}
	\Pi_T^{\mathrm{HTL}}(\omega_n,\bk)=m_D^2\left[\frac{(i\omega_n)^2}{\bk^2}+\frac{i\omega_n}{2\vert\bk\vert}\left(1-\frac{(i\omega_n)^2}{\bk^2}\right)\log\frac{i\omega_n+\vert\bk\vert}{i\omega_n-\vert\bk\vert}\right].
\end{equation}
From this equation it is easy to see that static transverse gluons do not develop a mass at the leading order in perturbation theory, since $\Pi_T^{\mathrm{HTL}}(0,\bk)=0$.\\
When integrating over the momentum region $k\sim gT$ the Hard Thermal Loops, such as these vacuum polarizations, have then to be resummed: this is best carried out in a systematic EFT approach, corresponding to integrating out from the QCD Lagrangian modes of energy and momenta $p\sim T$, obtaining an EFT that holds for momenta $k\sim gT\ll T$. To this end, two approaches are possible: they are called \emph{dimensional reduction} and \emph{Hard Thermal Loop effective theory}.\\
In the former approach one observes that, if the temperature is the largest energy scale available, then $1/T$, which is the length of the compactified imaginary time axis, is the shortest length scale and the theory can be mapped to an effective theory in three spatial dimensions. In this dimensionally-reduced theory only the zero-modes of the original theory survive, since all other  modes have an energy, given by their Matsubara frequency, of the order of the temperature. The original idea \cite{Gross:1980br,Appelquist:1981vg,Nadkarni:1982kb} was later turned into a full-fledged set of EFTs \cite{Braaten:1994na,Braaten:1995cm,Braaten:1995jr,Kajantie:1995dw,Kajantie:1997tt}. Integrating out the temperature from QCD leads to Electrostatic QCD (EQCD) \cite{Braaten:1994na,Braaten:1995cm,Braaten:1995jr,Kajantie:1995dw,Kajantie:1997tt}, whose Lagrangian is at the leading order a three-dimensional Yang-Mills Lagrangian minimally coupled to a $SU(\nc)$-adjoint scalar field $A_0$, whose mass (a matching coefficient of the EFT) is $m_D$ at the leading order in $g$. Fermions, being non-static due to their odd frequencies, are integrated out.\\
It is then clear that this approach, while simple and elegant, is limited to the computation of time-independent quantities, such as the partition function (the pressure), since it removes the time coordinate altogether. On the other hand the more 
intricate Hard Thermal Loop effective theory keeps the time (or energy) coordinates all along the procedure. The effective Lagrangian was developed in  \cite{Taylor:1990ia,Braaten:1991gm} and reads
\begin{equation}
	\lag_\mathrm{HTL}=\lag_\mathrm{QCD}+\lag_g+\lag_f+\ldots\,,
	\label{laghtl}
\end{equation}
where $\lag_\mathrm{QCD}$ is the standard QCD Lagrangian in~\eqref{lagrqcd} (real time) or in~\eqref{QCDimtime} (imaginary time). $\lag_g$ and $\lag_f$ encode the Hard Thermal Loop amplitudes in the gluon and light quark sectors of the theory respectively, and the dots stand for higher order, suppressed operators.\\
The effective term $\lag_g$ for the gluons reads, in Euclidean space-time
\begin{equation}
	\delta \lag_g=-\frac{m_D^2}{2}\mathrm{Tr}\int\frac{d\Omega_K}{4\pi}F_{\mu\alpha}\frac{\hat{K}_\alpha\hat{K}_\beta}{(\hat{K}\cdot D)^2}F_{\beta\mu}\,,
	\label{htllaggluon}
\end{equation}
where $\hat{K}\equiv(-i,\hat{\bk})$ is a light-like four-vector that represents the momentum of the hard, on-shell particle in the loop, hence its vanishing $\hat{K}^2$. Since $\hat{K}\cdot D$ appears in the denominator, this term is nonlocal. When performing an analytical continuation to Minkowski space-time, this term can possibly vanish, and the Lagrangian is understood to be valid only for fields such that $(\hat{K}\cdot D)$ does not vanish.\\
The light quark Lagrangian $\lag_f$ reads
\begin{equation}
	\delta\lag_f=im_f^2\,\overline{\psi}\int\frac{d\Omega_K}{4\pi}\frac{\hat{\slashed{K}}}{\hat{K}\cdot D}\,\psi\,,
	\label{htllagfermion}
\end{equation}
where again this Euclidean space-time expression is nonlocal, and the same considerations about the analytical continuation apply. The term $m_f^2=\cf/8g^2T^2$ is the thermal mass of the light quarks, similar to the Debye mass for gluons.\\
The HTL theory has been analyzed in the real-time framework in \cite{Carrington:1997sq}, where the intricacies due to the resummation of the HTL self-energies in the matrix formalism have been dealt with. The resulting Feynman rules for the gluon propagator in Coulomb gauge are listed in App.~\ref{sub_feyn_htl}. Here we remark that in Minkowski space the logarithms appearing in the HTL self-energies present imaginary parts in the spacelike region, that is for $k_0^2<\bk^2$. These imaginary parts are related, through the cutting rules and the optical theorem, to the cut polarization diagram. This is however different from what happens at zero temperature, where one has imaginary parts only in the forward timelike region, related to the branch cut for pair production, which is the single kinematically allowed process there. At finite $T$ on the other hand  the scattering of the gluon with the other light constituents of the plasma is allowed, since these particles already exist in the bath; furthermore the thermal particles are on shell and the scattering thus proceeds in the spacelike region. This feature of finite temperature is called \emph{Landau damping}, since it brings indeed to a damping of the gluon through these scattering processes.\\

A further remark involves the perturbative expansion in the HTL EFT and in EQCD: since the resummation gives a new scale $m_D\sim gT$ to the integrals, their results will depend on powers of it that are not necessarily even and positive, thus bringing the expansion parameter from the usual $g^2\sim\als$ to $g$. This can be understood in a qualitative way by looking again at Eq.~\eqref{boseexp}: if $k\sim gT$ then the leading IR behavior of the Bose--Einstein distribution is $\sim1/g$. This combines with the usual $g^2$ factor for each additional gluon to give a (larger) $g$ factor instead of $g^2$.\\ 
If we continue along this qualitative approach, we may wonder about what would happen if a scale $g^2T$ were to arise. In such a case, the IR Bose enhancement, now $1/g^2$, and the perturbative $g^2$ factor would cancel out. This would then imply that the scale $g^2T$ is non-perturbative, i.e. it cannot be explored by perturbative calculations, since all diagrams in principle contribute at the same order. It turns out that in QCD the scale $g^2T$ indeed exist, and is the so-called \emph{magnetic mass}. We have seen in Eq.~\eqref{pit} that, at the leading order, transverse static gluons, sometimes called \emph{magnetostatic gluons}, remain massless. However at the next order the self-interactions of the transverse gluons are indeed expected to cause a magnetic mass to appear. In QED, on the other hand, photons do not interact directly with each other and the theory is devoid of this issue.\\
The magnetic mass is in itself non-perturbative, since, for the reason mentioned above, an infinite number of diagrams contributes to it at order $g^2T$. However, in the framework of dimensional reduction, one can prove that such a mass must exist, and that it must be of order $g^2T$ \cite{Gross:1980br}. The issue of the magnetic mass and of the breakdown of the perturbative expansion was first pointed out in \cite{Linde:1980ts} in the context of the pressure, where the non-perturbative contribution was shown to arise at order $g^6$. The EFT approach of dimensional reduction renders this problem less severe  \cite{Braaten:1994na,Braaten:1995cm,Braaten:1995jr} than initially thought: by integrating out the scale $gT$ from EQCD one arrives at Magnetostatic QCD (MQCD) \cite{Braaten:1994na,Braaten:1995cm,Braaten:1995jr,Kajantie:1995dw,Kajantie:1997tt}, whose Lagrangian consists at leading order of a pure three-dimensional Yang-Mills term. While this last theory is inherently non-perturbative, this EFT approach nevertheless allows one to factorize the order-$g^6$ contribution to the pressure into contributions from the scales $T$ and $gT$, which can be calculated perturbatively in QCD and EQCD respectively, and a (dimensionless) contribution from the scale $g^2T$ that can be computed on the lattice within MQCD. The logarithmic contribution at order $g^6$, coming from QCD and EQCD, is known \cite{Kajantie:2002wa}. All higher order terms ($g^7$ and so on) are again free of magnetic non-perturbative effects.\\
For different observables, the non-perturbative contribution manifests itself at different orders, and should in principle be dealt with by analogous techniques. However in the calculations of this thesis the non-perturbative wall is never reached; we just remark that, for time-dependent observables, a time-dependent EFT of the HTL Lagrangian, obtained by integrating out the Debye mass in a non-static analogue of MQCD, has been developed in \cite{Bodeker:1998hm,Bodeker:2000da}. 


		\part{Real-time Effective Field Theories of QCD at finite temperature for heavy quarkonium}
		\label{part_realtime}
		\chapter{Overview}
		\label{chap_realtime}
\begin{figure}[ht]
	\begin{center}
		\includegraphics{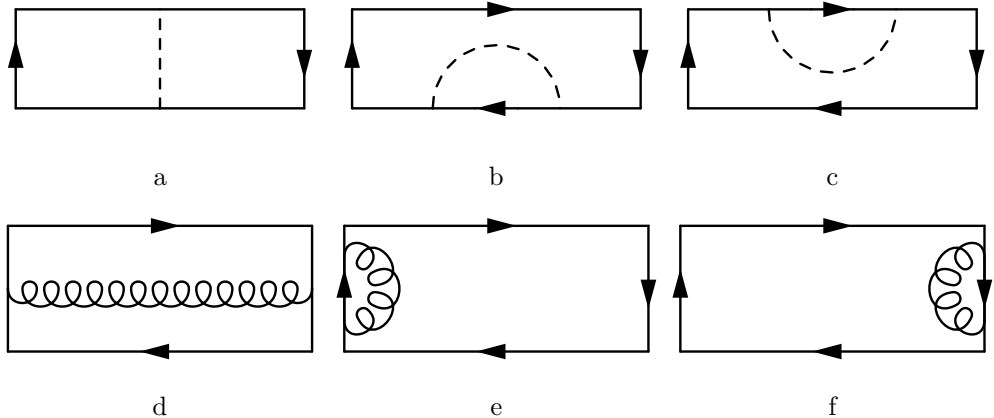}
	\end{center}
	\caption{The diagrams contributing to the tree-level rectangular Wilson loop in Feynman or Coulomb gauge. Imaginary time runs in the horizontal direction, space in the vertical one. As per the conventions of App.~\ref{app_feynrules}, dashed lines are longitudinal gluons and curly lines are transverse gluons. After analytical continuation to real times, it can be shown that in the large time limit only the diagrams in the first line contribute to the potential, both in Feynman gauge \cite{Laine:2006ns} and in Coulomb gauge \cite{Ghiglieri08}.}
	\label{fig_wloop}
\end{figure}

In Section~\ref{sec_exp_quarkonium}, the importance of heavy quarkonium as a \emph{hard probe} in heavy-ion collisions was emphasized and the most common theoretical tools for its finite-temperature study were introduced. As we mentioned, potential models were and still are a widespread tool; in these models  (see \cite{Karsch:1987pv,Digal:2001iu,Mocsy:2005qw,Kluberg:2009wc,Bazavov:2009us,Riek:2010fk}) it is assumed that all medium effects can be accounted for by solving a Schr\"odinger equation with a screened, temperature-dependent potential, often extracted from lattice calculations of correlation functions of Polyakov loops. However the connection between this simple Schr\"odinger picture and QCD was not established within the context of potential models.\\
A first step towards a QCD derivation of the $Q\overline{Q}$ potential at finite temperature was the calculation \cite{Laine:2006ns} of the rectangular Wilson loop in the imaginary-time formalism at order $\als$ (tree level). The diagrams contributing to the calculation are shown in Fig.~\ref{fig_wloop}. After analytical continuation to real time along the lines discussed in Eqs.~\eqref{spectralscalar} and~\eqref{defrho}, and taking the logarithm of the large-time limit, the calculation shows a real part, which is a screened Debye potential, and an imaginary part 
that may be traced back to the scattering of particles in the medium 
carrying momenta of order $T$ with space-like gluons, i.e. Landau damping, as mentioned in Sec.~\ref{sec_htl}. Some applications of this potential can be found in \cite{Laine:2007gj,Laine:2007qy,Burnier:2007qm}.\\
The logarithm of the large time limit of the Wilson loop, divided by the time extent, gives the singlet static potential in zero-temperature weakly-coupled pNRQCD\footnote{This is strictly true up to order $\als^3$. From order $\als^4$ and higher, ultrasoft effects start to contribute, as mentioned in Sec.~\ref{sec_pnrqcd}.}. In this Part we will show how the zero-temperature framework of NR EFTs can be generalized with the inclusion of the thermodynamical scales, enabling us to obtain a set of EFTs that can describe quarkonium in different temperature regimes, allowing a rigorous definition of the potential, the possibility of a systematic improvement, such as the calculation of non-static (i.e. $1/m$-suppressed) contributions and the inclusion of non-potential medium effects, i.e. effects that cannot be encoded in a potential. The results of this Part~\ref{part_realtime} are published in \cite{Brambilla:2008cx,Brambilla:2010vq,poincare}. A similar analysis, based on NR EFTs of QED for electromagnetic bound states (hydrogen and muonic hydrogen) can be found in \cite{Escobedo:2008sy,Escobedo:2010tu}.\\
As in any EFT, establishing a hierarchy and identifying the low-energy degrees of freedom is the first step in the construction of the effective theory, as we explained in Sec.~\ref{sec_princ_eft}. A crucial aspect of our EFT framework is thus the assumption of a \emph{scale hierarchy} between the non-relativistic and the thermodynamical scales. We remark that the aforementioned calculation \cite{Laine:2006ns} of the potential from the real-time continuation of the Wilson loop has been performed in the context of the HTL effective theory, with the Feynman rules derived from the Lagrangian~\eqref{laghtl}. As such it implicitly assumes a temperature $T$ much larger than the inverse spatial extent $1/r$ of the Wilson loop, i.e. $T\gg1/r$. The use of resummed HTL propagators furthermore implies $1/r\sim m_D$.\\ Sec.~\ref{sec_hierarchies} shall then be devoted to introducing the possible scale hierarchies that are relevant for $Q\overline{Q}$ bound states in the plasma, among which the one just discussed. These hierarchies will then be analyzed in detail in Chapters~\ref{chap_Tggr}, \ref{chap_rggT} and \ref{chap_poincare}.

\section{Scale hierarchies }
\label{sec_hierarchies}
Bound states at finite temperature are systems characterized by many energy scales. As we mentioned before, on one side there are the thermodynamical scales that describe the motion of the particles in the thermal bath: as discussed in Sec.~\ref{sec_htl}, one has
the temperature scale\footnote{\label{foot_piT}There is an ambiguity as to what is the effective scale between $T$, $\pi T$ and multiples thereof. The controversy arises because in the Matsubara formalism frequencies are even/odd multiples, according to the statistics, of $\pi T$. This in turn reflects itself on the running of $\als$, where the typical energy scale in the logarithms is again a multiple of $\pi T$ \cite{Kajantie:1997tt}. We will however not distinguish between $T$ and multiples of $\pi T$ in the text, always using $T$ to label this scale.} $T$, the Debye mass $m_D$, which is the screening scale of the chromoelectric interactions, and lower energy scales such as the magnetic mass. 
In the weak-coupling regime, which we will assume throughout this Part~\ref{part_realtime} and is the regime of validity of the HTL effective theory, 
one has $m_D \sim gT \ll T$. The magnetic mass would be smaller than $m_D$ in this hierarchy, and is consistently neglected in this Part, as its contribution would be smaller than the results reported here. This furthermore implies that, as we mentioned in the previous chapter, non-perturbative magnetic effects do not appear at the orders considered here.\\
On the other side there are the scales typical of a non-relativistic bound state, that have been widely discussed in Secs.~\ref{sec_nrqcd} and~\ref{sec_pnrqcd}. They are $m\gg mv\gg mv^2$ and, in the weak-coupling regime we are assuming, $v\sim\als(mv)$. Some of the results of this Part have been obtained in the \emph{static} limit only\footnote{In the context of heavy quarks the term ``static'' labels an infinitely heavy quark. This is not to be confused with the same term applied to gluons in the imaginary-time formalism of Thermal Field Theory, where it indicates the Matsubara zero mode, see also footnote~\ref{foot_static} of Chapter~\ref{chap_thermal}.}; in the context of non-relativistic EFTs, this corresponds to the zeroth order in the $1/m$ expansion, which is tantamount to infinitely massive, i.e. static, quarks. In such a case the scale $mv$ is naturally replaced by $1/r$ and $mv^2$ by $\als/r$, i.e. the Coulomb potential. We will often use $E$ to label the energy scale, both in the static and finite mass cases.\\
For what concerns the QCD scale $\lqcd$, we will always assume that $mv\gg\lqcd$ (or $1/r\gg \lqcd$ in the static limit). Since the potential arises when integrating out $mv$ ($1/r$), this amounts to concentrating on the short-distance, perturbative part of the potential, which may be the part of the potential relevant for the lowest quarkonium 
resonances like the $J/\psi$ or the $\Upsilon(1S)$, as we remarked in Sec.~\ref{sec_pnrqcd}. In some cases we will also assume that the energy scale is perturbative, i.e $mv^2\gg\lqcd$. On the other hand, the weak-coupling hierarchy of the thermodynamical scales implies $T\gg m_D>\lqcd$.\\
This set of assumptions (thermodynamical weak coupling, NR weak coupling) still leaves a sizeable freedom on the relative size of the thermodynamical and NR scales. We first restrict this freedom by the constraint $T\ll m$, since quarkonium is expected to exist in the medium if the temperature and the other thermodynamical scales are much smaller than the quark mass $m$. This, while leaving many possibilities on the relative size of $T$ and $m\als$, $m\als^2$, implies that $m$ will always be the largest scale and as such the first to be integrated out, leading to NRQCD. As we explained in Sec.~\ref{sec_princ_eft}, lower energy scales are customary put to zero in the matching procedure; hence, the resulting Lagrangian and matching coefficients are exactly the same one encounters at $T=0$, as they are laid out in Sec.~\ref{sec_nrqcd}. The only change with respect to the zero-temperature case lies in the Feynman rules:  in the next Section~\ref{sec_static} we will deal with the static quark propagator in the matrix representation of the real-time formalism. As we already remarked, the potential is related to large-time limits of correlation functions; hence the real-time formalism is better suited, not requiring any analytical continuation. All the calculations of this Part will then be performed within this formalism.\\
Once the mass has been integrated out, several scales, such as $m\als$ ($1/r$ in the static limit), $E$,  $T,\,m_D,\ldots$, remain dynamical in the resulting EFT (NRQCD). In Chapter~\ref{chap_Tggr} we will consider the case where the temperature is the next largest scale; this regime will be explored in the static limit only, corresponding to $T\gg1/r$. Different possibilities for the relative position of $1/r$ and $m_D$ will be explored, and for $1/r\simg m_D$ the results of \cite{Laine:2006ns} will be reobtained in our EFT context.\\
In Chapter~\ref{chap_rggT} we will instead consider smaller temperatures, such that $m\als\gg T$. The results of this Chapter apply to the finite mass case, giving the thermal corrections to the $T=0$ spectrum of pNRQCD, given in Sec.~\ref{sec_pnrqcd}, as well as the width caused by the thermal medium. We will in particular consider a hierarchy which, as we shall see, might be very relevant for the phenomenology of the $\Upsilon(1S)$ at the LHC. Finally, in Chapter~\ref{chap_poincare} we will consider, in the regime $m\als\gg T\gg m\als^2$, the spin-orbit potential, proving that thermal contributions thereto violate the \emph{Gromes relation}. As we discussed in Sec.~\ref{sec_pnrqcd}, this relation links a piece of the spin-orbit potential of pNRQCD to the derivative of the static potential and is a result of Poincar\'e invariance of the fundamental theory (QCD). The thermal bath explicitly breaking Lorentz invariance, a violation of the Gromes relation is in this context not unexpected. 

\section{The mass scale and the static quark propagator}
\label{sec_static}
As we just discussed, the first scale to be integrated out from QCD is, in all considered cases, the heavy quark mass $m$. 
In the matching procedure, smaller scales are expanded. 
Thus, at this stage, the presence of the thermal
scales does not affect the Lagrangian, which is the 
one of non-relativistic QCD (NRQCD), as in Eq.~\eqref{lagrnrqcd}. The matching coefficients are unmodified as well.\\
As we anticipated before, some care is required in deriving the heavy-quark propagator in the real-time formalism. To this end we start from the infinite mass (static) limit. The Lagrangian is then the NRQCD Lagrangian~\eqref{lagrnrqcd} at order $1/m^0$, i.e. 
\begin{equation}
{\cal L}  = 
- \frac{1}{4} F^a_{\mu \nu} F^{a\,\mu \nu} 
+ \sum_{i=1}^{n_f}\bar{q}_i\,iD\!\!\!\!/\,q_i 
+ \psi^\dagger( i D_0 -m) \psi  + \chi^\dagger( i D_0 +m) \chi,
\label{NRQCDstatic}
\end{equation}
where we have left explicit mass terms for the heavy quark fields, which are to be understood in a $m\to\infty$ limit. For reasons that will become clear in the next steps, the standard field redefinition~\eqref{fieldredef} has to be performed at a later stage in the derivation. We recall that the fermion rules for light quarks, gluons, as well as for their couplings, are given in the real-time formalism in App.~\ref{app_feyn_rt}. As a further remark, we observe that transverse gluons do not couple directly to static quarks.\\
We now set out on obtaining the static quark and antiquark propagators. Let us define  
\begin{equation} 
S_{\alpha\beta}^{>}(x) = \langle \psi_\alpha(x)\psi^\dagger_\beta(0)\rangle\,,\qquad S_{\alpha\beta}^{<}(x) = - \langle \psi^\dagger_\beta(0) \psi_\alpha(x)\rangle\,,
\end{equation}
where the minus sign in $S_{\alpha\beta}^{<}$ is a consequence of the fermionic statistics and the expectation values are, here and throughout the rest of this Part, understood to be thermal ones, as in Eq.~\eqref{defthermalaverage}. The free propagators, 
\begin{equation}
S_{\alpha\beta\,F}^{>} = \delta_{\alpha\beta}S^{>}_F, \qquad 
S_{\alpha\beta\,F}^{<} = \delta_{\alpha\beta}S^{<}_F,
\end{equation}
satisfy the equations (in momentum space)
\begin{eqnarray} 
k_0 S^{>}_F(k) = m S^{>}_F(k)\,, 
\label{eqmot}
\\
k_0 S^{<}_F(k) = m S^{<}_F(k)\,.
\label{eqmot1}
\end{eqnarray}
If the heavy quarks are part of the thermal bath, they have to satisfy the Kubo--Martin--Schwinger relation:
\begin{equation}
S^{<}(k) = - e^{-k_0/T}  S^{>}(k).
\label{KMSfermion}
\end{equation}
This relation is the fermionic counterpart of Eq.~\eqref{kms}, and it holds for full propagators as well as for free propagators. From the equal-time canonical anticommutation relation it follows the sum rule
\begin{equation}
\int \frac{dk_0}{2\pi}\left(S^{>}_F(k) -S^{<}_F(k)   \right) =1.
\label{sumrulefermion}
\end{equation}
The solutions of the equations (\ref{eqmot})-(\ref{sumrulefermion}) are 
\begin{eqnarray}
S^{>}_F(k) = (1-n_{\rm F}(k_0))\, 2\pi\delta(k_0-m),
\label{Spm}
\\
S^{<}_F(k) =  -n_{\rm F}(k_0)\, 2\pi\delta(k_0-m),
\label{Smm}
\end{eqnarray}
where $n_\mathrm{F}$ is the Fermi--Dirac distribution, as defined in Eq. \eqref{fermidistr}. The free spectral density $\rho^{(0)}_{F}$ is then given by
\begin{equation}
\rho^{(0)}_{F}(k) = S^{>}_F(k) - S^{<}_F(k) = 2\pi\delta(k_0-m),
\end{equation}
and the free time-ordered propagator, 
\begin{equation}
 S_F(x) = \theta(x^0) S^{>}_F(x) -  \theta(-x^0) S^{<}_F(x), 
\end{equation}
is given in momentum space by
\begin{equation}
 S_F(k) = \frac{i}{k_0-m+i\eta} -  n_{\rm F}(k_0) \, 2\pi \delta(k_0-m).
\end{equation}
We are now in the position to take the limit $m \to \infty$: the propagators simplify because 
$n_{\rm F}(m) \to 0$ for $m \to \infty$. Moreover, we may now 
get rid of the explicit mass dependence in the surviving part of the propagators and in the Lagrangian by means of the field redefinition
\eqref{fieldredef}, which 
amounts to change $k_0-m$ to $k_0$ in the expressions 
for the propagators and the spectral density; they read now 
\begin{eqnarray}
S^{>}_F(k) = 2\pi\delta(k_0),
\label{Sp}
\\
S^{<}_F(k) = 0,
\label{Sm}
\\
S_F(k) = \frac{i}{k_0+i\eta},
\label{SS}
\\
\rho_{F}(k) = 2\pi\delta(k_0).
\label{spectralS}
\end{eqnarray}
The free static propagators are the same as at zero temperature.
On the other hand, if we would have assumed from the beginning that  
$S^{<}_F(k) = 0$, i.e. that there is no backward propagation 
of a static quark (in agreement with  the Kubo--Martin--Schwinger formula 
in the $m\to\infty$ limit) then, together with the equations  of motion  $k_0 S^{>}_F(k) = 0$, 
$k_0 S^{<}_F(k) = 0$ (obtained after removing $m$ via field redefinitions) and  
the sum rule (\ref{sumrulefermion}), we would have obtained Eqs. (\ref{Sp}), (\ref{SS})
and (\ref{spectralS}).\\
The real-time free static propagator for the quark reads, with the conventions of Eq.~\eqref{genericrealprop} 
\begin{equation}
{\bf S}_{\alpha\beta\,F}(k) = 
\delta_{\alpha\beta}
\left(
\begin{matrix}
	\hspace{0mm}S_F(k)  
&&\hspace{-2mm}S^{<}_F(k) \\ 
\hspace{0mm}S^{>}_F(k)  
&&(S_{F}(k))^*
\end{matrix}
\right)
= 
\delta_{\alpha\beta}
\left(
\begin{matrix}
\displaystyle\frac{i}{k_0+i\eta}  
&&0 \\ 
2\pi\delta(k_0)  
&&\displaystyle\frac{-i}{k_0-i\eta}
\end{matrix}
\right),
\label{Squarkstatic}
\end{equation}
and for the antiquark
\begin{equation}
{\bf S}_{\alpha\beta\,F}(k) 
= 
\delta_{\alpha\beta}
\left(
\begin{matrix}
\displaystyle\frac{i}{-k_0+i\eta}  
&&0 \\ 
2\pi\delta(k_0)  
&&\displaystyle\frac{-i}{-k_0-i\eta}
\end{matrix}
\right).
\label{Santiquarkstatic}
\end{equation}
The main observation here is that, since the $[{\bf S}_{\alpha\beta\,F}(k)]_{12}$
component vanishes, the static  quark (antiquark) fields labeled ``2'' 
never enter in any physical amplitude, i.e. any amplitude that has the 
physical fields, labeled ``1'', as initial and final states. 
Hence, when considering physical amplitudes, the static fields ``2'' 
decouple and may be ignored.\\
For future convenience, we note that the propagator ${\bf S}_{\alpha\beta\,F}$ may be written in a diagonal form as 
\begin{equation}
{\bf S}_{\alpha\beta\,F}(k) = {\bf U}
\left(
\begin{matrix}
	\hspace{0mm}\displaystyle\frac{i\delta_{\alpha\beta}}{k_0+i\eta}  
&&0 \\ 
0  
&&\displaystyle\frac{-i\delta_{\alpha\beta}}{k_0-i\eta}
\end{matrix}
\right)
{\bf U}\,,
\label{diagS0}
\end{equation}
where
\begin{equation}
{\bf U} = 
\left(
\begin{matrix}
\hspace{0mm}1
&&0 \\ 
\hspace{0mm}1  
&&~1
\end{matrix}
\right)\,,
\quad \hbox{and for further use}\quad 
[{\bf U}]^{-1} = 
\left(
\begin{matrix}
\hspace{0mm}1
&&0 \\ 
\hspace{0mm}-1  
&&~1
\end{matrix}
\right)\,.
\label{matrixU0}
\end{equation}
In the next Chapters we will see how this diagonal representation will be very useful when resumming insertions of self-energies, potentials and kinetic energies.

Having integrated out the mass scale $m$ and obtained the expression for the static propagator, we are now ready to integrate out the subsequent, smaller scales. This is done in the following chapters, according to the different hierarchies we may assume.

\chapter{EFTs in the screening regime}
\label{chap_Tggr}

In this Chapter, we consider bound states made of a static quark and antiquark in a thermal 
bath at distances such that $1/r\ll T$. As we discussed previously, we still keep that $T$, $1/r$ and $m_D$ are 
perturbative scales and  we further neglect other thermodynamical scales. Under the above conditions, the first scale to integrate out from QCD after the mass is the temperature $T$. This first step will be done in Sec.~\ref{sec_NRQCDHTL}.\\
Even if in the static limit, the results we will show are extremely interesting. In Sec.~\ref{secVIA} we will study the regime $1/r\simg m_D$ and get to a rigorous EFT definition and derivation of the screened+imaginary potential first obtained in \cite{Laine:2006ns}, whereas in Sec.~\ref{secTrmD} we will consider the situation $1/r\gg m_D$, where a \emph{dissociation temperature} can be naturally defined. The importance of these imaginary parts, which lack in lattice-inspired potential models, will be strongly highlighted. In the conclusions we will also illustrate some phenomenological implications. A great body of the results of this Chapter have been published in \cite{Brambilla:2008cx}.

\section{Integrating out the temperature scale}
\label{sec_NRQCDHTL}
We now proceed to integrate out modes of energy and
momentum of the order of the temperature $T$. This corresponds to
modifying NRQCD into a new EFT where only modes with energies
and momenta lower than $T$ are dynamical. We label the new EFT $\textrm{NRQCD}_\mathrm{HTL}$ \cite{Vairo:2009ih}. This EFT can be used for $T\gg m\als\,(1/r\text{ in the static limit}),E,m_D$, no matter what the relation between $E$ and $m_D$ is. We remark that the opposite case, where $m\als\gg T$ ($1/r\gg T$), is discussed in the next Chapter.\\
The Lagrangian of  $\textrm{NRQCD}_\mathrm{HTL}$ will get additional
contributions with respect to NRQCD. For what concerns the gauge and light quark sectors, they are now described by the HTL effective Lagrangian~\eqref{laghtl}.  In the static quark-antiquark sector one can in principle expect the appearance of thermal mass terms and corrections to the couplings of the static quarks and antiquarks with gluons. While the thermal mass shift of the quark is expected to be gauge invariant, since it contributes to the pole of the propagator, the other corrections are in principle not gauge-invariant; in the following we analyze them in Coulomb gauge, which is the gauge we adopt for all real-time calculations. They will be relevant for calculations performed in App.~\ref{secpQCD}. At the zeroth-order in $\als$ in the static sector, the Lagrangian of static $\textrm{NRQCD}_\mathrm{HTL}$ is then
\begin{equation}
	\lag_{\mathrm{NRQCD}_\mathrm{HTL}}=\psi^\dagger( i D_0) \psi  + \chi^\dagger( i D_0 ) \chi\,+\lag_\mathrm{HTL}+\ldots\,,
	\label{lagnrqcdhtl}
\end{equation}
where $\lag_\mathrm{HTL}$ is the Lagrangian shown in Eq.~\eqref{laghtl}. An Abelian, non-static (with $1/m$ terms) version of $\textrm{NRQCD}_\mathrm{HTL}$ can be found in \cite{Escobedo:2008sy}.\\
As we just discussed, we now consider one-loop thermal 
contributions to the static quark propagator, quark-gluon vertices and gluon propagator. 
When the loop momenta and energies are taken at the scale $T$ and the external momenta 
are much lower, so that we may expand with respect to them, these correspond to the Hard Thermal Loop contributions in the static quark sector. 
\begin{figure}[ht]
	\begin{center}
		\includegraphics{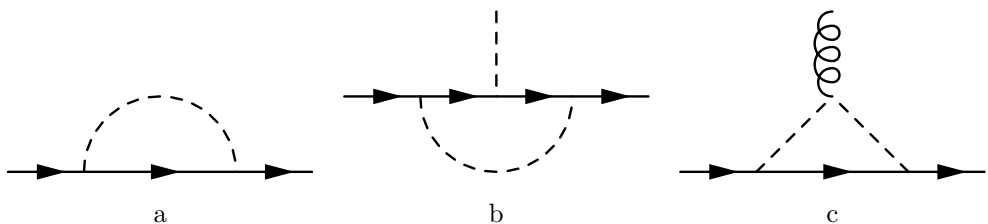}
	\end{center}
	\caption{Hard Thermal Loop contributions in the static quark sector. Diagram a is a static-quark self-energy, diagram b is a static quark-longitudinal gluon vertex and diagram c is the transverse gluon analogue. }
	\label{fig_statichtl}
\end{figure}\\
The one-loop contributions to the static-quark self energy, to the static-quark lon\-gitudinal-gluon vertex 
and to the  static-quark transverse-gluon vertex are displayed in Fig.~\ref{fig_statichtl} a, b and c respectively. In Coulomb gauge, longitudinal gluons do not depend on the temperature, as discussed in Sec.~\ref{sec_realtime} (see Eq. (\ref{longgluonpropreal}))
and the above diagrams do not give thermal contributions\footnote{Diagram a has been computed in QED in \cite{Donoghue:1984zz,Escobedo:2008sy}. In an inverse mass expansion the first contribution was found to be of order $T^2/m$, thus confirming our result. At one loop the QED fermion self-energy translates to the QCD one with the simple replacement $\alpha\to\cf\als$.}. Moreover, if evaluated 
in dimensional regularization, their vacuum contribution vanishes. 
At two-loop order, where transverse gluons, which do have a thermal part as per Eq.~\eqref{transgluonpropreal}, may appear, 
there may be effects. These would be of order $\als^2\, T$ and will be neglected in the following, where we shall concentrate 
on the leading contribution coming from the scale $m_D$. In this EFT, as we mentioned, several scales remain dynamical: the inverse distance $1/r$, the Debye mass $m_D$ and the (static) energy $E$, as well as $\lqcd$.\\
As discussed at length in Sec.~\ref{sec_pnrqcd}, the potential NR EFT picture appears naturally once the scale $mv$ ($1/r$ in this case) is integrated out. In Sec.~\ref{secVIA} we assume $1/r \simg m_D$ and proceed to integrate out both scales $1/r$ and $m_D$ at the same time. We shall specialize to the case 
$1/r \gg m_D$ in Sec.~\ref{secTrmD}, which is relevant for the definition of the \emph{dissociation temperature}. As the following steps will make clear, the opposite case $1/r\ll m_D$ is a region where no bound states are expected to survive.
\section{Integrating out the scales $1/r\simg m_D$}
\label{secVIA}
After integrating out the scale $1/r$, non-local four-fermion operators, i.e. the potentials, appear in the resulting EFT, as we discussed in Sec.~\ref{sec_pnrqcd} for the case of pNRQCD at zero temperature. The same happens in the case at hand; hence we name $\mathrm{pNRQCD}_{m_D}$ \footnote{In \cite{Vairo:2009ih} it was called instead $\textrm{pNRQCD}_\mathrm{HTL}$, along with an analogous, but different EFT arising when $1/r\gg T$. For clarity's sake, only the latter EFT is labelled $\textrm{pNRQCD}_\mathrm{HTL}$ in this thesis. $\textrm{pNRQCD}_\mathrm{HTL}$ will be introduced in Chapter~\ref{chap_rggT}.} the EFT we obtain from integrating out $1/r$ and $m_D$ from $\textrm{NRQCD}_\mathrm{HTL}$. In analogy with pNRQCD, we write its Lagrangian with colour-singlet and colour-octet quark-antiquark pairs as degrees of freedom in the static $Q\overline{Q}$ sector. In the gauge sector, integrating out the Debye mass leads to the effective theory developed in \cite{Bodeker:1998hm,Bodeker:2000da}, whereas light quarks, having a thermal mass of order $m_D$, are integrated out. Since we do not plan to use this EFT to perform calculations at scales lower than those we have just integrated out, the explicit form of the gauge Lagrangian is not relevant\footnote{The explicit form of this Lagrangian is, to the best of our knowledge, not mentioned in the literature. The corresponding effective Hamiltonian appears however in \cite{Bodeker:2000da}.}. We therefore write the Lagrangian of $\mathrm{pNRQCD}_{m_D}$ as
\begin{equation}
	{\cal L}_{\textrm{pNRQCD}_{m_D}} = 
{\cal L}_{\rm gauge}
+ \int d^3r \; {\rm Tr} \,  
\Biggl\{ {\rm S}^\dagger \left[ i\partial_0 - V_s -\delta m_s \right] {\rm S} 
+ {\rm O}^\dagger \left[ iD_0 - V_o -\delta m_o \right] {\rm O} \Biggr\}+\ldots\,,
\label{pNRQCDmd}	
\end{equation}
where $\mathrm{S}$ and $\mathrm{O}$ are the singlet and octet fields, normalized as in Sec.~\ref{sec_pnrqcd}, $V_{s,o}$ and $\delta m_{s,o}$ are respectively the potentials ($r$-dependent) and thermal mass shifts ($r$-independent) for the singlet and octet fields. The EFT is again organized as an expansion in $r$ and for our purposes here only the zeroth term suffices, corresponding to a (static) Schr\"odinger equation of motion for the singlet field. For what concerns the power counting, $\partial_0\sim E$ will be assigned the size of the potentials $V_{s,o}$ and mass terms $\delta m_{s,o}$, which one could expect to be of size $\als/r$ and $\als m_D$ respectively, with $\als/r\simg \als m_D$.\\
We now set out to match the potentials and mass terms. To this end, static quark-antiquark scattering diagrams in $\mathrm{NRQCD}_\mathrm{HTL}$ have to be matched to the bound state propagator in $\mathrm{pNRQCD}_{m_D}$, in the appropriate colour channel. The real-time quark-antiquark propagator, ${\bf S}(p)$, is  
a $2 \times 2$ matrix obtained by matching equal time quark and antiquark propagators 
such that $[{\bf S}(p)]_{ij}$ provides the propagator of a quark-antiquark pair of type 
``$i$'' into a quark-antiquark pair of type ``$j$''. The explicit expressions of the free ($V_{s,o}=\delta m_{s,o}=0$) colour singlet and colour octet quark-antiquark propagators are obtained analogously to the static quark ones in Sec.~\ref{sec_static}, since the free equations of motion are, colour and spin indices aside, identical, although singlet and octet are bosons. The propagators then read
\begin{equation}
{\bf S}^{\rm singlet}_F(p) =
\left(
\begin{matrix}
\displaystyle\frac{i}{p_0 +i\eta}  
&&0 \\ 
2\pi\delta(p_0)  
&&\displaystyle\frac{-i}{p_0 -i\eta}
\end{matrix}
\right)
= 
{\bf U}
\left(
\begin{matrix}
\displaystyle\frac{i}{p_0 +i\eta} 
&&0 \\ 
0  
&&\displaystyle\frac{-i}{p_0 -i\eta}
\end{matrix}
\right)
{\bf U}\,,
\label{SsingletstaticrT}
\end{equation}
\begin{equation}
{\bf S}^{\rm octet}_F(p)_{ab} = 
\delta_{ab}\,\left(
\begin{matrix}
\displaystyle\frac{i}{p_0 +i\eta}
&&0 \\ 
2\pi\delta(p_0)
&&\displaystyle\frac{-i}{p_0 - i\eta}
\end{matrix}
\right)
= 
\delta_{ab}\,{\bf U} 
\left(
\begin{matrix}
\displaystyle\frac{i}{p_0 +i\eta}
&&0 \\ 
0  
&&\displaystyle\frac{-i}{p_0-i\eta}
\end{matrix}
\right)
{\bf U}\,.
\label{SoctetstaticrT}
\end{equation}
Thermal and potential contributions from the scales $1/r \simg m_D$ modify the quark-anti\-quark propagator.
In particular, the singlet propagator gets the form 
\begin{eqnarray}
&&{\bf S}^{\rm singlet}(p) =  
{\bf S}^{\rm singlet}_F(p) + 
{\bf S}^{\rm singlet}_F(p) \left[ -i\,\delta {\bf m_s} -i\, {\bf V_s} \right] {\bf S}_F^{\rm singlet}(p) 
+ \dots \label{singletstaticexpand}\\
&&
\left(
\begin{matrix}
\displaystyle\frac{i}{p_0 - \delta m_s -V_s(r) +i\eta}  
&&0 \\ 
\displaystyle  
\frac{i}{p_0 - \delta m_s -V_s(r) +i\eta}  - \frac{i}{p_0 - \delta m_s^* -V_s^*(r) -i\eta}  
&&\displaystyle\frac{-i}{p_0  - \delta m_s^* -V_s^*(r) -i\eta}
\end{matrix}
\right),
\nn\\
\nn\\
\label{Ssingletstaticfull}
\end{eqnarray}
where in the first line, i.e. Eq.~\eqref{singletstaticexpand}, we have written the series which, when resummed, leads to the second line, Eq.~\eqref{Ssingletstaticfull}. The resummation is tantamount to the assumptions $p_0\sim\delta m_s\sim V_s$. The $2\times 2$ matrices introduced in Eq.~\eqref{singletstaticexpand} read
\begin{eqnarray}
\delta {\bf m_s}
&=& 
\left(
\begin{matrix}
\delta m_s
&&0 \\ 
 -2i \, {\rm Im} \, \delta m_s
&&\displaystyle -\delta m^*_s
\end{matrix}
\right)
=
[{\bf U}]^{-1}
\left(
\begin{matrix}
\delta m_s
&&0 \\ 
0
&&- \delta m_s^*
\end{matrix}
\right)
[{\bf U}]^{-1}\,,
\label{diagonalm}
\\
{\bf V_s}
&=& 
\left(
\begin{matrix}
 V_s
&&0 \\ 
 -2i \, {\rm Im} \, V_s
&&\displaystyle - V_s^*
\end{matrix}
\right)
=
[{\bf U}]^{-1}
\left(
\begin{matrix}
 V_s
&&0 \\ 
0
&&- V_s^*
\end{matrix}
\right)
[{\bf U}]^{-1}\,,
\label{diagonalVs}
\end{eqnarray}
where the corresponding propagator and matrices in the octet sector have the same structure. In \cite{Brambilla:2008cx} we have shown, through an explicit computation of the four components, that $\mathbf{\delta m_s}$ and $\mathbf{V_s}$ have, at the order considered, the structure shown in Eqs.~\eqref{diagonalm} and~\eqref{diagonalVs}. For clarity's sake we will write here only the results for the physical ``11'' component and unless otherwise specified all amplitudes are to be intended of this kind. 
\subsection{Matching the mass term $\delta m$}
\label{secVIB}
\begin{figure}[ht]
	\begin{center}
		\includegraphics{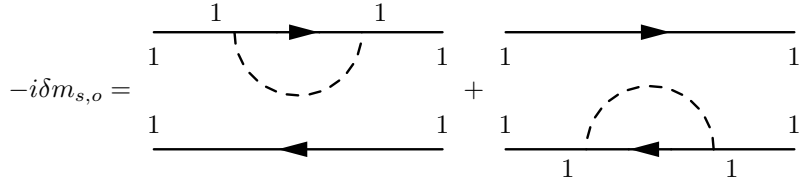}
	\end{center}
	\caption{Matching conditions for $\delta m_{s,o}$. Dashed lines represent Hard Thermal Loop-resummed longitudinal gluons. Numerical indices label the type (``1'' or ``2'') of the line or vertex.}
	\label{fig_realmass}
\end{figure}
The static quark and antiquark self-energies at one loop are shown in Fig.~\ref{fig_realmass}.
In the case considered here, the loop momentum is of order $m_D$ and the HTL resummed 
gluon propagator is used. We match in the real-time formalism the self energy diagram 
(normalized in colour space) with the second term in the series (\ref{singletstaticexpand}), 
${\bf S}^{\rm singlet}_F(p) \left[-i\, \delta {\bf m_s}\right]  {\bf S}^{\rm singlet}_F(p)$,  
obtaining:
\begin{eqnarray}
\delta  m_s &=& 
i\, (ig)^2\,C_F\, \int \frac{d^4k}{(2\pi)^4} 
\left[ \frac{i}{-k_0+i\eta} - \frac{i}{-k_0-i\eta} \right] \, [{\bf D}_{00}(k)]_{11} 
\nn\\
&=&
i\, (ig)^2 \,C_F\,\int \frac{d^4k}{(2\pi)^4} \, 2\pi \delta(-k_0)\,[{\bf D}_{00}(k)]_{11} 
= - C_F\, \als \left( m_D + i T \right)\,,
\label{deltam}
\end{eqnarray}
where $i/(-k_0+i\eta)$ and $ - i/(-k_0-i\eta)$ are the ``11'' components 
of the static quark and antiquark propagators respectively. The result follows from the static ($k_0\to 0$) limit of the ``11'' component of the longitudinal HTL propagator, which can be obtained from Eqs.~\eqref{11component} and~\eqref{prophtllong}. It reads
\begin{eqnarray}
	[{\bf D}_{00}(0,\bk)]_{11} &=&\lim_{k_0\to 0}\left[\frac{ D_{00}^{\rm R}(k) + D_{00}^{\rm A}(k)}{2}
	+ \left(\frac{1}{2} + n_{\rm B}(k_0)\right)\left(D_{00}^{\rm R}(k) - D_{00}^{\rm A}(k)\right)\right] \nn\\
	&=&\frac{i}{\bk ^2 + m_D^2}  +
\pi \,\frac{T}{|\bk|}\,\frac{m_D^2}{\left(\bk ^2 + m_D^2\right)^2}\,,
\label{D00HTLresumk0}
\end{eqnarray}
where the second term on the last line is obtained by employing the $k_0\ll T$ expansion of the Bose--Einstein distribution, as in Eq.~\eqref{boseexp}. As we observed there, the first term causes the temperature to appear at the numerator and the second cancels with the $1/2$ in round brackets.\\
For what concerns the matrix structure in Eq.~\eqref{diagonalm}, we just remark that $[\delta\mathbf{m_s}]_{12}$ vanishes because the component 
``12'' of the heavy quark and antiquark propagators vanishes, see Eqs.~(\ref{Squarkstatic}) 
and (\ref{Santiquarkstatic}).\\ 
The real part of  $\delta m_s$ corresponds to the free energy of two isolated 
static quarks in the imaginary-time formalism, which will be thoroughly dealt with in Part~\ref{part_imtime}. 
The imaginary part of $\delta m_s$ is minus twice the damping rate 
of an infinitely heavy fermion \cite{Pisarski:1993rf} and is caused by the Landau damping of the virtual longitudinal gluon in Fig.~\ref{fig_realmass}. We furthermore observe that the size of this imaginary part is of order $\als T$, whereas the real part, at size $\als m_D$, is smaller by a factor $m_D/T$. The imaginary part turns then out to be larger than our expectation: this is a product of the Bose enhancement, i.e. the first, singular term in the low momentum expansion of the Bose distribution, as in Eq.~\eqref{boseexp}. As we remarked before, it is this term that causes a factor of $T$ to appear in the numerator of the symmetric term of the $[{\bf D}_{00}(0,\bk)]_{11}$ propagator (i.e. the second term in the r.h.s in Eq.~\eqref{D00HTLresumk0}), effectively making it larger than the first term on the r.h.s.\\ 
For what concerns the octet sector, the calculation yields $\delta m_o=\delta m_s$ because, since no gluon is exchanged between the quark and antiquark, their colour state does not affect the result. Furthermore this contribution is guaranteed to be gauge invariant by being a contribution to the pole of the bound-state propagator. A covariant gauge calculation yields the same result, since the $k_0\to 0$ limit of the longitudinal propagator in these gauges \cite{Carrington:1997sq} is identical to the Coulomb gauge one in Eq.~\eqref{D00HTLresumk0}.

\subsection{Matching the static potential $V_{s,o}$}
\label{secVIC}
\begin{figure}[ht]
	\begin{center}
		\includegraphics{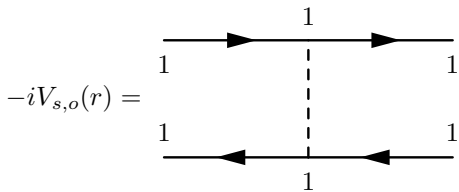}
	\end{center}
	\caption{Matching conditions for $V_{s,o}(r)$. The quark-antiquark pair is in the appropriate colour state, with the normalizations discussed in Sec.~\ref{sec_pnrqcd}.}
	\label{fig_realpot}
\end{figure}
We start with the colour singlet. The matrix elements $[{\bf V_s}]_{ij}$ are obtained by matching in real time one-gluon 
exchange diagrams that transform a colour-singlet quark-antiquark pair of type ``$i$'' into a colour-singlet quark-antiquark pair 
of type ``$j$'' with the third term in the expansion (\ref{singletstaticexpand}), 
${\bf S}^{\rm singlet}_F(p)$ $\left[-i\, {\bf V_s}\right]$  ${\bf S}^{\rm singlet}_F(p)$. We again only report here the calculation of the ``11'' component of the matrix in Eq.~\eqref{diagonalVs}, corresponding to matching the diagram of Fig.~\ref{fig_realpot} with 
$\left[{\bf S}^{\rm singlet}_F(p)\right]_{11}$ $\left[-i\, {\bf V_s}\right]_{11}  
\left[{\bf S}^{\rm singlet}_F(p)\right]_{11}$. 
We obtain  
\begin{eqnarray}
 V_s(r) &=& 
\int \frac{d^3k}{(2\pi)^3}e^{-i\bk \cdot \br}
\, i g^2\,C_F\, [{\bf D}_{00}(0,\bk)]_{11} 
\nn\\
&=&
\int \frac{d^3k}{(2\pi)^3}e^{-i\bk \cdot \br}
\left(
-C_F \frac{4\pi \als }{\bk^2+m_D^2} 
+ i \, C_F \, \frac{T}{|\bk|} \, m_D^2 \frac{4\pi^2\als }{(\bk^2+m_D^2)^2}
\right)\,,
\label{Vs11}
\end{eqnarray}
where the longitudinal HTL resummed gluon propagator, ${\bf D}_{00}(0,\bk)$,  
given in Eq.~(\ref{D00HTLresumk0}), comes from expanding in the external energy, 
which is much smaller than the typical momentum $\sim 1/r$.
The Fourier transform then yields
\begin{equation}
V_s(r) =  -C_F\,\frac{\als}{r}\,e^{-m_Dr} 
+ i2C_F\,\als\, T\,\int_0^\infty dx \,\frac{\sin(m_Dr\,x)}{m_Dr}\frac{1}{(x^2+1)^2}
\,,
\label{Vs}
\end{equation}
where the integral on the r.h.s is a monotonously decreasing function of $m_Dr$, from $1/2$ for $m_Dr=0$ to $0$ for $m_Dr\to\infty$. In the region $m_Dr\sim1$, which is the region of validity of Eq.~\eqref{Vs}, the function is still of the same order of $0.5$. The imaginary part of the potential is again due to Landau damping of the virtual gluon exchanged in Fig.~\ref{fig_realpot}.\\
The expression of $V_s(r)$, which we have derived here in the real-time formalism and in an EFT context,
agrees with the previously mentioned expression derived in the imaginary-time formalism, 
after analytical continuation of the sum of the amplitudes in Fig.~\ref{fig_wloop}, in \cite{Laine:2006ns}\footnote{A real-time derivation of that result, albeit not in an EFT context, is also available in \cite{Beraudo:2007ky}.}. It should be emphasized 
that under the condition $1/r \sim m_D$ the real part of (\ref{Vs}) is of order 
$\als m_D$, hence subleading with respect to the imaginary part, which is 
of order $\als T$: the quark-antiquark pair decays before forming the bound state, 
whose typical time scale is proportional to the inverse of the real part of the potential. The reason for this larger imaginary part can again be traced back to Bose enhancement, as in the previous Section.\\
Let us remark that the short-distance expansion of Eq.~(\ref{Vs}) would give, up to order $r^0$, 
the Coulomb potential and an $r$-independent term, $C_F\, \als \left( m_D + i T \right)$, which 
would cancel the mass term derived in (\ref{deltam}). In the following chapters we shall see other analogous cancellations.\\
For what concerns the matrix form of the potential, we refer again to \cite{Brambilla:2008cx}. As in the previous case for the mass term, we observe how the vanishing of the ``12'' component is guaranteed by the corresponding vanishing static-quark propagator.

The calculation of the octet static potential proceeds exactly like the one for its singlet counterpart. Since the potential is given by a one-gluon exchange, the only difference resides in the overall colour factor, which changes from $-C_F$ to $-C_F+C_A/2=1/(2\nc)$, causing the real part to become repulsive. The considerations on the gauge independence of $\delta m_o$ apply also for $V_o$, making it gauge invariant. It reads
\begin{equation}
	V_o(r)=\frac{1}{2\nc}\,\frac{\als}{r}\,e^{-m_Dr} 
	- \frac{i}{\nc}\,\als\, T\,\int_0^\infty dx \,\frac{\sin(m_Dr\,x)}{m_Dr}\frac{1}{(x^2+1)^2}
\,.
	\label{Vo}
\end{equation}
\subsection{Singlet static energy and width for $1/r \sim m_D$}
In Sec.~\ref{sec_pnrqcd} we discussed how the observable related to the potential is the spectrum or, in the static limit, the static energy. In this case the imaginary part of the potential gives rise to a second observable, the (static) width, defined as $\Gamma\equiv-2 \mathrm{Im}(V_s+\delta m_s)$. Furthermore the potential and the mass term are at this order free of divergences, so just adding the real parts of Eqs.~(\ref{deltam}) and (\ref{Vs}) gives the leading static quark-antiquark energy 
for $1/r \sim m_D$: 
\begin{equation}
E_s = -C_F\,\als\,m_D -C_F\,\frac{\als}{r}\,e^{-m_Dr} \,,
\label{energy2}
\end{equation}
and the imaginary parts of Eqs.~(\ref{deltam}) and (\ref{Vs}) provide the leading 
static quark-antiquark thermal decay width:
\begin{equation}
\Gamma = 2\,C_F\,\als\,T \left[
1 - \frac{2}{rm_D}\int_0^\infty dx \,\frac{\sin(m_Dr\,x)}{(x^2+1)^2} \right] \,.
\label{width2}
\end{equation}
The thermal width originates from the imaginary part of the Hard Thermal Loop gluon self energy, which has been discussed in Sec.~\ref{sec_htl}.\\ 
The static energy given by Eq. (\ref{energy2}) coincides with the leading-order result \cite{Petreczky:2005bd} 
of the so-called singlet free energy first introduced by Nadkarni \cite{Nadkarni:1986as} 
and also studied in lattice QCD (see e.g. \cite{Kaczmarek:2002mc,Kaczmarek:2004gv} 
and \cite{Brambilla:2004wf,Brambilla:2010cs} for reviews). The heavy quark-antiquark free energy, also called colour-averaged free energy, was defined by McLerran and Svetitsky in \cite{McLerran:1981pb} and will be the subject of Part~\ref{part_imtime}, where a comparison with the results of this Chapter and of Chapter~\ref{chap_rggT} will be carried out. Here we just anticipate that, while the free energy describes a thermodynamical property 
of the system and it is computed from the static quark-antiquark propagator evaluated at the imaginary 
time $1/T$ (for large temperatures this corresponds to small imaginary times), 
the static energy studied in this work describes the real-time evolution of a quark-antiquark 
pair and it is computed by evaluating the quark-antiquark propagator at infinite real times. In Part~\ref{part_imtime} we will present a case where the two energies no longer agree. Finally, the thermal decay width (\ref{width2}) coincides with the result of Ref.~\cite{Laine:2006ns}. 

\section{The $1/r \gg m_D$ case}
\label{secTrmD} 
In the $1/r \gg m_D$ case, but with $m_D$ still larger than the binding energy $E$, 
the scales $1/r$ and $m_D$ are integrated out in two subsequent matchings. We recall from our discussion in Sec.~\ref{sec_pnrqcd} that the potential receives contribution from all scales that are larger than the energy. So if $m_D$ were smaller than $E$ it would not contribute to the potential, but only to its related observable, the static energy.\\
As a first step we proceed to integrate out the inverse distance scale $1/r$. We label the resulting EFT $\mathrm{pNRQCD^\prime}_{m_D}$, where the apostrophe is to distinguish it from the previously discussed EFT. The Lagrangian reads
\begin{eqnarray}
	{\cal L}_{\mathrm{pNRQCD^\prime}_{m_D}}
&=& 
{\cal L}_{\rm HTL}
+ \int d^3r \; {\rm Tr} \,  
\Biggl\{ {\rm S}^\dagger \left[ i\partial_0 - V_s  \right] {\rm S} 
+ {\rm O}^\dagger \left[ iD_0 - V_o \right] {\rm O} \Biggr\}
\nonumber\\
&& \hspace{-2.8cm}
+ {\rm Tr} \left\{  {\rm O}^\dagger \br \cdot g\be\,{\rm S}
+ {\rm S}^\dagger \br \cdot g\be\,{\rm O} \right\} 
+ \frac{1}{2} {\rm Tr} \left\{  {\rm O}^\dagger \br\cdot g\be\, {\rm O} 
+ {\rm O}^\dagger {\rm O} \br \cdot g\be \right\}  + \dots\,,
\label{pNRQCDprimemd}	
\end{eqnarray}
where in the second line we have the order-$r$ dipole terms. Their matching coefficients are substituted by the tree level value, i.e. 1, which can be obtained by a multipole expansion of the Lagrangian~\eqref{lagnrqcdhtl} of static $\mathrm{NRQCD}_\mathrm{HTL}$, along the lines of Eqs.~\eqref{multipole0} and~\eqref{multipole1}.\\
For what concerns the gauge and light quark sectors, they are still described by the HTL Lagrangian~\eqref{laghtl}, the Debye mass having not been integrated out. For the same reason the mass terms $\delta m_s$ and $\delta m_o$ have been omitted, since the scale responsible for their generation at the tree level is $m_D$, as we have just shown in Eq.~\eqref{deltam}.  Furthermore, as we shall see, the leading contribution to the potential is in this regime the Coulomb potential, so that the power counting is that of (static) pNRQCD, i.e. $\partial_0\sim E\sim \als/r$.  \\
The matching at the scale $1/r$ can be done in close analogy with the discussion in the previous Section. However, since $\mbk \sim 1/r \gg m_D$ we expand $D_{00}(0,\bk)$ in powers of $m_D^2/\bk^2$. As in the previous Section, only the ``11'' component is considered in the matching. We furthermore need to regularize the integrals because after expansion they become infrared divergent. We employ dimensional regularization. Our conventions are $D\equiv 4-2\epsilon$, $d\equiv D-1$ and $\mu$ is the compensating scale. With those the matching yields
\begin{eqnarray}
V_s(r) &=& 
\mu^{4-D}\int \frac{d^{d}k}{(2\pi)^{d}}e^{-i\bk \cdot \br}
\left[-C_F \frac{4\pi \als }{\bk^2}\left(1 - \frac{m_D^2}{\bk^2} + \dots \right)\right.\nn\\
&&\hspace{4cm}\left.
+ i \, C_F \, \frac{T}{|\bk|} \, m_D^2 \frac{4\pi^2\als }{\bk^4}\left( 1 + \dots \right)
\right]
\nn\\
&=& -C_F\frac{\als}{r} - \frac{C_F}{2}\,\als\,r\, m_D^2 + \dots 
\nn\\
& & 
+ i \frac{C_F}{6} \, \als \, r^2 \, T \, m_D^2
\left(\frac{1}{\epsilon}+\gamma_E + \ln \pi + \ln (r\,\mu)^2 -1 \right) + \dots \;.
\label{VsTrmD1}
\end{eqnarray}
The dots stand for higher-order real and imaginary terms. In the Coulomb part,  
we have displayed only the leading term in $\als$. In the imaginary part, $\gamma_E$ is the Euler-Mascheroni constant and the divergence comes from the Fourier transform of $1/|\bk|^5$, which, in $d$ dimensions, reads \cite{Gelfand:1964}
\begin{equation}
\int\frac{d^dk}{(2\pi)^d}\frac{e^{-i\bk\cdot\br}}{\mbk^n}=
\frac{2^{-n}\pi^{-d/2}}{r^{d-n}}\frac{\Gamma\left(d/2-n/2\right)}{\Gamma\left(n/2\right)}.
\label{ftregdim}
\end{equation}
The divergence is of infrared origin and gives us a first example of an IR divergent potential at finite temperature. As we discussed in Chap.~\ref{chap_EFT}, we expect this divergence to cancel with an opposite UV divergence originating from a lower scale.\\
The octet counterpart is obtained as in the previous Section, since this is again a one-gluon exchange. In particular, the leading term is given by the (repulsive) Coulomb octet potential $V_o(r)=1/(2\nc)\als/r+\ldots$ .\\

Next, we integrate out the scale $m_D$, calling the resulting EFT  $\mathrm{pNRQCD^{\prime\prime}}_{m_D}$. The Lagrangian reads
\begin{equation}
	{\cal L}_{\mathrm{pNRQCD^{\prime\prime}}_{m_D}}
=
\int d^3r \; {\rm Tr} \,  
\Biggl\{ {\rm S}^\dagger \left[ i\partial_0 - V_s -\delta V_s \right] {\rm S} 
+ {\rm O}^\dagger \left[ iD_0 - V_o -\delta V_o\right] {\rm O} \Biggr\}+{\cal L}_{\rm gauge}
+  \dots
\label{pNRQCDsecondmd}	
\end{equation}
This EFT differs from the one of $\mathrm{pNRQCD^{\prime}}_{m_D}$, besides a general lowering of the UV cutoff, in the gauge sector, where $\lag_\mathrm{gauge}$ is again the EFT derived in \cite{Bodeker:1998hm,Bodeker:2000da}, and in the singlet and octet sectors, where  $\delta V_{s}$ and $\delta V_{o}$ are new matching coefficients that encode the contribution of the scale $m_D$. We display only the order $r^0$ terms, since we do not need to perform calculations within this theory, but only to match it to $\mathrm{pNRQCD^{\prime\prime}}_{m_D}$.\\ 
The calculation of $\delta V_s$ (we concentrate again only on the singlet) requires the matching of loop diagrams in  $\mathrm{pNRQCD^{\prime}}_{m_D}$ with the propagator of $\mathrm{pNRQCD^{\prime\prime}}_{m_D}$, since loop diagrams in the latter vanish when expanding for small external momenta.\\
At leading order (one-loop level) this corresponds to evaluating the contribution to the potential of the diagram generated 
by the singlet-octet vertex (dipole interaction) in the Lagrangian~\eqref{pNRQCDprimemd} of $\mathrm{pNRQCD^{\prime}}_{m_D}$, 
which is shown in Fig.~\ref{fig:hqself}; in it the colour singlet state emits and then reabsorbs a HTL chromoelectric gluon through an intermediate octet state. Let us call $\Sigma_s$ the amplitude of this diagram. The matching condition is then
\begin{equation}
	e^{-itV_s(r)}(1+\Sigma_s)=e^{-it(V_s+\delta V_s)}\,
	\label{matchingreal}
\end{equation}
where on the left we have the $\mathrm{pNRQCD^{\prime}}_{m_D}$ side of the matching and on the right the $\mathrm{pNRQCD^{\prime\prime}}_{m_D}$ side. The equality is to be understood for large time $t$. This then yields (see \cite{Brambilla:1999xf} for the $T=0$ case):
\begin{eqnarray}
 \delta { V_s}(r) &=& 
- i g^2 \, \frac{T_F}{N_c} \, \frac{r^2}{d}
\int_0^\infty \! dt^\prime \, e^{-it^\prime\Delta V} 
\left[\langle \be^a(t^\prime)\phi(t^\prime,0)^{\rm adj}_{ab} \be^b(0)\rangle\right]_{11}
 \nn\\
&=& 
- i g^2 \, C_F \, \frac{r^2}{d}
\mu^{4-D} \!\int \!\! \frac{d^Dk}{(2\pi)^D}
\frac{i}{-k_0 -\Delta V +i\eta}
\left[(k_0)^2 \,  D_{ii}(k) + \bk^2 \,  D_{00}(k) 
\right]\!,\nn\\
\label{EETfull11}
\end{eqnarray}
where $\frac{i}{-k_0-\Delta V+i\eta}$ is the octet propagator and $\displaystyle \Delta V\equiv V_o-V_s$ is the difference between the octet and singlet potentials of $\mathrm{pNRQCD^{\prime}}_{m_D}$. From the previous matching, we have at the leading order the difference between the Coulomb potentials, yielding $\Delta V= \frac{N_c\als}{2r}+\order{\als^2/r}$. The gluon propagators $D_{00}$ and $D_{ii}$ are the Hard Thermal Loop propagators.\\
Two scales, $m_D$ and $\Delta V$,  contribute to the amplitude~\eqref{EETfull11}. As we have stated at the beginning of this section, we assume $m_D\gg E\sim \Delta V$. Therefore, in order to single out the contribution of the Debye mass scale in this integral, corresponding to integrating over the momentum region $k_0\sim\mbk\sim m_D$, we expand the octet propagators for $\Delta V\ll k_0$. At the zeroth order this yields 
\begin{eqnarray}
 \delta { V_s}(r) &=& 
- i g^2 \, C_F \, \frac{r^2}{d}
\mu^{4-D} \int \frac{d^Dk}{(2\pi)^D}
\frac{i}{-k_0 +i\eta}
\left[(k_0)^2 \,  D_{ii}(k) + \bk^2 \,  D_{00}(k) 
\right]
\nn\\
&=&
- i g^2 \, C_F \, \frac{r^2}{d}
\mu^{4-D} \int \frac{d^Dk}{(2\pi)^D} \, \pi \delta(-k_0) \, 
\left[(k_0)^2 \,  D_{ii}(k) + \bk^2 \,  D_{00}(k) 
\right]\,,\nn\\
\label{EETmD11}
\end{eqnarray}
where the second line is justified by the even nature in $k_0$ of the propagators. After integration in $k_0$ only the longitudinal part, given by $D_{00}(0,\bk)$, contributes, the expression of which can be found in (\ref{D00HTLresumk0}).
Substituting and performing the dimensional integrals, we obtain 
\begin{equation}
\delta  V_s(r)=  \frac{C_F}{6} \, \als \, r^2m_D^3 -i \frac{C_F}{6} \, \als \, r^2 \, T \, m_D^2
\left(\frac{1}{\epsilon}-\gamma_E + \ln \pi + \ln \frac{\mu^2}{m_D^2} + \frac{5}{3} \right).
\label{VsmD}
\end{equation}
Equation (\ref{VsmD}) shows that the scale $m_D$ starts to contribute at order 
$g^2\, r^2\, m_D^3$ to the real part of the potential and at 
order  $g^2\, r^2\, T \, m_D^2$ to the imaginary one, the latter being larger than the smaller in our hierarchy by a factor $T/m_D$, which is again caused by Bose enhancement.\\
The Debye mass effectively plays the role of a gluon mass; in this sense, 
the real part of (\ref{VsmD}) agrees with a result that can be found in \cite{Brambilla:1999xf}, and which will be discussed in Chapter~\ref{chap_poincare}.
The imaginary part originates, as in the previous cases analyzed in this Chapter, from Landau damping. It corresponds, via the optical theorem, to a decay of the colour-singlet bound state to a colour-octet, through the scattering of the virtual gluon with the light constituents of the medium. \\
The imaginary part furthermore shows an ultraviolet divergence. This cancels against 
the infrared divergence of Eq.~(\ref{VsTrmD1}), as it was expected. Summing $V_s$, as given there, with $\delta V_s$ gives the full potential appearing in the Lagrangian~\eqref{pNRQCDsecondmd}. It reads 
\begin{eqnarray}
V_s(r)+\delta V_s(r) 
&=& -C_F\frac{\als}{r} - \frac{C_F}{2}\,\als\,r\, m_D^2 
+ \frac{C_F}{6} \, \als \, r^2m_D^3 + \dots 
\nn\\
& & 
-i \frac{C_F}{6} \, \als \, r^2 \, T \, m_D^2
\left(-2\gamma_E  - \ln (r m_D)^2 + \frac{8}{3} \right) + \dots \;.
\label{VsTrmD2}
\end{eqnarray}
We see that in the sum the divergences of Eqs.~(\ref{VsTrmD1}) and (\ref{VsmD}) cancel each 
other providing a finite physical result. The term $\displaystyle \frac{C_F}{6} \, \als \, r^2m_D^3$
in the real part is suppressed by a factor $r m_D$ with respect to $\displaystyle - \frac{C_F}{2}\,\als\,r\, m_D^2 $
and will be neglected in the following. Note the appearance of the logarithm 
$\ln (r m_D)^2$: it signals that divergences have been canceled 
when integrating out the scales $1/r$ and $m_D$. The real and imaginary parts of Eq.~(\ref{VsTrmD2}) can be also obtained by expansion 
in $r m_D$ of Eq.~(\ref{energy2}) and $-\Gamma/2$, as defined in Eq.~(\ref{width2}), respectively.
\subsection{Singlet static energy for $1/r \gg m_D$}
The real part of Eq.~(\ref{VsTrmD2}) provides the static quark-antiquark energy for $1/r \gg m_D$, whose 
leading thermal contribution is 
\begin{equation}
\delta E_s = - \frac{C_F}{2}\,\als\,r\, m_D^2 \,,
\label{energy3}
\end{equation}
and minus twice the imaginary part of Eq.~(\ref{VsTrmD2}) provides the static quark-antiquark thermal decay width
\begin{equation}
\Gamma =  \frac{C_F}{3} \, \als \, r^2 \, T \, m_D^2
\left(-2\gamma_E  - \ln (r m_D)^2 + \frac{8}{3} \right)\,.
\label{width3}
\end{equation}
\subsection{The dissociation temperature}
\label{sub_tdiss}
In \cite{Laine:2008cf} (see also \cite{Escobedo:2008sy} for the Abelian case), it was pointed out that the results of this section, or their Abelian counterparts, allow one to define and estimate a qualitative \emph{dissociation temperature} $T_d$ as the temperature where the width Eq.~\eqref{width3} becomes of the same order of the leading static energy, which in this regime is given by the Coulomb potential and hence of order $\als/r$. In order for the result to apply to physical quarkonium, rather than to static bound states, let us instate a kinetic term $\bp^2/m$ in the Lagrangians~\eqref{pNRQCDprimemd} and \eqref{pNRQCDsecondmd} and assume the power counting of standard, weakly-coupled pNRQCD (see Sec.~\ref{sub_weakpnrqcd}), i.e. $1/r\sim m\als$, which is justified by the potential being Coulombic at leading order in pNRQCD$_{m_D}^{\prime\prime}$. This then implies, equating Eq.~\eqref{width3} with the Coulomb potential and neglecting the logarithm in the former, $T_d\sim mg^{4/3}$.\\
Quantitative results are available for hydrogen in \cite{Escobedo:2008sy} and for muonic hydrogen and the $\Upsilon(1S)$ in \cite{Escobedo:2010tu}. 
\section{Conclusions}
\label{sec_concl_Tggr}
In this Chapter we have studied the real-time evolution of a static quark-antiquark pair 
in a medium of gluons and light quarks characterized by a temperature $T$ much larger than the inverse distance $1/r$.
We have addressed the problem of defining and deriving the potential 
between the two static sources, and of calculating their energy and 
thermal decay width. In the different ranges of temperature considered, 
we have set up and worked out a suitable sequence of effective field theories.
Our framework has been very close to the modern EFT treatment of non-relativistic and static bound states 
at zero temperature introduced in Chap.~\ref{chap_EFT}, but complicated by the existence of the thermal 
scales $T$ and $m_D$. We have assumed that all the energy scales are perturbative and worked in a strict 
weak-coupling framework. This had two consequences: first, we could exploit the hierarchy
$T \gg m_D$, second, the potential that we obtained is valid in the short range.
We recall that, in this EFT framework, the potential is the $r$-dependent matching coefficient that appears 
in front of the four-fermion operator that destroys and creates the bound state, after having 
integrated out all scales above the bound-state energy. Higher-order operators give lower 
energy contributions, entering into the computation of physical observables, but not 
in the Schr\"odinger equation that governs the motion of the bound state and hence are not 
of a potential type.

Our results pave the way for a systematic treatment of non-relativistic bound states in a thermal medium,
in an EFT framework and in real-time formalism. We have indeed devoted several parts of this Chapter and of the previous one to set up a proper real-time formalism for static sources.
The main outcome of this more formal aspect is in Eq.~(\ref{singletstaticexpand}), 
which expresses the real-time quark-antiquark propagator as an infinite sum of free propagators 
and potential or mass-shift insertions. In all the considered dynamical regimes, the structure 
of the potential is such to satisfy this equation, see Eqs.~(\ref{diagonalm}) 
and (\ref{diagonalVs}).

We have considered a wide range of temperatures and provided 
the leading thermal effects to the potential. The results may be summarized in the 
following way. 
\begin{enumerate}
\item If $T\gg1/r\simg m_D$ the static 
potential is given by Eq.~(\ref{Vs}): this result agrees with the earlier finding of \cite{Laine:2006ns}, but is now obtained in a modern and rigorous way as a matching coefficient of the EFT named pNRQCD$_{m_D}$, whose Lagrangian is given in Eq.~\eqref{pNRQCDmd}.
\item If the temperature is larger than $1/r$ but $m_D$ is smaller than $1/r$, the static potential is 
given by Eq.~(\ref{VsTrmD2}), as a matching coefficient of pNRQCD$^{\prime\prime}_{m_D}$. A dissociation temperature $T_d\sim mg^{4/3}$ can be estimated by imposing that the real and imaginary parts of the potential given in Eq.~(\ref{VsTrmD2}) be of the same order.\\
If $m_D$ is also smaller than or of the same order as $\Delta V$ 
then the potential is given by Eq.~(\ref{VsTrmD1}) only, the Debye mass cannot be integrated out and divergences cancel in physical observables 
against loop corrections from this scale. 
\end{enumerate}

We furthermore remark that equations (\ref{VsTrmD2}) and (\ref{Vs}) are finite because, in the kinematical regions of validity, 
they provide the leading thermal correction to the static energy
and the decay width (see Eqs.~(\ref{energy3}),(\ref{width3}), (\ref{energy2}) and (\ref{width2})). 
In the temperature ranges considered in this Chapter the thermal width comes from the imaginary part of the gluon self energy and is thus due to Landau damping. Moreover this thermal width is in both cases, i.e. Eqs.~\eqref{width2} and \eqref{width3}, larger by a factor $T/m_D$ than the corresponding thermal contributions to the static energies \eqref{energy2} and \eqref{energy3}. We have traced this fact back to the infrared behaviour of the Bose--Einstein distribution, which enters only in the symmetric part of the gluon propagator, that in turn is the one responsible for the imaginary parts. \\
Such large imaginary parts are then extremely important and alter the hypothesis of Matsui and Satz \cite{Matsui:1986dk}, which saw colour screening as the mechanism responsible for the dissociation of the bound state, whereas the real-time potential indicates that Landau damping in the imaginary parts is actually much stronger. Indeed, attempts at phenomenological analyses of the potential \eqref{Vs}, either by a numerical solution of the Schr\"odinger equation with this potential, as in \cite{Laine:2007gj,Burnier:2007qm}, or by adding its imaginary part only to a real part obtained from lattice inputs, as done recently in \cite{Miao:2010tk},  agree on a disappearance of charmonium and bottomonium states at temperatures lower than those obtained by standard potential models, which in general fail to take this imaginary part into account. However, for what concerns the ground states of bottomonium, it does not seem possible that the temperatures in current  heavy ion collision experiments may be larger than the typical momentum transfer or inverse distance, hence the results of this Chapter should not apply as they are to these systems. In the next Chapter we will indeed study the case where the temperature is smaller that the inverse distance and highlight its relevance for bottomonium phenomenology.\\

There are many possible developments of this work. First, the construction of a full EFT for non-relativistic bound states at finite temperature
requires to be completed in this regime. We have mostly focused  on the quark-antiquark 
colour-singlet state, but a complete identification and study of all relevant degrees 
of freedom that appear once the thermal energy scales have been integrated out 
is still to be done. This may require the usage of the EFT that includes  
the dynamics of gauge fields below the scale $m_D$ \cite{Bodeker:1998hm,Bodeker:2000da}.
Second, in the EFT framework presented here and in the temperature regime we have analyzed, the study of quark-antiquark states 
at large but finite mass, i.e. actual quarkonium in a thermal medium, 
should be addressed. As argued along this Chapter, the static limit provides the first piece of a  
$1/m$ expansion; higher-order corrections may be systematically implemented in the framework of 
NRQCD and pNRQCD, similarly to what will be done in the next Chapter for smaller temperatures. Finally, although the short-distance analysis performed in this work 
may provide a valuable tool for studying the thermal dissociation of the lowest 
quarkonium resonances, the inclusion  in the analysis of the non-perturbative scale $\lQ$ 
may become necessary for studying excited states.

		\chapter{Bound states for $m\als \gg T$}
\label{chap_rggT}
In this Chapter we will study the regime where the temperature is smaller than the typical momentum transfer scale $mv$. Most of the results will be obtained in the finite mass case, including the contributions of a large but finite quark mass, which is the case in physical quarkonia. Under the perturbative assumption $mv\sim m\als$, our results are expected to be applicable to the ground states of bottomonium ($\Upsilon(1S)$, $\eta_b$) and, to a lesser extent, of charmonium ($J/\psi$ and $\eta_c$).\\
For most of this Chapter we assume for definitiveness the following hierarchy between the thermodynamical 
and the non-relativistic scales:
\begin{equation}
m \gg m\als \gg T \gg m\als^2 \gg m_D.
\label{hierarchy}
\end{equation}
With this choice, the thermal bath affects the Coulombic bound state as a small perturbation, 
yet modifying the Coulomb potential. We remark that this temperature is below the dissociation temperature $T_d$, 
which is of order $mg^{4/3}$, as derived in \cite{Escobedo:2008sy,Laine:2008cf} and discussed in Sec.~\ref{sub_tdiss}. Moreover, this may indeed 
correspond to the situation of interest in present day colliders. For the ground states of bottomonium one has $m_bv\sim m_b\als\approx 1.5$ GeV, which is certainly larger than the highest temperatures reachable at colliders. At the LHC these are estimated to be of the order of 3--4 $T_c$, so one could have $m_b\approx 5\;\text{GeV}>m_b\als\approx1.5\;\text{GeV}>\pi T\approx 1\;\text{GeV}>m\als^2\approx 0.5\;\text{GeV}\simg m_D$.\\
As a consequence of~\eqref{hierarchy}, in the weak-coupling regime, we have
that $mg^3\gg T \gg mg^4$, corresponding to $m g^4\gg m_D\gg mg^5$. 
We furthermore assume that $\Lambda_{\rm QCD}$, the QCD scale, is smaller 
than $m_D$ (although results that do not involve a weak-coupling expansion at the scale 
$m_D$, which are all the results presented before Sec.~\ref{secmD}, 
are valid also for $m_D \sim \Lambda_{\rm QCD}$). 
In the Abelian case, a number of different inequalities has been addressed in \cite{Escobedo:2010tu}.\\
We will concentrate on the energy levels and decay widths. In the hiearchy~\eqref{hierarchy}, the former are given in a first approximation by $T=0$ pNRQCD, while the medium causes a perturbation to the spectrum and the appearance of a thermal width. Both observables will be computed with an accuracy 
of order $m\als^5$. In order to be definite, we will further assume $(m_D/E)^4 \ll g$, in this way keeping small the number of required  
corrections suppressed by powers of $m_D/E$.\\ 
A limitation for the practical application of our final
results to actual bottomonium and charmonium systems comes from the
fact that we use perturbation theory at the ultrasoft scale
$m\als^2$. Still, we expect them to be relevant for the ground states
of bottomonium and, to a lesser extent, charmonium.  Some intermediate
expressions, for which perturbation theory is only used at the scale
$T\gg m\als^2$ may have a wider range of applicability. We also assume a vanishing charm quark 
mass in the bottomonium case (effects of a non-vanishing mass are discussed in \cite{Escobedo:2010tu}).\\
This Chapter is organized in the following way. In Sec.~\ref{secm} the scales $m$ and $m\als$ are integrated out. This leads to NRQCD and pNRQCD, which have been presented in Chap.~\ref{chap_EFT}; the Section deals mostly with some subtleties related to the bound-state propagator in the real-time formalism. In Sec.~\ref{secT}, we integrate out the scale $T$ and calculate its contributions to the spectrum and the width, in Sec.~\ref{secE}, those 
coming from the scale $E$ and, finally, in Sec.~\ref{secmD}, those coming from the scale $m_D$.
In Sec.~\ref{secconclusions}, we summarize our results giving the thermal energy shifts and 
widths up to order $m\als^5$. Phenomenological implications of the results are also discussed.\\
The results of this Chapter have been published mostly in \cite{Brambilla:2010vq}, whose exposition we loosely follow. Some of the reported results had already been published, in the static limit only, in \cite{Brambilla:2008cx}. The leading static part of the results obtained here at the scale $T$ is re-derived in App.~\ref{app_realtime}, where the same results are obtained directly from QCD, i.e. without the EFT framework, highlighting the advantages introduced by the latter.  Some technical details of the calculations can be found as well in App.~\ref{app_realtime}.
\section{Integrating out the scales $m$ and $m\als$}
\label{secm}
As we mentioned before, the first scale to be integrated out from QCD is the heavy quark mass $m$. 
In the matching procedure, smaller scales are expanded. 
Thus, at this stage, the presence of the thermal
scales does not affect the matching of the Lagrangian, which is the 
Lagrangian~\eqref{lagrnrqcd} of non-relativistic QCD (NRQCD) (see Sec.~\ref{sec_nrqcd}).\\
The next scale to be integrated out is the inverse of the typical 
distance of the heavy quark and antiquark, which is of order $m\als$.
According to ~\eqref{hierarchy}, it is larger than the temperature. 
We are thus allowed to integrate out $m\als$ from NRQCD
setting to zero all thermodynamical scales. Furthermore, under the assumption that
$m\als \gg \Lambda_\mathrm{QCD}$, this integration can be carried out in
perturbation theory order by order in $\als$, yielding pNRQCD in the weak-coupling regime, which has been introduced in Sec.~\ref{sub_weakpnrqcd}. For our purposes and aimed accuracy, the Lagrangian shown in Eq.~\eqref{lagrpnrqcd} is sufficient, no other terms in the $1/m$ and multipole expansions shall be needed.\\
As we discussed in Sec.~\ref{sub_weakpnrqcd}, the discrete spectrum of the pNRQCD singlet field is customarily obtained by solving the Schr\"odinger equation (the EOM for the singlet field at the zeroth-order in the multipole expansion) with the leading Coulomb Hamiltonian $h^{(0)}_s$ in Eq.~\eqref{leadingham}, obtaining at order $m\als^2$ the QCD Bohr levels shown in Eq.~\eqref{coulomblevels}. Radiative and $1/m^{a\ge1}$ contributions to the singlet Hamiltonian are treated in quantum-mechanical perturbation theory, giving the $\order{m\als^3,m\als^4,m\als^5,\ldots}$ contributions to the spectrum from the scale $mv\sim m\als$. We remark again that at order $m\als^5$ this contribution is IR divergent due to the corresponding divergences in the potentials. While at $T=0$ this divergence cancels against a UV one from the scale $E\sim m\als^2$, we now have the presence of the temperature between $m\als$ and $m\als^2$. As we shall see, this will give a different pattern of cancellation.\\ 
We close this Section by analyzing the singlet and octet propagators in the real-time formalism, which will be very important in the following. We have shown in Eqs.~\eqref{SsingletstaticrT} and~\eqref{SoctetstaticrT} the free, static propagators, and how the potential can be resummed, with the help of the diagonal form introduced by the matrix $\mathbf{U}$ in Eq.~\eqref{matrixU0}. In this case we need to resum the leading, Coulombic Hamiltonians $h^{(0)}_s$ and $h^{(0)}_o$, as given in Eqs.~\eqref{leadingham} and~\eqref{leadinghamoctet}, because in the power counting of pNRQCD they are of size $m\als^2$. In matrix form  $h^{(0)}_s$ becomes
\begin{equation}
	\mathbf{h_s}^{(0)}=\left(\begin{matrix}
\hspace{0mm}h^{(0)}_s
&&0 \\ 
\hspace{0mm} 0 
&&-h^{(0)}_s
\end{matrix}
\right)=\mathbf{U}^{-1}\left(\begin{matrix}
\hspace{0mm}h^{(0)}_s
&&0 \\ 
\hspace{0mm}  0
&&-h^{(0)}_s
\end{matrix}
\right)\mathbf{U}^{-1}\,,
	\label{hammatrix}
\end{equation}
and the octet counterpart is obtained by replacing $h^{(0)}_s$ with $h^{(0)}_o$. The last equality, together with the diagonal form of the free propagators, allows to resum the geometric series of insertions to all orders, i.e.
\begin{equation}
{\bf S}^{\rm singlet}(E) = 
\left(
\begin{matrix}
\displaystyle\frac{i}{E +i\eta}  
&&0 \\ 
2\pi\delta(E) 
&&\displaystyle\frac{-i}{E -i\eta}
\end{matrix}
\right)
\sum_{n=0}^\infty 
\left[
\left(-i {\bf h_s}^{(0)}\right)
\left(
\begin{matrix}
\displaystyle\frac{i}{E +i\eta}  
&&0 \\ 
2\pi\delta(E) 
&&\displaystyle\frac{-i}{E -i\eta}
\end{matrix}
\right)
\right]^n,
\label{sumSinglet}
\end{equation}
where we are labeling the energy flowing in the propagator with $E$, in keeping with the usual notation at zero temperature (see App.~\ref{sub_feyn_pnrqcd}). The summation yields
\begin{equation}
{\bf S}^{\rm singlet}(E) = 
\left(
\begin{matrix}
	\displaystyle\frac{i}{E-h^{(0)}_s +i\eta}  
&&0 \\ 
2\pi\delta\left(E-h^{(0)}_s\right) 
&&\displaystyle\frac{-i}{E-h^{(0)}_s -i\eta}
\end{matrix}
\right),
\label{sumSingletexp}
\end{equation}
and analogously for the octet.\\
We notice that the ``12'' component of the quark-antiquark propagator keeps vanishing\footnote{This is true up to exponentially suppressed contributions. If we were to follow the rigorous method of Sec.~\ref{sec_static}, that is, keeping the mass term explicitly, we would end up with a thermal part, identical for all four components and proportional to $n_\mathrm{B}(\vert E \vert)\delta(E-2m-h^{(0)}_s)\sim n_\mathrm{B}(2m)$. Since $m\gg T$ this term is exponentially suppressed and its contribution is not considered here.} and the unphysical ``2'' component decouples again. It is thus conventient to drop the real-time formalism indices and
write only the ``11'' component of the propagator. For the rest of the Chapter, all amplitudes will be intended as the ``11'' components of the
real-time matrices unless otherwise specified.
In particular, for what concerns the singlet propagator, we thus have
\begin{equation}
	S^{\rm singlet}(E) =\frac{i}{E -h_s^{(0)}+i\eta},
	\label{hqprops}
\end{equation}
and similarly the octet propagator is
\begin{equation}
S^{\rm octet}(E)_{ab} =\frac{i\delta_{ab}}{E -h_o^{(0)}+i\eta},
\label{hqpropo}
\end{equation}
which are the same forms one encounters at $T=0$, as shown in App.~\ref{sub_feyn_pnrqcd}. We finally recall that subleading term in the singlet and octet Hamiltonians~\eqref{sinoctham}, being smaller than $m\als^2$, are treated as interaction terms, i.e. insertions in the propagator in the following way
\begin{eqnarray}
\hspace{-8mm} 
\frac{i}{E -h_{s,o} +i\eta} &=& \frac{i}{E -h_{s,o}^{(0)}+i\eta} 
\nonumber\\
&& \hspace{-10mm}
+ \frac{i}{E -h_{s,o}^{(0)}+i\eta} 
\left[ \frac{{\bf P}^2}{4m} + \frac{V^{(1)}_{s}}{m} + \frac{V^{(2)}_{s}}{m^2} +\ldots \right]
\frac{1}{E -h_{s,o}^{(0)}+i\eta} +\ldots\,.
\label{hqpropsozero}
\end{eqnarray}
\section{Integrating out the temperature}
\label{secT}
In this section, we proceed to integrate out modes of energy and
momentum of the order of the temperature $T$. This amounts to
modifying pNRQCD into a new EFT where only modes with energies
and momenta lower than $T$ are dynamical. We label the new EFT 
$\textrm{pNRQCD}_\mathrm{HTL}$ \cite{Vairo:2009ih}. The EFT can be used for $m\als\gg T\gg E,m_D$ 
no matter what the relation between $E$ and $m_D$ is. Its Lagrangian will get additional
contributions with respect to pNRQCD. For our purposes, we are
interested in the modifications to the singlet sector, corresponding
to a thermal correction $\delta V_s$ to the singlet potential, and to the Yang--Mills
sector, amounting to the Hard Thermal Loop (HTL) Lagrangian ${\cal L}_{\rm HTL}$, given in Eq.~\eqref{laghtl}. The  $\textrm{pNRQCD}_\mathrm{HTL}$ Lagrangian reads
\begin{eqnarray}
{\cal L}_{\textrm{pNRQCD}_\mathrm{HTL}} &=& 
{\cal L}_{\rm HTL}
+ \int d^3r \; {\rm Tr} \,  
\Biggl\{ {\rm S}^\dagger \left[ i\partial_0 - h_s - \delta V_s \right] {\rm S} 
+ {\rm O}^\dagger \left[ iD_0 -h_o - \delta V_o \right] {\rm O} \Biggr\}
\nonumber\\
&& \hspace{-1.5cm}
+ {\rm Tr} \left\{  {\rm O}^\dagger \br \cdot g\be\,{\rm S}
+ {\rm S}^\dagger \br \cdot g\be\,{\rm O} \right\} 
+ \frac{1}{2} {\rm Tr} \left\{  {\rm O}^\dagger \br\cdot g\be\, {\rm O} 
+ {\rm O}^\dagger {\rm O} \br \cdot g\be \right\}  + \dots\,,
\label{pNRQCDHTL}	
\end{eqnarray}
where we have set to one the matching coefficients of the dipole terms, whose 
quantum corrections are beyond the accuracy of the present calculation.\\
We calculate the correction $\delta V_s$ to the singlet potential. 
The leading thermal correction is again due to the dipole
vertices ${\rm O}^\dagger \br \cdot g\be\,{\rm S} + {\rm S}^\dagger
\br \cdot g\be\,{\rm O}$ in the pNRQCD Lagrangian
\eqref{lagrpnrqcd}. These terms induce the diagram depicted in
Fig.~\ref{fig:hqself}, where a colour-singlet state emits and
reabsorbs a chromoelectric gluon through the dipole vertex and an
intermediate colour-octet state. The amplitude reads (see
\cite{Brambilla:1999xj,Kniehl:1999ud,Kniehl:2002br} for the $T=0$
case and Eq.~\eqref{EETmD11} for the same diagram evaluated at the scale $m_D$ in the static limit)
\begin{equation}
	\Sigma_s^{(1\,\text{loop})}(E)=- i g^2\, C_F \, \frac{r^i}{d}
\mu^{4-D} \int \frac{d^Dk}{(2\pi)^D}
\frac{i}{E-h^{(0)}_o-k_0 +i\eta}\left[k_0^2 \,  D_{ii}(k) +  \bk^2 \,  D_{00}(k)
\right]r^i\,,
\label{defleading}
\end{equation}
where $E$ is the energy of the singlet and $k$ the four-momentum of the gluon. We remark that $h_o^{(0)}$ and $r^i$ do not commute, due to the operator $\bp^2=-{\boldsymbol \nabla}^2_\br$ appearing in the former. We furthermore recall that 
this expression corresponds to the ``11'' component in the real-time formalism. The pNRQCD matching coefficient of the dipole term,  $V_A=1+\order{\als^2}$, has been substituted with its leading value.   
Integrals over momenta are regularized in dimensional
regularization, again with $D\equiv4-2\epsilon$, $d\equiv D-1$ and $\mu$ being the subtraction point. 
In Coulomb gauge, with the free propagators given in
Eqs.~\eqref{longgluonpropreal} and~\eqref{transgluonpropreal}, the contribution of the
longitudinal gluon vanishes in dimensional regularization, whereas
that of the transverse gluon can be divided into a vacuum and a
thermal part:
\begin{eqnarray}
\nonumber\Sigma_s^{(1\,\text{loop})}(E)&=& -ig^2C_F\frac{d-1}{d} r^i\mu^{4-D}
\int\frac{\,d^Dk}{(2\pi)^D}\frac{i}{E-h^{(0)}_o-k_0+i\eta}k_0^2\left[\frac{i}{k_0^2-\bk^2+i\eta}\right.\\
&&\hspace{5.5cm}+2\pi \delta\left(k_0^2-\bk^2\right)n_\mathrm{B}\left(\vert k_0\vert\right)\bigg]r^i\,;
\label{transverseleading}
\end{eqnarray}
the first term in the square brackets is the vacuum part and the second term is the thermal part. 
The expression depends on the scales $T$ and $E$. In order to single out
the contribution from the scale $T$, which comes from the momentum regions $k_0\sim T$ 
and $\mbk\sim T$, we recall that $T\gg\left(E-h^{(0)}_o\right)$ and expand the octet propagator as
\begin{equation}
\frac{i}{E-h^{(0)}_o-k_0+i\eta}=
\frac{i}{-k_0+i\eta}-i\frac{E-h^{(0)}_o}{(-k_0+i\eta)^2}
+i\frac{\left(E-h^{(0)}_o\right)^2}{(-k_0+i\eta)^3}-i\frac{\left(E-h^{(0)}_o\right)^3}{(-k_0+i\eta)^4}+\ldots\,.
\label{octetexpand}
\end{equation}
The contribution of the vacuum part of the propagator is
scaleless for all the terms of the expansion and thus it vanishes. 
Conversely, in the thermal part, we have the Bose--Einstein
distribution giving a scale to the integration.\\
The zeroth-order term in the expansion~\eqref{octetexpand}, which would contribute at order $\als T^3 r^2$, gives however a vanishing
integral
\begin{equation}
	\Sigma_s^{\text{(zeroth)}}(E)= -ig^2C_F\frac{d-1}{d} r^2\mu^{4-D}
\int\frac{\,d^Dk}{(2\pi)^D}\frac{i}{-k_0+i\eta}k_0^2\,2\pi \delta\left(k_0^2-\bk^2\right)n_\mathrm{B}\left(\vert k_0\vert\right)=0.
	\label{vanishzero}
\end{equation}
The following terms instead do contribute to the potential. The linear and the cubic terms in $E-h^{(0)}_o$, i.e., after integration over $k_0$ 
\begin{equation}
	\Sigma_s^{\text{(linear)}}(E)=-g^2C_F\frac{d-1}{d}r^i\left(E-h^{(0)}_o\right)r^i\,\mu^{4-D}\int\frac{\,d^{d}k}{(2\pi)^{d}}\frac{n_\mathrm{B}(\mbk)}{\mbk}, 
\label{deflinear}
\end{equation}
and 
\begin{equation}
	\Sigma_s^{\text{(cubic)}}(E)=-g^2C_F\frac{d-1}{d}r^i\left(E-h^{(0)}_o\right)^3r^i\,\mu^{4-D}\int\frac{\,d^{d}k}{(2\pi)^{d}}\frac{n_\mathrm{B}(\mbk)}{\mbk^3},
\label{defcubic}
\end{equation}
can be shown to contribute to the real part of the potential.
Since in our counting~\eqref{deflinear} behaves as $mg^{8} \gg \als T^2 Er^2\gg mg^{10}$ 
and~\eqref{defcubic} as $\als E^3r^2\sim mg^{10}$, further terms
in the ${E}/{T}$ expansion~\eqref{octetexpand} are not needed. Similarly, insertions in the octet propagators of subleading terms of the octet Hamiltonian, as in~\eqref{hqpropsozero}, would result in a contribution smaller than $m\als^5$, since they are suppressed by at least a factor of $\als$ with respect to $h^{(0)}_o$.\\ 
Finally, the square term in the expansion, which would give an imaginary contribution to the potential, 
vanishes in dimensional regularization:
\begin{eqnarray}
	\nn\Sigma_s^{\text{(square)}}(E)&=&\frac{g^2C_F}{2}\frac{d-1}{d}r^i\left(E-h^{(0)}_o\right)^2r^i\,\mu^{4-D}
\int\frac{\,d^{d}k}{(2\pi)^{d}}\,\,n_\mathrm{B}(\mbk)\left[\frac{\mbk}{(-\mbk+i\eta)^3}\right.\\
&&\hspace{7cm}\left.+\frac{\mbk}{(\mbk+i\eta)^3}\right]
=0.
\label{defimag}
\end{eqnarray}
\subsection{The linear contribution}
\label{sub_linear}
We now evaluate the linear term defined in Eq.~\eqref{deflinear}. 
The integration yields
\begin{equation}
\Sigma_s^{(\text{linear})}(E)=-\frac{2\pi}{9}C_F\als T^2\,r^i\left(E-h^{(0)}_o\right)r^i\,.
\end{equation}
The singlet propagator in $\textrm{pNRQCD}_\mathrm{HTL}$ reads 
\begin{eqnarray}
&& \hspace{-5mm}
Z_s^{1/2} \frac{i}{E -h_s -\delta V_s  + i\eta} Z_s^{1/2\,\dagger} 
= \frac{i}{E -h_s + i\eta} 
\nonumber\\
&&
\hspace{12mm}
+ \frac{i}{E -h_s + i\eta} \delta V_s \frac{1}{E -h_s + i\eta} 
+ \left\{ \delta Z_s , \frac{i}{E -h_s + i\eta} \right\} + \dots\,.
\label{matchingHTL}
\end{eqnarray}
There is no self-energy contribution in~\eqref{matchingHTL}, 
because this would correspond to a scaleless integral eventually irrelevant 
(e.g. in dimensional regularization it would vanish).
$Z_s^{1/2} = 1 + \delta Z_s$ is the normalization of the singlet field in $\textrm{pNRQCD}_\mathrm{HTL}$; 
$\delta Z_s$ amounts then to a correction to the wavefunction. It is at least of order $\als$ and a function of $\br$, which implies that it does not 
commute with $h^{(0)}_s$. At our accuracy, $\delta Z_s$ is real.
Matching the singlet propagator in pNRQCD with the singlet propagator in $\textrm{pNRQCD}_\mathrm{HTL}$ then amounts to equating
\begin{eqnarray}
	&&\frac{i}{E-h^{(0)}_s+i\eta} + \frac{i}{E-h^{(0)}_s+i\eta}\Sigma_s^{\text{(linear)}}(E)\frac{1}{E-h^{(0)}_s+i\eta}
=\frac{i}{E -h^{(0)}_s + i\eta} 
\nonumber\\
&&
\hspace{-1cm}
+ \frac{i}{E -h^{(0)}_s + i\eta} \delta V^{(\text{linear})}_s \frac{1}{E -h^{(0)}_s + i\eta} 
+ \left\{ \delta Z_s^{(\text{linear})} , \frac{i}{E -h^{(0)}_s + i\eta} \right\} + \dots\,,
\label{matchcond}
\end{eqnarray}
where the left-hand part of the equality corresponds to the pNRQCD part of the 
matching and the right-hand side to the  $\textrm{pNRQCD}_\mathrm{HTL}$ part. We have expanded around the Coulomb Hamiltonian there as well.\\
In order to separate the contribution to $\delta V^{(\text{linear})}_s$ from that to $\delta Z_s^{(\text{linear})}$, we rewrite $E-h^{(0)}_o$ as 
\begin{equation}
	E-h^{(0)}_o=E-h^{(0)}_s-\left(h^{(0)}_o-h^{(0)}_s\right),
	\label{commute1}
\end{equation}
 where
$h^{(0)}_o-h^{(0)}_s$ is given by the difference between the octet and singlet Coulomb potentials:
\begin{equation} 
h^{(0)}_o-h^{(0)}_s\equiv\Delta V=\frac{\nc}{2}\frac{\als}{ r}\,.
\label{deltaV}
\end{equation}
Hence $r^i\left(E-h^{(0)}_o\right)r^i$ simplifies to $r^i\left(E-h^{(0)}_s\right)r^i- \nc\als r/2$; the second term is easily identified as contributing 
to $\delta V^{(\text{linear})}_s$, whereas the first term can be rewritten as
\begin{equation}
\label{commutexample}
r^i\left(E-h^{(0)}_s\right)r^i=
\frac{1}{2}\left(\left[\left[r^i,E-h^{(0)}_s\right],r^i\right]+\left\{r^2,\left(E-h^{(0)}_s\right)\right\}\right).
\end{equation}
The term $\left\{r^2,\left(E-h^{(0)}_s\right)\right\}$, when plugged in Eq.~\eqref{matchcond}, contributes to the normalization of the wave function  $\delta Z_s^{(\text{linear})}$, whereas the other contributes to the potential. We then obtain 
\begin{eqnarray}
\label{deltavslinear}
\delta V_s^{\rm (linear)}&=&\frac{\pi}{9}N_c C_F \als^2 T^2r+\frac{2\pi}{3m}C_F\als T^2\,,\\
\delta Z_s^{\rm (linear)}&=&-\frac{\pi}{9} C_F \als T^2r^2\,.
\label{deltazslinear}
\end{eqnarray}
The first term in Eq.~\eqref{deltavslinear} is the contribution of $\Delta V$ and was first obtained in \cite{Brambilla:2008cx}, where we considered the static limit only. The second term is the contribution of the kinetic term; a similar term appears in the
Abelian case of Ref.~\cite{Escobedo:2008sy}. We remark again that, since $r\sim1/(m\als)$, both terms are of the same size $mg^{8} \gg \delta V_s^{\rm (linear)}\gg mg^{10}$. Hence, in our power counting it happens that a static contribution and a $1/m$ contribution share the same size, thus highlighting the importance of the computation of finite-mass corrections, a feature which is in general missing from potential models, that do rely on static terms only.\\
As a final observation, let us point out that $\delta Z_s^{\rm (linear)}$ is not needed for the current calculation. We have computed it for future convenience in Chap.~\ref{chap_poincare}.

Using first-order quantum-mechanical perturbation theory and the
expectation values $\left\langle r \right\rangle_{n,l}$ on the eigenstates of the
Coulomb potential ($n$ and $l$ stand for the principal 
and angular momentum quantum numbers respectively, see, for instance, \cite{Titard:1993nn}) 
we obtain the following correction to the Coulomb energy levels
\begin{equation}
\delta E_{n,l}^{\rm (linear)}=\frac{\pi}{9}N_c C_F \als^2 T^2 \frac{a_0}{2}[3n^2-l(l+1)]+\frac{2\pi}{3m}C_F\als T^2\,.
\label{linear}
\end{equation}
where we recall that $a_0=2/(m\cf\als)$ is the QCD Bohr radius.\\
\subsection{The cubic contribution}
We now move to the cubic term, as defined in Eq.~\eqref{defcubic}. We have
\begin{eqnarray}
	\Sigma_s^{\text{(cubic)}}(E)&=&-g^2C_F\frac{d-1}{d}r^i\left(E-h^{(0)}_o\right)^3r^i\,\mu^{4-D}
\int\frac{\,d^{d}k}{(2\pi)^{d}}\frac{n_\mathrm{B}(\mbk)}{\mbk^3}\nn\\
&=&\frac{\als C_FI_T}{3\pi}r^i\left(E-h^{(0)}_o\right)^3r^i\,,
\label{cubic}
\end{eqnarray}
where $I_T$ comes from the evaluation of the integral. It reads (see also \cite{Escobedo:2008sy})
\begin{equation}
I_T=-\frac{1}{\epsilon}+\ln\frac{T^2}{\mu^2}-\gamma_E+\ln( 4\pi)-\frac{5}{3}\,.
\label{defit}
\end{equation}
The divergence of this expression is of infrared origin: it arises
when integrating over the Bose--Einstein distribution at momenta much
smaller than the temperature. Since we are integrating out the
temperature, i.e. getting the contribution for $\mbk\sim T$, this
divergence is an artifact of our scale separation. We identify two
possible schemes in which the cancellation of this divergence may be interpreted.
\begin{enumerate}
        \item In the first scheme, the divergence is cancelled by an
opposite ultraviolet divergence from a lower scale, in our case the
binding energy. In the next section, we will indeed show that
the thermal part of this very same diagram, when evaluated for
loop momenta of the order of the binding energy, yields 
an ultraviolet divergence that exactly cancels the one here, 
whereas the vacuum part of that diagram gives an
opposite UV divergence that cancels the previously mentioned IR divergence of the pNRQCD
potentials, yielding a finite spectrum.
	\item Alternatively one can observe that the pole of the
divergence is exactly opposite to the infrared pole of the pNRQCD potentials,
which can be read from \cite{Brambilla:1999xj} and the two therefore
cancel. More precisely, the scaleless, and hence vanishing in dimensional regularization, integral of 
the vacuum part of Eq.~\eqref{transverseleading}, with the octet propagator 
expanded at the cubic order, can be rewritten  
as the sum of an infrared and an ultraviolet divergent integral. 
The infrared pole cancels with the one in Eq.~\eqref{cubic} coming from the thermal
part, whereas the ultraviolet one cancels the IR divergence of the pNRQCD 
potentials. 
\end{enumerate}
The two interpretation schemes are equivalent and produce at the end a finite
spectrum, which is the relevant observable.\\ 
The evaluation of $r^i\left(E-h^{(0)}_o\right)^3r^i$ in (\ref{cubic}), in analogy to what has been performed previously in
Eqs.~\eqref{commutexample}, can be read from \cite{Kniehl:2002br}
\begin{eqnarray}
\nonumber &&\frac{1}{E-h^{(0)}_s}r^i\left(E-h^{(0)}_o\right)^3r^i\frac{1}{E-h^{(0)}_s}=
\frac{1}{E-h^{(0)}_s}\left(-\frac{N_c^3}{8}\frac{\als^3}{r}-(\nc^2+2\nc\cf)\frac{\als^2}{mr^2}\right.\\
&&\hspace{2.5cm}\left.+4(\nc-2\cf)\frac{\pi\als}{m^2}\delta^3(\br)+\nc\frac{\als}{m^2}
\left\{\nabla^2_\br,\frac{1}{r}\right\}\right)\frac{1}{E-h^{(0)}_s}+\cdots ,
\label{cubicexp}
\end{eqnarray}
where the dots stand for wave function renormalizations, which are again not relevant for the current calculation. 
Matching to the right-hand side of Eq.~\eqref{matchcond}, we obtain 
the corresponding contribution to the singlet potential 
$\delta V_s$ of $\textrm{pNRQCD}_\mathrm{HTL}$:
\begin{eqnarray}
\nonumber\delta V_s^{\rm (cubic)}&=&
\frac{\als C_FI_T}{3\pi}\left(-\frac{N_c^3}{8}\frac{\als^3}{r}-(\nc^2+2\nc\cf)\frac{\als^2}{mr^2}\right.\\
&& \hspace{2cm}
\left.+4(\nc-2\cf)\frac{\pi\als}{m^2}\delta^3(\br)+\nc\frac{\als}{m^2}\left\{\nabla^2_\br,\frac{1}{r}\right\}\right)\,.
\label{cubicpot}
\end{eqnarray}
We observe that the logarithmic $\mu$ dependence of the static part of this result, obtained from the first term in brackets, exactly cancels the IR logarithm appearing in the static potential~\eqref{staticpot}, as can be seen from the explicit expression of $\alVs$ in Eq.~\eqref{alvs}, thereby confirming our previous discussion on the cancellation of the divergences. We also remark that, as for the linear term, this potential is a sum of static, $1/m$ and $1/m^2$ terms, which however all share the same size $m\als^5$ in our power counting.
 
Using first-order quantum-mechanical perturbation theory 
and the value of the Cou\-lomb wave function at the origin, 
$\left\vert\psi_{n,l}(0)\right\vert^2=\delta_{l0}/(\pi n^3a_0^3)$, 
we obtain the shift of the energy levels
\begin{equation}
\delta E_{n,l}^{\rm (cubic)}=
\frac{E_nI_T\als^3}{3\pi}\left\{\frac{4 C_F^3\delta_{l0} }{n}+N_c \cf^2\left(\frac{8}{n (2l+1)}-\frac{1}{n^2} 
- \frac{2\delta_{l0}}{ n }   \right)+\frac{2N_c^2C_F}{n(2l+1)}+\frac{N_c^3}{4}\right\}\,.
\label{cubicshift}
\end{equation}

\subsection{Two-loop contribution}
\label{sub_twoloop}
As we have seen from Eq.~\eqref{vanishzero}, the zeroth-order term in the $E/T$ expansion $\Sigma_s^{(\text{zeroth})}$ vanishes. The size of this term would have been $mg^7\gg \als T^3 r^2\gg m g^{10}$. Hence, a radiative correction\footnote{Since we are evaluating radiative corrections from the temperature scale, the expansion parameter is still $g^2$.} to that diagram would be of size  $mg^9\gg \als^2 T^3 r^2\gg m g^{12}$, still contributing to the spectrum at order $m\als^5$ if $mg^3\gg T\ge mg^{10/3}$. In the same way it is easy to see that radiative corrections to the linear and higher terms in the $E/T$ expansion are smaller than $m\als^5$.\\ 
We thus consider radiative corrections to the diagram shown in Fig.~\ref{fig:hqself}. At the next order in $\als$,
corresponding to two loops, a sizable number of diagrams appears, corresponding to  next-to-leading order 
corrections to the chromoelectric field correlator. Their contribution to the static potential 
at zero temperature has been considered in \cite{Brambilla:2006wp}.\\
The chromoelectric correlator enters in the amplitude in the expression (see Eq.~\eqref{EETfull11})
\begin{displaymath}
\mu^{4-D} \int \frac{d^Dk}{(2\pi)^D} \, \frac{i}{-k_0+i\eta} \,  
\left[(k_0)^2 \, D_{ii}(k) + \bk^2 \,  D_{00}(k) \right]=
\int_0^\infty \! dt^\prime \,  
\frac{\left[\langle \be^a(t^\prime)\phi(t^\prime,0)^{\rm adj}_{ab} \be^b(0)\rangle\right]}{\nc^2-1}.
\end{displaymath}
Since $i/(-k_0 +i\eta) = -i\, {\rm P}\, (1/k_0) + \pi \delta(-k_0)$ and 
$\left[(k_0)^2 \,  D_{ii}(k) + \bk^2 \, D_{00}(k) \right]$ is even 
in $k_0$, only the $\pi \delta(-k_0)$ component of the static quark-antiquark propagator
contributes, therefore only the limit for $k_0\to 0$  of 
$\left[(k_0)^2 \,  D_{ii}(k) + \bk^2 \,  D_{00}(k)\right]$ matters.
In order to evaluate it, it is convenient to perform the calculation first in temporal-axial gauge 
$A^0 =0$. As we mentioned in Sec.~\ref{sec_imtime}, this is possible at finite temperature in the real-time formalism only  \cite{James:1990fd}. In this gauge, the chromoelectric field is simply $\be = - \partial_0 \ba$. Hence  
all corrections to the chromoelectric correlator are encoded in the spatial part of the gluon propagator alone: 
at one loop the correction is provided entirely by the gluon self energy. 
In temporal-axial gauge, from the transversality relation of the polarization tensor 
it follows that (compare with the explicit expressions of the propagators in \cite{Heinz:1986kz}):
\begin{equation}
\lim_{k_0\to 0} \; (k_0)^2 D_{ii}^{\rm R,A}(k)\bigg|_{\rm temporal-axial \; gauge} = 
\lim_{k_0\to 0} \; i \frac{\bk^2}{\bk^2 + \Pi_{00}^{\rm R,A}(k)}\bigg|_{\rm temporal-axial \; gauge} \,,
\end{equation}
where R and A stand again for retarded and advanced. Since in the $k_0\to 0$ limit $\Pi_{00}^{\rm R,A}(k)$ is equal in Coulomb and temporal-axial gauge, as shown in \cite{Heinz:1986kz}, we can also write that
\begin{equation}
\lim_{k_0\to 0} \; (k_0)^2 D_{ii}^{\rm R,A}(k)\bigg|_{\rm temporal-axial \; gauge} = 
\lim_{k_0\to 0} \; \bk^2 D_{00}^{\rm R,A}(k)\bigg|_{\rm Coulomb \; gauge}\,.
\end{equation}
\begin{figure}
	\begin{center}
\includegraphics{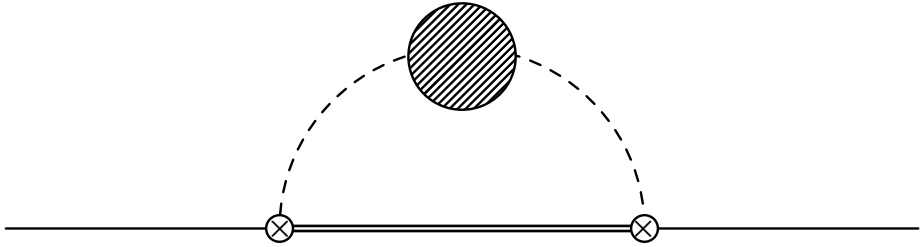}
\put(-110,21){$\boldsymbol\times$}
		\put(-246,21){$\boldsymbol\times$}
\end{center}
\caption {The only diagram contributing to the thermal part of the chromoelectric correlator at two loops in Coulomb gauge at the scale $T$. The dashed blob is the one-loop longitudinal self-energy, with light quarks and gluons in the loop. Ghosts do not couple to longitudinal gluons in Coulomb gauge; they furthermore do not have a thermal part.}  
\label{figEEmD}
\end{figure}The left-hand side is the only term of the chromoelectric correlator contributing to the potential 
in temporal-axial gauge: it may be evaluated by calculating the right-hand side in Coulomb gauge. 
At one loop, the right-hand side gets contribution from the gluon self-energy diagram 
shown in Fig.~\ref{figEEmD}; hence, at next-to-leading order we can write 
\begin{equation}
	\delta  V_s^{\rm (2 \, loops)}
 =
- i g^2 \, C_F \, \frac{r^2}{d}
\mu^{4-D} \int \frac{d^Dk}{(2\pi)^D} 
\, \pi \delta(-k_0) \,  
\bk^2 \, \left[\delta{\bf D}_{00}(k) \right]_{11}\,,
\label{Vs11Teq1}
\end{equation}
where we have already performed the matching, which is trivial due to the lack of dependence on $h^{(0)}_o$. Furthermore, from Eq. \eqref{11component} we have that
\begin{eqnarray} 
\left[\delta{\bf D}_{00}(k) \right]_{11} &=&  \frac{\delta D_{00}^{\rm R}(k)+ \delta D_{00}^{\rm A}(k)}{2}  
+ \left(\frac{1}{2} + n_{\rm B}(k_0)\right) \left(\delta D_{00}^{\rm R}(k) - \delta D_{00}^{\rm A}(k) \right), 
\label{D0011}
\\
\delta D_{00}^{\rm R,A}(k) &=& - \frac{i}{\bk ^4}  \Pi_{00}^{\rm R,A}(k) \,,
\label{DRAPi}
\end{eqnarray}
where the gluon polarization $\Pi_{00}^{\rm R,A}(k)$ in Coulomb gauge and its relevant limits are given by Eqs.~\eqref{Pi00full}, (\ref{RePi00k0}) and (\ref{ImPi00k0}) in App.~\ref{app_realtime}.
Finally, the correction to the real-time potential reads 
\begin{eqnarray}
\delta  V_s^{\rm (2 \, loops)}
&=& 
 - \frac{3}{2} \zeta(3)\,  C_F \, \frac{\als}{\pi} \, r^2 \, T \,m_D^2
+ \frac{2}{3} \zeta(3)\, N_c C_F \, \als^2 \, r^2 \, T^3 
\nonumber\\
&&
+ i \left[ \frac{C_F}{6} \als \, r^2 \, T \,m_D^2\, \left( 
\frac{1}{\epsilon} + \gamma_E + \ln\pi 
- \ln\frac{T^2}{\mu^2} + \frac{2}{3} - 4 \ln 2 - 2 \frac{\zeta^\prime(2)}{\zeta(2)} \right)\right.
\nonumber\\
&&  \quad \left.
+ \frac{4\pi}{9} \ln 2 \; N_c C_F \,  \als^2\, r^2 \, T^3 \right]
\,,
\label{VsTloop}
\end{eqnarray}
where $\zeta$ is the Riemann zeta function ($\zeta(2) = \pi^2/6$) and $m_D$ is the leading-order Debye mass, as given in Eq.~\eqref{mD}. This contribution to the potential was first evaluated in \cite{Brambilla:2008cx}.\\
Equation~\eqref{VsTloop} contains an imaginary part. It comes from the
imaginary part of the gluon self-energy, which is again related to the 
Landau-damping phenomenon. Furthermore, the imaginary part is infrared
divergent. In the EFT framework, this divergence has to be cancelled by an opposite
ultraviolet divergence coming from a lower scale. In the following section, we
will indeed show that the same diagram, when integrated over momenta
of the order of the binding energy, yields the desired UV divergence\footnote{\label{foot_mdggE}We observe that if we would have instead $m_D\gg E$, as we considered in \cite{Brambilla:2008cx}, the scale responsible for the cancellation of the divergence would be the Debye mass. The contribution to this diagram from that scale has been computed in Eq.~\eqref{VsmD} in the previous Chapter, and clearly shows an opposite UV divergence and $\mu$ dependence.}.\\ 
Finally, we remark that the result in Eq.~\eqref{VsTloop} comes from 
dimensionally regularizing only the integral over $k$ while keeping 
the thermal part of the gluon self energy, which is finite, 
in exactly four space-time dimensions. Using the same regularization when 
calculating the contribution coming from the binding-energy scale guarantees 
that the final result for the width is finite and scheme independent.
This is not the case for the potential, however, whose expression 
depends on the adopted scheme.

The contributions to the energy levels and to the thermal width can be
obtained easily from Eq.~\eqref{VsTloop} by using the expectation
value for $r^2$ on Coulombic states, i.e. $\left\langle
r^2\right\rangle_{n,l}= a_0^2n^2\left[5n^2+1-3l(l+1)\right]/2$:
\begin{eqnarray}
\hspace{-0.7cm} 
\delta  E_{n,l}^{\rm (2 \, loops)} &=& 
\left[ - \frac{3}{4} \zeta(3)\,  C_F \, \frac{\als}{\pi} \, T \,m_D^2
+ \frac{\zeta(3)}{3}  N_c C_F \, \als^2 \, T^3 \right]
a_0^2n^2\left[5n^2+1-3l(l+1)\right],\nn\\
\label{EnTloop}
\\
\hspace{-0.7cm} 
\Gamma_{n,l}^{\rm (2 \, loops)} &=& 
\left[ - \frac{C_F}{6} \als T m_D^2
\left( \frac{1}{\epsilon} + \gamma_E + \ln\pi 
- \ln\frac{T^2}{\mu^2}+ \frac{2}{3} - 4 \ln 2 - 2 \frac{\zeta^\prime(2)}{\zeta(2)} \right) \right.
\nonumber\\
&& \left.
-\frac{4\pi}{9} \ln 2 \; N_c C_F \,  \als^2\, T^3 \right] {a_0^2n^2}\left[5n^2+1-3l(l+1)\right]\,.
\label{GammanTloop}
\end{eqnarray}

\subsection{Summary}
Summing up Eqs.~\eqref{deltavslinear},~\eqref{cubicpot} and~\eqref{VsTloop} 
we obtain the thermal correction to the potential in $\textrm{pNRQCD}_\mathrm{HTL}$
up to terms whose contribution to the spectrum is smaller than $m\als^5$: 
\begin{eqnarray}
\nonumber
\delta V_s &=&\frac{\pi}{9}N_c C_F\, \als^2\, T^2\,r+\frac{2\pi}{3m}C_F\,\als \,T^2
+\frac{\als C_FI_T}{3\pi}\left[-\frac{N_c^3}{8}\frac{\als^3}{r}-(\nc^2+2\nc\cf)\frac{\als^2}{mr^2}\right.
\\
\nonumber&&
\hspace{5cm}
\left.+4(\nc-2\cf)\frac{\pi\als}{m^2}\delta^3(\br)+\nc\frac{\als}{m^2}\left\{\nabla^2_\br,\frac{1}{r}\right\}\right]
\\
&&- \frac{3}{2} \zeta(3)\,  C_F \, \frac{\als}{\pi} \, r^2 \, T \,m_D^2
+ \frac{2}{3} \zeta(3)\, N_c C_F \, \als^2 \, r^2 \, T^3
\nonumber\\
&&
+ i \left[ \frac{C_F}{6} \als \, r^2 \, T \,m_D^2\, \left( 
\frac{1}{\epsilon} + \gamma_E + \ln\pi 
- \ln\frac{T^2}{\mu^2} + \frac{2}{3} - 4 \ln 2 - 2 \frac{\zeta^\prime(2)}{\zeta(2)} \right)\right.
\nonumber\\
&&  \quad \left.
+ \frac{4\pi}{9} \ln 2 \; N_c C_F \,  \als^2\, r^2 \, T^3 \right]
\,,
\label{totalpotT}
\end{eqnarray}
where the first two terms come from the linear part of
Fig.~\ref{fig:hqself}, the terms in square brackets come from the
cubic term and the last three lines originate from the diagram in Fig.~\ref{figEEmD}. 
This correction to the potential can be used for $T\gg E,m_D$ 
no matter what the relative size between $E$ and $m_D$ is\footnote{\label{foot_potmdggE}For future convenience in Part~\ref{part_imtime}, we observe that, if we were to consider the case $m\als\gg T\gg m_D\gg E$, the potential one would have after integrating out the scale $m_D$ would be obtained by adding Eq.~\eqref{VsmD} to Eq.~\eqref{totalpotT}, as previously mentioned in footonote~\ref{foot_mdggE}. We remark that the real part of Eq.~\eqref{VsmD} would not contribute to the spectrum within our accuracy.}. We remark again that the first two terms have the same size, i.e. they contribute to the spectrum at the same order, in spite of their different countings in $1/m$. The same applies for all the terms in square brackets.\\
In App.~\ref{secpQCD} we will show how the static part of this potential can be obtained directly from perturbative QCD, without using the EFT framework. As we shall see, that derivation will be more cumbersome, highlighting the advantages of the EFT.

The total contribution to the energy levels coming from the scale $T$ is
\begin{eqnarray}
\nonumber \delta E_{n,l}^{(T)}&=&
\frac{\pi}{9}N_c C_F \,\als^2 \,T^2 \frac{a_0}{2}(3n^2-l(l+1))+\frac{2\pi}{3m}C_F\, \als\, T^2
\\
\nonumber&&
+\frac{E_nI_T\als^3}{3\pi}\left\{\frac{4 C_F^3\delta_{l0} }{n}
+N_c \cf^2\left(\frac{8}{n (2l+1)}-\frac{1}{n^2} - \frac{2\delta_{l0}}{ n }   \right)
+\frac{2N_c^2C_F}{n(2l+1)}+\frac{N_c^3}{4}\right\}
\\
\nonumber&& 
+\left(- \frac{3}{2} \zeta(3)\,  C_F \, \frac{\als}{\pi}  \, T \,m_D^2
+ \frac{2}{3} \zeta(3)\, N_c C_F \, \als^2 \, T^3\right) \frac{a_0^2n^2}{2}\left[5n^2+1-3l(l+1)\right].
\\
\label{totalenergyT}
\end{eqnarray}
The first and the second lines originate from the diagram in Fig.~\ref{fig:hqself}, 
and correspond to the linear and cubic terms in the expansion (\ref{octetexpand}). 
The last line originates from the gluon self-energy diagram in Fig.~\ref{figEEmD}, 
which also gives the full contribution of the scale $T$ to the width:
\begin{eqnarray}
\Gamma_{n,l}^{(T)}&=& \Gamma_{n,l}^{\rm (2 \, loops)}.
\label{totalwidthT}
\end{eqnarray}

\section{Contribution to the spectrum from the scale  $E$  \label{sec:energy}}
\label{secE}
After having integrated out the temperature in the previous section,
many different scales ($E$, $m_D$, $\Lambda_{\rm QCD}$, $\ldots$)
still remain dynamical in $\mathrm{pNRQCD}_\mathrm{HTL}$. 
In our hierarchy, the binding energy is much
larger than the Debye mass and $\Lambda_{\rm QCD}$ is smaller than all
other scales. Our purpose is to compute the correction to the
spectrum and the width coming from the scales $E$ and $m_D$. This is
achieved by computing loop corrections to the singlet propagator in 
$\mathrm{pNRQCD}_\mathrm{HTL}$. We recall that the gauge sector of $\mathrm{pNRQCD}_\mathrm{HTL}$ 
is described by the Hard Thermal Loop effective Lagrangian~\eqref{laghtl}.\\
We start by evaluating the one-loop dipole diagram shown in
Fig.~\ref{fig:hqself}, whose general expression is given in
Eq.~\eqref{defleading}, but now the longitudinal and transverse gluon propagators 
are the HTL ones, as given by Eqs.~\eqref{prophtllong} and~\eqref{prophtltrans}.
As we shall see, this is the only diagram we
need to consider to get the spectrum at order $m\als^5$.\\
At the energy scale, we have $k_0\sim\left(E-h^{(0)}_o\right)$ and therefore we have to keep 
the octet propagator unexpanded. However two expansions are still possible.
\begin{enumerate}
        \item Since $k_0\sim E\ll T$, the Bose--Einstein distribution can be expanded for $k_0\ll T$, as in Eq.~\eqref{boseexp}. 
        \item Moreover, since  $\mbk\sim E\gg m_D$, the Hard Thermal Loop propagators
can be expanded in $m_D^2/E^2\ll1$. At the zeroth order, this
corresponds to using the free propagators, given in Eqs.~\eqref{longgluonpropreal} and~\eqref{transgluonpropreal}. 
Some care is required in the expansion of the transverse gluons 
due to a collinear region, as we shall see later on.
\end{enumerate}
In the following, we will call $\delta \Sigma_s(E)$ the contribution 
of the diagram in Fig.~\ref{fig:hqself} to the singlet self energy; 
the corresponding energy shift and width for the state $\vert
n,l\rangle$ are given by 
$\delta E_{n,l}=\langle n,l\vert \mathrm{Re}\,\delta \Sigma_s(E_{n,l})\vert n,l\rangle$ and 
$\Gamma_{n,l}=-2\langle n,l\vert \mathrm{Im}\,\delta \Sigma_s(E_{n,l})\vert n,l\rangle$.

We now proceed to the evaluation of Eq.~\eqref{defleading} for loop
momenta of the order of the binding energy, with the HTL propagators
defined in Eqs.~\eqref{prophtllong} and~\eqref{prophtltrans}. We
find convenient to compute separately the contributions coming from the 
transverse and longitudinal gluons.

\subsection{Transverse gluon contribution}
\label{sub_trans}
The contribution of transverse gluons to Eq.~\eqref{defleading} is in $\mathrm{pNRQCD}_\mathrm{HTL}$
\begin{eqnarray}
\nonumber\delta \Sigma_s^{\rm (trans)}(E) &=&	- i g^2 \, C_F \, \frac{d-1}{d}r^i
\mu^{4-D} \!\! \int \!\! \frac{d^Dk}{(2\pi)^D}
\frac{i}{E-h^{(0)}_o-k_0 +i\eta}k_0^2 \left[ \frac{ \Delta_{\rm R}(k) + \Delta_{\rm A}(k)}{2}\right.\\
&&\hspace{3.2cm}\left.+ \left(\frac{1}{2} + n_{\rm B}(k_0)\right)\left(\Delta_{\rm R}(k) 
- \Delta_{\rm A}(k)\right)\right]r^i\,.
\label{deftransE}
\end{eqnarray}
We start by evaluating the contribution of the symmetric part, which we recall to be the one proportional to the difference between the retarded and advanced propagators. As we shall see, it turns out to be the leading one, the reason being again the leading, singular term in the infrared expansion of the Bose distribution (Bose enhancement), yielding
\begin{eqnarray}
\nonumber&& g^2 \, C_F \, \frac{d-1}{d}r^i
\mu^{4-D} \int \frac{d^Dk}{(2\pi)^D}
\frac{k_0^2}{E-h^{(0)}_o-k_0 +i\eta} \left(\frac{T}{k_0}+\order{\frac ET}\right)
\\
&&\hspace{7.5cm}
\times\left[\Delta_{\rm R}(k) - \Delta_{\rm A}(k)\right]r^i\,,
\label{deftranssym}
\end{eqnarray}
The expansion of the HTL propagators for $m_D\ll k_0,\mbk$
needs to be performed with care in the region around the light cone,
where the gluon propagator becomes singular. We refer to Appendix
\ref{app_trans} for details on the expansion and the evaluation of the
integral, whose final result reads
\begin{equation}
-i\frac{2}{3}\als\,\cf T r^i\left(E-h^{(0)}_o\right)^2r^i
+i\frac{\als\cf\,Tm_D^2\,r^2\,(\ln2-1/2)}{3}+\order{\frac{\als T m_D^4r^2}{E^2},\frac{\als r^2E^4}{T}}.
\label{finaltranssym}
\end{equation}
The suppressed term of order $\als r^2 E^4/T$ comes from the $k/(12\,T)$
term in the expansion of the thermal distribution, whereas the term of
order $\als T m_D^4r^2/E^2$ comes from subleading terms in the
expansion of the propagator\footnote{This term is of order
$m\als^5$ or bigger only in the very tiny window $mg^3\gg T\ge
mg^{3+1/5}$. For this reason, we will not include terms of order $\als T m_D^4r^2/E^2$
or smaller obtained from the expansion in $m_D^2/E^2$. \label{footpre}}.\\

We now consider the first term in the square brackets in
Eq.~\eqref{deftransE}; it does not depend on the Bose--Einstein
distribution and, when expanded for $k_0,\mbk\sim E\gg m_D$, gives
\begin{equation}
\frac{ \Delta_{\rm R}(k) + \Delta_{\rm A}(k)}{2}=i\mathrm{P}\frac{1}{k_0^2-\bk^2}+\order{m_D^2/E^4},
\label{r+atrans}
\end{equation}
where P stands for the principal value prescription.
Plugging Eq.~\eqref{r+atrans} back into Eq.~\eqref{deftransE} yields
\begin{eqnarray}
\nonumber&&	 g^2 \, C_F \, \frac{d-1}{d}r^i
\mu^{4-D} \int \frac{d^Dk}{(2\pi)^D}
\frac{1}{E-h^{(0)}_o-k_0 +i\eta}k_0^2 \left[ i\mathrm{P}\frac{1}{k_0^2-\bk^2}+\order{m_D^2/E^4}\right]r^i\\
&&=-i\frac{\als \, C_F \, }{3}r^i\left(E-h^{(0)}_o\right)^3r^i+\order{\als \,E \,m_D^2\,r^2}\,.
\label{impartr+a}
\end{eqnarray}
Summing up Eqs.~\eqref{finaltranssym} and~\eqref{impartr+a} we obtain the complete contribution of the transverse modes
\begin{eqnarray}
\nonumber\delta \Sigma_s^{\rm (trans)}(E)&=&
-i\frac{2}{3}\als\,\cf T r^i\left(E-h^{(0)}_o\right)^2r^i-i\frac{\als \, C_F \, }{3}r^i\left(E-h^{(0)}_o\right)^3r^i
\\
&&\hspace{-1cm}+i\frac{\als\cf\,Tm_D^2\,r^2\, }{3}\left(\ln2-\frac12\right)+\order{\als T m_D^4r^2/E^2,\als r^2E^4/T,\als \,E \,m_D^2\,r^2}.\nn\\
\label{tottrans}
\end{eqnarray}
We remark that the contribution of the transverse modes at the
energy scale is imaginary and finite, in contrast with what happens 
at zero temperature, where it is real and UV divergent, 
the divergence cancelling the infrared divergences appearing in the static, 
$1/m$ and $1/m^2$ potentials at the scale $m\als$. 
This is related to the discussion made in the previous section
regarding the cancellation of the IR divergence in Eqs.~\eqref{cubic}
and~\eqref{cubicshift} and can be understood in the following way.
For $E\gg m_D$, the Hard Thermal Loop transverse propagator can be
expanded for small $m_D$, giving, at the zeroth order, $(\Delta_{\rm R} +
\Delta_{\rm A})/2=i\mathrm{P} [{1}/{(k_0^2-\bk^2)}]$ and $(\Delta_{\rm R} - \Delta_{\rm
A})=2\pi\,\mathrm{sgn}(k_0)\delta(k^2_0-\bk^2)$. When plugged in
Eq.~\eqref{deftransE} we obtain Eq.~\eqref{transverseleading}. Evaluated at
the binding energy scale, the vacuum part is UV divergent and can be read from
\cite{Kniehl:1999ud,Brambilla:1999xj}: 
\begin{eqnarray}
&&g^2C_F\frac{d-1}{d} r^i\mu^{4-D}\int\frac{\,d^Dk}{(2\pi)^D}
\frac{k_0^2}{E-h^{(0)}_o-k_0+i\eta}\frac{i}{k_0^2-\bk^2+i\eta}r^i
\nonumber \\
&&\hspace{-1cm}=\frac{\als\cf}{3\pi}r^i\left(E-h^{(0)}_o\right)^3
\left(-\frac{1}{\epsilon}+2\ln\frac{-\left(E-h^{(0)}_o\right)-i\eta}{\mu}+\gamma_E-\frac53-\ln\pi\right)r^i\,,
\label{USvacuum}
\end{eqnarray}
where the logarithm of the Hamiltonian gives rise to the so-called QCD Bethe logarithm in the spectrum 
\cite{Kniehl:1999ud,Kniehl:2002br}. On the other hand, the temperature-dependent part gives
\begin{align}
\nonumber& 
g^2C_F\frac{d-1}{d} r^i\mu^{4-D}\int\frac{\,d^Dk}{(2\pi)^D}\frac{k^2_0}{E-h^{(0)}_o-k_0+i\eta}
\left(\frac{T}{\vert k_0\vert}-\frac12+\order{\frac kT}\right)2\pi \delta\left(k_0^2-\bk^2\right)r^i
\\
\nonumber
&=-i\frac{2}{3}\als\,\cf T r^i\left(E-h^{(0)}_o\right)^2r^i-\frac{\als\cf}{3\pi}r^i\left(E-h^{(0)}_o\right)^3
\left(-\frac{1}{\epsilon}+\gamma_E+2\ln\frac{\vert E-h^{(0)}_o\vert}{\mu}\right.
\\
&\hspace{7cm}
\left.-i\pi\mathrm{sgn}\left(E-h^{(0)}_o\right)-\frac53-\ln\pi\right)r^i\,,
\label{USmatter}
\end{align}
where the term proportional to $r^i\left(E-h^{(0)}_o\right)^2r^i$ comes from the first
term in the expansion of the Bose--Einstein distribution and the one
proportional to $r^i\left(E-h^{(0)}_o\right)^3r^i$ comes instead from the second term
in that expansion, see (\ref{boseexp}). In the sum of Eqs.~\eqref{USvacuum} and
\eqref{USmatter} the real parts, divergences included, cancel out and
the imaginary parts combine to give the two $m_D$-independent terms of
Eq.~\eqref{tottrans}. This shows  that the binding energy scale 
contribution produces two opposite UV divergences. 
In terms of the two interpretation schemes discussed in the previous section, 
we may understand the cancellation of divergences in two possible ways.
In the first way, the vacuum divergence in Eq.~\eqref{USvacuum}
cancels the IR divergences of the potentials, whereas the UV matter divergence 
in~\eqref{USmatter} cancels the IR matter divergence from the scale $T$ 
in~\eqref{cubic}. In the second way, we consider the real part of the potential 
in  $\mathrm{pNRQCD}_\mathrm{HTL}$ as finite, the IR divergences from the scales $m\als$ 
and $T$ cancelling each other, and  no UV divergences coming  from the
energy scale, which, as shown by Eq.~\eqref{tottrans}, is indeed the case. 
We stress that the cancellation of the divergences between the vacuum and thermal parts in
Eqs.~\eqref{USvacuum} and~\eqref{USmatter} is due to the second term
in the low-momentum expansion of the Bose--Einstein distribution,
i.e. $-1/2$, which, as we discussed in Sec.~\ref{sec_htl}, is known in Thermal Field Theory to cause
cancellations with the vacuum contribution. Finally, we observe that an
analogous cancellation is also obtained in the Abelian case \cite{Escobedo:2008sy}.\\

In order to obtain the contribution to the width from Eq.~\eqref{tottrans}, 
we need to evaluate $r^i\left(E-h^{(0)}_o\right)^2r^i$. 
We proceed as in the previous section and rewrite $\left(E-h^{(0)}_o\right)^2$ 
as $\left(E-h^{(0)}_s\right)^2-\left\{\left(E-h^{(0)}_s\right),\Delta V\right\}+\Delta V^2$. One then has
\begin{equation}
r^i\left(E-h^{(0)}_o\right)^2r^i =\left(\frac{\nc^2}{4}\als^2+\frac{2\nc\als}{mr}+\frac{4\bp^2}{m^2}\right)
+ ...\,,
\label{squareexp}
\end{equation}
where the dots stand for contributions that vanish on the physical state. The width thus reads
\begin{eqnarray}
\nonumber
\Gamma_{n,l}^{\rm (trans)}&=&
\frac{1}{3}N_c^2C_F\als^3T-\frac{16}{3m}C_F\als TE_n+\frac{4}{3}N_cC_F\als^2T\frac{2}{mn^2a_0}
\\
\nonumber&&
+\frac{2E_n\als^3}{3}\left\{\frac{4 C_F^3\delta_{l0} }{n}+N_c \cf^2\left(\frac{8}{n (2l+1)}-\frac{1}{n^2} 
- \frac{2\delta_{l0}}{ n }   \right)+\frac{2N_c^2C_F}{n(2l+1)}+\frac{N_c^3}{4}\right\}
\\
&&-\frac{\als\cf\, (\ln4-1)\,Tm_D^2}{3}\frac{a_0^2n^2}{2}[5n^2+1-3l(l+1)]\,,
\label{totaltranssym}
\end{eqnarray}
where the first line is the contribution from the term proportional to
$r^i\left(E-h^{(0)}_o\right)^2r^i$, the second line comes from the cubic term and has
been obtained using Eqs.~\eqref{cubicexp} and~\eqref{cubicshift}, and
the third line is the contribution from the last term in the first line of Eq.~\eqref{tottrans}.

The leading contribution to Eq.~\eqref{totaltranssym}  
is given by the first three terms, which are of the same size.
The first term comes from the static potential and agrees with the one  we first calculated in the static limit in \cite{Brambilla:2008cx}.
The second and third terms come from the kinetic energy;
the second one agrees with the one calculated in \cite{Escobedo:2008sy}.
This contribution to the thermal decay width originates from the possible break up 
of a quark-antiquark colour-singlet state into an unbound quark-antiquark colour-octet state: 
a process that is kinematically allowed only in a medium, the octet continuum having a higher energy than a discrete singlet state.
Clearly, the singlet to octet break up is a different phenomenon with respect 
to the Landau damping, which, in the previous section, 
provided another source for the in medium thermal width.  
In the situation $E \gg m_D$, which is the situation of 
interest for this work, the singlet to octet break up provides the dominant contribution 
to the thermal width, as one would expect from the fact that the former is caused by physics at the scale $E$ (a thermal gluon with enough energy to dissociate the singlet state) and the latter by the scale $m_D$. Indeed, comparing the Landau-damping width~\eqref{GammanTloop} with 
the singlet to octet break-up width (the first two lines of Eq.~\eqref{totaltranssym}), we see that 
the latter is larger than the former by a factor $(m\als^2/m_D)^2$.

The singlet-to-octet decay was first considered at zero temperature by Bhanot and Peskin in Refs.~\cite{Peskin:1979va,Bhanot:1979vb}. The authors computed the cross section for the process $g+\Phi(1S)\to (Q\overline{Q}_8$, where $\Phi(1S)$ is a $1S$ $Q\overline{Q}$ bound state being dissociated by the incoming gluon into an unbound octet $Q+\overline{Q}$. The calculation was performed in the framework of the Operator Product Expansion (OPE) \cite{Wilson:1969zs} and assuming a perturbative, Coulombic bound state, corresponding, as we know, to the assumption $mv\sim m\als(mv)$. The calculation was furthermore simplified by neglecting the octet potential altogether, corresponding in our formalism to having $h^{(0)}_o\to \bp^2/m$.\\
This calculation was, to the best of our knowledge, first considered in \cite{Kharzeev:1994pz} in the context of quarkonia in nuclear matter, either cold or hot. The authors concluded that in the latter case the thermal gluons, distributed along the Bose--Einstein distribution, have an average momentum $\langle p\rangle\sim 3T$ that is kinematically sufficient to lead to a dissociation of the bound state, i.e $\langle p\rangle\sim E$, whereas in the case of cold nuclear matter this was not the case. In \cite{Xu:1995eb} a more quantitative analysis was performed by convoluting the Bhanot-Peskin cross section, which had been computed as a function of the gluon momentum, with a thermal distribution. Since then this method has been widely used to model the width of the $J/\psi$ and of other quarkonium states in the medium. Within this context, this effect is often called \emph{gluo-dissociation}. It has been sometimes applied to non-Coulombic states, stretching the Bhanot-Peskin cross section by replacing its functional dependence on the Coulombic ground state energy with phenomenological binding energies extracted from potential models. We refer to \cite{Rapp:2008tf} for a review.\\
In Appendix~\ref{peskin} we show how our formalism can be brought to the form of a convolution of a cross section with the Bose--Einstein distribution. We show that, in the $h^{(0)}_o\to \bp^2/m$ approximation, we recover the Bhanot-Peskin cross section. We also perform the calculation with the full Coulombic Hamiltonian $h^{(0)}_o$ and we obtain the corresponding cross section, given by Eq.~\eqref{crossEFT}. By convoluting it with the Bose--Einstein distribution we show how the width obtained in this way is approximately $10\%$ larger than the one obtained from the Bhanot-Peskin cross section for $T>4\vert E_1\vert$, thereby evaluating the error introduced by neglecting the octet potential.

What our analysis should have furthermore made clear is that convoluting the thermal momentum distribution with the dissociation cross section, either the Bhanot-Peskin or the one given by Eq.~\eqref{crossEFT}, makes sense only for the ground states of bottomonium and, to a lesser extent, charmonium, which can be treated as Coulombic, whereas its application to $nS$ states with $n>1$ is no longer meaningful. Furthermore the Debye mass has to be much smaller than the the scale $m\als$, otherwise, as discussed in Chap.~\ref{chap_Tggr}, the potential becomes screened, changing the energy levels on which the cross section is based, and is furthermore overcome by its imaginary part. In the cases where the Bhanot-Peskin+thermal distribution approach is taken to non-Coulombic bound states with phenomenological binding energies, the connection to QCD appears dubious.

\subsection{Longitudinal gluon contribution}
The contribution of the longitudinal gluons to Eq.~\eqref{defleading} is
\begin{eqnarray}
\nonumber\delta \Sigma_s^{\rm (long)}(E)&=&	
- i g^2 \, C_F \, \frac{r^i}{d}
\mu^{4-D} \int \frac{d^Dk}{(2\pi)^D} \frac{i}{E-h^{(0)}_o-k_0 +i\eta}\bk^2 
\left[ \frac{ D_{00}^{\rm R}(k) + D_{00}^{\rm A}(k)}{2}\right.\\
&&\hspace{3.4cm}
\left.+ \left(\frac{1}{2} + n_{\rm B}(k_0)\right)
\left(D_{00}^{\rm R}(k) - D_{00}^{\rm A}(k)\right)\right]r^i\,,
\label{deflongE}
\end{eqnarray}
where $D_{00}^{\rm R,A}(k)$ is the HTL propagator in
\eqref{prophtllong}. The first term in square brackets,
i.e. $(D_{00}^{\rm R} + D_{00}^{\rm A})/2$, does not depend on the
Bose--Einstein distribution; therefore only the expansion in $m_D\ll E$, 
corresponding to $m_D\ll k_0,\mbk$, is possible. We then have
$(D_{00}^{\rm R} + D_{00}^{\rm
A})/2= {i}/{\bk^2}+\order{m_D^2/\bk^4}$, as in Eq.~\eqref{VsTrmD1}. The first term is the free 
propagator, which gives a scaleless integration, whereas the second
one can be shown to contribute at order $\als E m_D^2 r^2$, which is
smaller than $m\als^5$.

For what concerns the symmetric part of the
propagator, i.e. $(1/2 + n_{\rm B}(k_0))(D_{00}^{\rm R} - D_{00}^{\rm
A})$, it should be noted that the retarded and advanced propagators
depend on $k_0$ only through the HTL self-energy; therefore, imaginary
parts in their denominators can enter only through the logarithm
appearing in Eqs.~\eqref{pie} and~\eqref{prophtllong}. Hence, the symmetric propagator
is non-zero solely in the spacelike $\bk^2>k_0^2$ region, which is
related to the Landau-damping phenomenon. At leading order in the
expansions of the Bose--Einstein distribution and of the propagator for
$m_D^2/\bk^2\ll 1$,  we thus have 
\begin{equation}
\left(\frac{1}{2} + n_{\rm B}(k_0)\right)\left(D_{00}^{\rm R}(k) - D_{00}^{\rm A}(k)\right)
=\frac{2\pi T m_D^2}{\mbk^5}\theta\left(\bk^2-k_0^2\right)+\order{m_D^2/\bk^4,\,Tm_D^4/\mbk^7},
\label{deltashtllong}
\end{equation}
where the expansion of the thermal distribution provides again a Bose enhancement. 
The first term contributes to the spectrum at order $\als Tm_D^2r^2$, 
so further terms in Eq.~\eqref{deltashtllong} are not needed (see footnote~\ref{footpre} in the previous Subsection). 
We then have
\begin{eqnarray}
\nonumber\delta \Sigma_s^{\rm (long)}(E)&=&
g^2C_F\frac{2}{d}r^i\mu^{4-D}
\int\frac{\,d^Dk}{(2\pi)^D}\frac{k^2}{E-h^{(0)}_o-k_0+i\eta}\frac{Tm_D^2\pi}{\mbk^5}\theta\left(\bk^2-k_0^2\right)r^i\\
\nonumber&=&\frac{\als C_FTm_D^2}{6}r^i\bigg[2\pi\,\mathrm{sgn}\left(E-h^{(0)}_o\right)\\
&&\hspace{1.8cm}\left.+i\left(-\frac{1}{\epsilon}+\ln\frac{\left(E-h^{(0)}_o\right)^2}{\mu^2}+\gamma_E-\frac83-\ln\pi\right)\right]r^i.\label{tothtllong}
\end{eqnarray}
Equation~\eqref{tothtllong} translates into the following shift of the energy levels 
\begin{equation}
\label{longshiftE}
\delta E_{n,l}^{\rm (long)}=-\frac{\pi\als C_F\ Tm_D^2}{3}\frac{a_0^2n^2}{2}[5n^2+1-3l(l+1)].
\end{equation}
For what concerns the width, we observe that the divergence is of
ultraviolet origin and cancels the one in Eq.~\eqref{GammanTloop}, 
yielding a finite width; some care is, 
however, required in the handling of the logarithms of the energy, which
give rise to an analogue of the Bethe logarithm. We have
\begin{eqnarray}
\nonumber
\Gamma_{n,l}^{\rm (long)}&=&
-\frac{\als\cf Tm_D^2}{3}\left(-\frac{1}{\epsilon}
+\ln\frac{E_1^2}{\mu^2}+\gamma_E-\frac83-\ln\pi\right)\frac{a_0^2n^2}{2}[5n^2+1-3l(l+1)]\\
  &&+\frac{2\als\cf Tm_D^2}{3}\frac{C_F^2\als^2}{E_n^2}\,I_{n,l}\;,
\label{longwidhtE}
\end{eqnarray}
where $E_1=-m\cf^2\als^2/4$ is the energy of the Coulomb ground state and
\begin{equation}
I_{n,l}=\frac{E_n^2}{C_F^2\als^2}\int\frac{\,d^3k}{(2\pi)^3}
\left\vert\langle n,l\vert\br\vert\bk\rangle\right\vert^2\ln\frac{E_1}{E_n-{k^2}/{m}}\,.
\label{defin}
\end{equation}
$\langle n,l\vert\br\vert\bk\rangle$ is the matrix element between a
(bound) eigenstate $\vert n,l\rangle$ of $h^{(0)}_s$ and a continuum eigenstate
$\vert \bk\rangle$ of $h^{(0)}_o$.  This expression can be reduced to a single
integral using the techniques of \cite{Kniehl:1999ud,Kniehl:2002br}. 
We obtain for a singlet $nS$ state and an octet $P$ wave (the matrix element introduces a $\Delta l=1$
selection rule)
\begin{equation}
I_{n,0}\equiv\int_0^\infty d\nu\, Y_n^{m_D}(\nu)X^2_n(\nu)\,,
\label{explin}
\end{equation}
where
\begin{equation}
Y_n^{m_D}(\nu)=\frac{\nu^6}{(\nu^2+\rho_n^2)^3}Y_n^E\,.
\label{Ynmd}
\end{equation}
The definitions of $Y_n^E$, $X^2_n$ for $n=1,2,3$ and 
$\rho_n$ can be found in \cite{Kniehl:1999ud} and
\cite{Kniehl:2002br}, the latter reference correcting some misprints
in the former. A numerical evaluation of these integrals for the three
most tightly bound $l=0$ states yields:
\begin{equation}
I_{1,0}=-0.49673,\qquad I_{2,0}=0.64070,\qquad I_{3,0}=1.18970.		
\label{numin}
\end{equation}

\subsection{Summary}
In summary, the contribution to the energy levels coming from the binding energy scale
is entirely due to the longitudinal part of the chromoelectric correlator, 
\begin{equation}
\label{totalspectrumE}
\delta E_{n,l}^{(E)}= \delta E_{n,l}^{\rm (long)},
\end{equation}
which may be read from Eq.~\eqref{longshiftE}.
The contribution to the decay width coming from the binding energy scale
is the sum of $\Gamma_{n,l}^{\rm (trans)}$ and $\Gamma_{n,l}^{\rm (long)}$:
\begin{eqnarray}
\nonumber
\Gamma_{n,l}^{(E)}&=&
\frac{1}{3}N_c^2C_F\als^3T-\frac{16}{3m}C_F\als TE_n+\frac{8}{3}N_cC_F\als^2T\frac{1}{mn^2a_0}\\
\nonumber&&
+\frac{2E_n\als^3}{3}\left\{\frac{4 C_F^3\delta_{l0} }{n}+N_c \cf^2
\left(\frac{8}{n (2l+1)}-\frac{1}{n^2} - \frac{2\delta_{l0}}{ n }   \right)
+\frac{2N_c^2C_F}{n(2l+1)}+\frac{N_c^3}{4}\right\}\\
\nonumber&&
-\frac{\als\cf Tm_D^2}{6}\left(-\frac{1}{\epsilon}
+\ln\frac{E_1^2}{\mu^2}+\gamma_E-\frac{11}{3}-\ln\pi+\ln4\right) a_0^2n^2 [5n^2+1-3l(l+1)]\\
 &&+\frac{2\als\cf Tm_D^2}{3}\frac{C_F^2\als^2}{E_n^2}\,I_{n,l}\;,
\label{totalwidthE}
\end{eqnarray}
where the first two lines come from the first two in
Eq.~\eqref{totaltranssym} and the last two from Eq.~\eqref{longwidhtE}
and from the last term in~\eqref{totaltranssym}. $I_{n,l}$ is defined in
Eq.~\eqref{defin}.

\section{Contributions to the spectrum from the scale $m_D$}
\label{secmD}
In our hierarchy of energy scales, the next scale after the binding energy 
is the Debye mass. 
We thus have to evaluate Eqs.~\eqref{deftransE} and
\eqref{deflongE} for momenta of the order of $m_D$. In detail, we have
two regions to analyze: the first one is $k_0\sim E-h^{(0)}_o$, $\mbk\sim m_D$,
corresponding to having the octet propagator unexpanded and conversely
expanding the HTL propagators for $k_0\gg \mbk$. It can be easily shown
that both the transverse and the longitudinal parts result in a series
of scaleless integrations over $\bk$, which vanish in dimensional regularization.\\
The second region corresponds to having $k_0\sim
m_D$ and $\mbk\sim m_D$: the octet propagator then needs to be expanded,
whereas the HTL propagators are to be kept in their resummed form. The
resulting integrals are quite involved, however, by power counting
arguments, it can be easily seen from Eqs.~\eqref{deftransE} and
\eqref{deflongE} that, once the octet propagator is expanded, the
largest term comes again from the symmetric part of the gluon
propagator, due to the $T/k_0$ Bose enhancement factor. The size of this
term turns out to be of order $\als T m_D^3r^2/E$ and, 
since we have assumed $\left({m_D}/{E}\right)^4\ll g$, it is beyond $m\als^5$.

\section{Conclusions}
\label{secconclusions}
We have computed the heavy quarkonium energy levels and widths in a quark-gluon
plasma of temperature $T$ such that $m\als \gg  T \gg  m \als^2 \gg m_D$.
Assuming $(m_D/E)^4\ll g$, the spectrum is accurate up to order $m\als^5$. 

The thermal shift of the energy levels induced by the medium is obtained 
by summing the contribution from the scale $T$, given in
Eq.~\eqref{totalenergyT}, with the thermal part of the contribution from 
the energy scale. We remark that the contribution from the energy scale, 
given in Eq.~\eqref{totalspectrumE},  is the sum of both vacuum and thermal contributions, 
which, in the transverse sector, cancel. The thermal contribution of the transverse modes 
can be derived from  Eq.~\eqref{USmatter}. 
The complete thermal contribution to the spectrum up to order $m\als^5$ reads
(we recall that $E_n=-{mC_F^2\als^2}/{(4n^2)}$ and $a_0 = {2}/{(m\cf\als)}$)
\begin{eqnarray}
\nonumber
\delta E_{n,l}^{(\mathrm{thermal})}&=&
\frac{\pi}{9}N_c C_F \,\als^2 \,T^2 \frac{a_0}{2}\left[3n^2-l(l+1)\right] +\frac{\pi}{3}C_F^2\, \als^2\, T^2\,a_0
\\
\nonumber&&
+\frac{E_n\als^3}{3\pi}\left[\log\left(\frac{2\pi T}{E_1}\right)^2-2\gamma_E\right]
\left\{\frac{4 C_F^3\delta_{l0} }{n}+\frac{2N_c^2C_F}{n(2l+1)} +\frac{N_c^3}{4}
\right.
\\
\nonumber 
&& \hspace{5.7cm}
\left. +N_c \cf^2\left[\frac{8}{n (2l+1)}-\frac{1}{n^2} 
- \frac{2\delta_{l0}}{ n }   \right]\right\}
\\
\nonumber
&&
+\frac{2E_n\cf^3\als^3}{3\pi}L_{n,l}
\\
\nonumber
&&
+ \frac{a_0^2n^2}{2}\left[5n^2+1-3l(l+1)\right]
\left\{- \left[\frac{3}{2\pi} \zeta(3)+\frac{\pi}{3}\right]  C_F \, \als  \, T \,m_D^2
\right.
\\
&& \hspace{6.7cm}
+ \left. \frac{2}{3} \zeta(3)\, N_c C_F \, \als^2 \, T^3\right\},
\label{finalspectrum}
\end{eqnarray}
where $L_{n,l}$ is the QCD Bethe logarithm, defined as \cite{Kniehl:1999ud,Kniehl:2002br} 
\begin{equation}
L_{n,l}=\frac{1}{C_F^2\als^2E_n}\int\frac{d^3k}{(2\pi)^3}
\left\vert\langle n,l\vert\br\vert\bk\rangle\right\vert^2\left(E_n-\frac{k^2}{m}\right)^3\ln\frac{E_1}{E_n-{k^2}/{m}}.
\label{defln}
\end{equation}
We refer to \cite{Kniehl:1999ud,Kniehl:2002br} for details on the
numerical evaluation of this integral. We furthermore remark that the
thermal contribution to the spectrum is finite, the IR divergence in
Eq.~\eqref{totalenergyT} having cancelled against the UV divergence
coming from Eq.~\eqref{USmatter}.

The thermal width is obtained by summing the
contribution from the scale $T$, given in Eq.~\eqref{totalwidthT},
with the one coming from the energy scale as given in~\eqref{totalwidthE}, 
the IR divergence in the former canceling against the UV divergence in the latter. 
We then have
\begin{eqnarray}
\nonumber
\Gamma_{n,l}^{(\mathrm{thermal})}&=&
\frac{1}{3}N_c^2C_F\als^3T+\frac{4}{3}\frac{C_F^2\als^3 T}{n^2}(\cf+\nc)
\\
\nonumber&&
+\frac{2E_n\als^3}{3}\left\{\frac{4 C_F^3\delta_{l0} }{n}+N_c \cf^2
\left[\frac{8}{n (2l+1)}-\frac{1}{n^2} - \frac{2\delta_{l0}}{ n }   \right]
+\frac{2N_c^2C_F}{n(2l+1)}+\frac{N_c^3}{4}\right\}
\\
\nonumber&&
-a_0^2n^2\left[5n^2+1-3l(l+1)\right]\left[
\left(\ln\frac{E_1^2}{T^2}+ 2\gamma_E -3 -\log 4- 2 \frac{\zeta^\prime(2)}{\zeta(2)} \right)\right.\\
\nonumber
&& \hspace{5cm}\left.\times\frac{C_F}{6} \als T m_D^2+\frac{4\pi}{9} \ln 2 \; N_c C_F \,  \als^2\, T^3 \right]
\\
&&
+\frac{8}{3}\cf\als\, Tm_D^2\,a_0^2n^4
\,I_{n,l}\;,
\label{finalwidth}
\end{eqnarray}
where $I_{n,l}$ is defined in Eq.~\eqref{defin}. We remark that, up to the
order considered here, the thermal contribution to the spectrum and to
the width is independent of the spin. The spin-orbit contribution will be computed in the following Chapter.

Our results are expected to be relevant for the ground states of
bottomonium ($\Upsilon (1S)$ and $\eta_b$), and to a lesser extent to
those of charmonium ($J/\psi$ and $\eta_c$), for a certain range of
temperatures in the quark-gluon plasma for which (\ref{hierarchy}) is fulfilled, as we discussed in the introduction to this Chapter. Let us now try to figure out what our results imply for 
the electromagnetic decays to lepton pairs (vector states) or to two photons (pseudoscalars). First of all, the masses of
the heavy quarkonium states increase quadratically with the
temperature at leading order (first line of (\ref{finalspectrum})),
which would translate into the same functional increase in the
energy of the outgoing leptons and photons if produced by the quarkonium in the plasma. 
Second, since electromagnetic decays occur at short distances ($\sim 1/m \ll 1/T$), 
the standard NRQCD factorization formulas hold, and, at leading order, 
all the temperature dependence is encoded in the wave function
at the origin. The leading temperature correction to it comes from
first-order quantum-mechanical perturbation theory of the first term
of (\ref{totalpotT}). The size of this correction is $\sim n^4 T^2/(m^2\als)$. 
Hence, a quadratic dependence on the temperature should
also be observed in the frequency in which leptons or photons are
produced by the quarkonium in the plasma. However, due to the very short lifetime of the deconfined medium (up to $\sim10$ fm/c) compared to the inverse of the electromagnetic decay width in the vacuum ($\Gamma(\Upsilon(1S))\to e^+e^-=1.340\pm0.018$ keV $\approx\Gamma(\Upsilon(1S))\to \mu^+\mu^-$ \cite{Nakamura:2010zzi}), an overwhelmingly vast majority of electromagnetic decays happens after the hot medium has cooled down and its light constituents have hadronized, hence an experimental observation of this mass shift and production rate change does not appear to be possible in current experiments.\\

On the other hand, at leading order, a decay width linear with
temperature is developed (first line of (\ref{finalwidth})).
As we have discussed after Eq.~\eqref{totaltranssym}, this effect corresponds to a dissociation to an unbound colour-octet state due to the interaction with a sufficiently energetic, on-shell thermal gluon. This is related to the gluo-dissociation phenomenology \cite{Kharzeev:1994pz,Xu:1995eb,Rapp:2008tf}: as we mentioned, these calculations rely on convolutions of the $T=0$ $g+Q\overline{Q}$ dissociation cross section computed by Bhanot and Peskin \cite{Peskin:1979va,Bhanot:1979vb} with a thermal distribution for the gluon. In Appendix~\ref{peskin} we show how our EFT framework is equivalent to this gluo-dissociation approach and how we calculate the cross section~\eqref{crossEFT}, including the octet potential in the final state, which had been neglected by Bhanot and Peskin. Our approach furthermore clarifies the region of validity of the thermal gluo-dissociation approach, which is from temperature $T\sim E$ up to $m\als\simg T$.\\
A second mechanism, Landau damping, contributes to the thermal width. The last three lines of Eq.~\eqref{finalwidth} encode its contribution, which is parametrically of order $\als m_D^2 Tr^2$. In our power counting it is suppressed by a factor of $m_D^2/E^2$ with respect to the leading singlet-to-octet decay contribution, given by the first line of Eq.~\eqref{finalwidth}.\\
For what concerns the phenomenological implications, we observe that in the absence of screening it is this width that would cause a suppression of bound states satisfying the hierarchy we have assumed. So one expects that the pseudoscalar and vector ground states would have a tendency, linear with the temperature, to decay to the continuum of colour-octet states. A phenomenological analysis for the $\Upsilon(1S)$, based on the results presented here, appears certainly very important, also in the light of the recent LHC results mentioned in Sec.~\ref{sec_exp_quarkonium}.


		\chapter{Poincar\'e invariance and the spin-orbit potential}
\label{chap_poincare}
In the second half of Chapter~\ref{chap_EFT} we introduced the framework of non-relativistic EFTs of QCD, emphasizing how Poincar\'e invariance is realized in these EFTs through a set of relations between the matching coefficients. Eqs.~\eqref{poincnrqcd} and \eqref{poincpotentials} represent two examples of such relations.\\
In this Part we are analyzing these NR EFTs at finite temperature; in this context a thermal bath at equilibrium clearly introduces a preferred frame of reference, the one where the bath itself is at rest. One would then expect as a result that Poincar\'e invariance is broken, at the very least for what concerns Lorentz boosts. This should in turn reflect itself on the aforementioned relations in the EFTs.\\
In this Chapter we then concentrate on the spin-orbit potential. In zero-temperature pNRQCD it is part of $V^{(2)}$, i.e. it appears as a $1/m^2$ relativistic correction (see App.~\ref{app_potentials}). Its center-of-mass-momentum dependent part is related to the static potential $V^{(0)}$ by the so-called \emph{Gromes relation} \cite{Gromes:1984ma}, as we mentioned in Sec.~\ref{sec_pnrqcd}. We will now evaluate the temperature-dependent leading contribution to the spin-orbit potential assuming the following hierarchy:
\begin{equation}
m\gg m\als\gg T \gg m\als^2\,.
\label{poinchierarchy}
\end{equation}
Moreover, we will assume that all other thermodynamical scales as well as the typical hadronic scale, 
$\lQ$, are either of the same order as or smaller than the binding energy. With respect to the hierarchy of Chap.~\ref{chap_rggT}, there is then more freedom regarding the position of the Debye mass. The general considerations made in that Chapter regarding the relevance for bottomonium phenomenology still hold.\\
As in the previous Chapter, integrating out the first two scales, $m$ and $m\als$, leads to NRQCD and then pNRQCD. In Sec. \eqref{secpnrqcd} we briefly review how Poincar\'e invariance is realized in the spin-orbit sector in the latter EFT at zero temperature.\\
As a next step we integrate out the temperature $T$ from pNRQCD, obtaining again
$\mathrm{pNRQCD}_{\rm HTL}$, which has been introduced in Sec.~\ref{secT}.
Within this EFT we proceed to the calculation of the leading thermal contribution to the spin-orbit potential. In particular, in Sec.~\ref{secpnrqcdhtl} we calculate the part that depends on the center-of-mass momentum, and in Sec.~\ref{secpoi} we show that it violates the Gromes relation. In Sec.~\ref{secspinorb} we compute instead the relative-momentum-dependent part. In the conclusions of this Chapter we summarize the results and highlight the consequences for the phenomenology of bound states moving through the medium, as is the case in heavy-ion collision experiments. The results of this Chapter have been published in \cite{poincare}.

\section{The Gromes relation in pNRQCD}
\label{secpnrqcd}
As discussed in Sec.~\ref{secm} in the previous Chapter, integrating out the mass and the momentum transfer scale $m\als$ in the presence of a medium with temperature $T$ much smaller that $m\als$ leads to weakly-coupled pNRQCD, which has been introduced in Sec.~\ref{sub_weakpnrqcd}. There, we mentioned that the EFT is equivalent order by order to a fully Poincar\'e invariant theory (QCD). Therefore Poincar\'e invariance reflects itself in exact relations between the matching coefficients of the theory. By imposing the Poincar\'e algebra on the generators of Poincar\'e transformations in the EFT one can obtain such relations; this program was carried out in Ref.~\cite{Brambilla:2003nt}. One of these exact relations is indeed the Gromes relation, first derived by Gromes in Ref.~\cite{Gromes:1984ma} in the context of transformation properties of Wilson loops under Lorentz boosts (see also \cite{Brambilla:2001xk}).\\
We now set out to introduce this relation and some of the needed tools for the subsequent finite-temperature calculation. When we introduced the standard form of the pNRQCD Lagrangian in Eq.~\eqref{lagrpnrqcd}, we remarked that it contained all the operators needed for a calculation of the spectrum to order $m\als^5$. In the previous Chapter we saw that up to that order no thermal, spin-dependent contribution arises. We thus need to introduce higher-order operators in the multipole and $1/m$ expansions. To this end, we adopt the notation of \cite{Brambilla:2003nt}, which differs from the more standard one used in the rest of this thesis, but allows an easier bookkeeping. We now write the pNRQCD Lagrangian as
\begin{eqnarray}
&& \hspace{-8mm} {\mathcal L}_{\rm pNRQCD} = \int d^3r \, {\rm Tr} \,\left\{
  {\rm S}^\dagger \left( i\partial_0 - h_s \right) {\rm S}
+ {\rm O}^\dagger \left( iD_0 - h_o \right) {\rm O}
\right.
\nn\\
&&\qquad
- \left[ ({\rm S}^\dagger h_{so} {\rm O} + {\rm H.C.}) + {\rm C.C.} \right] \;
- \left[ {\rm O}^\dagger h_{oo} {\rm O} + {\rm C.C.}  \right]  \;
\nn \\
&&\qquad
\left.
 - \left[ {\rm O}^\dagger h_{oo}^A {\rm O}  h_{oo}^B + {\rm C.C.}  \right]
\right\}
- \frac{1}{4} F^a_{\mu \nu} F^{a\,\mu \nu}+ \sum_{i=1}^{n_f}\bar{q}_i\,iD\!\!\!\!/\,q_i\,,
\label{pnrpoinclag}
\end{eqnarray}
where C.C. stands for charge conjugation and H.C for Hermitean conjugation. The explicit form of the singlet-octet interaction term $h_{so}$, organized in the $1/m$ and multipole expansions, can be found in App.~\ref{app_multipole}. The various $h_{oo}$ terms can be read from \cite{Brambilla:2003nt}.\\
A very important remark for the following is that we assume to be in the laboratory reference frame, which we define as the frame where an infinitely heavy quarkonium would be at rest.

For what concerns the power counting, we observe that the center-of-mass momentum $\bpg$ appearing in the singlet and octet Hamiltonians $h_s$ and $h_o$ (see Eq. \eqref{sinoctham}) can be as large as $T$ (think about the $Q\overline{Q}$ state recoiling after interacting with a thermal gluon), contrarily to the zero-temperature case where it is assigned the size $m\als^2$. \\
The explicit form of the non-static potentials $V^{(1)}_s$ and $V^{(2)}_s$ appearing in $h_s$ and $h_o$ can be read from
App.~\ref{app_potentials}. In particular, let us consider the colour-singlet, $V_{LS\,s}$, and 
the colour octet, $V_{LS\,o}$, spin-orbit potentials, which
are part of $V^{(2)}_s$ and $V^{(2)}_o$ respectively. 
They can be conveniently split into a part that
depends on the center-of-mass momentum $\bpg$ and a part that depends
on the relative momentum $\bp$:
\begin{eqnarray}
V_{LS\,s}&=& \frac{({\bf r}\times {\bf P}) \cdot (\bfsigma^{(1)} - \bfsigma^{(2)})}{4 m^2} V_{LS\,s a}(r)
+ \frac{({\bf r}\times {\bf p}) \cdot (\bfsigma^{(1)} + \bfsigma^{(2)})}{2m^2} V_{LS\,s b}(r),
\label{defvls}
\\
V_{LS\,o}&=& \frac{({\bf r}\times {\bf P}) \cdot (\bfsigma^{(1)} - \bfsigma^{(2)})}{4 m^2} V_{LS\,o a}(r)
+ \frac{({\bf r}\times {\bf p}) \cdot (\bfsigma^{(1)} + \bfsigma^{(2)})}{2m^2} V_{LS\,o b}(r),
\label{defvlso}
\end{eqnarray}
where the notation differs from the more common one used in App.~\ref{app_potentials}. $V^{(2)}_{LS,\mathrm{CM}}$ there corresponds to $V_{LS\,s a}$ here, $V^{(2)}_{LS}$ to $V_{LS\,s b}$ and similarly for the octet. At leading order, it holds that (see Eqs.~\eqref{V2}, \eqref{Dsten2} and \eqref{poincpotentials})
\begin{equation}
V_{LS\,s a}(r) \! = \! -\frac{\cf}{2}\frac{\als}{r^3},\quad
V_{LS\,s b}(r) \! = \! \frac{3\cf}{2}\frac{\als}{r^3},\quad 
V_{LS\,o a}(r) \! = \! \frac{1}{4\nc}\frac{\als}{r^3},\quad
V_{LS\,o b}(r) \! = \! -\frac{3}{4\nc}\frac{\als}{r^3},
\label{spinorbitlo}
\end{equation}
which implies that $V_{LS\,s a}\sim V_{LS\,s b}\sim V_{LS\,o a}\sim V_{LS\,o b} \sim m^3\als^4$.

Poincar\'e invariance links the $\bpg$-dependent spin-orbit potentials $V_{LS\,s a}$ and $V_{LS\,o a}$ to the derivatives of the static potentials $V_s^{(0)}$ and $V^{(0)}_o$. In detail \cite{Brambilla:2003nt}
\begin{equation}
V_{LS\,sa}(r)=-\frac{V_s^{(0)}(r)^\prime}{2r}\,,\qquad  V_{LS\,oa}(r)=-\frac{V_o^{(0)}(r)^\prime}{2r}\,,
\label{gromes}
\end{equation}
where $f(r)'\equiv df(r)/dr$. The first relation is the aforementioned Gromes relation. We now set out to compute the leading thermal correction to $V_{LS\,sa}(r)$ and prove that it violates this relation.
\section{The $\bpg$-dependent spin-orbit potential in pNRQCD$_{\rm HTL}$}
\label{secpnrqcdhtl}
The EFT that results from integrating out the temperature scale from pNRQCD, under the assumptions $m\als\gg T\gg m\als^2$, is pNRQCD$_{\rm HTL}$, which has been introduced in Sec.~\ref{secT} in the previous Chapter. The contribution to the singlet potential from the scale $T$ is encoded in the matching coefficient $\delta V_s$ appearing in the Lagrangian \eqref{pNRQCDHTL} of this EFT. An important remark is that we will assume the thermal bath to be a quark-gluon plasma \emph{at rest with respect to the 
laboratory reference frame}. This implies, in particular, that the Bose--Einstein distribution,  which describes the distribution of the gluons in the bath, depends only on their energy, i.e. the propagators are those that have been used throughout this Part.\\
In Sec.~\ref{secT} we computed all the contributions to $\delta V_s$ necessary for the computation of the spectrum to order $m\als^5$ in the hierarchy assumed there; as we remarked, they do not depend on the hierarchy (if any) of the scales smaller than $T$, provided of course that $T\gg m\als^2$, so they apply in our case without any modification. These contributions are summarized in Eq. \eqref{totalpotT} and are a sum of static, $1/m$- and $1/m^2$-proportional terms. Up to that order, no spin-dependent corrections are relevant. Since our goal is to compute the leading thermal correction to $\delta V_{LS\,sa}$ and test it agains the Gromes relation \eqref{gromes}, we need the leading correction to the static potential as well. This has been computed in the previous Chapter in the so-called $\delta V_s^{(\text{linear})}$ term, given in Eq.~\eqref{deltavslinear}. Its static part reads
\begin{eqnarray}
\delta V_s^{(0)}(r) &=&
\frac{2\pi}{9}\,\cf\als\left(V_{so}^{(0,1)}(r)\right)^2 T^2r^2\left(V^{(0)}_o(r)-V^{(0)}_s(r)\right)\,,
\label{leadingstatic}
\\
\delta Z_s(r) &=& 
-\frac{\pi}{9}\,\cf\als \left(V_{so}^{(0,1)}(r)\right)^2 T^2 r^2 \,,
\label{leadingZs}
\end{eqnarray}
where $V^{(0)}_o(r)-V^{(0)}_s(r) = \Delta V =\nc\als/(2r)$. The normalization factor $\delta Z_s(r)$ has been included for future convenience. With respect to Eqs.~\eqref{deltavslinear} and \eqref{deltazslinear} we have reinstated the matching coefficient $V_{so}^{(0,1)}(r)=1+\order{\als^2}$ of the chromoelectric dipole operator. In the notation of this Chapter $V_{so}^{(0,1)}(r)$ corresponds to $V_A(r)$ in the standard notation of pNRQCD (see App. \ref{app_multipole}). Its unexpanded inclusion amounts to resumming all contributions from the scale $m\als$.\\
The power counting of pNRQCD and Eq.~\eqref{poinchierarchy} give the size of 
$\delta V_s^{(0)}$: $m\als^3 \gg \delta V_s^{(0)} \sim \als^2T^2r\gg m \als^5$. We notice that the upper limit $m\als^3$ is larger by a factor of $g^2$ with respect to the upper limit $mg^8$ for this term in the previous Chapter. This is a result of the laxer hierarchy adopted now.\\

We now set out to compute the leading spin-orbit terms in $\delta V_s$. 
In particular, we will compute the leading thermal correction, $\delta V_{LS\,s a}$, 
to the center-of-mass momentum dependent spin-orbit potential $V_{LS\,s a}$, 
defined in Eq.~\eqref{defvls}.
The computation follows the same line as the one carried out in Sec.~\ref{secT} for the spin-independent terms in $\delta V_s$: we calculate thermal spin-dependent corrections to the pNRQCD singlet propagator, 
and match it to the singlet propagator in $\mathrm{pNRQCD}_{\rm HTL}$.

We identify the following set of contributions to $\delta V_{LS\,s a}$:
\begin{equation}
\delta V_{LS\,s a}=\delta V_{LS,a}+\delta V_{LS,b}+\delta V_{LS,c}+\delta V_{LS,d}+\delta V_{LS,e}
\,,
\label{sumP}
\end{equation}
where 
\begin{itemize}
\item[(1)]{$\delta V_{LS,a}$ comes from inserting a $T=0$ spin-orbit potential 
in the singlet or octet propagators of the diagram in Fig.~\ref{fig:hqself};} 
\item[(2)]{$\delta V_{LS,b}$ comes from replacing one of the two chromoelectric dipole vertices 
in Fig.~\ref{fig:hqself} with the chromomagnetic vertex proportional 
to $c_FV_{so\,b}^{(1,0)}$ in Eq.~\eqref{1/m} and inserting a center-of-mass kinetic energy 
into the octet propagator;} 
\item[(3)]{$\delta V_{LS,c}$ comes from replacing one of the chromoelectric dipole vertices 
in Fig.~\ref{fig:hqself} with the vertex proportional to  $c_sV_{so\,a}^{(2,0)}$ in Eq.~\eqref{1/m2};}
\item[(4)]{$\delta V_{LS,d}$ comes from replacing one of the chromoelectric dipole vertices 
in  Fig.~\ref{fig:hqself} with the vertex proportional to  $V_{so\,b''}^{(2,0)}$ in Eq.~\eqref{1/m2};} 
\item[(5)]{$\delta V_{LS,e}$ comes from replacing one of the chromoelectric dipole vertices 
in  Fig.~\ref{fig:hqself} with the vertex proportional to  $c_FV_{so\,b}^{(1,0)}$ in Eq.~\eqref{1/m}  
and the other one with the vertex given by $h_{so}^{(1,1)}$ in Eq.~\eqref{1/m1m2}.} 
\end{itemize}
By explicit inspection, one sees that diagrams with vertices given by the terms 
proportional to $V^{(1,0)}_{so\,c}$, $V_{so\,b'}^{(2,0)}$ and $V_{so\,b'''}^{(2,0)}(r)$ 
in Eqs.~\eqref{1/m} and~\eqref{1/m2}, albeit spin-dependent, do not contribute 
to the spin-orbit potential.

\begin{figure}[ht]
\begin{center}
\includegraphics{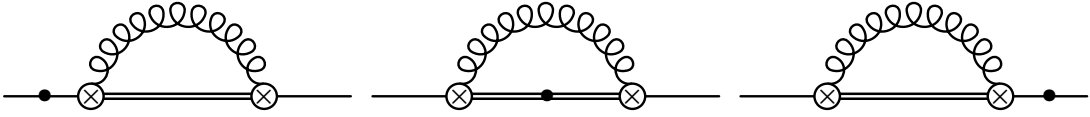}
\end{center}
\caption{Diagrams contributing to $\delta V_{LS,a}$. 
The dot stands for an insertion of the spin-orbit potential proportional to $V_{LS\,sa}$
(left and right diagram) or to $V_{LS\,oa}$ (middle diagram), 
all other symbols are as in Fig.~\ref{fig:hqself}.}
\label{diagram1}
\end{figure}

\subsubsection{Evaluation of $\delta V_{LS,a}$}
We evaluate the diagrams in Fig.~\ref{diagram1}.
As in the previous calculation in Sec.~\ref{secT} for the spin-independent terms, 
we expand the octet propagators for $k_0\gg E-h_o^{(0)}$ (see Eq.~\eqref{octetexpand}). 
The leading contribution comes again from the linear term (see Sec.~\ref{sub_linear} for the evaluation of the loop integral, which is the same as in Eq. \eqref{deflinear}) that we treat by means 
of the identities~\eqref{commute1} and \eqref{commutexample}. The left diagram of Fig.~\ref{diagram1} gives
\begin{eqnarray}
&& \hspace{-9mm}
-\frac{2\pi}{9}C_F\als T^2
\frac{i}{E-h_s^{(0)}} 
\left(V_{so}^{(0,1)}(r)\right)^2
\left[ \frac{1}{2}\left\{r^2,E-h^{(0)}_s \right\} - \left(V_o^{(0)}(r)-V_s^{(0)}(r)\right)r^2 \right]
\nonumber\\
&& \times
\frac{1}{E-h_s^{(0)}} \frac{({\bf r}\times {\bf P}) \cdot (\bfsigma^{(1)} - \bfsigma^{(2)})}{4 m^2} V_{LS\,s a}(r)
\frac{1}{E-h_s^{(0)}}\,,
\label{left}
\end{eqnarray}
the right one gives
\begin{eqnarray}
&& \hspace{-9mm}
-\frac{2\pi}{9}C_F\als T^2
\frac{i}{E-h_s^{(0)}} \frac{({\bf r}\times {\bf P}) \cdot (\bfsigma^{(1)} - \bfsigma^{(2)})}{4 m^2} V_{LS\,s a}(r)
\frac{1}{E-h_s^{(0)}} 
\nonumber\\
&& \times
 \left(V_{so}^{(0,1)}(r)\right)^2
\left[ \frac{1}{2}\left\{r^2,E-h^{(0)}_s \right\} - \left(V_o^{(0)}(r)-V_s^{(0)}(r)\right)r^2 \right] 
\frac{1}{E-h_s^{(0)}}\,,
\label{right}
\end{eqnarray}
and the middle one gives
\begin{eqnarray}
&& \hspace{-9mm}
\frac{2\pi}{9}C_F\als T^2
\frac{i}{E-h_s^{(0)}} 
 \left(V_{so}^{(0,1)}(r)\right)^2 r^2
\frac{({\bf r}\times {\bf P}) \cdot (\bfsigma^{(1)} - \bfsigma^{(2)})}{4 m^2} V_{LS\,o a}(r)
\frac{1}{E-h_s^{(0)}}\,,
\nonumber\\
\label{middle}
\end{eqnarray}
where we have kept only terms relevant at order $1/m^2$.
Matching to the $\textrm{pNRQCD}_\mathrm{HTL}$ propagator~\eqref{matchingHTL}, 
expanded according to~\eqref{hqpropsozero}, we observe that 
the terms proportional to $(V_o^{(0)}-V_s^{(0)})r^2$ in~\eqref{left} and~\eqref{right} 
cancel against one insertion of $\delta V_s^{(0)}(r)$ and one of the spin-orbit potential in 
the  $\textrm{pNRQCD}_\mathrm{HTL}$ propagator, while the term proportional to 
$(E-h^{(0)}_s) r^2/2$ in~\eqref{left} and the one proportional to 
$r^2 (E-h^{(0)}_s)/2$ in~\eqref{right} cancel against the term 
$\left\{ \delta Z_s, i/(E-h^{(0)}_s)\right.$ 
$\times [\hbox{spin-orbit potential}]$ $ \times$ $\left.1/(E-h^{(0)}_s) \right\}$
in ~\eqref{matchingHTL}. The expression of $\delta Z_s$ can be read from Eq.~\eqref{leadingZs}.
What is left gives the leading-order thermal correction, coming from the diagrams in Fig.~\ref{diagram1}, 
to the center-of-mass  momentum dependent spin-orbit potential:
\begin{equation}
\delta V_{LS,a}(r)=\frac{2\pi}{9}C_F\als \left(V_{so}^{(0,1)}(r)\right)^2 T^2r^2\left(V_{LS\,oa}(r)-V_{LS\,sa}(r)\right)\,.
\label{aspinorbit}
\end{equation}
According to the power counting of pNRQCD and Eq.~\eqref{poinchierarchy}, we have that 
$m\als^5\gg \delta V_{LS,a}\,({\bf r}\times {\bf P})\cdot
(\bfsigma^{(1)} - \bfsigma^{(2)})/m^2 \gg m \als^8$.

\begin{figure}[ht]
	\begin{center}
	\includegraphics[width=12cm]{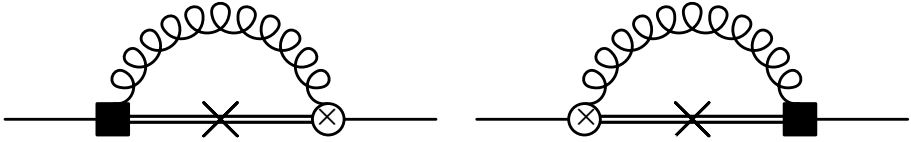} 
	\put(-128,13){$\boldsymbol\times$}
	\put(-225,13){$\boldsymbol\times$}
\end{center}
\caption{Diagrams contributing to $\delta V_{LS,b}$. 
The square stands for the chromomagnetic vertex proportional to $c_FV_{so\,b}^{(1,0)}(r)$, 
the cross for a center-of-mass kinetic energy insertion, and 
all other symbols are as in Fig.~\ref{fig:hqself}.}
\label{diagram3}
\end{figure}

\subsubsection{Evaluation of $\delta V_{LS,b}$}
We evaluate the diagrams in Fig.~\ref{diagram3}. Their thermal contribution to the amplitude reads 
\begin{eqnarray}
\Sigma_s(E)^{\hbox{\tiny Fig.\,\ref{diagram3}}}&=&
- 2 ig^2\cf V_{so}^{(0,1)}(r)V_{so\,b}^{(1,0)}(r)\frac{c_F}{2m}
\nonumber\\
&& \hspace{-1cm}\times
r^i \int\frac{\,d^4k}{(2\pi)^4}
\frac{i}{E-h_o^{(0)}-k_0+i\eta}
\frac{({\bf P}-{\bf k})^2}{4m}
\frac{1}{E-h_o^{(0)}-k_0+i\eta}
\nonumber\\
&&\hspace{-1cm}\times  
k_0\,\epsilon^{jln} k^l 
\left(\delta^{ni}-\frac{k^nk^i}{\bk^2}\right)\,2\pi\delta(k_0^2-\bk^2)
n_\mathrm{B}(\vert k_0\vert)\,
(\sigma^{(1)\,j}-\sigma^{(2)\,j})
\,,
\label{sigmab}
\end{eqnarray}
while the non-thermal part vanishes if regularized in dimensional regularization.
The factor $2$ follows from the fact that the two diagrams give the same contribution 
at order $1/m^2$. The octet propagators may be expanded according to Eq.~\eqref{octetexpand}; 
considering that, besides the two octet propagators, the integral in~\eqref{sigmab} is 
odd in $k_0$, the leading non-vanishing term coming from their expansion is $-2(E-h_o^{(0)})/(-k_0+i\eta)^3$.
The factor $E-h_o^{(0)}$ contains a part, $E-h_s^{(0)}$, that contributes to the singlet normalization, 
and a part, $V_s^{(0)}-V_o^{(0)}$, that contributes to the spin-orbit potential.
The octet center-of-mass kinetic energy, ${({\bf P}-{\bf k})^2}/(4m)$, contributes to the 
spin-orbit potential only through the term  $- {\bf P}\cdot{\bf k}/(2m)$.
With this in mind, we perform the matching analogously to what has been done in Eq.~\eqref{matchcond} and obtain
the leading-order thermal correction, coming from the diagrams in Fig.~\ref{diagram3}, 
to the center-of-mass  momentum dependent spin-orbit potential:
\begin{equation}
\delta V_{LS,b}(r) = 
-\frac{4\pi}{9}C_F \als  c_F V_{so}^{(0,1)}(r) V_{so\,b}^{(1,0)}(r) T^2 \left(V^{(0)}_o(r)-V^{(0)}_s(r)\right)\,.
\label{bspinorbit}
\end{equation}
Considering that the matching coefficients $c_F$, $V_{so}^{(0,1)}$ and  $V_{so\,b}^{(1,0)}$ are one 
at leading order (see App. \ref{app_multipole}), the size of the correction is 
$m\als^5\gg \delta V_{LS,b}\,({\bf r}\times {\bf P})\cdot (\bfsigma^{(1)} - \bfsigma^{(2)})/m^2 \gg m\als^8$.

\subsubsection{Evaluation of $\delta V_{LS,c}$}
\label{subsecc}
The calculation of $\delta V_{LS,c}$ is at this point simple: 
there are two contributing diagrams, which may be constructed 
by replacing one of the chromoelectric dipole vertices 
in Fig.~\ref{fig:hqself} with the vertex proportional 
to  $c_sV_{so\,a}^{(2,0)}$ in Eq.~\eqref{1/m2}.
Since this vertex contains a chromoelectric field as well, the integration is exactly the
same as the one performed in Eqs.~\eqref{transverseleading} and \eqref{deflinear}. The only
change is in the prefactor of the integral. Matching to the 
$\textrm{pNRQCD}_\mathrm{HTL}$ propagator, we obtain at leading order 
\begin{equation}
\delta V_{LS,c}(r) = \frac{2\pi}{9}C_F\als c_sV_{so}^{(0,1)}(r) V_{so\,a}^{(2,0)}(r) T^2\left(V^{(0)}_o(r)-V^{(0)}_s(r)\right)\,,
\label{cspinorbit}
\end{equation}
which, considering that  $c_s$, $V_{so}^{(0,1)}$ and $V_{so\,a}^{(2,0)}$ are one at leading order, 
has the same size as $\delta V_{LS,b}$.

\subsubsection{Evaluation of $\delta V_{LS,d}$}
The calculation of $\delta V_{LS,d}$ is similar to this last one, but with the vertices 
proportional to  $c_sV_{so\,a}^{(2,0)}$ replaced by the ones proportional to $V_{so\,b''}^{(2,0)}$ in Eq.~\eqref{1/m2}.
The leading-order result reads
\begin{equation}
\delta V_{LS,d}(r) = \frac{2\pi}{9}C_F\als V_{so}^{(0,1)}(r) V_{so\,b''}^{(2,0)}(r) T^2 \left(V^{(0)}_o(r)-V^{(0)}_s(r)\right)\,.
\label{dspinorbit}
\end{equation}
Considering that $V_{so}^{(0,1)}$ is one at leading order, but that $V_{so\,b''}^{(2,0)}$ is 
at least of order $\als^3$, as was shown at the end of App.~\ref{app_multipole}, $\delta V_{LS,d}(r)$ is suppressed with respect to  
$\delta V_{LS,a}$, $\delta V_{LS,b}$ and $\delta V_{LS,c}$ by, at least, a factor $\als^3$.

\begin{figure}[ht]
	\begin{center}
	\includegraphics[width=12cm]{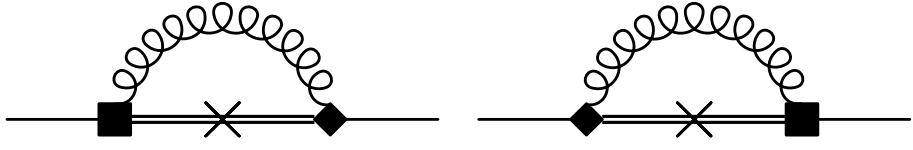} 
\end{center}
\caption{Diagrams contributing to $\delta V_{LS,e}$. 
The diamond stands for the vertex $h_{so}^{(1,1)}$, given in Eq.~\eqref{1/m1m2}, 
and all other symbols are as in Fig.~\ref{diagram3}.}
\label{diagram4}
\end{figure}

\subsubsection{Evaluation of $\delta V_{LS,e}$}
We evaluate the diagrams in Fig.~\ref{diagram4}. Their thermal contribution to the amplitude reads 
\begin{eqnarray}
\Sigma_s(E)^{\hbox{\tiny Fig.\,\ref{diagram4}}}&=&
 2ig^2\cf V_{so}^{(1,1)}(r)V_{so\,b}^{(1,0)}(r)\frac{c_F}{2m}
\left( \frac{({\bf P}\times{\bf r})^i}{2m} \right)
\int\frac{\,d^4k}{(2\pi)^4}
\frac{i}{E-h_o^{(0)}-k_0+i\eta}
\nonumber\\
&& \hspace{-1.7cm}\times
(ik^l)\epsilon^{jlr} \, \epsilon^{ins} (-ik^n) 
\left(\delta^{rs}-\frac{k^rk^s}{\bk^2}\right)\,2\pi\delta(k_0^2-\bk^2)
n_\mathrm{B}(\vert k_0\vert)\,
(\sigma^{(1)\,j}-\sigma^{(2)\,j})
\,,
\label{sigmae}
\end{eqnarray}
while the non-thermal part vanishes if regularized in dimensional regularization.
The factor $2$ follows from the fact that the two diagrams give the same contribution 
at order $1/m^2$. The octet propagators may be expanded according to Eq.~\eqref{octetexpand}: 
the linear term in $E-h_o^{(0)}$ contains a part, $E-h_s^{(0)}$, that contributes to the singlet normalization, 
and a part, $V_s^{(0)}-V_o^{(0)}$, that contributes to the spin-orbit potential.
This last contribution reads
\begin{equation}
\delta V_{LS,e}(r) = 
\frac{4\pi}{9}C_F \als  c_F V_{so}^{(1,1)}(r) V_{so\,b}^{(1,0)}(r) T^2 \left(V^{(0)}_o(r)-V^{(0)}_s(r)\right)\,.
\label{espinorbit}
\end{equation}
Considering that, according to~\eqref{poincare11}, 
the matching coefficient $V_{so}^{(1,1)}$  is equal to $V_{so}^{(0,1)}$, 
$\delta V_{LS,e}$ exactly cancels with $\delta V_{LS,b}$ in the sum~\eqref{sumP}.

\subsubsection{Summary}
In summary, the leading thermal correction to the center-of-mass momentum-depen\-dent 
spin-orbit potential,  
\begin{equation}
\delta V_{LS\,s}|_{{\bf P}\hbox{\tiny -dependent}} = 
\frac{({\bf r}\times {\bf P}) \cdot (\bfsigma^{(1)} - \bfsigma^{(2)})}{4 m^2} \delta V_{LS\,s a}(r)\,,
\end{equation}
is the sum of Eqs.~\eqref{aspinorbit},~\eqref{bspinorbit},~\eqref{cspinorbit},~\eqref{dspinorbit}
and ~\eqref{espinorbit}; it reads:
\begin{eqnarray}
\hspace{-7mm}
\delta V_{LS\,s a}(r) &=& \frac{2\pi}{9}C_F\als V_{so}^{(0,1)}(r) T^2
\left\{
V_{so}^{(0,1)}(r) r^2\left(V_{LS\,oa}(r)-V_{LS\,sa}(r)\right) \right.
\nonumber\\
&& 
\left. +
\left[ c_sV_{so\,a}^{(2,0)}(r) + V_{so\,b''}^{(2,0)}(r)       \right] 
\left(V^{(0)}_o(r)-V^{(0)}_s(r)\right) 
\right\}
\nonumber\\
&=& \frac{\pi}{6}\cf N_c \frac{\als^2}{r}T^2 + \hbox{higher orders}\,.
\label{summaryLS}
\end{eqnarray}
In the first equality of~\eqref{summaryLS}, the matching coefficients of NRQCD and pNRQCD have been kept unexpanded; 
this amounts to having provided an expression for the spin-orbit potential that resums 
contributions from the scales $m$ and  $m\als$, while it is of leading order in the temperature.
In the second equality, we have kept only the leading terms in the NRQCD and pNRQCD matching coefficients.
We note that the contribution coming from the term proportional to $V_{so\,b''}^{(2,0)}$ is negligible, 
of the same size or smaller than subleading thermal corrections that we have neglected throughout this Chapter, such as those coming from higher-order contributions in the expansion \eqref{octetexpand} or from radiative corrections, as in Sec.~\ref{secT}.

\section{Gromes relation at finite temperature}
\label{secpoi}
After having computed the leading contributions to $\delta V_{LS\,sa}$, 
we can now check whether these new terms fulfill the Gromes relation~\eqref{gromes} or not. 
This corresponds to verifying the equality
\begin{equation}
\label{testgromes}
\delta V_{LS\,sa}(r) \stackrel{?}{=} - \frac{\delta V^{(0)}_{s}(r)^\prime}{2r} \,.
\end{equation}
We use the expression of $\delta V_{LS\,sa}$ provided by the first equality in Eq.~\eqref{summaryLS}
that keeps unexpanded  the matching coefficients of NRQCD and pNRQCD. 
If we make use of the relations ~\eqref{gromes} and~\eqref{gromesmatching1}, which are exact, 
then $\delta V_{LS\,sa}$ may be rewritten as 
\begin{eqnarray}
\delta V_{LS\,s a}(r) &=&
- \frac{\delta V^{(0)}_{s}(r)^\prime}{2r} 
\nonumber\\
&& 
+ \frac{2\pi}{9}C_F\als V_{so}^{(0,1)}(r) T^2
\left( c_sV_{so\,a}^{(2,0)}(r) + V_{so}^{(0,1)}(r)\right)
\left(V^{(0)}_o(r) - V^{(0)}_s(r)\right),\nn\\
\label{nocancel}
\end{eqnarray}
which shows that the Gromes relation is violated by an amount, which at leading order is  
$\displaystyle \frac{2\pi}{9}\cf N_c \frac{\als^2}{r}T^2$.

\subsection{The spin-orbit potential in a covariant model}
In order to understand the origin of the observed violation of Poincar\'e invariance, it 
is useful to consider at zero temperature the case of a massive gluon, whose mass, $m_g$, is such that 
$m\als \gg m_g \gg m\als^2$. The massive gluon contributes to the potential, but 
clearly it does not break  Poincar\'e invariance. To see this let us evaluate the 
corrections to the spin-orbit potential. 
The diagrams contributing to the spin-orbit potential are the same of those 
considered in the previous section, only now the gluon propagator reads (in the unitary gauge, i.e. the limit $\xi\to\infty$ of the $R_\xi$ gauges)
\begin{equation}
-\frac{i}{k_0^2-\bk^2-m_g^2+i\eta}\left(g_{\mu\nu} - \frac{k_\mu k_\nu}{m_g^2}\right).
\label{massive}
\end{equation}

The contributions to the static potential, $\delta V_s^{(0)}$, and 
 $\delta V_{LS,a}$, $\delta V_{LS,c}$ and $\delta V_{LS,d}$  depend on the 
correlator of two chromoelectric fields. They are proportional to 
\begin{equation}
\int\frac{\,d^Dk}{(2\pi)^D}\frac{i}{E-h_o^{(0)}-k_0+i\eta} 
\frac{1}{k_0^2-\bk^2-m_g^2+i\eta} \left[(D-1)\,k_0^2-\bk^2\right],
\label{massiveEE}
\end{equation} 
where we have regularized the integral in dimensional regularization.
Integrating over the momentum region $k_0, k \sim m_g$ means that we are expanding the octet 
propagator according to Eq.~\eqref{octetexpand}. Contrary to what happens in the previous finite-temperature calculation (see Eq.~\eqref{vanishzero}), the leading order term does not vanish, but yields instead \cite{Brambilla:1999xf}
\begin{equation}
	\delta V_s^{(0)}(r)=\frac{\cf}{6}\als r^2m_g^3\,
	\label{leadingmassive}
\end{equation}
which, as we remarked after Eq.~\eqref{VsmD}, corresponds to the result one obtains when evaluating this diagram with the HTL-resummed propagators.\\
Three terms contribute to the spin-orbit potential $\delta V_{LS\,sa}$ at the leading order in the $(m\als^2)/m_g$ expansion and in $\als$. The first is obtained from the diagrams given in Fig. \ref{diagram3}. When evaluated with the propagator \eqref{massive} and at the zeroth order in the expansion \eqref{octetexpand}, it gives
\begin{equation}
	\delta V_{LS,sa}^{(i)}(r)=\frac{\cf}{3}c_F\als m_g^3\,.
	\label{massivei}
\end{equation}
The second contribution is analogous to $\delta V_{LS,c}$ discussed in Sec. \ref{subsecc}, evaluated now with the massive propagator and at the zeroth order in the octet energy. It yields
\begin{equation}
	\delta V_{LS,sa}^{(ii)}(r)=\frac{\cf}{6}c_s\als m_g^3\,.
	\label{massiveii}
\end{equation}
The third and last contribution comes from the diagrams shown in Fig.~\ref{diagram4}, computed with the prescriptions of the previous two terms. It reads
\begin{equation}
	\delta V_{LS,sa}^{(iii)}(r)=-\frac{2}{3}\cf c_F\als m_g^3\,.
	\label{massiveiii}
\end{equation}
It is now easy to verify that the sum of Eqs.~\eqref{massivei}--\eqref{massiveiii} satisfies the Gromes relation. Using Eq.~\eqref{poincnrqcd}, i.e $2c_F-c_s-1=0$, implies that the Gromes relation is satisfied also when resumming the contribution from the scale $m$ (encoded in these NRQCD matching coefficients) to all orders.\\

We now focus on the term linear in 
$E-h_o^{(0)}$, which is the relevant term in the finite temperature case analyzed  
in the rest of the Chapter.
It turns out that the linear term vanishes in the static potential in dimensional regularization (see \cite{Brambilla:1999xf}). 
The reason is that the contribution coming from the spatial components 
of the gluon propagator (proportional to $(D-1)\,k_0^2$ in Eq.~\eqref{massiveEE})
cancels against the contribution coming from the temporal component (proportional to $\bk^2$ in the equation). 
This is in sharp contrast with the finite temperature case, where the term linear in $E-h_o^{(0)}$ does not vanish
(in Coulomb gauge, this is due to the fact that only the spatial components of the gluon propagator get thermal 
contributions) and eventually generate a finite thermal contribution 
to $\delta V_s^{(0)}$, $\delta V_{LS,a}$, $\delta V_{LS,c}$ and $\delta V_{LS,d}$ 
(see Eqs.~\eqref{deflinear},~\eqref{aspinorbit},~\eqref{cspinorbit} and~\eqref{dspinorbit}
respectively).

The contributions to  $\delta V_{LS,b}$ and $\delta V_{LS,e}$ depend on the spatial components 
of the gluon propagator only. Both $\delta V_{LS,b}$ and $\delta V_{LS,e}$ get finite 
contributions from the massive gluon but the sum of the two terms linear in $E-h_o^{(0)}$
vanishes: the same happens in the finite temperature case discussed in the previous section.

The massive gluon example provides a simple case where Poincar\'e invariance is not broken.
The Gromes relation is realized both for the leading-order terms, that do not depend on the energy,  and for those that are linear in the energy. In the latter case the realization is trivial: such terms vanish for both the static and the spin-orbit potentials.  
In the finite temperature case, diagrams that depend on the 
correlator of two chromoelectric fields,  like the one shown in Fig.~\ref{fig:hqself}, 
do not vanish. This is a direct consequence of the fact that the thermal bath 
affects in a non-covariant way the gluon propagator.

\section{Singlet spin-orbit potential $\delta V_{LS\,s b}$ in $\mathrm{pNRQCD}_{\rm HTL}$}
\label{secspinorb}
In this section, we calculate the leading thermal corrections to the spin-orbit potential 
$\delta V_{LS\,sb}$, which is the spin-orbit potential experienced by the quarkonium when at rest with 
respect to the laboratory reference frame (we recall that, in our setup, this is also the reference frame of the 
thermal bath). This potential, even at zero temperature, is not constrained by Poincar\'e invariance.

In order to calculate $\delta V_{LS\,sb}$, we need to consider two new terms contributing to $h_{so}$ 
in the pNRQCD Lagrangian~\eqref{pnrpoinclag}: the term 
\begin{equation}
- \frac{c_F}{4m} \, \bfsigma^{(1)}\cdot r^i (\partial_i\,g{\bf B})\,,
\label{so2}
\end{equation}
and the term
\begin{equation}
\frac{c_s}{8m^2} \, \bfsigma^{(1)}\cdot [{\bf p} \times, g {\bf E}]\,,
\label{so1}
\end{equation}
where, for simplicity, we have put to their tree-level values the pNRQCD matching coefficients.

There are three classes of diagrams that contribute: 
\begin{equation}
\delta V_{LS\,sb} = \delta V_{LS\,(i)} + \delta V_{LS\,(ii)} + \delta V_{LS\,(iii)}.
\end{equation}
\begin{itemize}
\item[(1)]{The first class is similar to the one shown in Fig.~\ref{diagram1}, but now the 
dots stand for insertions of the spin-orbit potential proportional to $V_{LS\,sb}$
(left and right diagram) or to $V_{LS\,ob}$ (middle diagram). $V_{LS\,sb}$ and 
$V_{LS\,ob}$ have been defined in Eqs.~\eqref{defvls} and~\eqref{defvlso}, 
and given at leading order in Eq.~\eqref{spinorbitlo}. The result reads
\begin{equation}
\delta V_{LS\,(i)}(r)=\frac{2\pi}{9}C_F\als T^2r^2\left(V_{LS\,ob}(r)-V_{LS\,sb}(r)\right)\,.
\label{aprimespinorbit}
\end{equation}
}
\item[(2)]{The second class is similar to the one shown in Fig.~\ref{diagram3}, 
but now the squares stand for the vertex induced by~\eqref{so2} and the cross for a 
kinetic energy insertion, ${\bf p}^2/m$. The result reads
\begin{equation}
\delta V_{LS\,(ii)}(r) = 
-\frac{4\pi}{9}C_F \als  c_F T^2 \left(V^{(0)}_o(r)-V^{(0)}_s(r)\right)\,.
\label{bprimespinorbit}
\end{equation}
}
\item[(3)]{Finally, the third class of diagrams is similar to the ones evaluated in 
Sec.~\ref{subsecc}, but with the vertex proportional to $c_sV_{so\,a}^{(2,0)}$ in Eq.~\eqref{1/m2} 
replaced by the vertex induced by~\eqref{so1}. The result reads 
\begin{equation}
\delta V_{LS\,(iii)}(r) = 
\frac{2\pi}{9}C_F \als  c_s T^2 \left(V^{(0)}_o(r)-V^{(0)}_s(r)\right)\,.
\label{cprimespinorbit}
\end{equation}
}
\end{itemize}

Summing up all three contributions we obtain
\begin{eqnarray}
\hspace{-7mm}
\delta V_{LS\,s b}(r) &=& \frac{2\pi}{9}C_F\als T^2
\left[r^2\left(V_{LS\,ob}(r)-V_{LS\,sb}(r)\right) 
- \left(V^{(0)}_o(r)-V^{(0)}_s(r)\right) 
\right]
\nonumber\\
&=& - \frac{5\pi}{18}\cf N_c \frac{\als^2}{r}T^2 + \hbox{higher orders}\,,
\label{summaryLScm}
\end{eqnarray}
where, in the first equality, we have used Eq.~\eqref{poincnrqcd}, i.e $2c_F-c_s-1=0$.

\section{Conclusions}
\label{conclusions_poinc}
We have calculated the leading-order thermal corrections to the quarkonium spin-orbit 
potentials. These corrections go quadratically with the temperature and are proportional to 
$\als^2T^2/r$.

At zero temperature, the spin-orbit potential that depends on the center-of-mass momentum is 
protected by Poincar\'e invariance. We have computed its leading thermal correction 
in Eq.~\eqref{summaryLS}. In Eq.~\eqref{nocancel}, this correction has been shown 
to violate Poincar\'e invariance. This implies that order $\als^2T^2/r$ corrections to the
quarkonium potential will be experienced by the system differently in different 
reference frames, and, in particular, in a frame where the thermal bath is not at rest. Conversely, in the frame where the thermal bath is at rest, these corrections will be experienced differently by quarkonia moving with different momenta $\bpg$.\\
This is in general a very interesting aspect in the study of quarkonium suppression. Experimenters measure the dependence of the nuclear modification factor $R_{AA}$ on the transverse momentum $p_\mathrm{T}$ of the outgoing quarkonium state, which is in turn related to its momentum $\bpg$. While a quantitative description of this dependence is certainly not immediate, also due to effects caused by the finite size and short lifetime of the medium, effects due to the explicit breaking of Poincar\'e invariance are certainly to be taken into account and the results obtained here represent a first step in this direction. We remark that an analysis of NR QED bound states moving in a thermal bath was recently carried out in an analogous EFT formalism in \cite{Escobedo:2011ie} and a first study on the dependence of lattice spectral function on the center-of-mass momentum can be found in \cite{Nonaka:2010zz}.

We have also computed the leading thermal correction to the spin-orbit potential 
of a quarkonium state at rest with respect to the laboratory reference frame.
Its expression is in Eq.~\eqref{summaryLScm}.
The potential contributes to the spin-orbit splittings of the quarkonium levels. 
The thermal correction having an opposite sign with respect to the leading, $T=0$ term (see Eq. \eqref{defvlso}) implies 
a weakening of the spin-orbit interaction in the medium. For the reasons discussed in the conclusions of Chap.~\ref{chap_rggT}, i.e. the decays to leptons happening much after the deconfined phase, this spectral shift does not appear to be experimentally observable in current experiments.

		\part{Imaginary-time Effective Field Theories of QCD at finite temperature for thermodynamical quantities}
		\label{part_imtime}
		\chapter{Introduction to the Polyakov loop and to the Polyakov-loop correlator}
\label{chap_imtimeintro}
In Chap.~\ref{chap_thermal} we gave in Eq.~\eqref{defploop} the definition of the Polyakov loop, and we mentioned how it is one of the observables measured on the lattice to determine the pseudocritical temperature of QCD at $\mu_B=0$. In the absence of light fermions, the Polyakov loop and the correlator of two Polyakov loops become
the order parameters of the deconfinement phase transition in SU$(\nc)$ gauge
theories \cite{Kuti:1980gh,McLerran:1981pb}. The phase transition is signaled
by a non-vanishing expectation value of the Polyakov loop and a qualitative
change in the large-distance behaviour of the correlation function (from
confining to exponentially screened) \cite{McLerran:1981pb}.
In the deconfined phase, these quantities  provide information about 
the electric screening and can be calculated at sufficiently high temperatures $T$
in perturbation theory. For the correlation function of
Polyakov loops, the validity of the perturbative expansion is limited to
distances $r$ smaller than the magnetic screening length $r\ll 1/(g^2 T)$ 
\cite{Rebhan:1994mx,Arnold:1995bh} for the issues mentioned in Sec.~\ref{sec_htl}. 

From a phenomenological perspective, the Polyakov-loop correlator is
interesting because it provides an insight into the in-medium modifications of the 
quark-antiquark interaction, as we shall show in Sec.~\ref{sec_introloops}.
Indeed, in-medium modified heavy-quark potentials, inspired also by the behaviour of the 
Polyakov-loop correlator, have been used since long time in 
potential models, as we discussed in Sec.~\ref{sec_exp_quarkonium}. 
However, although the spectral decomposition of the Polyakov-loop correlator 
is known, its relation with the finite-temperature heavy-quark potential is still a matter of debate 
and in need of a clarifying analysis \cite{Philipsen:2008qx}. 
The issue has become particularly relevant in the light of the in-medium heavy-quark potentials we have rigorously defined in real-time in Part~\ref{part_realtime}. One of the aims of this Part is then to discuss, in the weak-coupling regime, the relation between the Polyakov-loop correlator and these findings.

The Polyakov-loop correlator is a gauge-invariant quantity, 
hence it is well suited for lattice calculations.
In fact, the correlator of two Polyakov loops has be calculated on the lattice for the pure gauge
theory \cite{Kaczmarek:1999mm,Petreczky:2001pd,Digal:2003jc,Bazavov:2008rw} 
as well as for full QCD \cite{Karsch:2000kv,Petreczky:2004pz}
(for a review see Ref.~\cite{Petreczky:2005bd}). As we mentioned, such calculations are often used as input for the form of the potential in potential models.
Surprisingly, not much is known instead about the correlator in perturbation theory. 
The correlator is known at leading order (LO)
since long time \cite{McLerran:1981pb,Gross:1980br}; beyond leading order, 
it was computed only for distances of the same order as the electric screening 
length in Ref.~\cite{Nadkarni:1986cz}.

The purpose of this Part is thus to evaluate the (connected) Polyakov-loop correlator in perturbation theory up to order $g^6$ at short distances, $rT \ll 1$, which corresponds to a next-to-next-to-leading order (NNLO) calculation. We also revisit the calculation of the expectation value of the Polyakov loop at order $g^4$, 
which corresponds also to a NNLO calculation, if we count $1$ as the leading-order 
result and $g^3$ as the NLO one. We will find a result that differs from the long-time accepted 
result of Gava and Jengo \cite{Gava:1981qd}. Finally, we will add on the discussion about 
the relation between the Polyakov-loop correlator and the in-medium heavy-quark potential, by rederiving the Polyakov-loop correlator in an EFT language that can be seen as an imaginary-time counterpart of the EFTs developed in the previous Part~\ref{part_realtime}.

This Part is organized as follows. In the remainder of this Chapter we will first introduce the Polyakov loop and the Polyakov-loop correlator, as well as their relation to the free energies of infinitely heavy quarks. This will be done in Sec.~\ref{sec_introloops}. Subsequently, in Sec.~\ref{self_energy}, we introduce the static gauge, which we adopt throughout this Part, and discuss the gluon propagator in this gauge at one-loop level. 
Chapter~\ref{chap_imtimepert} contains the NNLO calculations of the Polyakov loop and of the Polyakov-loop correlator, in Secs.~\ref{sec_ploop} and \ref{sec_corr} respectively. 
Finally, in Chap.~\ref{chap_imtimeEFT} we rederive the Polyakov-loop correlator in an effective 
field theory language. There, we also define a singlet and an octet free energy 
that we compute and we eventually present our conclusions in  Sec.~\ref{sec_concl}. The original results presented in this 
Part have been published in \cite{Brambilla:2010xn} (see also \cite{Ghiglieri:2010gi} for a brief summary).\\
\section{Polyakov loops and static-quark free energies}
\label{sec_introloops}
Our starting point is QCD with a static quark and a static antiquark, denoted 
in the following as {\em static QCD}. Its action in Euclidean space-time reads (see also Eq. \eqref{NRQCDstatic})
\begin{equation}
{\cal S}_{\rm QCD} = \int_0^{1/T} \!\! d\tau \int d^3x \; \left( \psi^\dagger D_0 \psi + \chi^\dagger D_0 \chi 
+ \frac{1}{4}F^a_{\mu\nu}F^a_{\mu\nu} + \sum_{l=1}^{n_f} \bar{q}_lD\!\!\!\!/
\, q_l \right),
\label{QCD}
\end{equation}
where $D_0 = \partial_0 + igA_0$, $\psi$ is the Pauli spinor field that annihilates 
a static quark, $\chi$ is the Pauli spinor field that creates a static antiquark, 
and $q_1$, ..., $q_{n_f}$ are the light quark fields, which we assume again to be massless. The infinitely heavy quark mass has instead been removed through a field redefinition.

We now set out to introduce the Polyakov loop and its correlator from this action, as was first done in Ref.~\cite{McLerran:1981pb}. Let us consider initially the free energy $F_{(1,0)}$ of a system with a single static quark located in $\bx$ and no static antiquark, hence the $(1,0)$ label. By standard thermodynamics $F_{(1,0)}$ is related to the logarithm of the partition function $Z_Q$ in the presence of a single static quark, so that
\begin{equation}
	e^{-\frac{F_{(1,0)}}{T}}=Z_Q=\frac{1}{\nc}\sum_s\langle s\vert e^{-\frac{H_\mathrm{QCD}}{T}}\vert s\rangle\,,
	\label{PL-QCD}
\end{equation}
where we have used the definition~\eqref{traceexplicit} for the partition function $Z_Q$ and $\vert s\rangle$ are the states with a single static quark in $\bx$. $H_\mathrm{QCD}$ is the Hamiltonian associated to the static QCD Lagrangian. The factor of $1/\nc$ is the colour normalization for a quark in the fundamental representation. This last equation can be rewritten as
\begin{equation}
	e^{-\frac{F_{(1,0)}}{T}}=\frac{1}{\nc}\frac{1}{\mathcal{N}_Q}\sum_{s^\prime}\langle s^\prime\vert \psi_i(\bx,0)e^{-\frac{H_\mathrm{QCD}}{T}}\psi_i^\dagger(\bx,0)\vert s^\prime\rangle\,,
	\label{PL-QCD-psi}
\end{equation}
where now the states $\vert s^\prime\rangle$ are all the states without static quarks and $\mathcal{N}_Q=\delta^3(\mathbf{0})$. The Boltzmann factor $e^{-H_\mathrm{QCD}/T}$ generates Euclidean time translations, so that
\begin{equation}
	e^{\frac{H_\mathrm{QCD}}{T}}\psi_i(\bx,0)e^{-\frac{H_\mathrm{QCD}}{T}}=\psi_i(\bx,\beta)\,.
	\label{timetransl}
\end{equation}
On the other hand, from the equation of motion for the field $\psi$ that follows from the action~\eqref{QCD} we have 
\begin{equation}
	\psi_i(\bx,\tau)=\left[P\exp\left(-ig\int_0^{\tau} d\tau^\prime A_0(\bx,\tau^\prime)\right)\right]_{ij}\psi_j(\bx,0)\,.
	\label{psievolution}
\end{equation}
Using these last two equations in Eq.~\eqref{PL-QCD-psi} and the fermionic equal-time anticommutation relation $\left\{\psi_i(\bx,\tau),\psi_j^\dagger(\by,\tau)\right\}=\delta_{ij}\delta^3(\bx-\by)$ we obtain
\begin{equation}
	e^{-\frac{F_{(1,0)}}{T}}=\frac{1}{\nc}\sum_{s^\prime}\langle s^\prime\vert e^{-\frac{H_\mathrm{QCD}}{T}}\left[P\exp\left(-ig\int_0^{1/T} d\tau A_0(\bx,\tau)\right)\right]_{ii}\vert s^\prime\rangle\,.
\end{equation}
This last equation can be brought to the form of a standard thermal average (see Eq.~\eqref{defthermalaverage}) by dividing by the partition function in the absence of static quarks, i.e. $Z_{(0,0)}=\sum_{s^\prime}\langle s^\prime\vert e^{-H_\mathrm{QCD}/T}\vert s^\prime\rangle$. This yields finally 
\begin{equation}
	e^{-\frac{F_Q}{T}}\equiv e^{-\frac{F_{(1,0)}-F_{(0,0)}}{T}}=\frac{1}{\nc}\langle \mathrm{Tr} L_F(\bx)\rangle\,,
	\label{finalPL}
\end{equation}
where we have defined $F_Q$ as the difference between the free energy of the system with one heavy quark $F_{(1,0)}$ and the free energy of the light degrees of freedom  only $F_{(0,0)}$. The expectation value on the right-hand side is a thermal average, as in Eq.~\eqref{defthermalaverage}, and the trace is a colour trace acting on the Polyakov line $L_F(\bx)$, as given by Eq.~\eqref{defploop}, in the fundamental representation $F$.\\
The operator on the right-hand side of Eq.~\eqref{finalPL} is the Polyakov loop. From the definition~\eqref{defploop} of the Polyakov line and the colour trace its gauge invariance follows immediately. For simplicity, let us introduce this notation
\begin{equation}
\langle L_R\rangle\equiv\langle\trt L_R\rangle,\quad\trt \equiv \frac{\mathrm{Tr}}{d(R)}\,,
\label{defploopvev}
\end{equation}
where $R$ labels a representation of dimension $d(R)$. $R$ shall be either the fundamental representation ($R=F$, $d(F)=\nc$) 
or the adjoint representation ($R=A$, $d(A)=\nc^2-1$), corresponding to having static quarks in the adjoint representation. The adjoint Polyakov loop $\langle L_A\rangle$ will play an important role in Chap.~\ref{chap_imtimeEFT}.

A completely analogous procedure can be applied to the free energy of a static quark and a static antiquark separated by a distance $\br$, yielding 
\begin{equation}
	e^{-\frac{F_{Q\overline{Q}}}{T}}\equiv e^{-\frac{F_{(1,1)}-F_{(0,0)}}{T}}=\frac{1}{\nc^2}\langle \mathrm{Tr} L_F^\dagger({\bf 0}) \mathrm{Tr} L_F(\br)\rangle\,,
	\label{PLC}
\end{equation}
where we have defined $F_{Q\overline{Q}}$, which is sometimes called the \emph{colour-averaged free energy}. The quantity on the right-hand side is the Polyakov-loop correlator, which, being a correlation function of two gauge-invariant operators, is also gauge invariant. We will only consider the fundamental representation for the correlator.\\
The Polyakov-loop correlator may furthermore be expressed in terms of the static QCD operators $\psi$ and $\chi$ as 
\begin{equation}
\langle\tilde{\mathrm{Tr}}L_F^\dagger({\bf 0})\tilde{\mathrm{Tr}}L_F(\br)\rangle 
= \frac{1}{\nc^2}\frac{1}{\mathcal{N}_{Q\overline{Q}}}
\langle \chi^\dagger_j({\bf 0},1/T)\psi_i({\bf r},1/T)\psi^\dagger_i({\bf r},0)\chi_j({\bf 0},0)\rangle,
\label{PLC-QCD}
\end{equation}
where ${\cal N}_{Q\overline{Q}} = [\delta^3({\bf 0})]^2$ and we have written explicitly the colour indices.\\
Since the Polyakov loop correlator is the partition function $Z_{Q\overline{Q}}$ in the presence of a static quark-antiquark pair divided by the partition function of the light degrees of freedom only, it follows that its spectral decomposition is  \cite{Luscher:2002qv,Jahn:2004qr}
\begin{equation}
\langle\tilde{\mathrm{Tr}}L_F^\dagger({\bf 0})\tilde{\mathrm{Tr}}L_F(\br)\rangle 
= \frac{1}{Z_{(0,0)}}\frac{1}{\nc^2} \sum_n e^{-E_n/T},
\label{PLC-spectrum}
\end{equation}
where $E_n$ are the eigenvalues of $H_\mathrm{QCD}$ relative to the eigenstates of the $Q\overline{Q}$ subspace of the Fock space, which we define as the set of all eigenstates with a static quark-antiquark pair and possibly other light degrees of freedom.

\section{The static gauge and the self energy}
\label{self_energy}
The Polyakov loop and the Polyakov-loop correlator are gauge-invariant quantities, as it was just shown.
We may exploit the gauge freedom by choosing the most suitable gauge.
A convenient gauge choice is the \emph{static gauge} \cite{D'Hoker:1981us}, as we mentioned in Sec.~\ref{sec_imtime}. This class of gauges is defined by the condition
\begin{equation}
\partial_0A_0(x)=0.
\end{equation}
The reason for using the static gauge is that in this gauge the Polyakov line has 
a very simple form
\begin{equation}
\label{ploop}
L(\bx) = {\rm P} \exp\left(-ig\int_0^{1/T} d\tau A_0(\bx,\tau)\right)=\exp\left(\frac{-igA_0(\bx)}{T}\right),
\end{equation}
where $\rm P$ stands for the path-ordering prescription.
The spatial part of the gluon propagator reads
\begin{equation}
\label{propnonstatic}
D_{ij}(\omega_n,\bk)=
\frac{1}{k^2}\left(\delta_{ij}+\frac{k_ik_j}{\omega_n^2}\right)(1-\delta_{n0})
+\frac{1}{\bk^2}\left(\delta_{ij}-(1-\xi)\frac{k_ik_j}{\bk^2}\right)\delta_{n0},
\end{equation}
where $\omega_n=2\pi T n$ are the bosonic Matsubara frequencies and
$k^2=\omega_n^2+\bk^2$. Throughout this Part italic letters will refer to Euclidean 
four-vectors and bold letters to the spatial components.
The parameter $\xi$ is a residual gauge-fixing parameter.
We call \emph{non-static modes} those propagating with nonzero Matsubara
frequencies and conversely we employ the term \emph{static mode} for the zero mode. 
The first term in the r.h.s. of Eq.~\eqref{propnonstatic}, proportional to $(1-\delta_{n0})$,
is then the non-static part, whereas the second, proportional to $\delta_{n0}$,  
is the static part. We stress again that, due to an unfortunate coincidence in the accepted terminology, the term static has, in this context, a different meaning from the the same term applied to heavy quarks, where it indicates an infinitely heavy quark, see also Footnote~\ref{foot_static} in Chap.~\ref{chap_thermal}.\\
The temporal part of the gluon propagator reads
\begin{equation}
\label{propstatic}
D_{00}(\omega_n,\bk)=\frac{\delta_{n0}}{\bk^2},
\end{equation}	
which is purely static.  Note that the gauge-fixing parameter affects only 
the static part of the spatial gluon propagator.
The complete set of Feynman rules in this gauge has been discussed in Refs. 
\cite{D'Hoker:1981us,Curci:1982fd,Curci:1984rd}. They are listed 
in App.~\ref{sub_stat_gauge} together with our Feynman diagram conventions.
We will adopt the static gauge in all the calculations of this Part, if not otherwise specified.

A necessary ingredient for the calculation of the Polyakov-loop expectation value and the Polyakov-loop
correlator at NNLO is the temporal component of the gluon self energy at LO. 
In the static gauge, due to the static nature of the temporal propagator in Eq.~\eqref{propstatic} only 
$\Pi_{00}(\bk) \equiv \Pi_{00}(0,\bk)$ enters. Furthermore, at LO  
static and non-static modes do not mix in $\Pi_{00}(\bk)$, which can thus be conveniently split into
\begin{equation}
\Pi_{00}(\bk)=\Pi_{00}^{\mathrm{NS}}(\bk)+\Pi_{00}^{\mathrm{S}}(\bk)+\Pi_{00}^{\mathrm{F}}(\bk),
\end{equation}
where the three terms correspond to the contribution of the non-static gluons, 
the static gluons and the fermion loops respectively. As discussed in App.~\ref{sub_stat_gauge}, ghosts do not couple to temporal gluon. They therefore do not contribute to $\Pi_{00}(\bk).$

\begin{figure}[ht]
\begin{center}
\includegraphics[width=10cm]{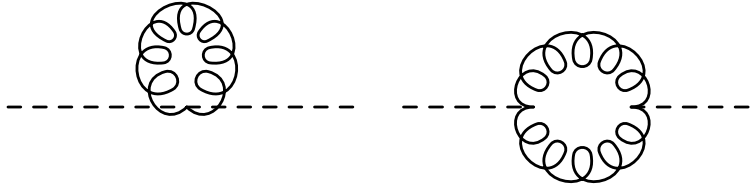}
\end{center}
\caption{Diagrams contributing to the non-static part of the gluon
  self-energy in the gluonic sector. 
  Dashed lines are temporal gluons, curly lines are spatial non-static gluons.}
\label{fig:ns}
\end{figure}

\begin{enumerate}
\item{\emph{$\Pi_{00}^{\mathrm{NS}}(\bk)$}}\\
In the gluonic sector, the non-static part of the self-energy receives contributions 
only from the two diagrams shown in Fig.~\ref{fig:ns}. Using the Feynman rules of 
appendix~\ref{sub_stat_gauge},  
it can be written in terms of five dimensionally-regularized master sum integrals 
\begin{equation}
\Pi_{00}^{\mathrm{NS}}(\bk)=
-2g^2\ca\left(\frac{d-1}{2}I_0 - (d-1) I_1 + I_2+\frac{1}{2}I_3 +\frac{1}{4}I_4\right),
\label{defp00}
\end{equation}
where $d\equiv 3-2\epsilon$ is the number of dimensions, 
\begin{equation}
\label{defmaster}
I_0=\int_p^\prime\frac{1}{p^2},\quad 
I_1=\int_p^\prime\frac{\bp^2}{p^2q^2},\quad 
I_2=\int_p^\prime\frac{\bk^2}{p^2q^2},\quad 
I_3=\int_p^\prime\frac{\bk^2}{\bp^2 p^2},\quad 
I_4=\int_p^\prime\frac{\bk^4}{p^2q^2\omega_n^2},
\end{equation}
$q=k-p$, $\displaystyle \int_p^\prime$ is a shorthand notation for the non-static, $n\ne0$, sum integral:
\begin{equation}
\label{shorthand}
\int_p^\prime \equiv T\sum_{n\ne0} \m2  \int\frac{d^dp}{(2\pi)^d},
\end{equation}
and $\mu$ is the scale in dimensional regularization.
The result~\eqref{defp00} can be conveniently cast in a sum of a vacuum part, a matter part, 
a part made of the subtracted zero modes and a part that we may call \emph{singular}, 
because it is singular for $T\to0$; the singular part is a peculiar 
feature of the static gauge. We then have
\begin{eqnarray}
\Pi_{00}^{\mathrm{NS}}(\bk)&=&
\Pi_{00}^{\mathrm{NS}}(\bk)_{\mathrm{vac}}+\Pi_{00}^{\mathrm{NS}}(\bk)_{\mathrm{mat}}
+\Pi_{00}^{\mathrm{NS}}(\bk)_{\mathrm{zero}}+\Pi_{00}^{\mathrm{NS}}(\bk)_{\mathrm{sing}},
\label{pi00nonstatic}
\end{eqnarray}
\begin{eqnarray}
\Pi_{00}^{\mathrm{NS}}(\bk)_{\mathrm{vac}}&=&
-\frac{g^2\bk^2}{(4\pi)^2}C_A\left[\frac{11}{3}
\left(\frac{1}{\epsilon}-\gamma_E+ \ln (4 \pi) -\ln\frac{\bk^2}{\mu^2}\right)
+\frac{31}{9}\right],
\label{pi00vacuum}\\
\nonumber
\Pi_{00}^{\mathrm{NS}}(\bk)_{\mathrm{mat}}&=&
g^2\ca\left\{\int_0^\infty d\mbp\frac{\mbp n_{\mathrm{B}}(\mbp)}{\pi^2}\left[1-\frac{\bk^2}{2\bp^2}
\right.\right.
\\
&& \left.\left. \hspace{1.4cm}
+\left(\frac{\mbp}{\mbk}-\frac{\mbk}{2\mbp}+\frac{\mbk^3}{8\mbp^3}\right)
\ln\left\vert\frac{\mbk+2\mbp}{\mbk-2\mbp}\right\vert\right]\right\}\!,
\label{pi00matter}\\
\Pi_{00}^{\mathrm{NS}}(\bk)_{\mathrm{zero}}&=&
g^2\ca\frac{T\mbk^{1-2\epsilon} \m2 }{4}\left[1+\epsilon(-1-\gamma_E + \ln (16\pi))\right], 
\label{pi00zero}\\
\Pi_{00}^{\mathrm{NS}}(\bk)_{\mathrm{sing}}&=&-g^2\ca\frac{\mbk^3}{192T},
\label{pi00sing}
\end{eqnarray}
where $n_\mathrm{B}$ is the Bose--Einstein distribution, see Eq.~\eqref{bosedistr}. 
We refer the reader to appendix~\ref{app_pi00} for details on the derivation of these equations. 
The vacuum part~\eqref{pi00vacuum} agrees with the static gauge computation in \cite{Curci:1982fd}. 
Furthermore, the vacuum part and the matter part are identical 
to the $k^0\to0$ limit of their Coulomb gauge counterparts, computed respectively 
in \cite{Duncan:1975kt,Appelquist:1977es,Andrasi:2003zf} and \cite{Heinz:1986kz}. The matter part in Coulomb gauge can also be read from App.~\ref{secvacuumpol}: Eq.~\eqref{RePi00k0} shows the $k_0\to0$ limit.  
$\Pi_{00}^{\mathrm{NS}}(\bk)_{\mathrm{zero}}$ consists of the subtracted zero modes.  
In the $\epsilon\to 0$ limit, it is $T\mbk/4$; we have kept the order $\epsilon$ corrections, 
because, in the Polyakov-loop correlator calculation of Sec.~\ref{sec_corr},
we will need to evaluate the Fourier transform of $\mbk^{1-2\epsilon}/\mbk^4$, coming 
from a self-energy insertion in a temporal-gluon propagator, which is divergent. \\ 

\item{\emph{$\Pi_{00}^{\mathrm{F}}(\bk)$}}\\
At leading order in the coupling, $\Pi_{00}^{\mathrm{F}}(\bk)$ may be written in terms 
of three di\-men\-sionally-regularized master sum integrals \cite{Heinz:1986kz} 
\begin{equation}
\Pi^{\mathrm{F}}_{00}(\bk) = 2g^2n_f\left(-\tilde{I}_0+2\tilde{I}_1+\frac{1}{2}\tilde{I}_2\right),
\label{pi00fermion}
\end{equation}
where
\begin{eqnarray} 
\nonumber
\tilde{I}_0  &=& T\sum_{n=-\infty}^{+\infty}\m2 \int\frac{d^dp}{(2\pi)^d}\frac{1}{p^2},  \quad
\tilde{I}_1   = T\sum_{n=-\infty}^{+\infty}\m2 \int\frac{d^dp}{(2\pi)^d}\frac{\tilde{\omega}_n^2}{p^2q^2}, 
\\
\tilde{I}_2 &=& T\sum_{n=-\infty}^{+\infty}\m2 \int\frac{d^dp}{(2\pi)^d}\frac{\bk^2}{p^2q^2},
\label{deffermionmaster}
\end{eqnarray}
$q=p+k$ and $\tilde{\omega}_n=(2n+1)\pi T$ are the fermionic Matsubara frequencies and
$n_f$ is the number of massless quarks contributing to the fermion loops.
 Since no fermionic Matsubara frequency vanishes, fermions are purely non-static.
The fermionic contribution can be cast into a sum of a vacuum and a matter part: 
$\Pi^{\mathrm{F}}_{00}(\bk)=\Pi_{00}^\mathrm{F}(\bk)_{\mathrm{vac}}+\Pi_{00}^\mathrm{F}(\bk)_{\mathrm{mat}}$. 
After the Matsubara frequencies summation, the matter part can be read from \cite{Kapusta:2006pm}
\begin{equation}
\label{matterfermion}
\Pi_{00}^\mathrm{F}(\bk)_{\mathrm{mat}}=
\frac{g^2}{2\pi^2}n_f\int_{0}^{\infty}d\mbp\,\mbp 
n_\mathrm{F}( \mbp)\left[2+\frac{4\bp^2-\bk^2}{2\mbp\mbk}
\ln\left\vert\frac{\mbk+2\mbp}{\mbk-2\mbp}\right\vert\right],
\end{equation} 
where $n_\mathrm{F}$ is the Fermi--Dirac distribution, see Eq.~\eqref{fermidistr}. This expression agrees with the fermionic part of Eq.~\eqref{RePi00k0}. 
The vacuum part is given by
\begin{equation}
\label{vacuumfermion}		
\Pi_{00}^\mathrm{F}(\bk)_{\mathrm{vac}}=
\frac{2}{3}\frac{g^2\bk^2}{(4\pi)^2}n_f
\left[\frac{1}{\epsilon}-\gamma_E +\ln (4\pi)-\ln\frac{\bk^2}{\mu^2}+\frac{5}{3}\right].
\end{equation}

\item{\emph{$\Pi_{00}^{\mathrm{NS}}(\bk) + \Pi_{00}^{\mathrm{F}}(\bk)$}}\\
Let us now consider the sum $\Pi_{00}^{\mathrm{NS}}(\bk) + \Pi_{00}^{\mathrm{F}}(\bk)$.
The divergences in the vacuum parts~\eqref{pi00vacuum} and~\eqref{vacuumfermion} are of 
ultraviolet origin and are accounted for by the charge renormalization.
In the $\MS$ scheme, the renormalized sum of vacuum parts reads 
\begin{equation}
\Pi_{00}^\mathrm{NS}(\bk)_{\mathrm{vac}}+\Pi_{00}^\mathrm{F}(\bk)_{\mathrm{vac}}
=-\frac{g^2\bk^2}{(4\pi)^2}\left[\beta_0\ln\frac{ \mu^2}{\bk^2}+\frac{31}{9}\ca-\frac{10}{9}n_f\right],
\label{vacuum}
\end{equation}
where $\beta_0$ is given by Eq.~\eqref{beta0}. 

Simple analytical expressions can be obtained for the renormalized 
sum $\Pi_{00}^{\mathrm{NS}}(\bk) + \Pi_{00}^{\mathrm{F}}(\bk)$ in the two limiting cases $\mbk\ll T$ and $\mbk\gg T$. 
In the former case, we have 
\begin{eqnarray}
\nonumber
\left(\Pi_{00}^{\mathrm{NS}}+\Pi_{00}^{\mathrm{F}}\right)(\mbk\ll T)
&=&\frac{g^2T^2}{3}\left(\ca+\frac{n_f}{2}\right)
\\
\nonumber
&-& \frac{g^2\bk^2}{(4\pi)^2}\left[\frac{11}{3} \ca\left( -\ln\frac{(4\pi
    T)^2}{\mu^2}+1+2 \gamma_E \right)\right.
\\
\nonumber
&&\left. \hspace{1.3cm} -\frac{2}{3}n_f\left(-\ln\frac{(4\pi T)^2}{\mu^2}-1+2 \gamma_E   + 4 \ln 2
\right)\right]
\\
&+& g^2\bk^2\mathcal{O}\left(\frac{\bk^2}{T^2}\right),
\label{kllt}
\end{eqnarray}
where the leading-order term is momentum independent and can be 
identified with the (square of the) Debye mass $m_D$, 
\begin{equation}
m_D^2 \equiv \frac{g^2T^2}{3}\left(\nc+\frac{n_f}{2}\right),
\label{fulldebye}
\end{equation}
which provides, in the weak-coupling regime, the inverse of an electric 
screening length.
We note that Eq.~\eqref{kllt} presents a logarithm of
the renormalization scale over the temperature rather than over the
momentum: this happens because in the limit $\mbk\ll T$ the matter part produces
a term proportional to $\bk^2\beta_0\ln(T^2/\bk^2)$ that combines
with the logarithm in the renormalized vacuum part~\eqref{vacuum} to
cancel its momentum dependence.

In the opposite limit $\mbk\gg T$, we have
\begin{eqnarray}
\nonumber
\left(\Pi_{00}^{\mathrm{NS}}+\Pi_{00}^{\mathrm{F}}\right)(\mbk\gg T)
&=&\Pi_{00}^\mathrm{NS}(\bk)_{\mathrm{vac}}+\Pi_{00}^\mathrm{F}(\bk)_{\mathrm{vac}}
+g^2 C_A \left(-\frac{T^2}{18} -\frac{\mbk^3}{192T}\right)
\\
&&+g^2\ca \frac{T\mbk^{1-2\epsilon} \m2 }{4}\left[1+\epsilon(-1-\gamma_E + \ln (16\pi))\right]
\nonumber\\
&&+g^2T^2\mathcal{O}\left(\frac{T^2}{\bk^2}\right).
\label{kggt}
\end{eqnarray}
We observe that, in this limit and at the considered order, fermions
enter only through their contribution to the vacuum part. It should
be also noted that, while the $-g^2\ca T^2/18$ term appears also in Coulomb
gauge, as shown in App.~\ref{secvacuumpolkoo}, the term proportional to $\mbk^3$ 
is instead a peculiar feature of the static gauge. 
The terms proportional to $\epsilon T \mbk^{1-2\epsilon}$, which appear in the second line,
come from the subtracted zero modes and contribute only when plugged into divergent amplitudes. 
Details on the derivation of these expressions can be found in appendix~\ref{app_exp}.

\begin{figure}[ht]
\begin{center}
\includegraphics{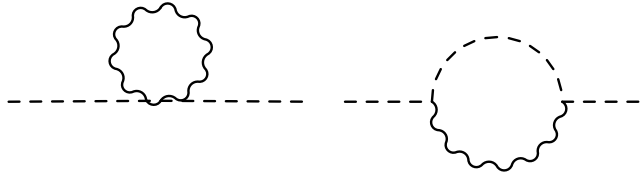}
\end{center}
\caption{Diagrams contributing to the static part of the self-energy: the
  dashed lines are temporal-gluon propagators,  the wavy lines are static
  spatial-gluon propagators. Loops made of two static spatial-gluon propagators  
  and of ghosts vanish.}
\label{fig_static}
\end{figure}

\item{\emph{$\Pi_{00}^{\mathrm{S}}(\bk)$}}\\
The diagrams contributing to the static part of the gluon self energy are shown
in Fig.~\ref{fig_static}. 
They are not sensitive to the scale $T$, since, by definition, 
static gluon propagators are just made of zero modes, however they 
are to the scale $m_D$. 
Hence, when evaluating the static contribution, it is important
to keep in mind that, if the incoming momentum is of the order of the Debye mass,
then insertions of gluon self-energies of the type of Eq.~\eqref{kllt} into the 
temporal-gluon propagator need to be resummed modifying the temporal-gluon propagator into
\begin{equation}
\label{propscreened}
D_{00}(\omega_n,\bk)=\frac{\delta_{n0}}{\bk^2+m_D^2}.
\end{equation}
We also remark that, as pointed out in Ref.~\cite{Nadkarni:1982kb}, the Feynman rules of the static modes in the static gauge with a resummed Debye mass are the same as those of EQCD up to a normalization.\\
The static part of the gluon self energy with resummed propagators 
reads, for all values of the gauge-fixing parameter $\xi$, 
\begin{eqnarray}
\Pi_{00}^{\mathrm{S}}(\bk)&=&
g^2C_AT \m2  \int\frac{d^dp}{(2\pi)^d}\left(\frac{1}{\bp^2+m_D^2}+\frac{d-2}{\bp^2}\right.
\nonumber\\
&&\left.+\frac{2(m_D^2-\bk^2)}{\bp^2(\bq^2+m_D^2)}
+(\xi-1)(\bk^2+m_D^2)\frac{\bp^2+2\bp\cdot\bk}{\bp^4(\bq^2+m_D^2)}\right),
\label{staticvacpol}
\end{eqnarray}
where $q=k+p$. The result agrees with Ref.~\cite{Rebhan:1993az,Rebhan:1994mx}.  
Note that Eq.~\eqref{staticvacpol} applies for all gauges sharing 
the same static propagator, among which the static and the covariant gauges. 
The expression is finite in three dimensions and reads
\begin{eqnarray}
\Pi_{00}^{\mathrm{S}}(\bk)&=&
\frac{g^2C_AT}{4\pi}\left[2\frac{m_D^2-\bk^2}{\mbk}\arctan\frac{\mbk}{m_D} -m_D+(\xi-1)m_D\right].
\label{staticvacpol2}
\end{eqnarray}
Finally for the static part $\mbk\gg T$ implies $\mbk\gg m_D$ and
\begin{equation}
\Pi_{00}^\mathrm{S}(\mbk\gg m_D)=-g^2C_A\left\{
\frac{T\mbk^{1-2\epsilon} \m2 }{4}\left[1+\epsilon(-\gamma_E + \ln (16\pi))\right]
+\mathcal{O}(m_DT)\right\},
\label{staticcontrhighk}
\end{equation}
where again we have kept up to order $\epsilon$ terms proportional to $T\mbk^{1-2\epsilon}$.

\item{\emph{$\Pi_{00}(\bk)$}}\\
$\Pi_{00}(\bk)$ is obtained by summing~\eqref{pi00vacuum},~\eqref{pi00matter},~\eqref{pi00zero}, 
\eqref{pi00sing},~\eqref{matterfermion},~\eqref{vacuumfermion} and~\eqref{staticvacpol} (or~\eqref{staticvacpol2}).
In particular, the asymptotic expression for the gluon polarization at high momenta is 
\begin{eqnarray}
\nonumber\Pi_{00}(\mbk\gg T)&=&
-\frac{g^2\bk^2}{(4\pi)^2}\left(\beta_0\ln\frac{\mu^2}{\bk^2}+\frac{31}{9}\ca-\frac{10}{9}n_f\right)
+g^2C_A\left(-\frac{T^2}{18}-\frac{\mbk^3}{192T}\right)\\
&&
-\epsilon g^2\ca\frac{T\mbk^{1-2\epsilon} \m2 }{4}
+\mathcal{O}\left(g^2\frac{T^4}{\bk^2},g^2m_DT\right).
\label{kggtfull}
\end{eqnarray}
Note that the term proportional to $T\mbk^{1-2\epsilon}\epsilon^0$ in Eq.~\eqref{staticcontrhighk} 
has canceled against the term proportional to $T\mbk^{1-2\epsilon}\epsilon^0$ in Eq.~\eqref{kggt}.
\end{enumerate}
In the following Chapter we will deal with the perturbative computation of the Polyakov loop and of the Polyakov-loop correlator. As we shall see, these expressions for the gluon self-energy will be extremely valuable.

\chapter{The Polyakov loop and the correlator of Polyakov loops in perturbation theory}
	\label{chap_imtimepert}
	The purpose of this Chapter is to evaluate the Polyakov loop and the Polyakov-loop correlator in perturbation theory. 
The former will be computed in Sec.~\ref{sec_ploop} at order $g^4$, 
	which corresponds to a next-to-next-to-leading order (NNLO) calculation, if we count $1$ as the leading-order 
	result and $g^3$ as the next-to-leading order (NLO) one. We will find a result that differs from the long-time accepted 
	result of Gava and Jengo \cite{Gava:1981qd}. We will also show part of the order-$g^5$ and $g^6$ results for their future relevance in the next Chapter.\\
	The (connected) Polyakov-loop correlator will be computed in Sec.~\ref{sec_corr} up to order $g^6$ at short distances, $rT \ll 1$. This corresponds to a NNLO calculation as well,	if we count the order $g^4$ as LO and the order $g^5$ as NLOß®.
\section{The Polyakov loop\label{sec_ploop}}
We shall evaluate the Polyakov loop both in the fundamental and adjoint representations, with the notation of Eq.~\eqref{defploopvev}. Expanding the Polyakov line in the static gauge~\eqref{ploop} up to order $g^4$ yields
\begin{equation}
\langle L_R\rangle=
1-\frac{g^2}{2!} \frac{\langle\trt A_0^2\rangle}{T^2}
+i\frac{g^3}{3!} \frac{\langle\trt A_0^3\rangle}{T^3}
+\frac{g^4}{4!} \frac{\langle\trt A_0^4\rangle}{T^4}
+\ldots \,.
\label{expandedlooptraces}
\end{equation}
In computing Eq.~\eqref{expandedlooptraces} perturbatively,  
each diagram can receive contributions from both scales $T$ and $m_D$, 
for which we assume a weak-coupling hierarchy:\footnote{
When discussing energy scales, we will consider $T$ and multiple of $\pi T$ to be 
parametrically of the same order. See also Footnote~\ref{foot_piT} in Chap.~\ref{chap_realtime}.
}
\begin{equation}
T\gg m_D.	
\label{wkhierarchy}
\end{equation}	
In the weak-coupling regime, the calculation of $\langle L_R\rangle$ may be organized in an 
expansion in the coupling $g$; our aim is to compute $\langle L_R\rangle$ up to order 
$g^4$. Sometimes, we will find it useful to keep $m_D/T$ as a separate expansion 
parameter with respect to $g$, in order to identify more easily
the origin of the various terms. We will call the $g^3$ term the NLO correction to the 
Polyakov loop and the $g^4$ term the NNLO correction.
We will also identify the source of some higher-order corrections of order $g^5$ and
$g^4\times(m_D/T)^2$ that will play a role in Chap.~\ref{chap_imtimeEFT}.

\begin{figure}[ht]
\begin{center}
\includegraphics{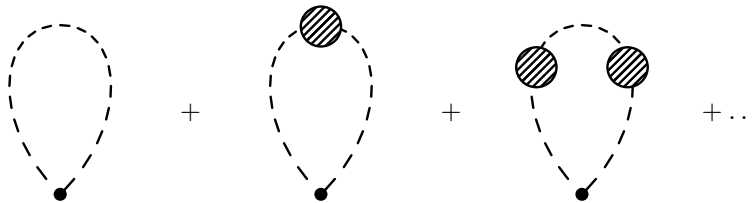}
\end{center}
\caption{Diagrams contributing to the perturbative expansion of $g^2\langle\trt A_0^2\rangle$. 
The dashed line is a temporal-gluon propagator, the dot is the point $\bx$ where the loop originates. 
The blob stands for the gluon self energy.}
\label{fig:square}
\end{figure}

\subsection{The order $g^3$ contribution\label{sub_lo}}
Let us start examining $g^2\langle\trt A_0^2\rangle$. Diagrams 
contributing to  $g^2\langle\trt A_0^2\rangle$ are shown in Fig.~\ref{fig:square}.
Summing up all these diagrams, $g^2\langle\trt A_0^2\rangle$ can be written as 
\begin{equation}
\delta\langle L_R\rangle
=-\frac{g^2}{2!} \frac{\langle\trt A_0^2\rangle}{T^2}
=-\frac{g^2C_R}{2T}\m2  \int \frac{d^dk}{(2\pi)^d}\frac{1}{\bk^2+\Pi_{00}(\bk)},
\label{loopb}
\end{equation}
where $C_R$ is the quadratic Casimir operator of the representation $R$, see Eqs.~\eqref{defcd} and \eqref{defca}. We observe that the integral
receives contributions from the scales $T$ and $m_D$. We set out to
separate the contributions from these two scales assuming the hierarchy~\eqref{wkhierarchy}. 
\subsubsection{Modes at the scale $T$}
We evaluate the integral~\eqref{loopb} for $\mbk\sim T\gg m_D$. 
In this momentum region, $\Pi_{00}(\mbk\sim T\gg m_D)\ll\bk^2$ and we may
expand the gluon propagator in $\Pi_{00}$. 
The LO term yields a scaleless integral
\begin{equation}
\delta\langle L_R\rangle=-\frac{g^2}{2T}C_R \m2  \int\frac{d^dk}{(2\pi)^d}\frac{1}{\bk^2}=0,
\label{tcontribnull}
\end{equation}
whereas the following term gives	
\begin{equation}
\delta\langle L_R\rangle_{T}=\frac{g^2C_R}{2T}\m2  \int \frac{d^dk}{(2\pi)^d}\frac{\Pi_{00}(\mbk\sim T\gg m_D)}{\bk^4}.
\label{loopbT}
\end{equation}
This term is of order $g^4$.

\subsubsection{Modes at the scale $m_D$}
We evaluate now the contribution from the scale $m_D$. 
We recall from Eqs.~\eqref{kllt} and~\eqref{staticvacpol2} that, 
for $\mbk\ll T$, $\Pi_{00}(\bk)=m_D^2(1 +\mathcal{O}(g))$. 
We then rewrite the propagator in Eq.~\eqref{loopb} as 
$1/(\bk^2+\Pi_{00}(\mbk\ll T))=1/(\bk^2+m_D^2+(\Pi_{00}(\mbk\ll T)-m_D^2))$ 
and expand in $\Pi_{00}(\mbk\ll T)-m_D^2$. The LO term yields 
\begin{equation}
\delta\langle L_R\rangle_{{\rm LO}\, m_D}
= -\frac{g^2C_R}{2T} \m2  \int \frac{d^dk}{(2\pi)^d}\frac{1}{\bk^2+m_D^2} 
=  \frac{\crr \als}{2} \frac{m_D}{T},
\label{loopbm0}
\end{equation}
whereas the following one gives 
\begin{equation}
\delta\langle L_R\rangle_{{\rm NLO}\,m_D}
= \frac{g^2C_R}{2T} \m2  \int \frac{d^dk}{(2\pi)^d}
\frac{\Pi_{00}(\mbk\sim m_D\ll T)-m_D^2}{(\bk^2+m_D^2)^2},
\label{loopbm}
\end{equation}
which is at least of order $g^4$.

Up to order $g^3$, we then have
\begin{equation}
\langle L_R\rangle=1+\frac{\crr \als}{2}\frac{m_D}{T}+\mathcal{O}\left(g^4\right).
\label{fleading}
\end{equation}
We remark that, since the static modes of the static gauge are, up to a normalization, the degrees of freedom of EQCD, the resummed contribution at the scale $m_D$ is the same contribution one would obtain when computing the Polyakov loop within EQCD.
\begin{figure}[ht]
\begin{center}
\includegraphics{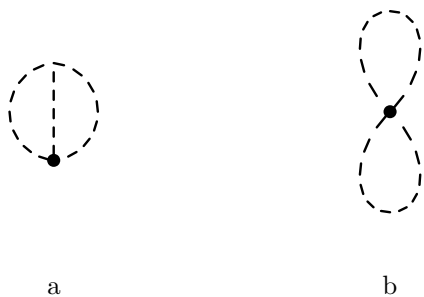}
\end{center}
\caption{Diagram a is the leading-order contribution to $\langle\trt A_0^3\rangle$: 
it vanishes because of the three-gluon vertex involving only temporal gluons.
Diagram b is the LO term of $g^4\langle\trt A_0^4\rangle$:  
it vanishes because scaleless.}
\label{fig:cubefourth}
\end{figure}

The LO contribution to the cubic term $g^3\langle\trt A_0^3\rangle$
is shown in Fig.~\ref{fig:cubefourth} a. 
It vanishes due to the structure of the three-gluon vertex.
This is just a LO manifestation of the charge-conjugation symmetry;
in fact, due to this symmetry, $g^3\langle\trt A_0^3\rangle$ vanishes
to all orders.
The quartic term $g^4\langle\trt A_0^4\rangle$ gets its LO
contribution from the diagram shown in Fig.~\ref{fig:cubefourth} b, which 
vanishes because scaleless. At higher order, a comparison with the analysis 
we have just performed for $\langle\trt A_0^2\rangle$ 
makes it clear that $g^4\langle\trt A_0^4\rangle$ 
starts to contribute at order $g^4\times(m_D/T)^2$, 
which is again beyond the accuracy of this analysis.
We can therefore identify as the only contributions to the Polyakov
loop at order $g^4$ the ones of Eqs.~\eqref{loopbT} and~\eqref{loopbm}. 
In Sec.~\ref{sub_nlo}, we will compute these contributions 
and, in Sec.~\ref{sub_nnlo}, we will analyze some sub-leading terms.

\subsection{The order $g^4$ contribution\label{sub_nlo}}
We now set out to compute Eqs.~\eqref{loopbT} and~\eqref{loopbm}. 
Following the discussion in Sec.~\ref{self_energy}, we
separate the non-static from the static modes in $\Pi_{00}(\bk)$.  We then
have four sources of contributions: non-static modes at the scale $T$,
non-static modes at the scale $m_D$, static modes at the scale $T$ and 
static modes at the scale $m_D$.  

\subsubsection{Non-static modes at the scale $T$}
The non-static contribution to Eq.~\eqref{loopbT} reads
\begin{equation}
\delta\langle L_R\rangle_{\mathrm{NS},\,T}=
\frac{g^2C_R}{2T}\m2  \int
\frac{d^dk}{(2\pi)^d}\frac{\Pi_{00}^{\mathrm{NS}}(\mbk\sim T)+\Pi_{00}^{\mathrm{F}}(\mbk\sim T)}{\bk^4},
\label{tcontribnlons}
\end{equation}
where $\Pi_{00}^{\mathrm{NS}}(\mbk\sim T)$ is the full
non-static contribution as defined in Eq.~\eqref{pi00nonstatic} and
similarly $\Pi_{00}^{\mathrm{F}}(\mbk\sim T)$ is the full fermionic
contribution as defined in Eqs.~\eqref{matterfermion} and~\eqref{vacuumfermion}. 
We can rewrite Eq.~\eqref{tcontribnlons} as
\begin{eqnarray}
\nonumber\delta\langle L_R\rangle_{\mathrm{NS},\,T}&=&
\frac{g^4C_R}{T}\left[-\ca\left(\frac{d-1}{2}J_0 - (d-1) J_1 +
  J_2+\frac{1}{2}J_3 +\frac{1}{4}J_4\right)\right.
\\
&&\left.+ n_f\left(-\tilde{J}_0+2\tilde{J}_1+\frac{1}{2}\tilde{J}_2\right)\right],
\label{tcontribnlonsj}
\end{eqnarray}
where we have defined the two-loop master sum-integrals $J_i$ and $\tilde{J}_i$ as
\begin{equation}
\label{defji}
J_i=\m2  \int \frac{d^dk}{(2\pi)^d} \frac{1}{\bk^4} I_i,\qquad 
\tilde{J}_i=\m2  \int \frac{d^dk}{(2\pi)^d} \frac{1}{\bk^4} \tilde{I}_i.
\end{equation}
These integrals are evaluated in appendix~\ref{app_integrals} and their sum yields
\begin{equation}
\delta\langle L_R\rangle_{\mathrm{NS},\,T}=
\frac{g^4\crr}{2(4\pi)^2}\left[\ca\left(\frac{1}{2\epsilon}-\ln\frac{4T^2}{\mu^2}+1-\gamma_E
+\ln (4\pi) \right)- n_f\ln2\right]. 
\label{finalcontribtnlons}
\end{equation}
The divergence stems from the $J_2$ integral and is expected 
to cancel against an opposite divergence coming from the scale $m_D$.

\subsubsection{Non-static modes at the scale $m_D$}
The non-static contribution to Eq.~\eqref{loopbm} reads 
\begin{equation}
\delta\langle L_R\rangle_{\mathrm{NS},\,m_D}=\frac{g^2C_R}{2T}\m2  
\int \frac{d^dk}{(2\pi)^d}\frac{\Pi_{00}^{\mathrm{NS}}(\mbk\sim m_D)
+\Pi_{00}^{\mathrm{F}}(\mbk\sim m_D)-m_D^2}{(\bk^2+m_D^2)^2}.
\label{mcontribnlons}
\end{equation} 
For $\mbk$ much smaller than the temperature, Eq.~\eqref{kllt} applies
and thus $\Pi_{00}^{\mathrm{NS}}(\bk) + \Pi_{00}^{\mathrm{F}}(\bk) =m_D^2+\mathcal{O}(g^2\bk^2)$.
Therefore, the contribution of Eq.~\eqref{mcontribnlons} is of order $g^4 \times (m_D/T) \sim g^5$.
More explicitly, plugging Eq.~\eqref{kllt} into Eq.~\eqref{mcontribnlons} gives
\begin{equation}
\delta\langle L_R\rangle_{\mathrm{NS},\,m_D}=
\frac{3g^4C_R}{4(4\pi)^3}\frac{m_D}{T}\left[\beta_0\ln\left(\frac{\mu }{4\pi T}\right)^2
+2 \beta_0\gamma_E +\frac{11}{3}\ca-\frac{2}{3}n_f\left(4\ln 2-1\right)\right].
\label{mcontribnlonsfinal}
\end{equation}
Although a term of order $g^5$ is beyond our accuracy, the contribution
\eqref{mcontribnlonsfinal} is of interest because 
it fixes the renormalization scale of $g^3$ in the LO term~\eqref{fleading} ($\als m_D/T \sim g^3$)
to $\mu = 4\pi T$. 

\subsubsection{Static modes at the scale $T$}
The static contribution at the scale $T$ to Eq.~\eqref{loopbT} reads
\begin{equation}
\delta\langle L_R\rangle_{\mathrm{S}\,T}=
\frac{g^2C_R}{2T}\m2  \int \frac{d^dk}{(2\pi)^d}\frac{\Pi_{00}^{\mathrm{S}}(\mbk\sim T)}{\bk^4}=0.
\label{tcontribnlos}
\end{equation}
It vanishes because $\Pi_{00}^{\mathrm{S}}(\mbk\sim T\gg m_D)\sim g^2T \mbk$ 
(see Eq.~\eqref{staticcontrhighk}) and thus the resulting integration over $\bk$ is scaleless. 

\subsubsection{Static modes at the scale $m_D$}
The static contribution to Eq.~\eqref{loopbm} is
\begin{equation}
\delta\langle L_R\rangle_{\mathrm{S}\,m_D}=
\frac{g^2C_R}{2T}\m2  \int \frac{d^dk}{(2\pi)^d}\frac{\Pi_{00}^{\mathrm{S}}(\mbk)}{(\bk^2+m_D^2)^2},
\label{mcontribnlos}
\end{equation}
where $\Pi_{00}^{\mathrm{S}}(\mbk)$ is the full static
contribution of Eq.~\eqref{staticvacpol}.  The computation
is carried out in detail in appendix~\ref{app_3d}; the result reads 
\begin{equation}
\delta\langle L_R\rangle_{\mathrm{S}\,m_D}=
\frac{g^4\crr\ca}{2(4\pi)^2}\left(-\frac{1}{2\epsilon}-\ln\frac{\mu^2}{4m_D^2}-\frac{1}{2}+\gamma_E -\ln (4\pi)\right).
\label{finalcontribm}
\end{equation}
The divergence cancels against the one of Eq.~\eqref{finalcontribtnlons} 
coming from non-static modes at the scale $T$.\footnote{
Both divergences in Eqs.~\eqref{finalcontribtnlons} and~\eqref{finalcontribm}
are of ultraviolet origin. This seems to contradict the expectation according to which 
infrared divergences from higher scales should cancel against ultraviolet divergences 
from lower scales. The contradiction is only apparent.
The static modes at the scale $T$ develop both an ultraviolet 
and an infrared divergence that cancel against each other if regularized by the same 
cutoff in dimensional regularization as assumed in Eq.~\eqref{tcontribnlos}.  
In general, however, the ultraviolet divergence of the static modes at the
scale $T$ cancels against the ultraviolet divergence of the non-static modes,
such that the sum of static and non-static modes at the scale $T$ ends up 
having only a residual infrared divergence. It is precisely this infrared
divergence coming from the scale $T$, formally identical to the divergence in Eq.~\eqref{finalcontribtnlons}, 
that cancels against the ultraviolet divergence in (\ref{finalcontribm}) coming from the scale $m_D$. Similar patterns of cancellations were observed in Chap.~\ref{chap_rggT}.} 
Note that the gauge-dependent part of Eq.~\eqref{staticvacpol} gives a vanishing integral, 
thus yielding the expected gauge-independent result.

\subsubsection{Final result at order $g^4$}
Summing all contributions (static and the non-static) 
from the scales $T$ and $m_D$ up to order $g^4$ thus gives
\begin{equation}
\langle L_R\rangle=1+\frac{\crr \als }{2}\frac{m_D}{T}+\frac{\crr\alpha^2_s}{2}
\left[\ca\left(\ln\frac{m_D^2}{T^2}+\frac{1}{2}\right)-n_f\ln2\right]+\mathcal{O}(g^5).
\label{finalg4loop}
\end{equation}

\subsection{Comparison with the literature} 
At order $g^4$, the Polyakov loop was first calculated in the pure gauge 
case ($n_f=0$) and in Feynman gauge, by Gava and Jengo (GJ) \cite{Gava:1981qd}, 
who find  
\begin{equation}
\langle L_R\rangle_{\mathrm{GJ}}=
1 +\frac{\crr \als }{2}\frac{m_D}{T}+\frac{C_RC_A\als^2}{2}
\left(\ln \frac{m_D^2}{T^2 }- 2\ln 2+\frac{3}{2}\right)+\mathcal{O}(g^5)\,.
\label{gavajengo}
\end{equation}
Their result disagrees with ours, given in Eq.~\eqref{finalg4loop}.

The disagreement may be traced back to an incorrect treatment of the 
static modes at the scale $m_D$ in \cite{Gava:1981qd}. In Feynman gauge, 
at order $g^4$, three terms contribute to the Polyakov loop: 
the non-static gluon self energy, whose dominant contribution comes from the scale $T$, the
static gluon self energy, getting contributions from the scale
$m_D$ only, and a third term coming from the fourth-order expansion of
the Polyakov line. The computation of Gava and Jengo correctly
reproduces the first and the third term. We show this with some detail 
in  appendix~\ref{app_feynman}. However, in the evaluation of the static 
gluon self energy, the Debye mass is not resummed in the temporal 
gluons, leading to an inconsistent treatment of the scale $m_D$.\footnote{
In \cite{Gava:1981qd}, some contributions coming from the resummation of the Debye mass 
seem to have been included in $\delta W(0)$.
} 
Indeed, they have 
\begin{equation}
\Pi_{00}^{\mathrm{S}}(\bk)_\mathrm{GJ}= g^2C_AT \m2  
\int\frac{d^dp}{(2\pi)^d}\left(\frac{d-1}{\bp^2}-\frac{2\bk^2}{\bp^2\bq^2}\right),
\label{staticvacpolgj}
\end{equation}
which is the static self energy in Feynman gauge but \emph{without} resumming the 
Debye mass in the internal propagators. If, instead, the Debye mass is resummed, the expression 
of the static self energy changes to Eq.~\eqref{staticvacpol} with $\xi=1$.
In this case, the calculation of the Polyakov loop in Feynman gauge 
leads to exactly the same result as in Eq.~\eqref{finalg4loop}.
 
Finally we observe that Burnier, Laine and Veps\"al\"ainen \cite{Burnier:2009bk}, in the context of a perturbative analysis of the singlet quark-antiquark free energy, independently obtained the same result for the Polyakov loop at order $g^4$, given by Eq.~\eqref{finalg4loop}. Their calculation was performed within a dimensionally reduced effective field theory framework in a covariant or Coulomb gauge\footnote{Ref.~\cite{Burnier:2009bk} was published while the next, final Chapter of this Part~\ref{part_imtime}, which also constitutes the second part of our publication~\cite{Brambilla:2010xn}, was still being completed.}.

\subsection{Higher-order contributions\label{sub_nnlo}}
In Sec.~\ref{sub_nlo}, we obtained in Eq.~\eqref{mcontribnlonsfinal} 
a term that is of order $g^4\times(m_D/T)\sim g^5$. Other contributions of order $g^5$ 
can only come from contributions from the scales $m_D$ and $g^2T$ to $\langle\trt A_0^2\rangle$.  Hence, they are encoded in the 
($n\ge2$)-loop expression of the gluon self energy.

At order $g^6$, we can expect other contributions from the two-loop self energy 
and contributions coming from the diagram in Fig.~\ref{fig:cubefourth} b. 
We explicitly calculate these last ones due to their relevance for  Chap.~\ref{chap_imtimeEFT}.
The computation is carried out by evaluating the colour trace of the
diagram in the representation $R$, whereas the loop integrations are
easily obtained by comparison with Eq.~\eqref{fleading}. Thus we obtain 
\begin{equation} 
\delta\langle L_\mathrm{R}\rangle=
\left(3\crr^2-\frac{\crr\ca}{2}\right)\frac{\als^2}{24}\left(\frac{m_D}{T}\right)^2.
\label{als2m2}
\end{equation} 
The colour structure of this quartic term is not linear in $C_R$, 
a fact that will play a role in Chap.~\ref{chap_imtimeEFT}.
We recall here that the linear dependence of $\ln \langle L_\mathrm{R}\rangle$ 
on the Casimir operator $C_R$ is called \emph{Casimir scaling} of the Polyakov loop.
Equation~\eqref{als2m2} provides the leading perturbative correction that breaks the 
Casimir scaling. It is a tiny correction of order $g^6$, 
which may explain, at least in the weak-coupling regime, the approximate Casimir scaling 
observed in lattice calculations \cite{Gupta:2007ax}.

\section{The Polyakov-loop correlator at order $g^6$ for $rT\ll 1$\label{sec_corr}}
The spatial correlator of Polyakov loops in the fundamental representation 
has been defined in Eq.~\eqref{PLC} as
\begin{equation}
\label{defcorr}
\langle\tilde{\mathrm{Tr}}L_F^\dagger({\bf0})\tilde{\mathrm{Tr}}L_F(\br)\rangle.
\end{equation}
Following the notation of \cite{Nadkarni:1986cz}, we define 
$C_{\mathrm{PL}}(r,T)$ as the connected part of the correlator 
\begin{equation}
\label{defcpl}
C_{\mathrm{PL}}(r,T)\equiv
\langle\tilde{\mathrm{Tr}}L_F^\dagger({\bf0})\tilde{\mathrm{Tr}}L_F(\br)\rangle_\mathrm{c}
= \langle\tilde{\mathrm{Tr}}L_F^\dagger({\bf0})\tilde{\mathrm{Tr}}L_F(\br)\rangle-\langle L_F\rangle^2.
\end{equation}
Expanding Eq.~\eqref{defcpl} up to order $g^6$ yields\footnote{
We adopt a slightly different definition of
$C_\mathrm{PL}(r,T)$ with respect to \cite{Nadkarni:1986cz}, in that we
consider the zeroth-order term in the perturbative expansion, i.e $1$,
as part of $\langle L_F\rangle^2$ rather than of $C_\mathrm{PL}$.}
\begin{eqnarray}
C_{\mathrm{PL}}(r,T)&=&
\frac{g^4}{(2!)^2}\frac{\langle\trt A_0^2({\bf0})\trt A_0^2(\br)\rangle_\mathrm{c}}{T^4}
+\frac{g^6}{(3!)^2}\frac{\langle \trt A_0^3({\bf0})\trt A_0^3(\br)\rangle_\mathrm{c}}{T^6}
\nonumber\\
&&
-\frac{2g^6}{2!\,4!}\frac{\langle \trt A_0^4({\bf0})\trt A_0^2(\br)\rangle_\mathrm{c}}{T^6}
+\mathcal{O}(g^8).
\label{pertcpl}
\end{eqnarray}
Since the generators of SU$(\nc)$ are traceless, the first term in the
expansion, which is $g^2\langle\trt A_0({\bf0})\trt A_0(\br)\rangle_\mathrm{c}$, 
vanishes and thus the correlator starts in perturbation theory with
a two-gluon exchange term. Terms with an odd number of gauge fields have been omitted from Eq.~\eqref{pertcpl} since they vanish for charge-conjugation symmetry.

We will perform a complete calculation of the Polyakov-loop correlator 
for distances $r T \ll 1$. This situation corresponds to temperatures lower 
than the inverse distance of the quark-antiquark pair, hence it is the right one to make 
contact with known zero-temperature results. We assume the following hierarchy: 
\begin{equation}
\frac{1}{r}\gg T\gg m_D\gg \frac{g^2}{r}.
\label{defhierarchy}
\end{equation}
The scales $1/r$ and $g^2/r$ are the typical scales appearing in any perturbative 
static quark-antiquark correlator calculation, as discussed at length in the previous Parts of this thesis.
The scales $T$ and $m_D$ are associated to the thermodynamics of the system.
We assume that they are smaller than $1/r$, because we are interested in short 
distances. We assume that they are larger than $g^2/r$, because 
we would like to study a situation where both thermodynamical scales affect the 
quark-antiquark potential. A similar hierarchy has been studied in real-time in the previous Part~\ref{part_realtime}, see footnotes~\ref{foot_mdggE} and \ref{foot_potmdggE} in Chap.~\ref{chap_rggT}. In the weak-coupling regime, 
as discussed above, $T \gg m_D$, where $m_D$ is given by Eq.~\eqref{fulldebye}. 
Equation~\eqref{defhierarchy} amounts 
to having two largely unrelated small parameters, $g$ and $rT$, the hierarchy 
only requiring $rT \gg g$. Differently from the Polyakov-loop calculation where we had only $g$, 
the perturbative expansion of the Polyakov-loop correlator is, therefore, organized as a double 
expansion in $g$ and $rT$. We will stop the expansion for the Polyakov-loop correlator 
at order $g^6(rT)^0$, meaning that, given a term of order  $g^k(rT)^n$, 
we will display it only if $k<6$, for any (positive or negative) $n$, or if $k=6$, for $n\le 0$; 
we will not display it elsewhere. We should note here that,  
as in any double expansion whose expansion parameters are unrelated, 
undisplayed terms may, under some circumstances, turn out to be numerically 
as large as or larger than some of the displayed ones.\footnote{ 
A posteriori (see the final result in Eq.~\eqref{finalcpltot}), this may be avoided, in our case,  
by further requiring that $rT \gg \sqrt{g}$.}

In \cite{Nadkarni:1986cz}, Nadkarni computed the Polyakov-loop correlator
up to order $g^6$ within EQCD, using resummed temporal-gluon propagators throughout the
computation, which amounts to calculating the Polyakov-loop correlator for
distances $rm_D\sim 1$. Our calculation will differ from Nadkarni's one in that 
we adopt the different hierarchy~\eqref{defhierarchy}. Nevertheless, some of our results can be obtained 
by expanding Nadkarni's result for $rm_D\ll 1$; we refer to Sec.~\ref{sub_nadkarni} 
for a detailed comparison between the two results.

The calculation of the different contributions to Eq.~\eqref{pertcpl} will proceed 
similarly to the calculation of the Polyakov loop performed in the previous 
section. We will consider the different Feynman diagrams contributing to each of the terms 
in~\eqref{pertcpl}, separate the contributions from the different energy scales and, in case, 
distinguish between static and non-static modes. A similar calculation, albeit in real time, with a separation of the energy scales has been performed in App.~\ref{secpQCD}.

\begin{figure}[ht]
\begin{center}
\includegraphics{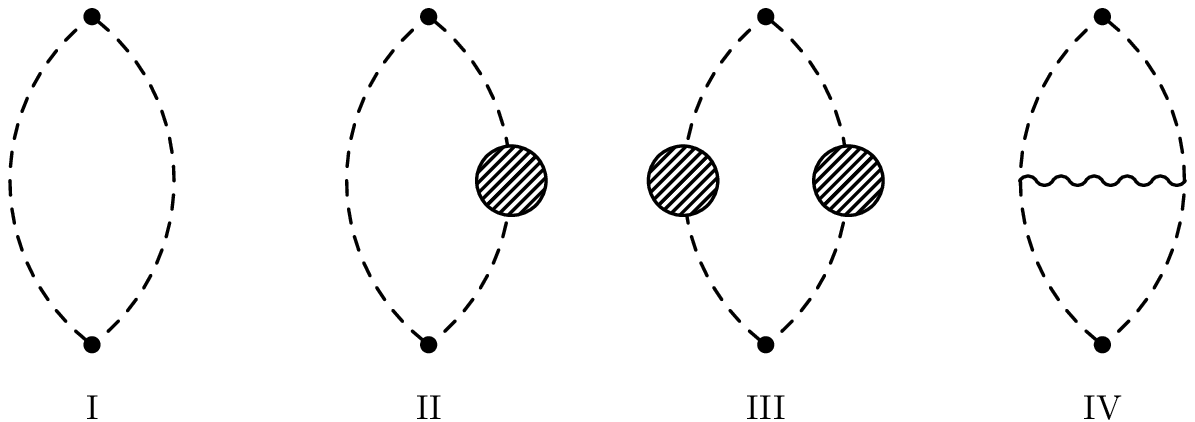}
\end{center}
\caption{Diagrams contributing to $\langle\trt A_0^2({\bf0})\trt A_0^2(\br)\rangle_\mathrm{c}$.}
\label{fig:corr-lo}
\end{figure}

\subsection{The leading-order contribution: diagram I}
We start by evaluating the four-field correlation function: its
leading-order contribution is given by diagram I in Fig.~\ref{fig:corr-lo}. 
It does not vanish only for momenta of order $1/r$, giving 
\begin{equation}
\delta C_\mathrm{PL}(r,T)_\mathrm{I}
=\frac{\nc^2-1}{8\nc^2}\frac{g^4}{T^2}\left(\m2
\int\frac{d^dk}{(2\pi)^d}\frac{e^{-i\bk\cdot\br}}{\bk^2}\right)^2
=\frac{\nc^2-1}{8\nc^2}\frac{\als^2}{ (rT)^2}.
\label{rg4contrib}
\end{equation}

\subsection{The contribution from diagrams of type II}
As we go beyond leading order, the first class of diagrams that we consider are 
those with gluon self-energy insertions in one temporal-gluon line, whose 
first example is diagram II in Fig.~\ref{fig:corr-lo}.
They give 
\begin{equation}
\delta C_\mathrm{PL}(r,T)_\mathrm{II} =
2\frac{\nc^2-1}{8\nc^2}\frac{g^4}{T^2}\frac{1}{4\pi r}\m2 
\int\frac{d^dk}{(2\pi)^d}e^{-i\bk\cdot\br}
\left(\frac{1}{\bk^2+\Pi_{00}(\bk)} - \frac{1}{\bk^2}\right),
\label{defselfenergyinsert}
\end{equation}
where the factor $2$ comes from the symmetric diagrams and 
$\Pi_{00}$ is the sum of bosonic and fermionic contributions to the gluon self
energy, as in the Polyakov-loop case. This diagram receives contributions from all scales and
depends on the gauge parameter $\xi$. However it can be shown that the gauge
dependence cancels with diagram IV \cite{Nadkarni:1986cz}, so, for
simplicity, here we write our results in static Feynman gauge, $\xi=1$.

\subsubsection{Contribution from the scale $1/r$}
We start by evaluating the contribution from the scale $1/r$ in the integral.
If $\mbk\sim 1/r\gg T$, then we have 
\begin{equation}
\delta C_\mathrm{PL}(r,T)_{\mathrm{II}\,1/r} =
-\frac{\nc^2-1}{4\nc^2}\frac{g^4}{T^2}\frac{1}{4\pi r}\m2 
\int\frac{d^dk}{(2\pi)^d} e^{-i\bk\cdot\br} \frac{\Pi_{00}(\mbk\gg T)}{\bk^4}
\left[1 + {\cal O}\left( \frac{g^2}{rT} \right) \right],
\label{defselfenergyinsertr}
\end{equation}
where $\Pi_{00}(\mbk\gg T)$ is given by Eq.~\eqref{kggtfull}. 
The Fourier transform of the vacuum part corresponds to the one-loop
static QCD potential and can be read from Eq.~\eqref{alvs}.
Using Eq.~\eqref{ftregdim} for the Fourier transform of $1/\mbk^n$ in dimensional regularization, we have
\begin{eqnarray}
\delta C_\mathrm{PL}(r,T)_{\mathrm{II}\,1/r}&=&
\frac{\nc^2-1}{8\nc^2}\frac{\als^3}{ (rT)^2}
\left\{\frac{1}{2\pi}
\left[2\beta_0(\ln(\mu r)+\gamma_E)+ \frac{31}{9}\ca-\frac{10}{9}n_f \right]\right.
\nonumber\\
&&\left.+\ca\left(\frac{1}{12rT} -rT -\frac{2 }{9}\pi(rT)^2\right)\right\}
+\mathcal{O}\left(g^6(rT)^2,g^7\right).
\label{selfenergyinsertr}
\end{eqnarray}
The term in the first line comes from the Fourier transform of the vacuum
contribution, whereas the terms in the second line come respectively from the
singular part, the (zero mode) order $\epsilon$ term\footnote{
The dimensionally-regularized Fourier transform of the order $\epsilon$ term 
in Eq.~\eqref{kggtfull}  yields a $1/\epsilon$ pole, eventually leading to a finite contribution.}  
and the $T^2$ term in Eq.~\eqref{kggtfull}. 
Higher-order corrections to Eq.~\eqref{kggtfull} contribute at order 
$g^6(rT)^2$ or $g^7$. Higher order radiative corrections to the gluon
self energy contribute at order $g^8$. Note that the $(\als^3/\pi)  \beta_0 \ln(\mu r)$ term in 
Eq.~\eqref{selfenergyinsertr} fixes the natural scale of  $\als^2$ in the LO 
term $\delta C_\mathrm{PL}(r,T)_\mathrm{I}$ to be $1/r$.

\subsubsection{Contributions from the scales $T$ and $m_D$}
We now consider the contributions from the thermal scales. For what
concerns the temperature, $\mbk\sim T$ translates into $r \mbk\ll 1$
and $m_D \ll \mbk$. Integrating out the temperature leads to the following 
contribution 
\begin{eqnarray}
\delta C_\mathrm{PL}(r,T)_{\mathrm{II}\,T}&=&
-\frac{\nc^2-1}{4\nc^2}\frac{g^4}{T^2}\frac{1}{4\pi r}
\m2 \int\!\frac{d^dk}{(2\pi)^d}
\left[1+  {\cal O}((\bk \cdot \br)^2) \right]
\frac{\Pi_{00}(\mbk\sim T)}{\bk^4}
\nonumber\\
&& \hspace{5cm}
\times
\left[ 1 +  {\cal O}(g^2)\right],
\label{defselfenergyinsertt}
\end{eqnarray}
where we have implemented the condition  $r\mbk\ll 1$ by expanding the Fourier exponent.
Integrating out the Debye-mass scale leads to the following contribution
\begin{eqnarray}
\nonumber\delta C_\mathrm{PL}(r,T)_{\mathrm{II}\,m_D}&=&
\frac{\nc^2-1}{4\nc^2}\frac{g^4}{T^2}\frac{1}{4\pi r}\m2 \int\frac{d^dk}{(2\pi)^d}
\left[1+  {\cal O}((\bk \cdot \br)^2) \right]
\left[\frac{1}{\bk^2+m_D^2}   \right.
\\
&& \hspace{2cm} \left.-\frac{\Pi_{00}(\mbk\sim m_D)-m_D^2}{(\bk^2+m_D^2)^2}
+ {\cal O}\left( \frac{g^4}{m_D^2}\right) 
\right].
\label{defselfenergyinsertm}
\end{eqnarray}
The integrals to be evaluated are the same needed to evaluate 
Eqs.~\eqref{loopbT},~\eqref{loopbm0} and~\eqref{loopbm}.  
Thus, summing the $T$ and $m_D$ contributions, we obtain
\begin{eqnarray}
\nonumber\delta C_\mathrm{PL}(r,T)_{\mathrm{II}\,T+m_D}&=&
-\frac{\nc^2-1}{4\nc^2}\frac{\als^2}{rT}\left\{\frac{m_D}{T}
+\als\left[\ca\left(\ln\frac{m_D^2}{T^2}+\frac12\right)-n_f\ln2\right]\right\}
\\
&&+\mathcal{O}\left(\frac{g^7}{rT},g^6(rT)\right).
\label{selfenergyinserttm}
\end{eqnarray}
The term of order $g^5/(rT)$ comes from the first term in
\eqref{defselfenergyinsertm}, the terms of order $g^6/(rT)$ come
from the non-static modes in~\eqref{defselfenergyinsertt} and from the 
static ones in the second term of Eq.~\eqref{defselfenergyinsertm}, the 
appearance of the logarithm $\ln{m_D^2}/{T^2}$ signals the cancellation between 
divergences at the scale $T$ and $m_D$, the suppressed term $g^7/(rT)$ 
comes from the non-static modes in the second term of
Eq.~\eqref{defselfenergyinsertm} (see Eqs.~\eqref{mcontribnlons} and 
\eqref{mcontribnlonsfinal} for the analogous case in the Polyakov-loop
calculation), whereas the suppressed term $g^6(rT)$ comes from the
$(\bk\cdot\br)^2$ term in Eq.~\eqref{defselfenergyinsertt}.

\subsection{The contribution from diagrams of type III}
Diagram III in Fig.~\ref{fig:corr-lo} is the first example of the class of
diagrams with gluon self-energy insertions in both temporal-gluon lines.
They may be evaluated from the diagrams of type II:
\begin{eqnarray}
\delta C_\mathrm{PL}(r,T)_\mathrm{III} 
&=&\frac{\nc^2-1}{8\nc^2}\frac{g^4}{T^2}\left[\m2
\int\frac{d^dk}{(2\pi)^d} e^{-i\bk\cdot\br}
\left( \frac{1}{\bk^2+\Pi_{00}(\bk)} - \frac{1}{\bk^2} \right) \right]^2
\nonumber\\
&=&\frac{8\nc^2}{\nc^2-1}\frac{T^2}{g^4}\left(4\pi r\frac{\delta C_\mathrm{PL}(r,T)_\mathrm{II}}{2}\right)^2 .
\label{defIII}
\end{eqnarray}
The leading-order term in~\eqref{selfenergyinserttm} gives a $g^6$ contribution 
to $\delta C_\mathrm{PL}(r,T)_\mathrm{III}$, all other contributions being at
least of order $g^7/(rT)^2$, 
\begin{equation}
\delta C_\mathrm{PL}(r,T)_{\mathrm{III}}=
\frac{\nc^2-1}{8\nc^2}\als^2\frac{m_D^2}{T^2}+\mathcal{O}\left(\frac{g^7}{(rT)^2}\right).
\label{finalIII}
\end{equation}

\subsection{The contribution from diagrams of type IV}
The transverse static-gluon exchange between the two temporal-gluon lines
(diagram IV and the diagrams derived from IV by inserting gluon self energies 
in each of the gluon lines) receives the following contributions. 

\subsubsection{Contribution from the scale $1/r$}
The contribution from the scale $1/r$ reads at leading order (with $\xi=1$) 
\begin{eqnarray}
\delta C_\mathrm{PL}(r,T)_\mathrm{IV\,1/r}
&=& \frac{g^6}{4T}\frac{\nc^2-1}{2\nc^2}\ca 
\mu^{6\epsilon} \int \frac{d^dk_1}{(2\pi)^d} \int \frac{d^dk_2}{(2\pi)^d} \int \frac{d^dp}{(2\pi)^d}
e^{-i(\bk_1-\bk_2)\cdot\br}
\nonumber\\
&&
\hspace{3.5cm} \times 
\frac{(2\bk_1+\bp)\cdot(2\bk_2+\bp)}{\bk_1^2\bk^2_2(\bk_1+\bp)^2(\bk_2+\bp)^2\bp^2}.
\label{nadkarnicontrib}
\end{eqnarray}
Gluon self-energy insertions are suppressed by $g^2$.

\subsubsection{Contribution from the scale $T$}
The contribution from the scale $T$ vanishes, because scaleless, if no 
self-energy insertions are considered. Hence, the leading contribution from the
scale $T$ is of order $g^6/T \times g^2T \sim g^8$. 

\subsubsection{Contribution from the scale $m_D$}
The contribution from the scale $m_D$ reads 
\begin{eqnarray}
&&\hspace{-10mm}
\delta C_\mathrm{PL}(r,T)_\mathrm{IV\,m_D} =
\nonumber\\
&& 
\frac{g^6}{4T}\frac{\nc^2-1}{2\nc^2}\ca
\mu^{6\epsilon} \int \frac{d^dk_1}{(2\pi)^d} \int\frac{d^dk_2}{(2\pi)^d}  \int\frac{d^dp}{(2\pi)^d}
\left[1+  {\cal O}(((\bk_1-\bk_2) \cdot \br)^2) \right]
\nonumber\\
&&
\hspace{0.5cm} 
\times 
\frac{(2\bk_1+\bp)\cdot(2\bk_2+\bp)}{(\bk_1^2+m_D^2)(\bk^2_2+m_D^2)((\bk_1+\bp)^2+m_D^2)((\bk_2+\bp)^2+m_D^2)\bp^2}
\left[1+  {\cal O}(g) \right].\nn\\
\label{nadkarnicontribm}
\end{eqnarray}
This corresponds to a contribution  of order $g^6(m_D/T)\sim g^7$, which is beyond our accuracy.

The leading contribution to $\delta C_\mathrm{PL}(r,T)_\mathrm{IV}$ comes, therefore, from  $\delta
C_\mathrm{PL}(r,T)_\mathrm{IV\,1/r}$, which can be computed in dimensional
regularization with the help of Eq.~\eqref{ftregdim}. Our final result reads
\begin{equation}
\delta C_\mathrm{PL}(r,T)_\mathrm{IV}
=\frac{\nc^2-1}{2\nc}\frac{\als^3 }{rT}\left(1-\frac{\pi^2}{16}\right)+\mathcal{O}\left(g^7\right).
\label{corr-IV}
\end{equation}
The same result follows from \cite{Nadkarni:1986cz} by expanding in $rm_D\ll 1$.

\begin{figure}[ht]
\begin{center}
\includegraphics{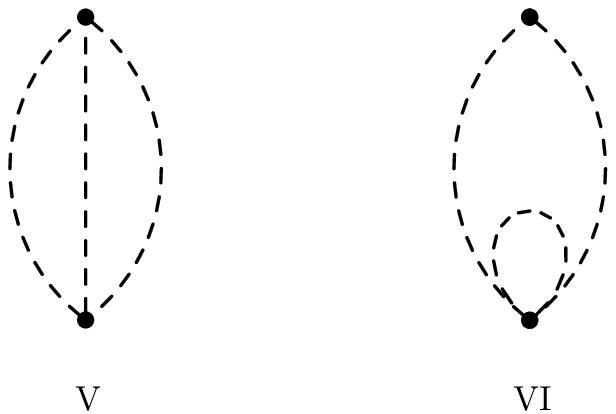}
\end{center}
\caption{Diagram V is the first contribution to $\langle \trt
  A_0^3({\bf0})\trt A_0^3(\br)\rangle_\mathrm{c}$, whereas diagram VI is the
  first contribution to $\langle \trt A_0^4({\bf0})\trt A_0^2(\br)\rangle_\mathrm{c}$.}
\label{fig:corr-nlo}
\end{figure}

\subsection{The contribution from diagrams of type V}
Diagrams contributing to the correlators of six $A_0$ fields in
Eq.~\eqref{pertcpl} are shown in Fig.~\ref{fig:corr-nlo}. 
The LO diagram  contributing to $\langle \trt A_0^3({\bf0})\trt
A_0^3(\br)\rangle$ is diagram V, which gives 
\begin{equation}
\delta C_\mathrm{PL}(r,T)_\mathrm{V}
=\frac{(\nc^2-4)(\nc^2-1)}{96\nc^3}\frac{\als^3}{(rT)^3}.
\label{corr-33}
\end{equation}
If we consider diagram V with gluon self-energy insertions in one of the
temporal lines, in analogy to~\eqref{defselfenergyinsertm},  then this starts 
contributing at order $g^7/(rT)^2$, which is beyond our accuracy.

\subsection{The contribution from diagrams of type VI}
Diagrams contributing to $\langle \trt A_0^4({\bf0})\trt
A_0^2(\br)\rangle$ are like diagram VI in Fig.~\ref{fig:corr-nlo}
and diagrams derived from VI by inserting gluon self energies 
and other radiative corrections. 
Colour factors aside, their leading contribution may be estimated 
by simply multiplying the contribution of the diagrams of
Fig.~\ref{fig:corr-lo} to the Polyakov-loop correlator 
with the contribution of the diagrams of Fig.~\ref{fig:square}
to the Polyakov loop. Hence, diagrams of type VI
contribute at LO to order $g^4/(rT)^2 \times g^2m_D/T \sim g^7/(rT)^2$, 
which is beyond our accuracy.

\subsection{The Polyakov-loop correlator up to order $g^6$}
Summing up all contributions, we then have 
\begin{eqnarray}
C_{\mathrm{PL}}(r,T)&=&
\frac{\nc^2-1}{8\nc^2}\left\{ 
\frac{\als(1/r)^2}{(rT)^2}
-2\frac{\als^2}{rT}\frac{m_D}{T} \right.
\nonumber\\
&&
\hspace{1.5cm}
+\frac{\als^3}{(rT)^3}\frac{\nc^2-2}{6\nc}
+ \frac{1}{2\pi }\frac{\als^3}{(rT)^2}\left(\frac{31}{9}\ca-\frac{10}{9}n_f +2\gamma_E\beta_0\right)
\nonumber\\
&&
\hspace{1.5cm}
+ \frac{\als^3}{rT}\left[
\ca\left(-2 \ln\frac{m_D^2}{T^2} + 2-\frac{\pi^2}{4}\right) + 2n_f\ln 2\right]
\nonumber\\
&&
\hspace{1.5cm}
\left.
+\als^2\frac{m_D^2}{T^2} -\frac{2}{9}\pi \als^3 \ca
\right\}
+\mathcal{O}\left(g^6(rT),\frac{g^7}{(rT)^2}\right),
\label{finalcpltot}
\end{eqnarray}
where we have made explicit the scale dependence of $\als$ in the leading term. 
Note that the $r$, $T$ and $m_D$ independent term proportional to 
$-2\pi \als^3 \ca/9$ comes from Eq.~\eqref{selfenergyinsertr}, so it is 
actually a contribution from the scale $1/r$ that accounts for the matter part of the 
gluon self energy. The term proportional 
to $\als^3/(rT)^3$ comes from diagram V, Eq.~\eqref{corr-33}, 
and from the singular part of the gluon self energy in the static gauge, 
Eq.~\eqref{selfenergyinsertr}.

\subsection{Comparison with the result of Nadkarni\label{sub_nadkarni}}
We compare here with Nadkarni's (N) computation of the Polyakov-loop correlator \cite{Nadkarni:1986cz}. 
The regime of validity of Nadkarni's computation is $T \gg 1/r \sim  m_D$, while ours is 
$1/r \gg T \gg m_D$. Therefore, we may only compare results obtained here that 
do not involve the hierarchy $rT \ll 1$, with Nadkarni's results that do not 
involve the hierarchy $rT \gg 1$, expanded for $rm_D \ll 1$.

In \cite{Nadkarni:1986cz}, the tree-level expression of 
$g^4\langle\trt A_0^2({\bf0})\trt A_0^2(\br)\rangle_\mathrm{c}/(4T^4)$ reads 
$(\nc^2-1)/(8\nc^2)$ $\als^2\exp(-2rm_D)/(rT)^2$, which expanded for $rm_D \ll 1$ 
gives $\delta C_\mathrm{PL}(r,T)_\mathrm{I}$, the LO of 
$\delta C_\mathrm{PL}(r,$ $T)_{\mathrm{II}\,m_D}$ (to be read from Eq.~\eqref{selfenergyinserttm})
and $\delta C_\mathrm{PL}(r,T)_{\mathrm{III}}$. Also, the tree-level expression of 
$g^6\langle \trt A_0^3({\bf0}) \trt A_0^3(\br)\rangle_\mathrm{c}/(36T^6)$
in \cite{Nadkarni:1986cz} agrees with $\delta C_\mathrm{PL}(r,T)_\mathrm{V}$
once expanded for $rm_D \ll 1$.

Diagram IV in Fig.~\ref{fig:corr-lo} also contributes 
to Nadkarni's calculation. The diagram does not involve gluon self-energy 
insertions and therefore its calculation does not rely on the hierarchy 
between $1/r$ and $T$. As already remarked, $\delta C_\mathrm{PL}(r,T)_\mathrm{IV}$ 
agrees with  Nadkarni's result once expanded for $rm_D\ll 1$.\footnote{
In Nadkarni's paper this contribution is called  $f_{II}$.}

Let us now consider the NLO contribution to $\delta C_\mathrm{PL}(r,T)_{\mathrm{II}\,m_D}$.
This contribution is given by the static part of Eq.~\eqref{defselfenergyinsertm}:
\begin{equation}
\delta C_\mathrm{PL}(r,T)_{\mathrm{II}\,{\rm N}\,m_D}=
-\frac{\nc^2-1}{4\nc^2}\frac{g^4}{T^2}\frac{1}{4\pi r}\m2
\int\frac{d^dk}{(2\pi)^d}\frac{\Pi_{00}^\mathrm{S}(\mbk)}{(\bk^2+m_D^2)^2}.
\label{defselfenergyinsertmnad}
\end{equation}
The integral is divergent. In our case, i.e. assuming $1/r \gg T \gg m_D$, 
the divergence cancels against $\delta C_\mathrm{PL}(r,T)_{\mathrm{II}\,T}$, eventually 
leading to a finite result in $\delta C_\mathrm{PL}(r,T)_{\mathrm{II}\,T+m_D}$.
The $\ln m_D/T$ term in Eq.~\eqref{selfenergyinserttm} signals precisely that 
a divergence at the scale $m_D$ has canceled against a divergence at the scale $T$.
In  Nadkarni's case,  i.e. assuming $T \gg 1/r \gg m_D$, we get, along with 
$\delta C_\mathrm{PL}(r,T)_{\mathrm{II}\,{\rm N}\,m_D}$, a contribution from the scale $1/r$, which is 
\begin{equation}
\delta C_\mathrm{PL}(r,T)_{\mathrm{II}\,{\rm N}\,1/r} =
-\frac{\nc^2-1}{4\nc^2}\frac{g^4}{T^2}\frac{1}{4\pi r}\m2 
\int\frac{d^dk}{(2\pi)^d}e^{-i\bk\cdot\br}\frac{\Pi_{00}^\mathrm{S}(\mbk\gg m_D)}{\bk^4}.
\label{defselfenergyinsertrnad}
\end{equation}
This is like Eq.~\eqref{defselfenergyinsertr},   
but involves only the static part of the self energy~\eqref{staticcontrhighk},
since non-static modes have been already integrated out at the larger scale $T$.
According to Eq.~\eqref{staticcontrhighk}, we have $\Pi_{00}^\mathrm{S}(\mbk\gg m_D)\sim T\mbk^{1-2\epsilon}$.
The Fourier transform of $1/\mbk^{3+2\epsilon}$ originates a $1/\epsilon$ pole.
It is this divergence that in Nadkarni's hierarchy cancels against the divergence in 
$\delta C_\mathrm{PL}(r,T)_{\mathrm{II}\,{\rm N}\,m_D}$ leading to the finite result
\begin{equation}
\delta C_\mathrm{PL}(r,T)_{\mathrm{II}\,{\rm N}\,m_D} + \delta C_\mathrm{PL}(r,T)_{\mathrm{II}\,{\rm N}\,1/r} =
-\frac{\nc^2-1}{2\nc}\frac{\als^3}{rT}\left[\ln (2 m_D r)
  +\gamma_E-\frac{3}{4} +\mathcal{O}(rm_D)\right],
\label{nadkarniII}
\end{equation}
which agrees with the result in  \cite{Nadkarni:1986cz}.\footnote{
In Nadkarni's paper this contribution is called  $f_I$.} 
In this case, the $\ln m_D r$ term signals that a divergence at the scale $m_D$ 
has canceled against a divergence at the scale $r$.
\section{Summary}
\label{sec_summary_pert}
In the weak-coupling regime, we have calculated the Polyakov loop up to order
$g^4$ and the correlator of two Polyakov loops up to order $g^6(rT)^0$, 
assuming the hierarchy of scales $\displaystyle \frac{1}{r}\gg T\gg m_D\gg \frac{g^2}{r}$. The former may be read from Eq.~\eqref{finalg4loop} and the latter from Eq.~\eqref{finalcpltot}.

The Polyakov-loop calculation differs from the result of Gava and Jengo
\cite{Gava:1981qd} by a finite contribution at order $g^4$. We have analyzed
in detail the origin of the difference and shown in Appendix~\ref{app_feynman} that our
result may be reproduced also performing the calculation in Feynman gauge.
Our calculation agrees with the recent finding of Ref.~\cite{Burnier:2009bk}.

The calculation of the Polyakov-loop correlator is new in the considered
regime, although some partial results may be deduced from a previous work of
Nadkarni, who studied distances $r \sim 1/m_D$ \cite{Nadkarni:1986cz}, as pointed out in Sec.~\ref{sub_nadkarni}. The significance of our result will be made clearer in the next chapter, where it will be understood in terms of colour-singlet and colour-octet contributions.

\chapter{The Polyakov-loop correlator in an EFT language}
\label{chap_imtimeEFT}
The calculation of the Polyakov-loop correlator discussed in the previous chapter can be 
conveniently rephrased in an Effective Field Theory language that exploits 
at the Lagrangian level the hierarchy of energy scales in Eq.~\eqref{defhierarchy} and is in close contact with the EFT framework introduced in real-time in Part~\ref{part_realtime}.
The EFT framework has the advantage to allow more easily for systematic improvements of the 
calculation and to make more transparent its physical meaning, bringing to a better understanding of the relation between the free energies extracted from the correlator and the real-time potentials governing the behaviour of quarkonium in media.

The Chapter is organized as follows: in Sec.~\ref{secpNRQCD} we deal with the scale $1/r$, the first in the hierarchy~\eqref{defhierarchy}. Integrating it out we shall obtain pNRQCD in imaginary time and we will show how to write the correlator in terms of the pNRQCD degrees of freedom. In the subsequent Sections~\ref{secpNRQCDT} and \ref{secpNRQCDmD} we will compute the contributions from the thermal scales $T$ and $m_D$, in the end reobtaining the result of the previous Chapter for the correlator, as given by Eq.~\eqref{finalcpltot}.\\
In Sec.~\ref{sec_free} we will show how gauge-invariant singlet and octet free energies can be naturally introduced within our formalism, and we will compare the singlet free energy with the static energies computed in the previous Part~\ref{part_realtime} in real time. In the following Sec.~\ref{sec_compare} we will compare our results with other calculations in the literature and finally in Sec.~\ref{sec_concl} we shall draw our conclusions.

\section{The scale $1/r$: pNRQCD}
\label{secpNRQCD}
Our starting point is the action of static QCD, as shown in Eq.~\eqref{QCD}. By integrating out from the static quark-antiquark sector gluons of energy or momentum that scale like 
the inverse of the distance $r$ between the quark and the antiquark we obtain pNRQCD. Since $1/r$ is the largest scale, we may again set to zero all other scales, the thermal ones in particular, in the matching of the pNRQCD Lagrangian;
as in Chap.~\ref{chap_rggT}, the Lagrangian is then identical to the one derived at zero 
temperature and discussed in Sec.~\ref{sec_pnrqcd}.
In Euclidean space-time, the action reads
\begin{eqnarray}
{\cal S}_{\rm pNRQCD} &=&  
\int_0^{1/T} \!\! d\tau \int d^3R \int d^3r\, {\rm Tr} \Bigg\{ {\rm S}^\dagger (\partial_0+V_s){\rm S}
+ {\rm O}^\dagger (D_0+V_o){\rm O}
\nonumber\\
&& \hspace{1.5cm}
- iV_A \left( {\rm S}^\dagger {\bf r}\cdot g{\bf E} {\rm O} + {\rm O}^\dagger {\bf r}\cdot g{\bf E} {\rm S}\right)
- \frac{i}{2}V_B \left({\rm O}^\dagger {\bf r}\cdot g{\bf E} {\rm O} + {\rm O}^\dagger{\rm O} {\bf r}\cdot g{\bf E} \right) 
\nonumber\\
&&  
\hspace{1.5cm} 
+ \frac{i}{8} V_C \left(r^i r^j  {\rm O}^\dagger D^igE^j {\rm O} - r^i r^j  {\rm O}^\dagger {\rm O} D^igE^j \right)
+ \delta {\cal L}_{\rm pNRQCD}
\Bigg\}
\nonumber\\
&&
+ \int_0^{1/T} \!\! d\tau \int d^3x \; \left(
\frac{1}{4}F^a_{\mu\nu}F^a_{\mu\nu} + \sum_{l=1}^{n_f} \bar{q}_lD\!\!\!\!/\,q_l
\right).
\label{pNRQCDeuc}
\end{eqnarray}
The terms at order $r^2$ in the multipole expansion were first derived in \cite{Brambilla:2002nu}. The normalization of the singlet and octet fields $\mathrm{S}$ and $\mathrm{O}$ is the same as in Sec.~\ref{sub_weakpnrqcd}, i.e ${\rm S} = 1\hspace{-1.1mm}{\rm l}_{\nc \times \nc} S/\sqrt{\nc}$,  ${\rm O} = \sqrt{2} T^a O^a$. The trace is over the colour indices.
In Euclidean space-time we have again $D_0{\rm O} = \partial_0 + ig[A_0,{\rm O}]$, 
${\bf D} = \bfnabla +ig{\bf A}$ and $E^i = F_{i0}$. 
The quantities $V_s$, $V_o$, $V_A$, $V_B$ and $V_C$ are 
the matching coefficients of the EFT. These are non-analytic functions of $r$. Since, as discussed in Sec.~\ref{sub_weakpnrqcd},  
$V_A(r) = 1 + {\cal O}(\als^2)$, 
$V_B(r) = 1 + {\cal O}(\als^2)$ and  $V_C(r) = 1 + {\cal O}(\als)$ it will suffice to our 
purposes to put $V_A(r)=V_B(r)=V_C(r)=1$ from now on. $V_s$ and $V_o$ are the singlet and octet 
potentials\footnote{$V_s$ and $V_o$ label in this Chapter the static potentials only, as the action~\eqref{pNRQCDeuc} is entirely devoid of non-static, $1/m$ suppressed corrections.} in pNRQCD: they are shown in Eq.~\eqref{staticpot}. For the purpose of obtaining the Polyakov-loop 
correlator at NNLO accuracy it is sufficient to know $V_s$ and $V_o$ at one-loop accuracy and their 
difference at two-loop accuracy: these three elements correspond to Eqs.~\eqref{a1} and \eqref{a2o} in App.~\ref{app_potentials} and can be summarized in
 \begin{eqnarray}
 && V_s(r) = -C_F\frac{\als(1/r)}{r}\left[ 1 
 + \left( \frac{31}{9}\ca-\frac{10}{9}n_f +2\gamma_E\beta_0 \right)\frac{\als}{4\pi} + {\cal O}(\als^2)\right],
 \label{Vspnrqcd}\\
 && V_o(r) = \hspace{0.9mm}\frac{1}{2\nc} \frac{\als(1/r)}{r}\left[ 1 
 + \left( \frac{31}{9}\ca-\frac{10}{9}n_f +2\gamma_E\beta_0 \right)\frac{\als}{4\pi} + {\cal O}(\als^2)\right],
 \label{Vopnrqcd}\\
 && (\nc^2-1)V_o(r) + V_s(r) =
 \frac{\nc(\nc^2-1)}{8}\frac{\als^3}{r}\left(\frac{\pi^2}{4}-3\right)\left[ 1 + {\cal O}(\als)\right].
 \label{Vo-Vs}
 \end{eqnarray}
Finally, $\delta {\cal L}_{\rm pNRQCD}$ includes all operators that are of
order $r^3$ or smaller. At tree-level, they may be read from the multipole expansion
of the quark and antiquark coupling to the temporal gluon in the static QCD
Lagrangian~\eqref{QCD}, hence they just involve covariant derivatives acting on a chromoelectric field: 
the leading-order operator being
$- i r^i r^j r^k {\rm Tr} \{ {\rm O}^\dagger D^iD^jgE^k {\rm S}$ $+{\rm S}^\dagger D^iD^jgE^k {\rm O} \}/24$ 
\cite{Brambilla:2002nu}. As we will argue in the next section, these terms 
contribute in principle to the correlator at order $g^4$, however, their contribution eventually cancels 
up to order $g^6(rT)^0$. For this reason, 
we do not need to specify them further here.

Matching the connected Polyakov-loop correlator to pNRQCD gives 
\begin{eqnarray}
C_\mathrm{PL}(r,T)
&=& 
\frac{1}{\nc^2}\Bigg[
Z_s \frac{\langle S(\br,{\bf 0},1/T)S^\dagger(\br,{\bf 0},0)\rangle}{\mathcal{N}_{Q\overline{Q}}}
+
Z_o \frac{\langle O^a(\br,{\bf 0},1/T)O^{a \, \dagger}(\br,{\bf 0},0)\rangle}{\mathcal{N}_{Q\overline{Q}}}
\nonumber\\
&& \hspace{0.8cm}
+ {\cal O}\left(\als^3(rT)^4\right)\Bigg] - \langle L_F \rangle^2.
\label{PLC-pNRQCD}
\end{eqnarray}
We recall that $\mathcal{N}_{Q\overline{Q}}=[\delta^3(\mathbf{0})]^2$. The right-hand side is the pNRQCD part of the matching.
It contains the singlet and octet correlators, 
$\langle S(\br,{\bf 0},1/T)S^\dagger(\br,{\bf 0},0)\rangle$ and 
$\langle O^a(\br,{\bf 0},1/T)O^{a \, \dagger}(\br,{\bf 0},0)\rangle$, not surprisingly 
because in the $r\to 0$ limit the tensor fields $\chi^\dagger_j({\bf 0},1/T)\psi_i({\bf r},1/T)$ 
and $\psi^\dagger_i({\bf r},0)\chi_j({\bf 0},0)$,  
appearing in the right-hand side of Eq.~\eqref{PLC-QCD}, decompose into the direct sum 
of a colour-singlet and a colour-octet component.
The colour-singlet and colour-octet correlators may be read from the Lagrangian~\eqref{pNRQCDeuc}:
\begin{eqnarray}
\frac{\langle S(\br,{\bf 0},1/T)S^\dagger(\br,{\bf 0},0)\rangle}{\mathcal{N}_{Q\overline{Q}}} &=& e^{-V_s(r)/T}(1+ \delta_s),
\label{SSpNRQCD}
\\
\frac{\langle O^a(\br,{\bf 0},1/T)O^{a\,\dagger}(\br,{\bf 0},0)\rangle}{\mathcal{N}_{Q\overline{Q}}} &=&
e^{-V_o(r)/T}\left[(\nc^2-1)\, \langle L_A \rangle  +  \delta_o\right],
\label{OOpNRQCD}
\end{eqnarray}
where $\delta_s$ and $\delta_o$ stand for loop corrections to the singlet and octet correlators respectively.
The factor $\langle L_A \rangle$
comes from the covariant derivative $D_0$ acting on the octet field in~\eqref{pNRQCDeuc}. 
The adjoint Polyakov loop $\langle L_A \rangle$ factorizes the contribution
coming from the gluons in the thermal bath that bind with the colour-octet quark-antiquark states to
form part of the spectrum appearing in the right-hand side of Eq.~\eqref{PLC-spectrum}.
In pNRQCD at zero temperature, a similar expression
factorizes the non-perturbative gluonic contribution to the gluelumps masses \cite{Brambilla:1999xf}.\\
Note that at finite temperature, for $T\simg g^2/r$, 
the octet correlator is not suppressed with respect to the singlet one, while 
in the opposite limit, $T \ll g^2/r$, the Polyakov-loop correlator 
is dominated by the singlet contribution.
Higher-dimensional operators have not been
displayed, because they are negligible with respect to our present accuracy, which is 
of order $g^6(rT)^0$.  The reason is that higher-dimensional operators involve the coupling with 
at least two field-strength tensors, hence the corresponding matrix elements
are at least of order $(rT)^4$; moreover, as can be seen by adding two
external gluons to diagram I of Fig.~\ref{fig:corr-lo}, the matrix element of an operator coupled
with two external gluons is at least of order $g^6$.
The normalization factors $Z_s$ and $Z_o$ have to be determined from the
matching condition~\eqref{PLC-pNRQCD}. While $V_s$ and $V_o$ 
are the same at zero and finite temperature, the normalization
factors are not for they depend on the boundary conditions. 

In order to determine the normalization factors $Z_s$ and $Z_o$, let us consider in Eq.~\eqref{PLC-pNRQCD}
only contributions coming from the scale $1/r$. 
In dimensional regularization, all loop corrections vanish in the pNRQCD part
of the matching and the Polyakov loops $\langle L_F \rangle$ and $\langle L_A \rangle$ 
reduce to one; therefore, the matching condition reads
\begin{equation}
C_\mathrm{PL}(r,T)_{1/r} = 
\langle\tilde{\mathrm{Tr}}L_F^\dagger({\bf 0})\tilde{\mathrm{Tr}}L_F(\br)\rangle_{1/r} - 1= 
\frac{1}{\nc^2}\left[ Z_se^{-V_s(r)/T} + Z_o(\nc^2-1)e^{-V_o(r)/T} \right] -1.
\label{pNRQCDrmatching}
\end{equation}
We may now proceed in different ways. A way consists in matching 
with the spectral decomposition~\eqref{PLC-spectrum}. 
By noting that at the scale $1/r$ the spectrum is just given by a singlet state 
of energy $V_s(r)$ and $\nc^2-1$ degenerate octet states of energy $V_o(r)$ and that $Z_{(0,0)}=1$ at the scale $1/r$,
the matching condition  implies that $Z_s = Z_o = 1$.
Another way consists in taking advantage of the Polyakov-loop correlator 
calculation done in Sec.~\ref{sec_corr} and matching to it.
$C_\mathrm{PL}(r,T)_{1/r}$ is the sum of Eq.~\eqref{rg4contrib}, 
Eq.~\eqref{selfenergyinsertr}  without the contribution from the matter part 
of the gluon self energy, Eq.~\eqref{corr-IV} and Eq.~\eqref{corr-33}; it reads
\begin{eqnarray}
C_\mathrm{PL}(r,T)_{1/r} &=& 
\frac{\nc^2-1}{8\nc^2}\left\{ \frac{\als(1/r)^2}{(rT)^2} 
+ \frac{\als^3}{(rT)^3}\frac{\nc^2-2}{6\nc} 
\right.
+ \frac{1}{2\pi }\frac{\als^3}{(rT)^2}\left(\frac{31}{9}\ca-\frac{10}{9}n_f\right. 
\nonumber\\
&& \hspace{1.7cm}
\left.+2\gamma_E\beta_0\bigg)
+ \frac{\als^3}{rT}\ca \left(3-\frac{\pi^2}{4}\right)
+{\cal O}\left(\frac{\als^4}{(rT)^4}\right)
\right\}.
\label{CPLr}
\end{eqnarray}
A direct inspection shows that this expression satisfies 
\begin{equation}
C_\mathrm{PL}(r,T)_{1/r} = 
\frac{1}{\nc^2}\left[ e^{-V_s(r)/T} + (\nc^2-1)e^{-V_o(r)/T} \right] -1,
\label{pNRQCDrverification}
\end{equation}
up to order $\als^3$, for $V_s(r)$ and $V_o(r)$ given by
Eqs.~\eqref{Vspnrqcd}-\eqref{Vo-Vs}.\footnote{
More precisely, the matching to~\eqref{CPLr} fixes $Z_s=Z_o=1$ up to order
$\als^2$ and $Z_s + (\nc^2-1)Z_o= \nc^2$ up to order $\als^3$.
}
We note that Eqs. (\ref{CPLr}) and (\ref{pNRQCDrverification}) are equivalent for $rT \gg g^2$,  
however, in Eq.~\eqref{pNRQCDrverification}, we resum some contributions
that would become large for $r T \siml g^2$. Equation~\eqref{pNRQCDrverification} is therefore valid also in that regime.
We furthermore observe that the combination of the two procedures provides a
non-trivial verification of Eq.~\eqref{Vo-Vs}, i.e. of the two-loop difference between the octet 
and the singlet potentials, known, so far, only from the direct calculation 
of the two-loop octet potential in a covariant gauge, done in Ref.~\cite{Kniehl:2004rk}. This method has recently been used by the authors of Ref.~\cite{Pineda:2011db} to check the two-loop calculation of the octet potential in an arbitrary number of spatial dimensions.

Loop corrections to the singlet and octet correlators in Eqs.~\eqref{SSpNRQCD} and~\eqref{OOpNRQCD} 
get contributions from the scales $T$, $m_D$ and lower ones. 
We now proceed to evaluate these corrections, separating the contributions 
of the temperature from the ones of the Debye mass.

\section{The temperature scale\label{secpNRQCDT}}
In the hierarchy~\eqref{defhierarchy}, the next scale after the inverse
distance is the temperature. Our aim is thus to compute the
temperature contributions to loop corrections in pNRQCD. These loop
corrections are the terms $\delta_s$ and $\delta_o$ 
that were introduced in Eqs.~\eqref{SSpNRQCD} and~\eqref{OOpNRQCD}. 
We call $\delta_{s,T}$ and $\delta_{o,T}$ the parts of $\delta_{s}$ and
$\delta_{o}$ respectively that encode the contributions coming from the scale $T$; they
may be obtained by expanding $\delta_{s}$ and $\delta_{o}$ in $m_D$,
$V_s$, $V_o$ and in any lower energy scale. Similarly, 
$\delta \langle L_R \rangle_T$ is the part of $\langle L_R \rangle$ that encodes 
the contributions coming from the scale $T$.
Different terms contribute to  $\delta_{s,T}$, $\delta_{o,T}$ and $\delta \langle L_R \rangle_T$; 
we examine them in the following.


\subsection{The singlet $r^2$ contributions}
We start considering the one-loop, order $r^2$ in the multipole expansion, correction 
to the singlet correlator induced by the standard one-loop diagram shown in Fig.~\ref{fig:hqself}, where the singlet emits and reabsorbs a chromoelectric gluon through an intermediate octet state. It yields
\begin{eqnarray}
\delta_{s}^{\,{\cal O}(r^2)} &=& \left( ig\sqrt{\frac{1}{2\nc}} \right)^2 r^ir^j T\sum_n
\int \frac{d^dk}{(2\pi)^d} \int_0^{1/T} \!\! d\tau \int_0^{\tau} \!\! d\tau'
e^{\tau V_s} \, e^{-(\tau-\tau')V_o} \, e^{-\tau' V_s}
\nonumber\\
&& \hspace{4cm} \times e^{-i(\tau-\tau')\omega_n}\,
\langle E^{i\,a} U_{ab}E^{j\,b}\rangle(\omega_n,\bk).
\label{EEVa}
\end{eqnarray}
In the sum integral, we may distinguish between contributions coming from the non-zero modes and from the zero modes.

For the contribution coming from the non-zero modes, only 
the leading-order chromoelectric correlator in momentum space $\langle E^{i\,a}
U_{ab}E^{j\,b}\rangle(\omega_n,\bk)$ ($U_{ab}$ stands for a Wilson straight line in the adjoint representation
connecting $E^{i\,a}$ with $E^{j\,b}$; at leading order $U_{ab} = \delta_{ab}$) is relevant at our accuracy:
\begin{equation}
\langle E^{i\,a} U_{ab}E^{j\,b}\rangle(\omega_n,\bk) = 
(\nc^2-1)\left[ \frac{k^ik^j}{\bk^2} + (\delta_{ij}-\hat{k}^i\hat{k}^j)\frac{\omega_n^2}{\omega_n^2+\bk^2}\right].
\label{EEleading}
\end{equation}
Loop corrections to the chromoelectric correlator contribute to the Polyakov-loop correlator 
at order $g^6(rT)$ or smaller.
Because of the hierarchy~\eqref{defhierarchy}, we  can  expand the
right-hand side of~\eqref{EEVa} in $V_o-V_s$. The longitudinal part of
the chromoelectric correlator, i.e. the first term in square brackets,
vanishes in dimensional regularization, whereas the transverse part is
sensitive to the scale $T$ through the Matsubara frequencies. After
performing the sum integral over the non-zero modes, we obtain
\begin{eqnarray}
\delta_{s,T}^{\, {\cal O}(r^2)\, {\rm NS}} &=& 
- g^2C_F \frac{r^2T}{9}(V_o-V_s) + g^2C_F \frac{r^2}{36}(V_o-V_s)^2 +  {\cal  O}\left(g^6(rT),\frac{g^8}{rT}\right)
\nonumber\\
&=& - \frac{2}{9} \pi \nc C_F\als^2rT + \frac{\pi}{36}\nc^2 C_F\als^3 
+  {\cal  O}\left(g^6(rT),\frac{g^8}{rT}\right).
\label{deltasEa}
\end{eqnarray}

The contribution coming from the zero modes reads
\begin{equation}
\delta_{s,T}^{\, {\cal O}(r^2)\, {\rm S}} = 
\left( ig\sqrt{\frac{1}{2\nc}} \right)^2  \frac{r^ir^j}{2T}
\int \frac{d^dk}{(2\pi)^d} \langle E^{i\,a} U_{ab}E^{j\,b}\rangle(0,\bk)\vert_{|\bk|\sim T}
 + {\cal  O}\left(g^6(rT)\right).
\label{EEVb}
\end{equation}
Here, the first non-vanishing contribution in dimensional regularization
comes from the one-loop correction to the chromoelectric correlator.
The integral with $\langle E^{i\,a} $ $U_{ab}E^{j\,b}\rangle(0,\bk)$ at one loop 
has been calculated in Coulomb gauge in Sec.~\ref{sub_twoloop}. The chromoelectric correlator however is gauge invariant.
In static gauge, corrections from the scale $T$ arise only from the non-static part of the spatial 
gluon propagator. Hence, at one loop, only gluon self-energy diagrams may provide 
thermal corrections; we have 
\begin{equation}
\langle E^{i\,a} U_{ab}E^{j\,b}\rangle(0,\bk)\vert_{|\bk|\sim T} = 
\langle \partial_i A_0^{a} \, \partial_j A_0^{a}\rangle(0,\bk)\vert_{|\bk|\sim T} = 
(\nc^2-1)\frac{k^ik^j}{\bk^2+\Pi_{00}^{\rm NS}(\bk)_{\rm mat}},
\label{staticcoulomb}
\end{equation}
where $\Pi_{00}^{\rm NS}(\bk)_{\rm mat}$ is the matter part of the gluon self-energy's temporal component  
calculated in static gauge, which can be read from Eq.~\eqref{pi00matter} and is the same in static gauge and in Coulomb gauge, as we discussed in Sec.~\ref{self_energy}. Hence, the result of the integration is given by the real part of Eq.~\eqref{VsTloop} and reads
 \begin{equation}
\delta_{s,T}^{\, {\cal O}(r^2)\, {\rm S}} = 
\frac{3}{2}\zeta(3)C_F\frac{\als}{\pi}(rm_D)^2 
- \frac{2}{3}\zeta(3)\nc C_F \als^2(rT)^2 
 + {\cal  O}\left(g^6(rT)\right).
\label{deltasEb}
\end{equation}

\subsection{Higher multipole terms}
Our aim is to calculate in the EFT the Polyakov-loop correlator at order $g^6$, neglecting terms of order 
$g^6(rT)$ or smaller. Contributions coming from the $\delta {\cal L}_{\rm pNRQCD}$ part of the pNRQCD Lagrangian, 
which includes terms of order $r^3$ or smaller coming from the multipole expansion, share, at leading order,
the same colour structure and the same order in $\als$ as Eqs.~\eqref{deltasEa} and~\eqref{deltasEb} but are suppressed 
by powers of $rT$. We may write these contributions as
\begin{equation}
\delta_{s,T}^{\,\delta {\cal L}_{\rm pNRQCD}} 
= \delta_{s,T}^{\, {\cal O}(r^2)\, {\rm NS}}
\sum_{n=0}^\infty c_n^{\rm NS}(rT)^{2n+2}  
+
\delta_{s,T}^{\, {\cal O}(r^2) \, {\rm S}}
\sum_{n=0}^\infty c_n^{\rm S}(rT)^{2n+2}  
+  {\cal  O}\left(g^6(rT)^3\right),
\label{Vsmult}
\end{equation}
where the unknown coefficients $c_n^{\rm NS}$ and $c_n^{\rm S}$ are, as we will see, 
irrelevant for the purpose of calculating the Polyakov-loop correlator at order 
$g^6(rT)^0$.

\subsection{The octet contributions}
As in the singlet case, one loop-corrections to the octet correlator may be divided into 
order $r^2$ non-zero mode contributions ($\delta_{o,T}^{\, {\cal O}(r^2)\, {\rm NS}}$), 
order $r^2$ zero-mode contributions ($\delta_{o,T}^{\, {\cal O}(r^2)\, {\rm S}}$), 
and higher multipole terms ($\delta_{o,T}^{\,\delta {\cal L}_{\rm pNRQCD}}$). 
It turns out that 
\begin{eqnarray}
\delta_{o,T}^{\, {\cal O}(r^2)\, {\rm NS}}   &=&  \delta_{s,T}^{\, {\cal O}(r^2)\, {\rm NS}}\vert_{V_s\leftrightarrow V_o},
\label{deltaoEa}
\end{eqnarray}
and, up to order $g^6(rT)^0$, 
\begin{eqnarray}
\delta_{o,T}^{\, {\cal O}(r^2)\, {\rm S}}   &=&  - \delta_{s,T}^{\, {\cal O}(r^2)\, {\rm S}},
\label{deltaoEb}\\
\delta_{o,T}^{\,\delta {\cal L}_{\rm pNRQCD}} &=& -\delta_{s,T}^{\,\delta {\cal L}_{\rm pNRQCD}}.
\label{Vomult}
\end{eqnarray}
These equalities are proved in Appendix~\ref{app_octet}.

\subsection{$\delta \langle L_R \rangle_T$}
Finally, we need to calculate the contributions to the Polyakov loop coming from the scale $T$.
The order $g^4$ contribution may be read from Eq.~\eqref{finalcontribtnlons}. 
Since we do not know the order $\crr \, g^6$ contribution, we write $\delta \langle L_R \rangle_T$ as 
\begin{equation}
\delta\langle L_R\rangle_{T}=
\frac{\crr\als^2}{2}\left[\ca\left(\frac{1}{2\epsilon}-\ln\frac{4T^2}{\mu^2}+1-\gamma_E
+\ln (4\pi) \right)- n_f\ln2 + a\,\als \right]
+ {\cal  O}\left(\als^4\right),
\label{LRT}
\end{equation}
where the explicit value of the coefficient $a$ does not matter.
Instead, what matters here is that this coefficient is common to all colour representations.
The first correction from the scale $T$ not of the type $\crr \, \als^n$ 
appears at order $\als^4$ and comes from diagram~b in Fig.~\ref{fig:cubefourth} 
with two self-energy insertions, one in each temporal gluon.
Note that, while Eq.~\eqref{als2m2} provides the first correction not of the type $\crr \, \als^n$, it is however 
coming from the scale $m_D$.

\subsection{Summary}
In summary, we obtain the contribution of the scale $T$ to the singlet and octet correlators:
\begin{eqnarray}
&&
e^{-V_s(r)/T} \delta_{s,T} =
\nonumber\\
&&\hspace{6mm}
 e^{-V_s(r)/T}\Bigg\{  
- \frac{2}{9} \pi \nc C_F\als^2rT\left[1 + \sum_{n=0}^\infty c_n^{\rm NS}(rT)^{2n+2}\right]
+ \frac{\pi}{36}\nc^2 C_F\als^3 
\nonumber\\
&&\hspace{25mm}
+ \left(\frac{3}{2}\zeta(3)C_F\frac{\als}{\pi}(rm_D)^2 
- \frac{2}{3}\zeta(3)\nc C_F \als^2(rT)^2 \right)
\left[1 + \sum_{n=0}^\infty c_n^{\rm S}(rT)^{2n+2}\right]
\nonumber\\
&&\hspace{25mm}
+ {\cal  O}\left(g^6(rT),\frac{g^8}{rT}\right) 
\Bigg\},
\label{VsE}
\end{eqnarray}
\begin{eqnarray}
&&
 e^{-V_o(r)/T}\left[(\nc^2-1)\, \delta \langle L_A \rangle_T + \delta_{o,T}\right] =
\nonumber\\
&&\hspace{2mm}
(\nc^2-1)e^{-V_o(r)/T}\Bigg\{
\frac{\ca}{2}  \als^2 \left[\ca\left(\frac{1}{2\epsilon}-\ln\frac{4T^2}{\mu^2}+1-\gamma_E
+\ln (4\pi) \right)- n_f\ln2 + a\,\als \right]
\nonumber\\
&&\hspace{33mm}
+ \frac{1}{9} \pi \als^2rT\left[1 + \sum_{n=0}^\infty c_n^{\rm NS}(rT)^{2n+2}\right]
+ \frac{\pi}{72}\nc \als^3 
\nonumber\\
&&\hspace{33mm}
- \left(\frac{3}{4}\zeta(3)\frac{1}{\nc}\frac{\als}{\pi}(rm_D)^2 
- \frac{1}{3}\zeta(3) \als^2(rT)^2 \right)
\left[1 + \sum_{n=0}^\infty c_n^{\rm S}(rT)^{2n+2}\right]
\nonumber\\
&&\hspace{33mm}
+ {\cal  O}\left(g^6(rT),\frac{g^8}{rT}\right) 
\Bigg\}.
\label{VoE}
\end{eqnarray}

Inserting Eqs.~\eqref{LRT}-\eqref{VoE} into Eq.~\eqref{PLC-pNRQCD} and expanding, 
we obtain that the connected Polyakov-loop correlator is given by
\begin{eqnarray}
C_\mathrm{PL}(r,T) &=& C_\mathrm{PL}(r,T)_{1/r}
\nonumber\\
&&
-\frac{\pi}{18} C_F\als^3 + \frac{\nc^2-1}{8\nc^2}\frac{\als^3}{rT}
\bigg[
\ca\left( - \frac{1}{\epsilon}- 2\ln\frac{\mu^2}{4T^2} -2 +2\gamma_E
-2\ln (4\pi) \right) 
\nonumber\\
&&
\hspace{6.2cm}
+2 n_f\ln 2 \bigg]
+{\cal  O}\left(g^6(rT),\frac{g^8}{(rT)^4}\right)
\nonumber\\
&& + \;\hbox{loop corrections at the scale }m_D\hbox{ or lower}\,,
\label{CPLT}
\end{eqnarray}
where  $C_\mathrm{PL}(r,T)_{1/r}$ may be read from Eq.~\eqref{CPLr}.
We observe that, in the connected Polyakov-loop correlator, terms proportional to the 
unknown coefficients  $c_n^{\rm NS}$, $c_n^{\rm S}$ and $a$ have canceled.
The thermal corrections in~\eqref{CPLT} agree with those calculated 
in Sec.~\ref{sec_corr}; in particular, they correspond to the sum of the 
gluon self-energy matter-part contribution in Eq.~\eqref{selfenergyinsertr} with 
Eq.~\eqref{defselfenergyinsertt}. The result in Eq.~\eqref{CPLT} has an infrared 
divergence that originates at the scale $T$. This divergence shall cancel against 
an opposite ultraviolet one at the scale $m_D$, which will be the subject of the next section.

\section{The Debye mass scale}
\label{secpNRQCDmD}
Here we compute the contributions to the singlet correlator, 
the octet correlator and the Polyakov loop coming from loop momenta sensitive 
to the Debye mass scale. We call these contributions  
$\delta_{s,m_D}$, $\delta_{o,m_D}$ and $\delta \langle L_R\rangle_{m_D}$
respectively. They may be computed by evaluating the loop integrals in $\delta_{s}$, 
$\delta_{o}$ and $\delta \langle L_R\rangle$ over momenta of the order 
$m_D$ and expanding with respect to any other scale.
The Debye mass scale is the lowest scale we need to consider here; 
contributions coming from scales lower than $m_D$ are beyond our accuracy.
Different terms contribute to  $\delta_{s,m_D}$, $\delta_{o,m_D}$ and $\delta \langle L_R \rangle_{m_D}$; 
we examine them in the following.

\subsection{The singlet and octet contributions}
The leading-order contribution to $\delta_{s,m_D}$ comes from the self-energy diagram 
shown in Fig.~\ref{fig:hqself} when evaluated over loop momenta of order $m_D$.
The contribution reads
\begin{eqnarray}
\delta_{s,m_D} &=&
\left( ig\sqrt{\frac{1}{2\nc}} \right)^2 r^ir^j T\sum_n
\int \frac{d^dk}{(2\pi)^d} \int_0^{1/T} \!\! d\tau \int_0^{\tau} \!\! d\tau'
e^{\tau V_s} \, e^{-(\tau-\tau')V_o} \, e^{-\tau' V_s}
\nonumber\\
&& \hspace{3cm} \times e^{-i(\tau-\tau')\omega_n}\,
\langle E^{i\,a} U_{ab}E^{j\,b}\rangle(\omega_n,\bk)\vert_{\mbk\sim m_D}.
\label{defEEVsmE}
\end{eqnarray}
The chromoelectric correlator evaluated over the region 
$\mbk\sim m_D$ gives rise to scaleless momentum integrals 
unless for the temporal part of the zero mode, $n=0$, which is at leading order 
$\langle E^{i\,a} U_{ab}E^{j\,b}\rangle(0,\bk)\vert_{|\bk|\sim m_D} = (\nc^2-1)\,k^ik^j\,/(\bk^2+m_D^2)$.
We obtain
\begin{eqnarray}
\delta_{s,m_D}&=&-g^2 \cf\frac{r^ir^j}{2T}\,\int \frac{d^dk}{(2\pi)^d} \frac{k^ik^j}{\bk^2+m_D^2}
\left[1 + {\cal  O}\left(\frac{g^2}{rT}\right)\right]\nn\\
 &=&
- \cf \frac{\als}{6} r^2\frac{m_D^3}{T} + {\cal  O}\left(g^7(rT)\right).
\label{EEVsmE}
\end{eqnarray}

The leading-order contribution to $\delta_{o,m_D}$ comes from the octet self-energy diagrams 
shown in Fig.~\ref{fig:octetEEV}, when evaluated over the region $\mbk\sim m_D$.
Also in this case, the only non-vanishing contribution comes from the zero mode
of the temporal gluon propagator, which is $1/(\bk^2+m_D^2)$ (see Eq.~\eqref{propscreened}). 
For the same argument developed in appendix~\ref{app_octet}, we find that 
\begin{equation}
\delta_{o,m_D} = -\delta_{s,m_D}.
\label{EEVomE}
\end{equation}
Higher multipole terms are of order $\displaystyle \als r^2\frac{m_D^3}{T} (rm_D)^2 \sim g^7 (rT)^4$ or smaller 
and, therefore, beyond our accuracy.

\subsection{$\delta\langle L_{R} \rangle_{m_D}$}
We need to calculate the contribution to the Polyakov loop coming from the scale $m_D$.
It may be read from Eqs.~\eqref{loopbm0},~\eqref{finalcontribm} and~\eqref{als2m2}.
Since we do not know the order $\crr \,g^5$ and $\crr \,g^6$ contributions, 
we write $\langle L_{R} \rangle_{m_D}$ as 
\begin{eqnarray}
\delta\langle L_{R}\rangle_{m_D} &=& 
\frac{\crr\als}{2}\frac{m_D}{T} 
\nonumber\\ 
&& 
+ \frac{\crr\als^2}{2}
\left[\ca\left(- \frac{1}{2\epsilon}-\ln\frac{\mu^2}{4m_D^2}
-\frac{1}{2}+\gamma_E - \ln (4\pi) \right) + b_1\,g + b_2\,g^2 \right]
\nonumber\\
&& 
+ \left(3\crr^2-\frac{\crr\ca}{2}\right)\frac{\als^2}{24}\left(\frac{m_D}{T}\right)^2
+ {\cal  O}\left(g^7\right),
\label{LRmE}
\end{eqnarray}
where the explicit values of the coefficients $b_1$  and $b_2$ do not matter.
Instead, what matters here is that these coefficients are common to all colour representations. The last line comes instead from Eq.~\eqref{als2m2}.
\subsection{Summary}
In summary, we obtain the contribution of the scale $m_D$ to the singlet and octet correlators:
\begin{eqnarray}
&& e^{-V_s(r)/T} \delta_{s,m_D} = 
e^{-V_s(r)/T} \Bigg\{
- \cf \frac{\als}{6} r^2 \frac{m_D^3}{T} 
+   {\cal  O}\left(g^7(rT)\right) \Bigg\},
\label{VsM}
\\
&& 
e^{-V_o(r)/T}\left[(\nc^2-1)\, \delta \langle L_A \rangle_{m_D} + \delta_{o,m_D}\right] =
(\nc^2-1)e^{-V_o(r)/T}\Bigg\{
\frac{\ca\als}{2}\frac{m_D}{T} 
\nonumber\\
&&
\hspace{-5mm}
+ \frac{5}{48} \ca^2\als^2 \left(\frac{m_D}{T}\right)^2 + \frac{\ca\als^2}{2}
\left[\ca\left(- \frac{1}{2\epsilon}-\ln\frac{\mu^2}{4m_D^2}
-\frac{1}{2}+\gamma_E - \ln (4\pi)\right) 
+ b_1\,g + b_2\,g^2 \right]
\nonumber\\
&& 
\hspace{-5mm}
+ \frac{1}{\nc} \frac{\als}{12} r^2 \frac{m_D^3}{T} 
+  {\cal  O}\left(g^7\right) \Bigg\}.
\label{VoM}
\end{eqnarray}

Inserting Eqs.~\eqref{LRmE}-\eqref{VoM} into Eq.~\eqref{CPLT} and expanding\footnote{
In terms of $\delta_{s,T}$, $\delta_{s,m_D}$, $\delta_{o,T}$, $\delta_{o,m_D}$, 
$\delta \langle L_F \rangle_T$, $\delta \langle L_F \rangle_{m_D}$, 
$\delta \langle L_A \rangle_T$ and $\delta \langle L_A \rangle_{m_D}$,  
$C_\mathrm{PL}(r,T)$ reads 
\begin{eqnarray*}
C_\mathrm{PL}(r,T) &=& 
\frac{1}{\nc^2}\left\{ e^{-V_s(r)/T}\left(1+\delta_{s,T}+\delta_{s,m_D}\right) \right.
\\
&& \hspace{5mm} 
\left.
+ e^{-V_o(r)/T} \left[ (\nc^2-1)\left(1 + \delta \langle L_A \rangle_T + \delta \langle L_A \rangle_{m_D}\right)
+\delta_{o,T}+\delta_{o,m_D} \right]
\right\} 
\\
&& - \left(1 + \delta \langle L_F \rangle_T + \delta \langle L_F \rangle_{m_D}\right)^2.
\end{eqnarray*}
}, 
we obtain that the connected Polyakov-loop correlator is given by
\begin{eqnarray}
C_\mathrm{PL}(r,T) &=& C_\mathrm{PL}(r,T)_{1/r}
\nonumber\\
&&
-\frac{C_F}{18}\pi \als^3 + \frac{\nc^2-1}{8\nc^2} \als^2 \left(\frac{m_D}{T}\right)^2
\nonumber\\
&&
+ \frac{\nc^2-1}{\nc^2}\frac{\als}{rT}\left\{
-\frac{\als}{4}\frac{m_D}{T} - \frac{\als^2}{4}\left[
\ca\left( - \ln\frac{T^2}{m_D^2} +\frac{1}{2}\right) 
- n_f\ln 2 \right]\right\}
\nonumber\\
&&
+{\cal  O}\left(g^6(rT),\frac{g^7}{(rT)^2}\right),
\label{CPLmE}
\end{eqnarray}
where  $C_\mathrm{PL}(r,T)_{1/r}$ may be read from Eq.~\eqref{CPLr}.
We observe that, in the Polyakov-loop correlator, terms proportional to the 
unknown coefficients $b_1$ and $b_2$, as well as the divergences, have canceled.
The origin of the thermal corrections to the Polyakov-loop correlator 
in the situation $1/r \gg T \gg m_D \gg g^2/r$ is clear. 
The term $-C_F\pi \als^3/18$ arises from the dipole interaction contributions and from 
their interference with the zero-temperature potentials.
The other thermal corrections arise from the interference of the 
adjoint Polyakov loop with the zero-temperature potentials.

The result coincides with Eq.~\eqref{finalcpltot}, obtained in
Sec.~\ref{sec_corr} after a direct calculation.
The differences in the way the two results were achieved illustrate 
well the typical differences between a direct computation and a computation 
in an EFT framework. In the EFT framework, some more conceptual work was
necessary in order to identify the relevant contributions.
Once this was done, we could take advantage of previously done calculations 
(in particular for $V_s(r)$ and $V_o(r)$) and reduce the calculation to essentially 
one diagram, shown in Fig.~\ref{fig:hqself}, evaluated in different momentum regions. 
In the EFT framework, we will also gain some new insight by reconstructing 
the spectral decomposition of the Polyakov-loop correlator and 
by providing two new quantities: the colour-singlet and the colour-octet quark-antiquark correlators.

\section{Singlet and octet free energies}
\label{sec_free}
Potential NRQCD at finite temperature allows to define 
a colour-singlet correlator, $\langle S(\br,{\bf 0},1/T)S^\dagger(\br,{\bf 0},0)\rangle$,  
and a colour-octet correlator, 
$\langle O^a(\br,{\bf 0},1/T)O^{a \, \dagger}(\br,{\bf 0},0)\rangle$, which 
are both gauge-invariant quantities.
We may associate to them 
a  \emph{colour-singlet free energy}, 
$f_{s}(r,T,m_D)$, and a \emph{colour-octet free energy}, 
$f_{o}(r,T,m_D)$, such that 
\begin{eqnarray}
\langle S(\br,{\bf 0},1/T)S^\dagger(\br,{\bf 0},0)\rangle 
&=& e^{-V_s(r)/T}\left(1+\delta_{s,T}+\delta_{s,m_D}\right) 
\nonumber\\
&\equiv&e^{-f_{s}(r,T,m_D)/T},
\label{deffs}\\
\langle O^a(\br,{\bf 0},1/T)O^{a \, \dagger}(\br,{\bf 0},0)\rangle 
&=& e^{-V_o(r)/T} \left[(\nc^2-1)\left(1 + \delta \langle L_A \rangle_T + \delta \langle L_A \rangle_{m_D}\right)\right.\nn\\
&&\hspace{5.3cm}+\delta_{o,T}+\delta_{o,m_D}\big]
\nonumber\\
&\equiv&(\nc^2-1)e^{-f_{o}(r,T,m_D)/T}.
\label{deffo}
\end{eqnarray}
Using the results of the previous sections, we have that 
\begin{align}
f_{s}(r,T,m_D) 
=& 
V_s(r) 
\nonumber\\
&
+ \frac{2}{9} \pi \nc C_F\als^2rT^2\left[1 + \sum_{n=0}^\infty c_n^{\rm NS}(rT)^{2n+2}\right]
- \frac{\pi}{36}\nc^2 C_F\als^3T 
\nonumber\\
&
- \left(\frac{3}{2}\zeta(3)C_F\frac{\als}{\pi}(rm_D)^2T 
- \frac{2}{3}\zeta(3)\nc C_F \als^2r^2T^3 \right)
\left[1 + \sum_{n=0}^\infty c_n^{\rm S}(rT)^{2n+2}\right]
\nonumber\\
&
+ \cf \frac{\als}{6} r^2 m_D^3 
+ T {\cal  O}\left(g^6(rT),\frac{g^8}{rT}\right),
\label{freesinglet}
\end{align}
and
\begin{eqnarray}
f_{o}(r,T,m_D) 
&=&
V_o(r) 
\nonumber\\
&&
- \frac{\ca\als}{2}m_D 
+ \frac{1}{48} \ca^2\als^2 \frac{m_D^2}{T}
\nonumber\\
&&
-\frac{\ca\als^2}{2}T\left[\ca\left(-\ln\frac{T^2}{m_D^2}+\frac{1}{2}\right)-
  n_f\ln2 + b_1\,g + b_2\,g^2 + a\,\als \right]
\nonumber\\
&&
- \frac{\pi}{9}\als^2rT^2\left[1 + \sum_{n=0}^\infty c_n^{\rm NS}(rT)^{2n+2}\right]
- \frac{\pi}{72}\nc \als^3T 
\nonumber\\
&&
+ \left(\frac{3}{4\nc}\zeta(3)\frac{\als}{\pi}(rm_D)^2T - \frac{1}{3}\zeta(3) \als^2r^2T^3 \right)
\left[1 + \sum_{n=0}^\infty c_n^{\rm S}(rT)^{2n+2}\right]
\nonumber\\
&&
- \frac{1}{\nc} \frac{\als}{12} r^2 m_D^3 
+ T {\cal  O}\left(g^6(rT),\frac{g^8}{rT}\right).
\label{freeoctet}
\end{eqnarray}
We note that $f_{s}(r,T,m_D)$ and $f_{o}(r,T,m_D)$ are both finite and gauge invariant.
They also do not depend on some special choice of Wilson lines connecting 
the initial and final quark and antiquark states.

In Part~\ref{part_realtime}, the static, colour-singlet quark-antiquark potential was calculated 
in real-time formalism in the same thermodynamical situation considered 
here and specified by Eq.~\eqref{defhierarchy}. The result may be obtained from Chapters~\ref{chap_Tggr} and \ref{chap_rggT} by summing the static part of Eq.~\eqref{totalpotT} with Eq.~\eqref{VsmD}, as observed in footnotes~\ref{foot_mdggE} and \ref{foot_potmdggE} in Chap.~\ref{chap_rggT}. It reads
\begin{eqnarray}
	V_s^{(\text{real-time})}(r) &=& -\cf\frac{\als}{r}\left\{1+\left(a_1+ 2 {\gamma_E \beta_0}\right) {\als(r) \over 4\pi}\right.
	\nonumber\\
	&&\qquad\qquad 
	+
	\left[\gamma_E\left(4 a_1\beta_0+ 2{\beta_1}\right)+\left( {\pi^2 \over 3}+4 \gamma_E^2\right) 
	{\beta_0^2}+a_{2\,s,0}\right] {\als^2(r) \over 16\,\pi^2}\nn\\
	&&\qquad\qquad\qquad\qquad+\left.\left[\frac{16\pi^2}{3}\nc^3\left(\ln(4\pi T\,r)-\gamma_E\right)+\tilde{a}_{3\,s,o}\right] {\als^3(r) \over 64\,\pi^3}\right\}
	\nn\\
	& &
	\hspace{-12mm}
	+ \frac{\pi}{9} \, N_c C_F \, \als^2 \, r \, T^2\,\nn\\
	&&\hspace{-12mm}- \frac{3}{2} \zeta(3)\,  C_F \, \frac{\als}{\pi} \, r^2 \, T \,m_D^2
	+ \frac{2}{3} \zeta(3)\, N_c C_F \, \als^2 \, r^2 \, T^3
	+ \frac{C_F}{6} \, \als \, r^2m_D^3  + \dots
	\nn\\
	& &
	\hspace{-12mm}
	+ i \left[
	- \frac{N_c^2 C_F}{6} \, \als^3\, T\,
	+ \frac{C_F}{6} \als \, r^2 \, T \,m_D^2\, \left( 2 \gamma_E
	- \ln\frac{T^2}{m_D^2} -1 - 4 \ln 2 - 2 \frac{\zeta^\prime(2)}{\zeta(2)} \right)
	\right.
	\nn\\
	&& \quad \left.
	+ \frac{4\pi}{9} \ln 2 \; N_c C_F \,  \als^2\, r^2 \, T^3
	\right]
	+ \dots
	\,,
	\label{sumV11}
\end{eqnarray}
where we have used Eq.~\eqref{alvs} in App.~\ref{app_potentials} to write the zero-temperature part of this potential and see how it combines at three loops with the divergent part coming from the scale T in Eq.~\eqref{totalpotT}, yielding a finite expression. The logarithm in the third line of this equation signals the the cancellation of the divergence. The dots stand for higher orders in the real and imaginary parts. Comparing terms of the same order, the real part of the real-time potential 
differs from $f_{s}(r,T,m_D)$ by  
\begin{equation}
	f_{s}(r,T,m_D)-\mathrm{Re}V_s^{(\text{real-time})}(r)=\frac{1}{9} \pi \nc C_F\als^2rT^2 - \frac{\pi}{36}\nc^2 C_F\als^3T +\ldots\,.
\end{equation}
The origin of the difference may be traced back to terms in Eq.~\eqref{EEVa} that 
would vanish for large real times. Indeed, performing the calculation of 
$\langle S(\br,{\bf 0},\tau)S^\dagger(\br,{\bf 0},0)\rangle$ for an imaginary time $\tau \le 1/T$,  
along the lines of Secs.~\ref{secpNRQCDT} and~\ref{secpNRQCDmD}, 
and then continuing analytically $\tau$ to large real times, one gets 
back exactly both the real and the imaginary parts of the real-time colour-singlet potential 
given by Eq.~\eqref{sumV11} at the corresponding order.\\
It is then important to remark that this difference between the singlet free energy and the real part of the real-time 
colour-singlet potential appears to be a relevant finding to be considered when using free-energy 
lattice data for the quarkonium in media phenomenology.

\section{Comparison with the literature} 
\label{sec_compare}
An EFT approach for the calculation of the correlator of Polyakov loops 
was developed in \cite{Braaten:1994qx} for the situation $m_D  \simg 1/r$ and 
in \cite{Nadkarni:1986cz} for $T \gg 1/r$. In neither of the two cases, the
scale $1/r$ was integrated out: the Polyakov-loop correlator was
described in terms of dimensionally reduced effective field theories of QCD, MQCD in the former and EQCD in the latter calculation,
while the complexity of the bound-state dynamics remained implicit in the correlator. 
The description developed in  \cite{Nadkarni:1986cz,Braaten:1994qx} is valid for 
largely separated Polyakov loops. Under that condition, the correlator turns out 
to be screened either by the Debye mass, for $rm_D \sim 1$, or by the mass of the
lowest-lying glueball in 2+1-dimensional QCD, for $rm_D \gg 1$.

In \cite{Jahn:2004qr}, the spectral decomposition of the Polyakov-loop correlator was analyzed. 
It was concluded that the quark-antiquark component of an allowed intermediate state, 
i.e. a field $\varphi$ describing a quark located in $\bx_1$  and an antiquark located in $\bx_2$,
should transform as $\varphi(\bx_1,\bx_2) \to g(\bx_1) \varphi(\bx_1,\bx_2) g^\dagger(\bx_2)$
under a gauge transformation $g$. 
Equation~\eqref{PLC-pNRQCD}  is in accordance with that result for, 
in pNRQCD, both the singlet field S and the octet field O transform in that way, as shown in Eqs.~\eqref{decomposesingoct} and \eqref{gaugesingoct}. 
We remark, however, a difference in language:
in our work, singlet and octet refer to the gauge transformation properties of 
the quark-antiquark fields, while, in \cite{Jahn:2004qr}, they refer to the gauge 
transformation properties of the physical states.

In \cite{Burnier:2009bk}, a weak-coupling calculation of the untraced
Polyakov-loop correlator in Coulomb gauge and of the cyclic Wilson loop was
performed up to order $g^4$. Each of these objects contributes to
the correlator of two Polyakov loops through a Fierz transformation that also generates some 
octet counterparts. It is expected that large cancellations occur between 
those correlators and their octet counterparts in
order to reproduce the Polyakov-loop correlator given in Eq.~\eqref{finalcpltot}.
Such large cancellations should occur at the level of the scales $1/r$, 
$T$ and $m_D$ as we have already experienced in this work.
Note that in the case of the untraced Polyakov-loop correlator, 
the octet contribution shall also restore gauge invariance.

\section{Summary and outlook}
\label{sec_concl}
In this Chapter we have performed the calculation of the Polyakov-loop correlator in a suitable EFT that exploits the hierarchy of scales in the problem, reobtaining the results obtained previously in a direct perturbative computation. In this EFT approach, we have used pNRQCD at finite temperature and subsequently integrated 
out lower momentum regions. The advantages of this EFT approach are that the 
calculations do not rely on any specific choice of gauge 
and the systematics is clearer. Moreover, it makes explicit the quark-antiquark colour-singlet and 
colour-octet contributions to the Polyakov-loop correlator.
In particular, we have shown in Eqs.~\eqref{PLC-pNRQCD}, \eqref{deffs} and \eqref{deffo} that at leading order in the multipole expansion
the Polyakov-loop correlator can be written as the colour average of a colour-singlet 
correlator, which defines a gauge-invariant colour-singlet free energy, and a 
colour-octet correlator, which defines a gauge-invariant colour-octet free energy.
This is in line with some early intuitive arguments given in 
\cite{McLerran:1981pb,Gross:1980br,Nadkarni:1986cz}. 
In general, however, such a decomposition does not hold and higher-order terms in the multipole expansion do contribute at higher orders.

We have furthermore shown that the colour-singlet free energy we have defined and computed differs from the real-time potential and the corresponding static energy obtained in Part~\ref{part_realtime}. Not only does the real-time potential have an imaginary part, it also differs in the real part by an amount that we have traced back to the different boundary conditions in the two cases. In the present Chapter we have an imaginary time extent $\tau=1/T$ with periodic boundary conditions, whereas in the real-time calculation we have a large, real time $t\to\infty$. This difference between the free energy derived from the Polyakov-loop correlator and the real-time potential governing the evolution of quark-antiquark pairs in the medium should be considered when using lattice calculations of correlation functions of Polyakov loops as input for phenomenological potential models.

In the weak-coupling regime, the degrees of freedom of pNRQCD are quark-antiquark 
colour-singlet fields, quark-antiquark colour-octet fields, gluons and light quarks.
The obtained result for the Polyakov-loop correlator 
is consistent with its spectral decomposition.
In the strong-coupling regime, the degrees of freedom are expected to change 
when the typical energy of the bound state is smaller than the 
confinement scale $\Lambda_{\rm QCD}$. In that situation, the bound state 
would become sensitive to confinement and give rise to a new 
spectrum of gluonic excitations (hybrids, glueballs). 
In the present work, we have not discussed this situation, which 
surely deserves investigation.

Possible further extensions of this work also include the study of the
Polyakov-loop correlator in different scale hierarchies, in particular at
temperatures of the same order as or higher than $1/r$, where the present
analysis should smoothly go over the ones performed in \cite{Nadkarni:1986cz,Braaten:1994qx}. 
In particular, the EFT treatment of the correlator in the region $T\gg 1/r\sim m_D$ should be related to the real-time EFTs developed in Chap.~\ref{chap_Tggr}. As mentioned above, also analyses that involve the strong-coupling scale
should be addressed.

An interesting completion of the results presented in this last Chapter would be the recasting of the contributions from the scales $T$ and $m_D$ in the form of matching coefficients of two subsequent, dimensionally-reduced EFTs of pNRQCD. In this way the contributions from the scale $T$, displayed in Sec.~\ref{secpNRQCDT}, would arise when integrating out the temperature from pNRQCD, thus leading to modified matching coefficients in the singlet-singlet and octet-octet sectors, as well as to the Lagrangian of EQCD in the gauge sector. As a last step, integrating out the Debye mass would lead to MQCD in the gauge sector, to further modifications in the order-$r^0$ singlet-singlet and octet-octet sectors and to the disappearance of all other terms in the multipole expansion, since chromoelectric fields are absent from MQCD.

Finally, the present study could be extended by the study of correlators 
different from the Polyakov-loop one. Among these, the most studied in lattice 
gauge theories are the untraced Polyakov-loop correlator and the cyclic Wilson
loop. Also the octet Wilson loop should be included for its role in the
Polyakov-loop correlator. Since some partial perturbative results are already available for some 
of these correlators, it would be interesting to see how they can be reproduced 
in the EFT framework introduced here and how they combine to give back the Polyakov-loop 
correlator.

		\part{Conclusions}
		\label{part_concl}
	\chapter{Conclusions and outlook}
		\label{chap_conclusions}
		In this final Chapter we first draw our conclusions in Sec.~\ref{conclusions} and then we present perspectives for future activities based on the results of this thesis in the outlook in Sec.~\ref{outlook}.
\section{Conclusions}
\label{conclusions}
Let us try to summarize here the most relevant results of this thesis. After the introductory Part~\ref{part_intro} we have set out to generalize in Part~\ref{part_realtime} the successful $T=0$ EFT framework to finite temperatures. The more formal part of the results obtained there is contained in the EFT formalism itself, and how it allows to derive rigorously the potential as the matching coefficient of the non-local four-fermion operator that arises after having integrated out all scales higher than the binding energy, eventually yielding the EFTs that we have called pNRQCD$_{m_D}$, pNRQCD$^{\prime\prime}_{m_D}$ and pNRQCD$_\mathrm{HTL}$. We have furthermore shown how these theories, coherently with the general properties of EFTs described in Sec.~\ref{sec_princ_eft}, can be systematically improved and easily allow to keep track of all effects contributing to a given order in the power counting.\\
Another important formal result is the real-time formalism for heavy quarks and heavy-quark bound states in the medium, whose main outcomes are Eqs.~\eqref{singletstaticexpand} and \eqref{sumSinglet}, which express the bound-state propagator as an infinite sum of free propagators and insertions of the Hamiltonian.

These results pave the way for a rigorous QCD description of $Q\overline{Q}$ bound states in heavy ion collisions. In particular, in Chap.~\ref{chap_Tggr} we have studied the region $T\gg mv$, where screening effects become important. We have however noted how the imaginary parts, which represent a novel feature of the real-time potentials derived from QCD in perturbation theory, as first done in \cite{Laine:2006ns}, become even more important than the real parts and eventually lead to dissociation at lower temperatures than screening alone would imply. By imposing that the real and imaginary part of the potentials be equal one can define a \emph{dissociation temperature} $T_d\sim m \als^{2/3}$. A quantitative dissociation temperature for the $\Upsilon(1S)$, considering also the charm quark mass dependence, has been calculated in \cite{Escobedo:2010tu}.\\
The region $mv\gg T\gg mv^2$ has been dealt with in Chap.~\ref{chap_rggT}. In this case the temperature is below $T_d$ and, as we argue, this is the region relevant for the phenomenology of the $\Upsilon(1S)$ in current collision experiments, especially at the LHC, where the $b\overline{b}$ cross section is relevant and the resolution is very good in the bottomonium mass region. The recent experimental results from CMS \cite{Collaboration:2011pe,CMS-PAS-HIN-10-006}, which point to a modest direct suppression of this state, are in agreement with this picture. In this region we are able to compute the correction induced by the thermal medium to the energy levels and to the width of the bound state, which are summarized in Eqs.~\eqref{finalspectrum} and \eqref{finalwidth}. The latter in particular has the most phenomenological relevance, since it is the one responsible for the suppression of the bound state in this region, screening being absent from the real part of the potential. We have shown how two mechanisms, colour-singlet-to-colour-octet thermal decay and Landau damping, contribute to this width, the first being dominant. We have also shown (see App.\ref{peskin}) how the first can be seen as a rigorous EFT derivation of the previous results in the literature going under the name of \emph{gluo-dissociation} \cite{Kharzeev:1994pz,Xu:1995eb,Rapp:2008tf}, based on an old OPE calculation at zero temperature of the $Q\overline{Q}$ gluo-dissociation cross section by Bhanot and Peskin \cite{Peskin:1979va,Bhanot:1979vb}. Our calculation includes the octet potential and the final state effects it introduces, which had been neglected by Bhanot and Peskin. In the region $T\gg mv^2$ we are considering we are able to quantify the error introduced by this approximation as $\approx10\%$. These results will appear in \cite{uspeskin}.

In this $mv\gg T\gg mv^2$ region we also consider the Lorentz-invariance breaking effects that the thermal medium introduces in the potentials. In particular we concentrate on the spin-orbit part of the potential and show that the Gromes relation, which realizes Lorentz invariance in pNRQCD at zero temperature, is broken at finite temperature in pNRQCD$_\mathrm{HTL}$.  This breaking happens at the leading order in the thermal contribution, corresponding to $\als^2 T^2 r$ in the static potential and $\als^2 T^2/(m^2r)$ in the spin-orbit. As such, it appears to be an important element for the understanding of the $p_\mathrm{T}$ distribution of the suppression factor measured in experiments and is certainly a subject worth further investigations.

In Part~\ref{part_imtime} we have studied the Polyakov loop and its correlator, which are associated to the free energies of a static quark and of a static quark-antiquark pair in the medium. For both quantities we have first performed a perturbative NNLO calculation. For what concerns the Polyakov loop, our result differs from the long-time accepted result \cite{Gava:1981qd} and agrees instead with another determination contemporary to ours \cite{Burnier:2009bk}. We have shown in detail the origin of the discrepancy in App.~\ref{app_feynman}. For what concerns the correlator our short-distance result is new, although parts of it agree with a short-distance expansion of a previous calculation by Nadkarni \cite{Nadkarni:1986cz}, which was based on a different hierarchy. Part~\ref{part_imtime} is concluded by Chap.~\ref{chap_imtimeEFT}, where the correlator is analyzed in the framework of finite-temperature pNRQCD in the imaginary-time formalism. This allows us to understand the origin of the contributions to the correlator in terms of singlet and octet free energies, which we define in a new, gauge-invariant way, thereby giving a rigorous footing to the previous statements in the literature \cite{McLerran:1981pb,Gross:1980br,Nadkarni:1986cz}. We also show that this gauge-invariant colour-singlet free energy differs from the real part of the real-time potential computed in the same scale setting in the previous Part and we remark how this fact should be considered when using lattice calculations of correlation functions of Polyakov loops as non-perturbative input for potential models. In the Outlook we will show how our framework allows for a rigorous non-perturbative approach as well.

From a technical/computational standpoint, some of the results presented in the main text and in the appendices are also of relevance, such as the complete one-loop expression of the longitudinal gluon propagator in the static gauge presented in Sec.~\ref{self_energy}.

\section{Outlook}
\label{outlook}
In this Section we will concentrate on possible future extensions of the results exposed in this thesis, as well as possible applications of the EFT methodology that has been widely discussed and applied throughout this work.\\
As we mentioned before, an extension to the non-perturbative regime would be of great importance. As discussed in Sec.~\ref{sec_pnrqcd}, at zero temperature pNRQCD can be formulated in the strong-coupling regime as well, corresponding to $mv\simg\lqcd$. In this case the momentum transfer and the confinement scales are integrated out at the same time, resulting in a Lagrangian where only colourless singlet states can appear. In the absence of light fermions and due to the mass gap of QCD the only state with energy $E\ll\lqcd$ is the $Q\overline{Q}$ colour-singlet, while glueballs and hybrids are both integrated out. The Lagrangian is then a simple Schr\"odinger Lagrangian for the singlet field only \cite{Brambilla:1999xf,Brambilla:2004jw},\footnote{In the presence of light quarks the present picture still applies for states far from the open heavy flavour threshold. However Goldstone bosons (pions), which have a mass of the order of $\lqcd$, should appear in the Lagrangian.} and the potentials can be written as large-time limits of expectation values of chromoelectric and chromomagnetic fields inserted along rectangular Wilson loops \cite{Brambilla:2000gk,Pineda:2000sz,Brambilla:2004jw}. In particular, the static potential is given by the logarithm of the large-time limit of the Wilson loop divided by time, as per the pre-EFT definition \cite{Wilson:1974sk,Susskind:1976pi}, and the $1/m$ potential by the insertion of two chromoelectric fields along one of the timelike Wilson lines.\\
The extension of this approach to finite temperatures is currently underway \cite{pnrstrong}. On one hand it requires the identification of the Wilson-loop operators giving the right potentials at finite temperature. In perturbation theory we have checked both at the static level (see App.~\ref{secpQCD}) and at the $1/m$ level that the zero-temperature operators, when evaluated in perturbation theory, give the same results obtained in Part~\ref{part_realtime} in the EFT framework at the orders considered. Whether this statement extends to the non-perturbative levels still needs to be proved. On the other hand such operators need to be reliably evaluated non-perturbatively. At zero temperature the static term (the Wilson loop) has been extensively computed on the lattice and recently also the $1/m$ and $1/m^2$ terms have been evaluated (see for instance \cite{Koma:2006fw,Koma:2009ws}). At finite temperature the non-perturbative measurement is more complicated due to the presence of the important imaginary parts. A large Euclidean-time limit is also not possible, because it would correspond to a zero-temperature limit, and one has to resort to the extraction of the spectral function of the considered operator from a discrete set of data points, then obtaining the real and imaginary part of the potential from the position and width of the peaks of this function. A preliminary work in this direction, based on the Maximum Entropy Method \cite{Asakawa:2000tr}, is underway for the static potential \cite{Rothkopf:2009pk}.

Other possible extensions of the results presented here are, for what concerns Part~\ref{part_realtime}, the study of more hierarchies and the improvement of the matching in the existing ones. For instance in Chap.~\ref{chap_Tggr} we only considered the static limit: the computation of $1/m$-suppressed terms appears necessary, also in the lights of the results of Chap.~\ref{chap_rggT}, that show how static and $1/m$-suppressed terms can have the same size in the power counting of the theory. A complete analysis of the octet sector could also be interesting and would be necessary for higher-order calculations.\\
A phenomenological analysis of the results of this Part is also very interesting: as we discussed, the width computed in Chap.~\ref{chap_rggT} is relevant for the understanding and quantitative description of the suppression of the $\Upsilon(1S)$ at the LHC. To this end, the calculation of Chap.~\ref{chap_rggT} should be extended to the case of a plasma with a momentum-space anisotropy, which resembles more closely the medium produced in heavy-ion collision, where the large pressure gradients given by the geometry of the collision (Lorentz contraction in the collision axis and impact parameter in the transverse plane) indeed cause such anisotropies. In the real-time formalism the anisotropy is implemented by a simple modification of the thermal distribution. In the $T\gg mv$ region the potential has been computed in the presence of anisotropies in \cite{Dumitru:2007hy,Dumitru:2009ni,Burnier:2009yu,Dumitru:2009fy,Philipsen:2009wg}, whereas in the region $mv\gg T\gg mv^2$ the anisotropic extension still remains to be done.

For what concerns Part~\ref{part_imtime}, the EFT presented there could be improved, as discussed at the end of Chap.~\ref{chap_imtimeEFT}, by making use of the framework of dimensional reduction introduced in Sec.~\ref{sec_htl}. Other extensions include the analysis of different correlators, such as the untraced Polyakov-loop correlator and the cyclic Wilson loop, which are now being studied \cite{matthias} and contribute to the correlator through Fierz identities together with their octet counterparts, as well as the analysis of different hierarchies. A strong-coupling analysis, where the correlator is no longer given at the first orders by colour singlet and colour-octet degrees of freedom, but instead by the singlet ground state and its gluonic excitations, is certainly interesting and would also overlap with the strong-coupling investigation of the real-time EFTs.

The EFT framework that has been introduced in this thesis, and which can be considered one of its most important outcomes, can also be suitably generalized and extended, applying it to different physical problems at finite temperature. In particular, the problem of heavy quark thermalization/energy loss in heavy ion collisions appears suited to an EFT treatment.\\
In the very early stages of an heavy ion collision several heavy quark-antiquark pairs are produced, as has been explained in Sec.~\ref{sec_exp_quarkonium}. In this thesis we have in a way only considered the fate of the pairs that remain correlated and form, at least temporarily, a quarkonium bound state.
However the fate of uncorrelated heavy quarks or antiquarks is equally interesting. Due to their large mass $m\gg T$, $T$ being the typical temperature in heavy ion collisions, charm and bottom quarks can be expected to thermalize much more slowly
than the light constituents of the plasma, which is itself very short-lived in heavy-ion collision experiments. As we have mentioned, the lifetime of the plasma is indeed estimated to be of a few fm/c at RHIC and at most $\sim10$ fm/c at the LHC. These facts would then cause one to expect a little variation between the observed yields of the decay products
of heavy-light mesons, which are eventually formed when the heavy quark hadronizes, with respect to those observed in proton-proton collisions, suitably scaled to the appropriate number of binary collisions. Experimental data from RHIC \cite{Adare:2006nq,Abelev:2006db,Adare:2010de} and the early LHC data \cite{Dainese:2011vb} show however a significant
suppression of high-$p_\mathrm{T}$ electrons and positrons coming from the weak decay of heavy quarks with respect to the pp and pA baselines, as well as a significant elliptic flow of these electrons, thus hinting at a somewhat larger than expected thermalization of the heavy quarks. The ALICE collaboration has also measured directly the suppression of D mesons \cite{Dainese:2011vb}.\\
On the theory side, the \emph{energy loss} $-dE/dx$ of a heavy quark was first computed in perturbation theory by Braaten and Thoma \cite{Braaten:1991jj,Braaten:1991we}, representing the first computation of a heavy quark \emph{transport coefficient}. Recent efforts have focussed on a formalism based on the Langevin equation (or relativistic generalizations thereof), first introduced in this context in \cite{Moore:2004tg}. In this formalism the heavy quarks are assumed to obey a classical Langevin equation
\begin{equation}
	\frac{dp_i}{dt}=-\eta_Dp_i(t)+\xi_i(t)\,,
	\label{langevin}
\end{equation}
where $\eta_D$ is the \emph{drag coefficient}, representing the dissipative response of the medium, and $\xi_i(t)$ is a stochastic noise term, defined  by its moments $\langle \xi_i(t)\rangle\equiv0$ and $\langle \xi_i(t)\xi_j(t')\rangle\equiv\kappa\delta_{ij}\delta(t-t')$, where $\kappa$ is the \emph{momentum diffusion coefficient}. Classically $\kappa=2mT\eta_D$ through equipartition and the fluctuation-dissipation theorem.\\
It is then clear that the assumption at the basis of the application of Eq.~\eqref{langevin} to heavy quarks is that the typical timescale of the medium is much smaller that the typical timescale of the heavy quarks, thus allowing in a first approximation to consider interactions with the medium as completely uncorrelated momentum kicks.\\
In the past years many efforts, starting from \cite{Moore:2004tg}, went into the determination of one of the Langevin transport coefficients (mostly $\kappa$), the other being determined by the classical relation. The Langevin equation can then be solved numerically and, together with a parametrization of the evolution of the medium, allows for predictions on the phenomenology of heavy quarks through the above-mentioned spectra of their decay products.\\
For what concerns the evaluation of the transport coefficients in the Langevin picture, it was shown in \cite{CasalderreySolana:2006rq} that if one identifies the stochastic noise with the Lorentz force for a static quark $\xi_i(t)=\int d^3x\,\psi^\dagger(t,\bx)E_i(t,\bx)\psi(t,\bx)$ and then integrates out the static quarks along the Schwinger-Keldysh contour (see Sec.~\ref{sec_realtime} and Fig.~\ref{fig_sk}), one obtains an expression for $\kappa$ as a correlator of two chromoelectric fields inserted into Wilson lines spanning said contour. In \cite{CaronHuot:2009uh}, a similar analysis was performed making a more systematic use of HQET/NRQCD, claiming a better control of suppressed terms in the inverse mass  and perturbative expansions and obtaining an imaginary-time analogue of the real-time correlator in \cite{CasalderreySolana:2006rq}. A first exploratory lattice study of the imaginary-time correlator has recently been performed in \cite{Meyer:2010tt}, whereas the real-time correlator was used in \cite{CaronHuot:2007gq,CaronHuot:2008uh} for a perturbative NLO determination of $\kappa$.

The link between the dynamics of heavy quarks in the medium and the simple Langevin picture does not appear however to have been fully justified conceptually in the literature; an EFT treatment of the problem seems a promising path in establishing whether the Langevin picture represents an effective description at the leading order in some expansion and, if yes, if the size and relevance of the sub-leading corrections can be estimated, allowing in case for a calculation of these corrections.\\ 
The way to proceed would be to start from an HQET/NRQCD Lagrangian~\eqref{lagrnrqcd} at some fixed order in the $1/m$ expansion. Since $m\gg T$, one must then proceed to integrate out in succession the temperature and the other relevant thermodynamical scales, in analogy to what has been done in Part~\ref{part_realtime}. We have already seen in Sec.~\ref{secVIB} (see Eq.~\eqref{deltam} in particular) that integrating out the temperature  and the Debye mass at the static level already introduces a thermal mass shift and a damping rate, respectively through the real and imaginary parts of the static quark self-energy. Extending this approach to non-static corrections will introduce new matching coefficients in the effective Lagrangian obtained by integrating out the thermal scales from HQET/NRQCD; in particular one would expect for dimensional reason a dissipative response to appear already at order $1/m$, which should be related to the drag coefficient in the Langevin picture. An analysis along these basic principles is currently ongoing.

An altogether similar analysis could also be applied to study the transport properties of $Q\overline{Q}$ bound states, also in the light of the recent experimental results (\cite{Atomssa:2009ek} from PHENIX and \cite{Masui:2011qi} from STAR) that show an elliptic flow $v_2$ for the $J/\psi$ at RHIC that is compatible with zero and much smaller than that attributed to the heavy quarks. In our EFT framework the interactions of the colourless bound state with the medium are described at leading order by the chromoelectric dipole operator, so that our theory seems ready to be extended to the study of transport properties of the bound state.

	\part*{Appendices}
	\label{part_appendices}
	\addcontentsline{toc}{part}{{}Appendices}
	\appendix
	\chapter{Feynman rules}
		\label{app_feynrules}
	 	\section{Feynman rules in real time}
\subsection{Feynman rules of QCD at zero temperature}
\label{sub_feyn_qcd_realtime}
\begin{fmffile}{qcdrealtimetime}
\setlength{\unitlength}{1mm}
The Feynman rule for the three-gluon vertex reads 
\begin{equation}
	\label{3gretime}
\parbox{30mm}{
\begin{fmfchar*}(30,30)
\fmftop{up}
\fmfbottom{down1,down2}
\fmf{gluon,label=$p\,,\mu\,,a$,label.dist=10}{up,center}
\fmf{gluon,label=$k\,,\rho\,,c$,label.dist=10}{down1,center}
\fmf{gluon,label=$q\,,\nu\,,b$,label.dist=10}{down2,center}
\end{fmfchar*}}\quad=\;-gf^{abc}\left[g^{\mu\nu}(p-q)^\rho +g^{\nu\rho}(q-k)^\mu +g^{\rho\mu}(k-p)^\nu \right],
 \end{equation}
 where all momenta are understood as inflowing in the vertex and the curly line is taken to represent either a longitudinal or a transverse gluon. For the four-gluon vertex we have instead
\begin{equation}
	\label{4gretime}
	\hspace{10mm}\parbox{30mm}{
\begin{fmfchar*}(30,30)
\fmftop{up1,up2}
\fmfbottom{down1,down2}
\fmf{gluon}{up1,center}
\fmf{gluon}{up2,center}
\fmf{gluon}{down1,center}
\fmf{gluon}{down2,center}
\fmfv{label=$\mu\,,a$}{up1}
\fmfv{label=$\nu\,,b$}{up2}
\fmfv{label=$\sigma\,,d$}{down1}
\fmfv{label=$\rho\,,c$}{down2}
\end{fmfchar*}}=\begin{array}{l}
	-ig^2\left[f^{eab}f^{ecd}(g^{\mu\rho}g^{\nu\sigma}-g^{\mu\sigma}g^{\nu\rho})+f^{eac}f^{edb}(g^{\mu\sigma}g^{\nu\rho}\right.\\
	\\
	\hspace{8mm}\left.-g^{\mu\nu}g^{\rho\sigma})+f^{ead}f^{ebc}(g^{\mu\nu}g^{\sigma\rho}-g^{\mu\rho}g^{\sigma\nu}) \right].
\end{array}
\end{equation}
The gluon-ghost vertex is, in covariant gauges
\begin{equation}
	\label{gluonghostretime}
\parbox{30mm}{
\begin{fmfchar*}(30,30)
\fmftop{up}
\fmfbottom{down1,down2}
\fmf{gluon}{up,center}
\fmf{ghost}{down1,center}
\fmf{ghost,label=$p$}{center,down2}
\fmfv{label=$\mu\,,a$,lab.angle=0}{up}
\fmfv{label=$c$}{down1}
\fmfv{label=$b$}{down2}
\end{fmfchar*}}\quad=\;-gf^{abc}p^\mu\,,
 \end{equation}
where the momentum $p$ is understood as inflowing in the vertex and thus in the opposite direction of the arrow.\\
The quark-gluon vertex is simply
 \begin{eqnarray}
	 \nonumber \\
	\label{gluonquarkretime}
	\parbox{30mm}{
\begin{fmfchar*}(30,30)
\fmftop{up}
\fmfbottom{down1,down2}
\fmf{gluon}{up,center}
\fmf{quark}{down1,center}
\fmf{quark}{center,down2}
\fmfv{label=$\mu\,,a$}{up}
\end{fmfchar*}}\quad=\;-ig\gamma^\mu T^a\,.
 \end{eqnarray}
 \end{fmffile}

 \subsection{Feynman rules of pNRQCD}
 \label{sub_feyn_pnrqcd}
\begin{fmffile}{prnqcd}
\setlength{\unitlength}{1mm}
In this Section we list the basic Feynman rules of pNRQCD. The free singlet propagator is
\begin{equation}
	\label{singletprop}
	\parbox{30mm}{
\begin{fmfchar*}(30,10)
\fmfleft{in}
\fmfright{out}
\fmf{plain}{in,out}
\end{fmfchar*}}\quad=
\displaystyle{{i \over 
\displaystyle{E-h_s^{(0)}+i\eta}}
=
{i \over \displaystyle{E-{\bf p}^2/m+C_F\als/ r+i\eta}}}\,.
\end{equation}
The free octet propagator reads
\begin{equation}
	\label{octetprop}
	\parbox{30mm}{
\begin{fmfchar*}(30,10)
\fmfleft{in}
\fmfright{out}
\fmf{dbl_plain}{in,out}
\fmfv{label=$b$,lab.angle=90}{in}
\fmfv{label=$a$,lab.angle=90}{out}
\end{fmfchar*}}\quad=
\displaystyle{{i \delta_{ab}\over 
\displaystyle{E-h_o^{(0)}+i\eta}}
=
{i \delta_{ab}\over \displaystyle{E-{\bf p}^2/m-1/(2\nc)\als/ r+i\eta}}}\,.
\end{equation}
For illustration purposes we display here the singlet-octet chromoelectric dipole vertex, which is be the most used throughout the thesis. It reads
\begin{equation}
	\label{singoctvertex}
	\parbox{30mm}{
\begin{fmfchar*}(30,30)
\fmfleft{in}
\fmfright{out}
\fmftop{glue}
\fmf{plain,tension=100}{in,middle}
\fmf{dbl_plain,tension=100}{middle,out}
\fmf{gluon}{middle,glue}
\fmfv{label=$b$,lab.angle=0,lab.dist=8}{glue}
\fmfv{decor.shape=circle,decor.filled=empty,decor.size=10}{middle}
\fmfv{label=$a$,lab.angle=90}{out}
\end{fmfchar*}
\put(-16.5,14){$\boldsymbol\times$}
}\;=-g V_A\sqrt\frac\tf\nc\delta_{ab}\br k^0\,,\qquad \parbox{30mm}{
\begin{fmfchar*}(30,30)
\fmfleft{in}
\fmfright{out}
\fmftop{glue}
\fmf{plain,tension=100}{in,middle}
\fmf{dbl_plain,tension=100}{middle,out}
\fmf{dashes}{middle,glue}
\fmfv{label=$b$,lab.angle=0,lab.dist=8}{glue}
\fmfv{decor.shape=circle,decor.filled=empty,decor.size=10}{middle}
\fmfv{label=$a$,lab.angle=90}{out}
\end{fmfchar*}
\put(-16.5,14){$\boldsymbol\times$}
}\;=g V_A\sqrt\frac\tf\nc\delta_{ab}\br\cdot\bk\,,
\end{equation}
where the curly line represents a transverse gluon and the dashed line a longitudinal gluon. The rules for the non-Abelian part of the chromoelectric field, as well as for the other vertices, can be obtained in a similar fashion and are summarized in \cite{Brambilla:2004jw}.


\end{fmffile}
\subsection{Feynman rules at finite temperature in real time}
\label{app_feyn_rt}
\begin{fmffile}{rtprops}
\setlength{\unitlength}{1mm}
The free light quark propagator in the real-time formalism reads, neglecting colour indices
\begin{eqnarray}
	\nonumber {\bf S}(p)&=&(\slashed{p}+m)\left[\left(\begin{matrix}\displaystyle\frac{i}{p_0^2-\bp^2-m^2+i\eta}&\theta(-p_0)2\pi\delta(p_0^2-\bp^2-m^2)\\ \theta(p_0)2\pi\delta(p_0^2-\bp^2-m^2)&\displaystyle-\frac{i}{p^2_0-\bp^2-m^2-i\eta}\end{matrix}\right)\right.\\
	\label{fermionpropreal}
	&&\left.-2\pi\delta(p^2_0-\bp^2-m^2)\,n_\mathrm{F}(\vert p_0\vert)\left(\begin{matrix}1&1\\1&1\end{matrix}\right)\right]=\quad\parbox{30mm}{
\begin{fmfchar*}(30,10)
\fmfleft{in}
\fmfright{out}
\fmf{fermion}{in,out}
\end{fmfchar*}}\;,
\end{eqnarray}
where we notice that, due to the fermionic statistics, there is a minus sign in front of the thermal part, whose distribution is now the Fermi--Dirac distribution.\\
We now present the free gluon propagator in Coulomb gauge, which is the gauge we will use for all perturbative calculations in real time. The longitudinal gluon propagator reads
\begin{equation}
	\label{longgluonpropreal}
	{\bf D}_{00}(\bk)=\quad\parbox{30mm}{
\begin{fmfchar*}(30,10)
\fmfleft{in}
\fmfright{out}
\fmf{dashes}{in,out}
\end{fmfchar*}}\;=\;
\left(\begin{matrix}\displaystyle\frac{i}{\bk^2}&0\\0&\displaystyle-\frac{i}{\bk^2}\end{matrix}\right).
\end{equation}
The longitudinal propagator contains no thermal part; this is a consequence of the frequency-independent nature of the propagator, which causes the spectral density to vanish.\\
In the transverse sector we have instead
\begin{eqnarray}
	\nonumber {\bf D}_{ij}(k)&=&\left(\delta_{ij}-\frac{k_ik_j}{\bk^2}\right)\left[\left(\begin{matrix}\displaystyle\frac{i}{k^2_0-\bk^2+i\eta}&\theta(-k_0)2\pi\delta(k^2_0-\bk^2)\\\theta(k_0)2\pi\delta(k^2_0-\bk^2)&\displaystyle-\frac{i}{k^2_0-\bk^2-i\eta}\end{matrix}\right)\right.\\
	\label{transgluonpropreal}
	&&\left.+2\pi\delta(k^2_0-\bk^2)\,n_\mathrm{B}(\vert k_0\vert)\left(\begin{matrix}1&1\\1&1\end{matrix}\right)\right]=\quad\parbox{30mm}{
\begin{fmfchar*}(30,10)
\fmfleft{in}
\fmfright{out}
\fmf{gluon}{in,out}
\end{fmfchar*}}\;.
\end{eqnarray}
In Coulomb gauge ghosts couple only to transverse gluons; hence they never enter any of the calculations of the thesis and for this reason we omit their propagator.\\
For what concerns the vertices, the rules are those of Sec.~\ref{sub_feyn_qcd_realtime}, bearing in mind the opposite sign for vertices of type ``2''. \end{fmffile}
\subsection{Feynman rules in the Hard Thermal Loop effective theory}
\label{sub_feyn_htl}
The longitudinal and transverse gluon propagators in Coulomb gauge are the only necessary ingredients for the calculation in the HTL-resummed theory performed in this thesis. Their form in the real-time formalism can be read from \cite{Carrington:1997sq}\footnote{The transverse 
propagator given there contains a misprint: a factor of $p_0/(2p)$ should be multiplying 
the logarithm in Eq. (27), as follows from the transverse HTL self-energy given in Eq. (17) 
of the same paper.}
\begin{equation}
D^{\mathrm{R,A}}_{00}(k)=
\frac{i}{\bk^2+m_D^2\left(1-\displaystyle\frac{k_0}{2\mbk}\ln\frac{k_0+\mbk\pm i\eta}{k_0-\mbk\pm i\eta}\right)}\,,
\label{prophtllong}
\end{equation}
and 
\begin{equation}
D^{\mathrm{R,A}}_{ij}(k)=\left(\delta_{ij}-\frac{k_ik_j}{\bk^2}\right)\Delta_\mathrm{R,A}(k)\,,
\label{prophtltrans}
\end{equation}
respectively,  where
\begin{equation}
\Delta_\mathrm{R,A}(k)=
\frac{i}{k_0^2-\bk^2-\displaystyle\frac{m_D^2}{2}
\left(\displaystyle\frac{k_0^2}{\bk^2}-(k_0^2-\bk^2)\displaystyle\frac{k_0}{2\mbk^3}
\ln\left(\displaystyle\frac{k_0+\mbk\pm i\eta}{k_0-\mbk\pm i\eta}\right)\right)\pm i\,\mathrm{sgn}(k_0)\,\eta}\,,
\label{defdelta}
\end{equation}
and the upper sign refers to the retarded propagator (R), the lower sign to the advanced one (A). 
The ``11'' component can be obtained from the relation~\eqref{11component}.
 \section{Feynman rules in imaginary time}
\label{sec_im_time}
\subsection{Interaction vertices and the quark propagator}
\label{sub_ferm_vertex}
\begin{fmffile}{imtime}
\setlength{\unitlength}{1mm}
The Feynman rule for the three-gluon vertex reads 
\begin{equation}
	\label{3gimtime}
\parbox{30mm}{
\begin{fmfchar*}(30,30)
\fmftop{up}
\fmfbottom{down1,down2}
\fmf{gluon,label=$p\,,\mu\,,a$,label.dist=10}{up,center}
\fmf{gluon,label=$k\,,\rho\,,c$,label.dist=10}{down1,center}
\fmf{gluon,label=$q\,,\nu\,,b$,label.dist=10}{down2,center}
\end{fmfchar*}}\quad=\;igf^{abc}\left[\delta_{\mu\nu}(p-q)_\rho +\delta_{\nu\rho}(q-k)_\mu +\delta_{\rho\mu}(k-p)_\nu \right],
 \end{equation}
 where all momenta are understood as inflowing in the vertex and the curly line is taken to represent either a longitudinal or a transverse gluon. For the four-gluon vertex we have instead
\begin{equation}
	\label{4gimtime}
	\hspace{10mm}\parbox{30mm}{
\begin{fmfchar*}(30,30)
\fmftop{up1,up2}
\fmfbottom{down1,down2}
\fmf{gluon}{up1,center}
\fmf{gluon}{up2,center}
\fmf{gluon}{down1,center}
\fmf{gluon}{down2,center}
\fmfv{label=$\mu\,,a$}{up1}
\fmfv{label=$\nu\,,b$}{up2}
\fmfv{label=$\sigma\,,d$}{down1}
\fmfv{label=$\rho\,,c$}{down2}
\end{fmfchar*}}=\begin{array}{l}
	-g^2\left[f^{eab}f^{ecd}(\delta_{\mu\rho}\delta_{\nu\sigma}-\delta_{\mu\sigma}\delta_{\nu\rho})+f^{eac}f^{edb}(\delta_{\mu\sigma}\delta_{\nu\rho}-\delta_{\mu\nu}\delta_{\rho\sigma})\right.\\
	\\
	\hspace{8mm}\left.+f^{ead}f^{ebc}(\delta_{\mu\nu}\delta_{\sigma\rho}-\delta_{\mu\rho}\delta_{\sigma\nu}) \right].
\end{array}
 \end{equation}
The gluon-ghost vertex is, in covariant gauges
\begin{equation}
	\label{gluonghostimtime}
\parbox{30mm}{
\begin{fmfchar*}(30,30)
\fmftop{up}
\fmfbottom{down1,down2}
\fmf{gluon}{up,center}
\fmf{ghost}{down1,center}
\fmf{ghost,label=$p$}{center,down2}
\fmfv{label=$\mu\,,a$,lab.angle=0}{up}
\fmfv{label=$c$}{down1}
\fmfv{label=$b$}{down2}
\end{fmfchar*}}\quad=\;igf^{abc}p_\mu\,,
 \end{equation}
where the momentum $p$ is understood as inflowing in the vertex and thus in the opposite direction of the arrow.\\
The quark-gluon vertex is simply
 \begin{eqnarray}
	 \nonumber \\
	\label{gluonquarkimtime}
	\parbox{30mm}{
\begin{fmfchar*}(30,30)
\fmftop{up}
\fmfbottom{down1,down2}
\fmf{gluon}{up,center}
\fmf{quark}{down1,center}
\fmf{quark}{center,down2}
\fmfv{label=$\mu\,,a$}{up}
\end{fmfchar*}}\quad=\;g\gamma_\mu T^a\,.
 \end{eqnarray}
The fermion propagator in imaginary time reads 
 \begin{equation}
 	\label{matsubarafermion}
	S_\mathrm{F}(\tilde\omega_n,\bk)=\quad\parbox{30mm}{
\begin{fmfchar*}(30,10)
\fmfleft{in}
\fmfright{out}
\fmf{fermion}{in,out}
\end{fmfchar*}}\;=\; \frac{m-\slashed{k}}{\tilde\omega_n^2+\bk^2+m^2}\,,
 \end{equation}
where $\tilde\omega_n=(2n+1)\pi T$ is the fermionic Matsubara frequency 
and in Euclidean space-time
 \begin{equation}
 	\label{euclslash}
 	\slashed{k}=-\gamma_4\tilde\omega_n+\gamma\cdot\bk,\qquad\gamma_4=i\gamma_0\,.
 \end{equation}
 \end{fmffile}
\subsection{Feynman rules in the static gauge}
\label{sub_stat_gauge}
\begin{fmffile}{statgauge}
\setlength{\unitlength}{1mm}
In the following, we list the Feynman rules in Euclidean space-time  
under the gauge condition $\partial_0A_0=0$. We refer to \cite{D'Hoker:1981us,Nadkarni:1982kb,Curci:1982fd,Curci:1984rd} for a thorough derivation.\\
The temporal propagator reads (dropping colour indices)  
\begin{equation}
\label{temporal}
D_{00}(\omega_n,\bk)=\quad\parbox{30mm}{
\begin{fmfchar*}(30,10)
\fmfleft{in}
\fmfright{out}
\fmf{dashes}{in,out}
\end{fmfchar*}}\;=\; \frac{\delta_{n0}}{\bk^2},
\end{equation}
where, as usual, $\omega_n=2\pi nT$ and the Kronecker delta fixes
$n=0$, making this propagator purely static. The spatial propagator
can be divided into a non-static ($n\ne0$) and a static ($n=0$)
part. The former reads
\begin{equation}
\label{nonstatic}
D_{ij}(\omega_n\ne0,\bk)=\quad\parbox{30mm}{
\begin{fmfchar*}(30,10)
\fmfleft{in}
\fmfright{out}
\fmf{curly}{in,out}
\end{fmfchar*}}\;=\; 
\frac{1}{\omega_n^2+\bk^2}\left(\delta_{ij}+\frac{k_ik_j}{\omega_n^2}\right)(1-\delta_{n0}),
\end{equation}
and thus mixes longitudinal and transverse components. The static part
has a residual gauge dependence on the parameter $\xi$; it reads
\begin{equation}
\label{magnetostatic}
D_{ij}(\omega_n=0,\bk)=\quad\parbox{30mm}{
\begin{fmfchar*}(30,10)
\fmfleft{in}
\fmfright{out}
\fmf{photon}{in,out}
\end{fmfchar*}}\;=\; \frac{1}{\bk^2}\left(\delta_{ij}-(1-\xi)\frac{k_ik_j}{\bk^2}\right)\delta_{n0}.
\end{equation}
Finally the ghost propagator reads
\begin{equation}
\label{ghost}
D_{\mathrm{ghost}}(\omega_n,\bk)=\quad\parbox{30mm}{
\begin{fmfchar*}(30,10)
\fmfleft{in}
\fmfright{out}
\fmf{ghost}{in,out}
\end{fmfchar*}}\;=\; \frac{\delta_{n0}}{\bk^2},
\end{equation}
and is thus purely static. It couples to spatial gluons only according to the Feynman rule in Eq.~\eqref{gluonghostimtime}. The non-static ghost can be shown to decouple \cite{D'Hoker:1981us}. The gluon-gluon and quark-gluon interaction vertices are the usual ones shown in Eqs.~\eqref{3gimtime},~\eqref{4gimtime} and~\eqref{gluonquarkimtime}.
\end{fmffile}

	\chapter{The pNRQCD Lagrangian at higher orders in the expansions}
	\label{app_pnrqcd}
	In this Appendix we give more details on the matching coefficients of pNRQCD, as we mentioned in Sec.~\ref{sec_pnrqcd}. In the first part, Sec.~\ref{app_potentials}, the potentials appearing in the singlet and octet Hamiltonians $h_s$ and $h_o$ (see Eq.~\eqref{lagrpnrqcd}) are given up to order $1/m^2$.\\
In Sec.~\ref{app_multipole} the singlet-octet interaction terms up to order $1/m^2$ in the inverse mass and $r$ in the multipole expansion are given and their matching coefficients are reported. Some of these terms have been used for the finite-temperature calculation of the spin-orbit potential in Chap.~\ref{chap_poincare}.
\section{Matching of the potentials}
\label{app_potentials}
We recall from Eq.~\eqref{sinoctham} that the potentials are organized in a $1/m$ expansion, i.e. $V^{(0)}_{s,o}
+\frac{V^{(1)}_{s,o}}{m}+\frac{V^{(2)}_{s,o}}{m^2}+\ldots$. For what concerns $V^{(0)}$, the matching coefficients $\alVs$ and $\alVo$ appearing in the singlet and octet static potentials respectively (see Eq.~\eqref{staticpot}) read
\begin{eqnarray}
	\alpha_{V_{s,o}} &=&\als(r)
	\left\{1+\left(a_1+ 2 {\gamma_E \beta_0}\right) {\als(r) \over 4\pi}\right.
	\nonumber\\
	&&\qquad\qquad 
	+
	\left[\gamma_E\left(4 a_1\beta_0+ 2{\beta_1}\right)+\left( {\pi^2 \over 3}+4 \gamma_E^2\right) 
	{\beta_0^2}+a_{2\,s,0}\right] {\als^2(r) \over 16\,\pi^2}\nn\\
	&&\qquad\qquad\qquad\qquad+\left.\left[\frac{16\pi^2}{3}\ca^3\ln(r\mu)+\tilde{a}_{3\,s,o}\right] {\als^3(r) \over 64\,\pi^3}+\order{\als^4}\right\},
	\label{alvs}
\end{eqnarray}
where $\beta_i$ are the coefficients of the $\beta$-function~\eqref{betaexpand}, the one-loop coefficient $a_1$
\begin{equation}
	a_1=\frac{31}{9}\ca-\frac{10}{9}n_f\,,
	\label{a1}
\end{equation}
is identical for both colour states and was first computed in \cite{Fischler:1977yf,Billoire:1979ih}. Starting from $a_2$ the two matching coefficients differ. The singlet was computed in \cite{Peter:1997ig,Schroder:1998vy} and the octet in \cite{Kniehl:2004rk}, yielding
\begin{equation}
	a_{2\,o}=a_{2\,s}+\left(\pi^4-12\pi^2\right)\ca^2\,.
	\label{a2o}
\end{equation}
Finally, at order $\als^4$ we notice the logarithm that accompanies the IR divergence mentioned in Sec.~\ref{sec_pnrqcd}. Only the singlet coefficient $\tilde{a}_{3\,s}$ is known \cite{Smirnov:2008pn,Smirnov:2009fh,Anzai:2009tm}. At order $\als^5$ only the logarithmic, IR divergent part is known \cite{Brambilla:2006wp,Brambilla:2009bi}.

The $1/m$ potential $V^{(1)}$ reads 
\begin{equation}
	{V^{(1)}_s } (r)	= -{C_FC_A D^{(1)}_s \over 2r^2}\,
	\label{v1}
\end{equation}
where from now on we concentrate on the singlet sector only. At the leading order (one loop) $D^{(1)}_s=\als^2(r)$ \cite{Duncan:1975kt}, and the NLO (two-loop) contribution was calculated by \cite{Kniehl:2001ju} (the logarithmic corrections were computed by
\cite{Kniehl:1999ud,Brambilla:1999xj}).

$V^{(2)}$ can be written as a sum of spin-dependent ($SD$) and spin-independent ($SI$) terms:
\begin{eqnarray}
	V^{(2)}&=&V^{(2)}_{SD}+V^{(2)}_{SI},\label{v2}\\
V^{(2)}_{SI}
&=&
{1\over 8}\{ {\bf P}^2, V^{(2)}_{{\bf p}^2,{\rm CM}}(r)\}
+
{({\bf r}\times {\bf P})^2\over 4 r^2} V^{(2)}_{{\bf L}^2,{\rm CM}}(r)
\nn
\\
&&
\nn+{1 \over 2}\left\{{\bf p}^2,V_{{\bf p}^2}^{(2)}(r)\right\}
+{V_{{\bf L}^2}^{(2)}(r)\over r^2}{\bf L}^2 + V_r^{(2)}(r),
\\
V^{(2)}_{SD} &=&
{({\bf r}\times {\bf P})\cdot ({\bf S}_1 - {\bf S}_2) \over 2}
V^{(2)}_{LS,{\rm CM}}(r)
+
V_{LS}^{(2)}(r){\bf L}\cdot{\bf S} + V_{S^2}^{(2)}(r){\bf S}^2
 + V_{{\bf S}_{12}}^{(2)}(r){\bf S}_{12}({\hat {\bf r}}), \nn
\end{eqnarray}
where ${\bf S}_1=\bfsigma_1/2$, ${\bf S}_2=\bfsigma_2/2$, ${\bf L}_1 \equiv {\bf r} \times {\bf p}_1$, ${\bf L}_2 \equiv {\bf
  r} \times {\bf p}_2$ and ${\bf S}_{12}({\hat {\bf r}}) \equiv 
3 {\hat {\bf r}}\cdot \bfsigma_1 \,{\hat {\bf r}}\cdot \bfsigma_2 - \bfsigma_1\cdot \bfsigma_2$, ${\bf S}={\bf S}_1+{\bf S}_2$ and ${\bf L}={\bf r}\times {\bf p}$.
Other forms of the potential can be brought to the one above by using 
unitary transformations, or the relation
\begin{equation}
- \left\{ {1 \over r},{\bf p}^2 \right\} + 
{1 \over r^3}{\bf L}^2 + 4\pi\delta^{(3)}({\bf r})
= - {1 \over r} \left( {\bf p}^2 + 
{ 1 \over r^2} {\bf r} \cdot ({\bf r} \cdot {\bf p}){\bf p} \right).
\end{equation}
By dimensional analysis $ V^{(2)}_{{\bf p}^2}$ scales like $1/r$, 
$V^{(2)}_r$ like $1/r^3$ or $\delta^{(3)} ({\bf r})$, and so on. They read
\begin{displaymath}
V_{{\bf p}^2,s}^{(2)}(r)=- { C_F D^{(2)}_{1,s} }\frac{1}{r},\qquad
V_{{\bf L}^2,s}^{(2)}(r)= { C_F D^{(2)}_{2,s} \over 2 }{1 \over r},
\end{displaymath}
\begin{displaymath}
V_{r,s}^{(2)}(r)={\pi C_F D^{(2)}_{d,s} }
\delta^{(3)}({\bf r}),\qquad
 V_{S^2,s}^{(2)}(r)={4\pi C_F D^{(2)}_{S^2,s} \over 3}
 \delta^{(3)}({\bf r}),
\end{displaymath}
\begin{equation}
V_{LS,s}^{(2)}(r)={ 3 C_F D^{(2)}_{LS,s} \over 2 }{1 \over r^3},\qquad
V_{{\bf S}_{12},s}^{(2)}(r)= { C_F D^{(2)}_{S_{12},s} \over 4 }{1 \over r^3},
\label{V2}
\end{equation}
where the various $D$s depend logarithmically on $r$ and the
renormalization scale $\nu_\mathrm{pNR}$.
They read
\begin{eqnarray}
D^{(2)}_{1,s}=
D^{(2)}_{2,s}=
D^{(2)}_{d,s}= 
D^{(2)}_{S^2,s}=
D^{(2)}_{LS,s}=
D^{(2)}_{S_{12},s}=\als(r)
\,.
\label{Dsten2}
\end{eqnarray}
The complete $V_s^{(2)}$ have been
computed over the years
\cite{Gupta:1981pd,Gupta:1982qc,Buchmuller:1981aj,Pantaleone:1985uf,
Titard:1993nn,Pineda:1998kn,Brambilla:1999xj,Manohar:2000hj,Kniehl:2002br} 
and can be found to one-loop in \cite{Kniehl:2002br}.\\
The center-of-mass (CM) $1/m^2$ potentials are linked by Poincar\'e invariance to the static potential and derivatives thereof \cite{Brambilla:2003nt}:
\begin{equation}
	{V_{LS,{\rm CM}} \over V^{(0)\prime}} =  -{1\over 2r},
	\qquad
	V_{{\bf L}^2,{\rm CM}} + {r \, V^{(0)\prime} \over 2} = 0,
	\qquad
	V_{{\bf p}^2,{\rm CM}} + V_{{\bf L}^2,{\rm CM}} + {V^{(0)} \over 2}= 0\,,
	\label{poincpotentials}
\end{equation}
where $f(r)'\equiv df(r)/dr$ and we notice that the first relation is the \emph{Gromes relation} \cite{Gromes:1984ma}, which has been at the center of Chap.~\ref{chap_poincare}, where it was introduced in Eq.~\eqref{gromes}.\\
We furthermore remark that, in order to obtain the spectrum at order $m\als^ 4$, $\alpha_{V_s}$ has to be 
calculated to order $\als^ 3$ (two loops), $V_s^{(1)}$ to order $\als^ 2$ (one loop) 
and the remaining potentials to order $\als$ (tree level). If one wishes to have the spectrum to one order higher, namely $m\als^ 5$, all these potentials must be calculated
to one more power in $\als$.\\
At order $1/m^2$ the potentials present also an imaginary part governing the annihilation to gluons and photons. It can be found in \cite{Brambilla:2004jw}.

\section{Matching of higher-order terms in the multipole expansion}
\label{app_multipole}
In this section we deal with the terms appearing in the multipole expansion at higher orders in $r$ and $1/m$. Only the terms necessary for the calculations of Chap.~\ref{chap_poincare} are shown here. For a full treatmen we refer to \cite{Brambilla:2003nt}, whose notation we adopt in this section.\\
In this new notation the Lagrangian of pNRQCD becomes
\begin{eqnarray}
&& \hspace{-8mm} {\mathcal L}_{\rm pNRQCD} = \int d^3r \, {\rm Tr} \,\left\{
  {\rm S}^\dagger \left( i\partial_0 - h_s \right) {\rm S}
+ {\rm O}^\dagger \left( iD_0 - h_o \right) {\rm O}
\right.
\nn\\
&&\qquad
- \left[ ({\rm S}^\dagger h_{so} {\rm O} + {\rm H.C.}) + {\rm C.C.} \right] \;
- \left[ {\rm O}^\dagger h_{oo} {\rm O} + {\rm C.C.}  \right]  \;
\nn \\
&&\qquad
\left.
 - \left[ {\rm O}^\dagger h_{oo}^A {\rm O}  h_{oo}^B + {\rm C.C.}  \right]
\right\}
- \frac{1}{4} F^a_{\mu \nu} F^{a\,\mu \nu}+ \sum_{i=1}^{n_f}\bar{q}_i\,iD\!\!\!\!/\,q_i\,,
\label{pnrqcd}
\end{eqnarray}
which differs from  Eq.~\eqref{lagrpnrqcd} in the second and third lines. C.C. stands
for charge conjugation and H.C. stands for Hermitian conjugation. In the third line we have a new kind of octet-octet operator.\\ 
The operators $h_{so}$, $h_{oo}$, $h_{oo}^A$ and $h_{oo}^B$ describe the singlet-octet and octet-octet interactions. For our purposes only the former is needed, corresponding to the operator $h_{so}$. We now set out to detail this term: it may be ordered in powers of $1/m$ and $r$ as 
\begin{equation} 
h_{so} = h_{so}^{(0,1)} + h_{so}^{(0,2)} + h_{so}^{(1,0)} + h_{so}^{(1,1)} + h_{so}^{(2,0)}+\ldots\,,
\end{equation}
where the indices $(i,j)$ refer to the order in powers of $1/m$ and $r$ respectively. 
The dots stand for higher orders in those expansions
The explicit expressions of $ h_{so}^{(i,j)}$ may be taken from \cite{Brambilla:2003nt} and read
\begin{eqnarray}
&&h_{so}^{(0,1)} = - \frac{V_{so}^{(0,1)}(r)}{2} {\bf r}\cdot g {\bf E}\,,
\\
&&h_{so}^{(1,0)} = - \frac{c_F}{2m} \, V_{so\,b}^{(1,0)}(r) \, \bfsigma^{(1)}\cdot g {\bf B}
\nn\\
\label{1/m}
&&
\qquad-\frac{1}{2m} \, \frac{V_{so\,c}^{(1,0)}(r)}{r^2}
\, ({\bf r}\cdot\bfsigma^{(1)})\, ({\bf r}\cdot g {\bf B})
- \frac{1}{m}\frac{V_{so\,d}^{(1,0)}(r)}{2 r} {\bf r}\cdot g {\bf E}\,,
\\
&&h_{so}^{(1,1)}  = \frac{1}{8m} \, V_{so}^{(1,1)}(r) \,
\{ {\bf P} \cdot ,{\bf r}\times g {\bf B} \} +\ldots\,,
\label{1/m1m2}
\end{eqnarray}
\begin{eqnarray}
&&h_{so}^{(2,0)} = \frac{c_s}{16m^2} V_{so\,a}^{(2,0)}(r) \, \bfsigma^{(1)}\cdot [{\bf P} \times, g {\bf E}]
\nn\\
&&\qquad
+ \frac{1}{16m^2} \frac{V_{so\,b'}^{(2,0)}(r)}{r^2} \,
({\bf r}\cdot\bfsigma^{(1)}) \,
\{ {\bf P} \cdot, (g {\bf E} \times {\bf r}) \}
\nn\\
&&\qquad
+ \frac{1}{16m^2} \frac{V_{so\,b''}^{(2,0)}(r)}{r^2} \,
\{ ({\bf r}\cdot g {\bf E}),
{\bf P} \cdot ({\bf r} \times \bfsigma^{(1)})\}
\nn\\
&&\qquad
+ \frac{1}{16m^2} \frac{V_{so\,b'''}^{(2,0)}(r)}{r^2} \,
\{({\bf r}\cdot{\bf P}),
\bfsigma^{(1)} \cdot ({\bf r} \times g {\bf E}) \}
\nn\\
\label{1/m2}
&&\qquad
+ \frac{1}{8m^2} \, \frac{V_{so\, e}^{(2,0)}(r)}{r}\,
\{ {\bf P} \cdot ,{\bf r}\times g {\bf B} \} +\ldots\,.
\end{eqnarray}
Charge conjugation invariance requires that $h_{so}^{(0,2)} = 0$. $[\bpg\times,g\be]=\bpg\times g\be-g\be\times\bpg$ and similarly for the anticommutators. 
For $h^{(1,1)}_{so}$ and $h^{(2,0)}_{so}$ only the ${\bf P}$-dependent terms have been displayed.
The coefficients $c_F$ and $c_s$ are inherited from NRQCD (see Sec.~\ref{sec_nrqcd}) and encode non-analytical 
contributions in $1/m$, whereas the various $V_{so}^{(i,j)}(r)$ come from the matching to pNRQCD and 
encode non-analytical contributions in $r$. In the notation of Eq.~\eqref{lagrpnrqcd}, $V_{so}^{(0,1)}(r)$ corresponds to $V_A(r)$. At leading order in the coupling, the matching gives $c_F=c_s=1$ and
$V_{so}^{(0,1)}(r) = V_{so\,b}^{(1,0)}(r) = V_{so\,a}^{(2,0)}(r) = V_{so}^{(1,1)}(r) = 1$, 
while all other matching coefficients are of order $\als$ or smaller.

Poincar\'e invariance imposes further constraints on the matching coefficients. Beside the relation linking $c_F$ and $c_s$ in NRQCD, given by Eq.~\eqref{poincnrqcd}, we have that the following exact relations \cite{Brambilla:2003nt}  hold at the level of pNRQCD 
\begin{eqnarray}
&&  V_{so}^{(1,1)}(r) = V_{so}^{(0,1)}(r)\,,
\label{poincare11}
\\
&&2\, c_F V_{so\,b}^{(1,0)}(r) - c_s V_{so\,a}^{(2,0)}(r) = V_{so}^{(0,1)}(r) \,, 
\label{gromesmatching0}
\\
&&2\, c_F V_{so\,b}^{(1,0)}(r) - c_s V_{so\,a}^{(2,0)}(r) - V_{so\,b''}^{(2,0)}(r) = \left(r \, V_{so}^{(0,1)}(r)\right)^\prime\,.
\label{gromesmatching}
\end{eqnarray}
Combining the last two it follows that 
\begin{equation}
V_{so\,b''}^{(2,0)}(r) = -r V_{so}^{(0,1)}(r)^\prime\,.
\label{gromesmatching1}
\end{equation}
An interesting consequence of this relation is that, since $V_{so}^{(0,1)}(r)$ is at least 
of order $\als^2$ \cite{Brambilla:2006wp} but has not infrared divergences at that order 
\cite{Brambilla:2009bi}, $V_{so\,b''}^{(2,0)}(r)$ is at least of order $\als^3$.

	\chapter{Details on the real-time calculations}
	\label{app_realtime}
	In this Appendix we give some technical details on the calculation of Part~\ref{part_realtime}. In Sec.~\ref{secvacuumpol} we give the one-loop expression of the gluon polarization tensor in real time, in Sec.~\ref{secpQCD} we reobtain the static subset of the results of Sec.~\ref{secT}, while in Sec.~\ref{app_trans} we lay out the detailed calculation of the transverse gluon contribution in Sec.~\ref{sub_trans}.
\section{The longitudinal gluon polarization tensor}
\label{secvacuumpol}
The gluon polarization tensor is obtained by summing up all thermal contributions from the diagrams of  Fig.~\ref{figvacpol}. In Coulomb gauge this yields
(for details see \cite{Ghiglieri08}):
\begin{figure}
	\begin{center}
\includegraphics{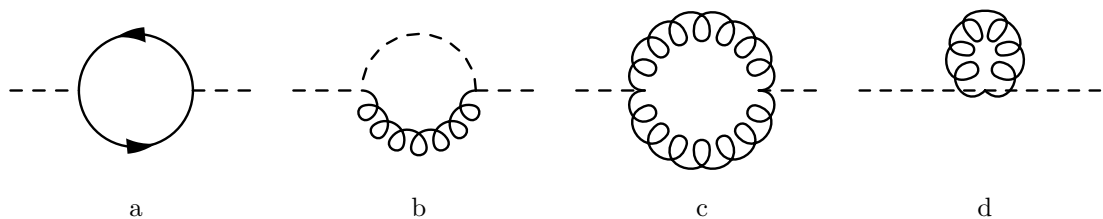}
\end{center}
\caption{Diagrams contributing to the longitudinal component of the 
gluon polarization tensor at one-loop order. Dashed lines for longitudinal gluons and 
curly lines for transverse gluons. Ghosts do not contribute to the thermal part of the gluon polarization tensor 
\cite{Landshoff:1992ne}. Furthermore, they do not couple to longitudinal gluons in Coulomb gauge.}
\label{figvacpol}
\end{figure}
\begin{eqnarray} 
\left[\Pi_{00}^{\rm R}(k)\right]_{\rm thermal} &=& 
\left[\Pi_{00}(k_0+i\epsilon,\bk)\right]_{\rm thermal}\,, 
\label{Pi00fullR} \\
\left[\Pi_{00}^{\rm A}(k)\right]_{\rm thermal} &=& 
\left[\Pi_{00}(k_0-i\epsilon,\bk)\right]_{\rm thermal}\,, 
\label{Pi00fullA} 
\end{eqnarray}
\begin{equation}
\left[\Pi_{00}(k)\right]_{\rm thermal} 
= \left[\Pi_{00,\,{\rm F}}(k)\right]_{\rm thermal} 
+ \left[\Pi_{00,\,{\rm G}}(k)\right]_{\rm thermal}\,,  
\label{Pi00full}
\end{equation}
\begin{eqnarray} 
\left[\Pi_{00,\,{\rm F}}(k)\right]_{\rm thermal} 
&=& 
\frac{g^2 \, T_F \, n_f}{2\pi^2}\int_{-\infty}^{+\infty}dq_0\,|q_0|\, n_{\rm F}(\vert q_0\vert)
\nn \\
&& \hspace{-2cm}
\times\left[2-\left(\frac{4q_0^2+k^2-4q_0k_0}{4|q_0||{\bk}|}\right)
\ln\frac{k^2-2q_0k_0+2|q_0||{\bk}|}{k^2-2q_0k_0-2|q_0||{\bk}|}\right.
\nn \\
&& \hspace{-2cm} 
\qquad 
\left.
+\left(\frac{4q_0^2+k^2+4q_0k_0}{4|q_0||{\bk}|}\right)
\ln\frac{k^2+2q_0k_0-2|q_0||{\bk}|}{k^2+2q_0k_0+2|q_0||{\bk}|}\right]\,,
\label{Pi00fermion}
\\
\left[\Pi_{00,\,{\rm G}}(k)\right]_{\rm thermal} 
&=&
\frac{ g^2 \, N_c}{2\pi^2}\int_{-\infty}^{+\infty}dq_0\,|q_0| n_{\rm B}(|q_0|)
\nn\\
&& \hspace{-2cm} 
\times 
\left\{1+\frac{(2q_0-k_0)^2}{8q_0^2}-\frac{1}{2}-\frac{\bk^2}{2q_0^2}\right.
\nn\\
&& \hspace{-2cm} 
\qquad
+2\left[\frac{|{\bk}|}{2|q_0|}
-\frac{(\bk^2+q_0^2)^2}{8|q_0|^3|{\bk}|}
-\frac{(2q_0-k_0)^2}{4(q_0-k_0)^2}
\left(-\frac{(\bk^2+q_0^2)^2}{8|q_0|^3|{\bk}|}
+\frac{|{\bk}|}{2|q_0|}\right)\right]
\nn\\ 
&& \hspace{-2cm} 
\qquad\qquad\qquad\qquad\qquad\qquad
\times \ln\left\vert\frac{|{\bk}|-|q_0|}{|{\bk}|+|q_0|}\right\vert
\nn\\
&& \hspace{-2cm} 
\qquad
-\frac{(2q_0-k_0)^2}{4} \left[\frac{1}{(q_0-k_0)^2}
\left(\frac{(k^2-2q_0k_0)^2}{8|q_0|^3|{\bk}|}
+\frac{k^2-2q_0k_0}{2|q_0||{\bk}|}+\frac{|q_0|}{2|{\bk}|}\right)\right.
\nn\\
&& \hspace{-2cm} 
\qquad\qquad\qquad\qquad\qquad\qquad
\left.\left. +\frac{1}{2|q_0||{\bk}|}\right]\ln\frac{k^2-2q_0k_0+2|q_0||{\bk}|}{k^2-2q_0k_0-2|q_0||{\bk}|}\right\}\,,
\label{Pi00gluon}
\end{eqnarray}
where ``R'' stands for retarded, ``A'' for advanced, ``F'' labels the 
contribution coming from the loops of $n_f$ massless quarks (first diagram of Fig.~\ref{figvacpol}) 
and ``G'' labels the contribution from the second, third and fourth diagram of Fig.~\ref{figvacpol}.
In the context of the imaginary-time formalism, Eqs.~(\ref{Pi00fermion}) 
and (\ref{Pi00gluon}) can be found also in textbooks like \cite{Kapusta:2006pm}. 
The original derivation of (\ref{Pi00gluon}) is in \cite{Heinz:1986kz}.\\
The retarded and advanced gluon self energies contribute to the retarded 
and advanced gluon propagators. From the retarded and advanced gluon propagators 
we may derive the full propagator, the spectral density and finally all components 
of the $2 \times 2$ matrix of the real-time gluon propagator along the lines 
of Eqs.~\eqref{kms} and~\eqref{11component}. In the following, we study 
Eqs.~(\ref{Pi00fullR})-(\ref{Pi00gluon}) in different kinematical limits which will be useful for the calculations of Chap.~\ref{chap_rggT} and Sec.~\ref{secpQCD}.
\subsection{The longitudinal gluon polarization tensor for $k_0 \ll T\sim|\bk|$}
\label{secvacuumpolk0}
The typical loop momentum $q_0$ is of order $T$. 
If we expand $\left[\Pi_{00}^{\rm R}(k)\right]_{\rm thermal}$ 
and\\
$\left[\Pi_{00}^{\rm A}(k)\right]_{\rm thermal}$ in $k_0 \ll$ $T \sim |\bk|$ 
and keep terms up to order $k_0$, the result is 
\begin{eqnarray}
{\rm Re} \, \left[\Pi_{00}^{\rm R}(k)\right]_{\rm thermal} = 
{\rm Re} \, \left[\Pi_{00}^{\rm A}(k)\right]_{\rm thermal} &=& 
\nn\\
&& \hspace{-7cm}
\frac{g^2 \, T_F \, n_f}{\pi^2} 
\int_0^{+\infty}dq_0\,q_0\,n_{\rm F}(q_0)\left[
2 +  \left( \frac{|\bk|}{2q_0} - 2 \frac{q_0}{|\bk|} \right) 
\ln \left| \frac{|\bk|-2q_0}{|\bk|+2q_0} \right|
\right]
\nn\\
&& \hspace{-7cm}
+ \frac{g^2 \, N_c}{\pi^2}
\int_0^{+\infty}dq_0\,q_0\,n_{\rm B}(q_0)\left[
1 - \frac{\bk^2}{2q_0^2}
+ \left( - \frac{q_0}{|\bk|} + \frac{|\bk|}{2q_0} - \frac{|\bk|^3}{8q_0^3} \right)
\ln \left| \frac{|\bk|-2q_0}{|\bk|+2q_0} \right|
\right] \,,
\nn\\
\label{RePi00k0}
\end{eqnarray}
\begin{eqnarray}
{\rm Im} \, \left[\Pi_{00}^{\rm R}(k)\right]_{\rm thermal} = 
- {\rm Im} \, \left[\Pi_{00}^{\rm A}(k)\right]_{\rm thermal} &=& 
\nn\\
&& \hspace{-4cm}
\frac{2\, g^2 \, T_F \, n_f}{\pi}\,\frac{k_0}{|\bk|} 
\int_{|\bk|/2}^{+\infty}dq_0\,q_0\,n_{\rm F}(q_0)
\nn\\
&& \hspace{-4cm}
+ \frac{g^2 \, N_c}{\pi} \, \frac{k_0}{|\bk|}
\left[\frac{\bk^2}{8}\,n_{\rm B}(|\bk|/2) + 
\int_{|\bk|/2}^{+\infty}dq_0\,q_0\,n_{\rm B}(q_0)\left( 1 - \frac{\bk^4}{8q_0^4} \right) 
\right]
\,.\nn\\
\label{ImPi00k0}
\end{eqnarray}
Equation (\ref{ImPi00k0}) and the gluonic part of (\ref{RePi00k0})
are in agreement with \cite{Heinz:1986kz}.

\subsection{The longitudinal gluon polarization tensor for $|\bk| \gg T \gg k_0$}
\label{secvacuumpolkoo}
If we assume that  $|\bk| \gg T \gg k_0$, then the expression for the longitudinal 
gluon polarization tensor may extracted from Eqs.~(\ref{RePi00k0}) and (\ref{ImPi00k0})
by expanding for large $|\bk|/T$. At leading order, we obtain 
\begin{equation}
\left[\Pi_{00}^{\rm R}(k)\right]_{\rm thermal} = \left[\Pi_{00}^{\rm A}(k)\right]_{\rm thermal}
= - \frac{N_c\,g^2\,T^2}{18}\,.
\label{Pi00koo}
\end{equation}
The result is real and does not depend on $k$. Moreover, only the gluonic part of 
the polarization tensor contributes in this limit and at this order.
Higher-order real corrections are suppressed by $T^2/\bk^2$, while higher-order imaginary 
corrections are exponentially suppressed.

\section{Short-distance thermal corrections to the potential in perturbative QCD for $1/r\gg T\gg \als/r\gg m_D$}
\label{secpQCD}
\begin{figure}[ht]
	\begin{center}
\includegraphics{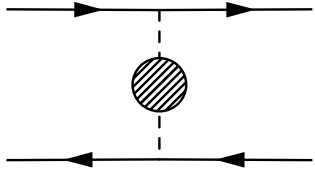}
\end{center}
\caption{Longitudinal gluon exchange between a static quark and a static antiquark; 
the dashed blob stands for the gluon self energy.}  
\label{figgluon00}
\end{figure}

In this Section, we ask the question of what would be the origin of 
the static part of the potential if we would not introduce any EFT treatment, but simply perform a calculation in 
perturbative QCD under the condition that $1/r \gg T\gg \als/r\gg m_D$. The static potential in this regime can be obtained from Eq.~(\ref{totalpotT}). It reads
\begin{eqnarray}
	\nonumber
	\delta V_s &=&\frac{\pi}{9}N_c C_F\, \als^2\, T^2\,r -\frac{\als^4 C_F\nc^3I_T}{24\pi r}
	\\
	&&- \frac{3}{2} \zeta(3)\,  C_F \, \frac{\als}{\pi} \, r^2 \, T \,m_D^2
	+ \frac{2}{3} \zeta(3)\, N_c C_F \, \als^2 \, r^2 \, T^3
	\nonumber\\
	&&
	+ i \left[ \frac{C_F}{6} \als \, r^2 \, T \,m_D^2\, \left( 
	\frac{1}{\epsilon} + \gamma_E + \ln\pi 
	- \ln\frac{T^2}{\mu^2} + \frac{2}{3} - 4 \ln 2 - 2 \frac{\zeta^\prime(2)}{\zeta(2)} \right)\right.
	\nonumber\\
	&&  \quad \left.
	+ \frac{4\pi}{9} \ln 2 \; N_c C_F \,  \als^2\, r^2 \, T^3 \right]
	\,,
	\label{staticT}
\end{eqnarray}
The answer is that the leading static part of (\ref{totalpotT}) would originate 
from the longitudinal gluon exchange, with a self-energy insertion, between 
a static quark and a static antiquark shown in Fig.~\ref{figgluon00}. 

We first consider the diagram in Fig.~\ref{figgluon00}, which 
contributes to the physical ``11'' component of the static potential by 
\begin{equation}
\left[ \delta {\bf V_s}(r)\right]_{11} = 
\mu^{4-D} \int \frac{d^{d}k}{(2\pi)^{d}} \, e^{-i \bk \cdot \br}
\,g^2 \, C_F \, \left[ i \delta {\bf D}_{00}(0,\bk)\right]_{11}\,,
\label{VsQCD}
\end{equation}
where $\delta {\bf D}_{00}(k)$ is defined in Eqs.~(\ref{D0011})-(\ref{DRAPi}) and 
depends on the gluon polarization $\Pi_{00}^{\rm R,A}$.
Note that we have set to zero the fourth-component of the momentum in the longitudinal gluon: 
corrections would be suppressed by powers of $k_0/|\bk| \sim V_s\,r$ or $V_s/T$ or $V_s/m_D$.
Equation (\ref{VsQCD}) gets contributions from different momentum regions. 

{\bf (1)} The first  momentum region is $|\bk| \sim 1/r$. The thermal contribution to the 
longitudinal gluon polarization tensor when $|\bk| \sim 1/r \gg T$ 
is provided by Eq.~(\ref{Pi00koo}), which, substituted in 
Eq.~(\ref{VsQCD}), gives (the integral is finite, hence $d=3$)
\begin{equation}
 \delta  V_s(r) = 
\int \frac{d^3k}{(2\pi)^3}e^{-i\bk \cdot \br} 
\left(-C_F \frac{4\pi\als}{\bk^4}\right) \frac{N_c\,g^2\,T^2}{18}
= \frac{\pi}{9} \, N_c C_F \, \als^2 \, r \, T^2\,,
\label{VsTbis}
\end{equation}
where we have used that the Fourier transform of $4\pi/\bk^4$ is $-r/2$.
Equation (\ref{VsTbis}) agrees with the first term of Eq.~(\ref{staticT}).

\begin{figure}[ht]
	\begin{center}
\includegraphics{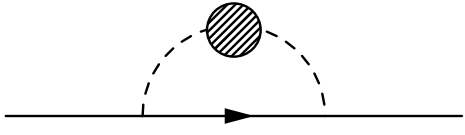}
\end{center}
\caption{Gluon self-energy correction to the one-loop self-energy diagram of a static quark.}  
\label{figself00}
\end{figure}

{\bf (2)} A second momentum region is $|\bk|\sim T$. Since $T \ll 1/r$, under the condition 
$|\bk| \sim T$ we may expand the exponential 
$\displaystyle e^{-i \bk \cdot \br}$ in (\ref{VsQCD}):
\begin{equation}
\delta  V_s(r) =  
\mu^{4-D} \int \frac{d^{d}k}{(2\pi)^{d}} \, \left(1 - \frac{(\bk \cdot \br)^2}{2} + \dots \right) 
\,g^2 \, C_F \, \left[ i \delta {\bf D}_{00}(0,\bk)\right]_{11}\,.
\label{VsQCDmultipole}
\end{equation}
The first term in the expansion corresponds to a mass correction and cancels against twice the 
thermal contribution of the static quark self energy with a gluon self-energy insertion, 
see Fig.~\ref{figself00}. The second term coincides with the expression in Eq.~(\ref{Vs11Teq1}) 
and gives the same result as (\ref{VsTloop}), corresponding to the last three lines of Eq.~\eqref{staticT}.

\begin{figure}[ht]
	\begin{center}
\includegraphics{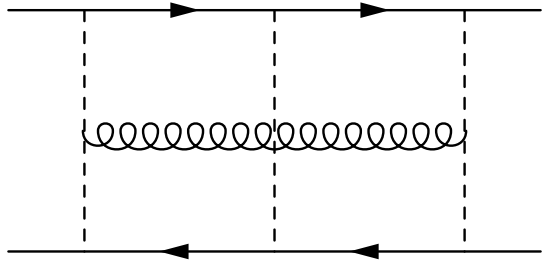}
\end{center}
\caption{The diagram contributing to the divergent term at order $\als^4$ in Coulomb gauge.}  
\label{fig_adm}
\end{figure}

No other diagrams contribute to the thermal part of the potential at order $\als^2$, since bare longitudinal gluons do not have a thermal part and vertex corrections to the quark-gluon vertex vanish at one loop in Coulomb gauge, as it was shown in Sec.~\ref{sec_NRQCDHTL}. As we can see from Eq.~\eqref{staticT}, no terms contribute to the static potential at order $\als^3$ at our accuracy of $m\als^5$, while at order $\als^4$ there is the IR divergent term on the first line of Eq.~\eqref{staticT}, which can be traced back in Coulomb gauge to the diagram shown in Fig.~\ref{fig_adm}, which is also responsible for the divergence at the scale $1/r$ at zero temperature discussed in Sec.~\ref{sub_weakpnrqcd}, first identified in \cite{Appelquist:1977es} and then treated in the context of pNRQCD in \cite{Brambilla:1999qa}.

\section{Details on the evaluation of the transverse HTL contribution}
\label{app_trans}
Our aim is the evaluation of Eq.~\eqref{deftranssym}. Owing to the
symmetries of the retarded and advanced propagators and of the
Bose--Einstein distribution we can restrict the integration in
\eqref{deftranssym} to positive values of $k_0$. We then have
\begin{eqnarray}
\nonumber\delta \Sigma_s^{\rm (trans,\, symm)}(E)&=&g^2C_F\frac{d-1}{d}r^i\mu^{4-D}
\int\frac{\,d^{d}k}{(2\pi)^{d}}\int_0^\infty\frac{\,dk_0k_0^2}{2\pi} \left(\frac{T}{k_0}+\order{\frac ET}\right)\\
&& \hspace{-2.5cm}
\times\left(\Delta_{\rm R}(k) 
- \Delta_{\rm A}(k)\right)\left(\frac{1}{E-h_o^{(0)}-k_0+i\eta}+\frac{1}{E-h_o^{(0)}+k_0+i\eta}\right)r_i\,.
\label{pot}
\end{eqnarray}

Let us define the quantity $\lambda\equiv k_0-\mbk$.
There exist two momentum regions 
that contribute to the integral~\eqref{pot} for $k_0\sim \mbk\sim E-h_o^{(0)}$.
We call the first region the \emph{off-shell region}. It is defined by
\begin{equation}
\label{defhard}
\lambda\sim \left(E-h^{(0)}_o\right)\,,\qquad \mbk\sim \left(E-h^{(0)}_o\right)\,,
\end{equation}
i.e. the region where the gluon is far from being on shell.
The second region is called the \emph{collinear region}. In this region, we have
\begin{equation}
\label{defcoll}
\lambda\sim\frac{m_D^2}{E-h_o^{(0)}}\,,\qquad \mbk\sim \left(E-h^{(0)}_o\right)\,.
\end{equation}
We observe that the collinear scale $m_D^2/\left(E-h^{(0)}_o\right)$ has,
in our energy scale hierarchy, a magnitude in between $mg^4$ and $mg^6$. 
It is, therefore, smaller that the Debye
mass by a factor of $m_D/E\ll 1$ and still larger than the
non-perturbative magnetic mass, which is of order $g^2 T$, by a factor $T/E\gg1$.
For simplicity, we separate the two regions by a cut-off $\Lambda$, such that
\begin{equation}
\left(E-h^{(0)}_o\right)\gg \Lambda\gg\frac{m_D^2}{\left(E-h^{(0)}_o\right)}\,.
\end{equation}

We start by analyzing the off-shell region. Here $k_0^2-\bk^2=\lambda(2\mbk+\lambda)\gg m_D^2$ 
and we can thus expand the retarded propagator propagator in Eq.~\eqref{prophtltrans} as
\begin{equation}
\Delta_\mathrm{R}(k_0>0)=
\frac{i}{k_0^2-\bk^2+i\eta}+\frac{i\frac{m_D^2}{2}\left(\frac{k_0^2}{\bk^2}-(k_0^2-\bk^2)
\frac{k_0}{2\mbk^3}\ln\left(\frac{k_0+\mbk+i\eta}{k_0-\mbk+i\eta}\right)\right)}{(k_0^2-\bk^2+i\eta)^2}
+\mathcal{O}\left(\frac{m_D^4}{E^6}\right).
\end{equation}
Terms contributing to the real part of this propagator and hence to
$\Delta_\mathrm{R}-\Delta_\mathrm{A}$ can come either from the poles
of the denominators, yielding a $\delta(k_0^2-\bk^2)$, or from the
imaginary part of the logarithm. However, $\delta(k_0^2-\bk^2)=0$
over the whole off-shell region. We can safely discard these terms and obtain
\begin{equation}
(\Delta_\mathrm{R}-\Delta_{A})(k_0>0)=-\frac{m_D^2k_0\pi\theta(\mbk-k_0)}{2\mbk^3}\mathrm{P}\frac{1}{k_0^2-\bk^2}.
\end{equation}
Note that the principal value prescription is irrelevant since our integration region excludes the poles. 
From Eq.~\eqref{pot}, we get
\begin{eqnarray}
\nonumber
\delta \Sigma^{\rm (trans,\, symm)}_{s,\mathrm{off \; shell}}(E)&=&
-\frac{g^2C_Fm_D^2T\pi r^i(d-1)}{2d}\int\frac{\,d^{d}k}{(2\pi)^{d}}\frac{1}{\mbk^3}
\\
&&\times\int_0^{k-\Lambda}\! \frac{dk_0}{2\pi}\mathrm{P}\frac{k_0^2}{k_0^2-\bk^2}
\frac{2(E-h_o^{(0)}+i\eta)}{(E-h_o^{(0)}+i\eta)^2-k_0^2}r^i\,.
\end{eqnarray}
This integral does not need to be dimensionally regularized, 
so we can set $D=4$ at this point and obtain 
\begin{equation}
\delta \Sigma^{\rm (trans,\, symm)}_{s,\mathrm{off \; shell}}(E)=-\frac{g^2C_Fm_D^2T r^i}{12\pi}
\int_0^\infty\!\frac{dk_0}{2\pi}\frac{2(E-h_o^{(0)}+i\eta)}{(E-h_o^{(0)}+i\eta)^2-k_0^2}
\left[\ln\frac{2\Lambda}{k_0}+\mathcal{O}\left(\frac{\Lambda}{k_0}\right)\right]r^i\,.
\label{hardtrans}
\end{equation}

We consider, now, the collinear region.
We start again from the retarded propagator introduced in
Eq.~\eqref{prophtltrans}. We perform the change of variables
$k_0-\mbk=\lambda$ and we expand for $\lambda\sim {m_D^2}/{\mbk}\ll \mbk$,
thereby implementing the collinear hierarchy. We then have
\begin{equation}
(\Delta_\mathrm{R}-\Delta_\mathrm{A})(k_0>0)=\Delta_1+\Delta_2+\Delta_3+\Delta_4+\Delta_5+\order{\frac{\lambda}{\mbk^3}},
\end{equation}
where the $\Delta_i$ are defined as
\begin{equation}
\Delta_1=\frac{i}{2\mbk}\left(\frac{1}{\lambda-\frac{m_D^2}{4\mbk}+i\eta}-\frac{1}{\lambda-\frac{m_D^2}{4\mbk}-i\eta}\right),
\end{equation}
\begin{equation}
\Delta_2=\frac{3im_D^4}{64\bk^4}
\left(\frac{1}{(\lambda-\frac{m_D^2}{4\mbk}+i\eta)^2}-\frac{1}{(\lambda-\frac{m_D^2}{4\mbk}-i\eta)^2}\right),
\end{equation}
\begin{equation}
\Delta_3=-\frac{im_D^2}{8\mbk^3}\left(\frac{\ln\left(\frac{2\mbk}{\lambda+i\eta}\right)}
{\lambda-\frac{m_D^2}{4\mbk}+i\eta}-\frac{\ln\left(\frac{2\mbk}{\lambda-i\eta}\right)}{\lambda-\frac{m_D^2}{4\mbk}-i\eta}\right),
\end{equation}
\begin{equation}
\Delta_4=-\frac{im_D^4}{32\bk^4}\left(\frac{\ln\left(\frac{2\mbk}{\lambda+i\eta}\right)}
{(\lambda-\frac{m_D^2}{4\mbk}+i\eta)^2}
-\frac{\ln\left(\frac{2\mbk}{\lambda-i\eta}\right)}{(\lambda-\frac{m_D^2}{4\mbk}-i\eta)^2}\right),
\end{equation}
\begin{equation}
\Delta_5=\frac{im_D^2}{8\mbk^3}\left(\frac{1}{\lambda-\frac{m_D^2}{4\mbk}+i\eta}
-\frac{1}{\lambda-\frac{m_D^2}{4\mbk}-i\eta}\right).
\end{equation}
We start by plugging $\Delta_1$ in Eq.~\eqref{pot}. We then have 
\begin{eqnarray}
\nonumber
\delta \Sigma^{\rm (trans,\, symm)}_{s,1}(E)&=&
\frac{g^2C_F}{6}r^i\int\frac{\,d^3k}{(2\pi)^3}\int_{-\Lambda}^{\Lambda}\! d\lambda(\mbk+\lambda)\frac{T}{k}
\delta\left(\lambda-\frac{m_D^2}{4\mbk}\right)
\\
&& \hspace{5cm}
\times \frac{2(E-h_o^{(0)}+i\eta)}{(E-h_o^{(0)}+i\eta)^2-(\mbk+\lambda)^2}r^i
\nonumber
\\
&=&-i\frac{2}{3}\als\,\cf\,T\,r^i\left(E-h^{(0)}_o\right)^2r^i+\order{\frac{\als T m_D^4r^2}{E^2}}.
\label{tot1}
\end{eqnarray}
The contribution of $\Delta_2$ is 
\begin{eqnarray*}
\delta \Sigma^{\rm (trans,\, symm)}_{s,2}(E)&=&\frac{g^2C_FTr^i}{3}\int\frac{\,d^3k}{(2\pi)^3}
\int_{-\Lambda}^\Lambda\frac{\,d\lambda}{2\pi}\frac{3im_D^4}{32\mbk^3}
\\
&& \hspace{-1.5cm}\times 
\left(\frac{1}{(\lambda-\frac{m_D^2}{4\mbk}+i\eta)^2}-\frac{1}{(\lambda-\frac{m_D^2}{4\mbk}-i\eta)^2}\right) 
\frac{2(E-h_o^{(0)}+i\eta)}{(E-h_o^{(0)}+i\eta)^2-(\mbk+\lambda)^2}r^i\\
&=&\order{\als T m_D^4r^2/E^2};
\end{eqnarray*}
the leading order term in the expansion of $((E-h_o^{(0)}+i\eta)^2-(k+\lambda)^2)^{-1}$, 
which would contribute at order $\als T m_D^2 r^2$, vanishes because the integral over $\lambda$ is zero.
The contribution of $\Delta_3$ is 
\begin{eqnarray*}
\delta \Sigma^{\rm (trans,\, symm)}_{s,3}(E)&=&
-\frac{ig^2C_FTm_D^2r^i}{12}\int\frac{\,d^3k}{(2\pi)^3}
\frac{1}{\bk^2} \int_{-\Lambda}^\Lambda\frac{\,d\lambda}{2\pi}
\frac{2(E-h_o^{(0)}+i\eta)}{(E-h_o^{(0)}+i\eta)^2-(\mbk+\lambda)^2} \\
&&\times
\left[\ln\left\vert\frac{2\mbk}{\lambda}\right\vert\left(-2i\pi\delta
\left(\lambda-\frac{m_D^2}{4\mbk}\right)\right)
-2i\pi\theta(-\lambda)\mathrm{P}\frac{1}{\lambda-\frac{m_D^2}{4\mbk}}\right]r^i\,.
\end{eqnarray*}
We then have
\begin{equation}
\delta \Sigma^{\rm (trans,\, symm)}_{s,3}(E)=
-\frac{g^2 C_F Tm_D^2r^i}{12\pi}\int_0^\infty\! \frac{dk}{2\pi} \frac{2(E-h_o^{(0)}+i\eta)}{(E-h_o^{(0)}+i\eta)^2-\bk^2}
\left[\ln\left(\frac{2\mbk}{\Lambda}\right)+...\right]r^i\,,
\end{equation}
where the dots mean terms suppressed by ${1}/{\Lambda}$. 
We now combine this result with the contribution from the off-shell region in Eq.~\eqref{hardtrans} to obtain
\begin{align}
\nonumber\delta \Sigma^{\rm (trans,\, symm)}_{s,\mathrm{off \; shell}}(E)+\delta \Sigma^{\rm (trans,\, symm)}_{s,3}(E)=&
\frac{\als \cf Tm_D^2r^i}{3\pi}\ln4\int_0^\infty\! dk\frac{-(E-h_o^{(0)}+i\eta)}{(E-h_o^{(0)}+i\eta)^2-\bk^2}r^i \\
=&i\frac{\als \cf Tm_D^2r^2}{3}\ln2+\ldots\,,
\label{c+hard}
\end{align}
where the dots stand for higher orders. We remark that the dependence on 
the cut-off scale $\Lambda$ has disappeared.
The contribution of $\Delta_4$ is
\begin{eqnarray*}
&& \hspace{-0.5cm}\delta \Sigma^{\rm (trans,\, symm)}_{s,4}(E)=
-\frac{ig^2Tm_D^4C_Fr^i}{48}\int\frac{\,d^3k}{(2\pi)^3}\frac{1}{\mbk^3}
\int_{-\Lambda}^\Lambda\frac{\,d\lambda}{2\pi}\frac{2(E-h_o^{(0)}+i\eta)}{(E-h_o^{(0)}+i\eta)^2-(\mbk+\lambda)^2}
\\
&& \hspace{3cm}  \times 
\left[\left(\frac{\ln\left\vert\frac{2\mbk}{\lambda}\right\vert}
{(\lambda-\frac{m_D^2}{4\mbk}+i\eta)^2}-\frac{\ln\left\vert\frac{2\mbk}{\lambda}\right\vert}
{(\lambda-\frac{m_D^2}{4\mbk}-i\eta)^2}\right)
-\frac{2i\pi\theta(-\lambda)}{(\lambda-\frac{m_D^2}{4\mbk}-i\eta)^2}\right]r^i\,.
\end{eqnarray*}
The needed $\lambda$ integrals are 
\begin{eqnarray*}
\int_{-\Lambda}^\Lambda\frac{\,d\lambda}{2\pi}
\ln\left\vert\frac{2\mbk}{\lambda}\right\vert\left(\frac{1}{(\lambda-\frac{m_D^2}{4\mbk}+i\eta)^2}
-\frac{1}{(\lambda-\frac{m_D^2}{4\mbk}-i\eta)^2}\right)=i\frac{4\mbk}{m_D^2}\,,
\end{eqnarray*}
and
\begin{equation*}
-i\int_{-\Lambda}^\Lambda\,d\lambda\frac{\theta(-\lambda)}{(\lambda-\frac{m_D^2}{4\mbk}-i\eta)^2}=
-\frac{i4\mbk}{m_D^2}+...\,,
\end{equation*}
so that $\delta \Sigma^{\rm (trans,\, symm)}_{s,4}(E)$ has only contributions that are suppressed 
by powers of ${1}/{\Lambda}$.

Finally, the contribution of $\Delta_5$ is 
\begin{eqnarray}
\nonumber
\delta \Sigma^{\rm (trans,\, symm)}_{s,5}(E)&=&
\frac{g^2C_Fr^i(d-1)}{2d}
\int\frac{\,d^{d}k}{(2\pi)^{d}}\int_{-\Lambda}^\Lambda\frac{\,d\lambda}{2\pi}
\frac{T\pi m_D^2}{2\bk^2}\delta(\lambda-\frac{m_D^2}{4\mbk})
\\
\nonumber
&& \hspace{5cm}
\times \frac{2(E-h_o^{(0)}+i\eta)}{(E-h_o^{(0)}+i\eta)^2-(\mbk+\lambda)^2}r^i 
\\
&=&-i\frac{\als \cf Tm_D^2r^2}{6} +\ldots\,,
\label{finale}
\end{eqnarray}
where the dots stand for higher orders.
The contribution 
of the symmetric part of the transverse propagator is then given 
by the sum of Eqs.~\eqref{tot1},~\eqref{c+hard} and~\eqref{finale}.

	\section{The thermal width in pNRQCD and its relation with the gluo-dissociation cross-section}
\label{peskin}
In Chapter~\ref{chap_rggT} we computed the thermal width for $m\als\gg T\gg E\gg m_D$. It is given by Eq.~\eqref{finalwidth}. As we mentioned in Sec.~\ref{sub_trans}, the first two lines of that equation, which are also the leading ones, are caused by the process of singlet-to-octet thermal decay, where the bound singlet becomes a colour octet after absorbing a sufficiently energetic thermal gluon. As we discussed there, an altogether similar process has been considered in the literature under the name of \emph{gluo-dissociation} \cite{Kharzeev:1994pz,Xu:1995eb,Rapp:2008tf,Park:2007zza,Zhao:2010nk}. In this case the thermal width is obtained by convoluting the cross section for $g+\Phi(1S)\to (Q\overline{Q})_8$ in the vacuum, where $\Phi(1S)$ is a $1S$ quarkonium state, computed by Bhanot and Peskin (BP) \cite{Peskin:1979va,Bhanot:1979vb} in 1979, with a thermal distribution for the gluon. 
In detail one has (see for instance Eq. (23) of \cite{Rapp:2008tf})
\begin{equation}
	\label{gluodiss}
	\Gamma_{gd}=\int\frac{d^3q}{(2\pi)^3}n_\mathrm{B}(q)\,\sigma_\mathrm{BP}(q)\,
\end{equation}
where we have assumed the bound state to be at rest in the preferred frame where the bath is at rest and $\sigma_\mathrm{BP}(k)$ is the  $g+\Phi(1S)\to (Q\overline{Q})_8$ BP cross section as a function of the gluon momentum $q$. It reads
\begin{equation}
	\label{sigmapeskin}
	\sigma_\mathrm{BP}(q)=\frac{2}{3}\pi\left(\frac{32}{3}\right)^2\left(\frac{16\pi}{3g^2}\right)\frac{1}{m^2}\frac{(\mbq/\epsilon_0-1)^{3/2}}{(\mbq/\epsilon_0)^5},
\end{equation}
where $\epsilon_0$ is the absolute value of the binding energy of the $1S$ ground state. As we remarked in Chap.~\ref{chap_rggT}, this cross section was computed in an operator product expansion, treating the gluon-quarkonium interaction at leading order as a chromoelectric dipole interaction, which corresponds to our EFT treatment, but neglecting the (repulsive) octet potential, which is tantamount to neglecting final state interactions. This has been achieved in a large-$\nc$ limit, where $V_s(r)\to-(\nc/2)\als/r$ and the octet potential, which is proportional to $1/(2\nc)$, vanishes. As a result, after taking this limit and then reinstating $\nc=3$, the Bohr radius changes to $\tilde{a}_0=4/(m\nc\als)=4/(3m\als)$ and the absolute value of the binding energy is $\epsilon_0=1/(m\tilde{a}_0^2)=9/16 m\als^2$. The cross section can then be rewritten as
\begin{equation}
	\label{peskinrewrite}
	\sigma_\mathrm{BP}(q)=\frac{2^9\pi\als}{9}\frac{\epsilon_0^{5/2}}{m}\frac{(\mbq-\epsilon_0)^{3/2}}{\mbq^5}.
\end{equation}
As we know the kinetic term and the octet potential in the octet Hamiltonian $h_o^{(0)}$ have the same size $m\als^2$ in our power counting. As such, neglecting the octet potential is in contrast with the power counting of a Coulombic bound state.  We will now show how our EFT formalism is analogous to Eq.~\eqref{gluodiss}; we will see how, neglecting the octet potential, we reobtain the BP cross section and finally we will perform the calculation with the entire Coulomb Hamiltonian $h_o^{(0)}$. These results will appear in a forthcoming publication \cite{uspeskin}.

In Sec.~\ref{sub_trans} we have obtained the singlet-to-octet contribution to the width in the case $m\als\gg T\gg E\gg m_D$. Let us now relax the hierarchy a bit, in that we do not assume the temperature to be much larger than the energy, i.e. $T\sim E\gg m_D$. The singlet-to-octet thermal width is still obtained from Eqs.~\eqref{deftransE} and \eqref{deftranssym}, but now we do not expand the Bose--Einstein distribution for $T\gg E$. We then have
\begin{equation}
	\label{widthdef}
\delta \Sigma_s^{(\mathrm{trans,\,symm})}(E) =	- i g^2 \, C_F \, \frac{2}{3}r^i
	 \!\! \int \!\! \frac{d^4k}{(2\pi)^4}
	\frac{i}{E-h^{(0)}_o-k_0 +i\eta}k_0^2\, 2\pi n_{\rm B}(\vert k_0\vert)\delta(k_0^2-\bk^2)\,r^i\,,
\end{equation}
where the bare propagators are again used for gluons, coherently with our hierarchy, and the number of dimensions has been set to 4, the integral being convergent. Evaluating it yields
\begin{equation}
	\label{ampliwidth}
	\delta \Sigma_s^{(\mathrm{trans,\,symm})}(E) =-i\frac{g^2\cf}{6\pi}r^i\left\vert E-h_o^{(0)}\right\vert^3n_\mathrm{B}(\vert E-h_o^{(0)}\vert)r^i\,,
\end{equation}
so that the width for the $1S$ state reads
\begin{equation}
	\label{width}
	\Gamma_{1S}=\frac{g^2\cf}{3\pi}\langle1S\vert r^i\left\vert E-h_o^{(0)}\right\vert^3n_\mathrm{B}(\vert E-h_o^{(0)}\vert)r^i\vert 1S\rangle\,,
\end{equation}
where $\vert 1S\rangle=1/\sqrt{\pi}a_0^{-3/2}\exp(r/a_0)$ is the Coulomb $1S$ wavefunction. The difficulty in the evaluation stems from the Bose--Einstein distribution and its nontrivial dependence on $h_o^{(0)}$. If we had expanded it for $(E-h_o^{(0)})\ll T$, as in Chapter~\ref{chap_rggT}, we would have obtained the simpler matrix elements of the first two terms in Eq.~\eqref{tottrans}.\\
The matrix element in Eq.~\eqref{width} can be evaluated analogously to how the QCD Bethe logarithms discussed in Sec.~\ref{secE} are dealt with in \cite{Kniehl:1999ud,Kniehl:2002br}, i.e. a by introducing a complete set of octet states, which are the only ones that can contribute for colour reasons. If we label them $\vert \bp\rangle$ according to their energy $\mbp^2/m$ and require that they obey
\begin{equation}
	\label{defnormp}
	\int\frac{d^3p}{(2\pi)^3}\langle\bx\vert\bp\rangle\langle\bp\vert\by\rangle=\delta^3(\bx-\by)\,,
\end{equation}
we have
\begin{equation}
		\label{widthinsert}
		\Gamma_{1S}=\frac{g^2\cf}{3\pi}\int\frac{d^3p}{(2\pi)^3}\left\vert\langle1S\vert r^i\vert \bp\rangle\right\vert^2 \left( \frac{\mbp^2}{m}-E_1\right)^3\,n_\mathrm{B}\left(\frac{\mbp^2}{m}-E_1\right),
	\end{equation}
where we have used the fact that the continuum octet states obey $h_o^{(0)}\vert\bp\rangle=\mbp^2/m\vert\bp\rangle$ in the presence of the potential as well. We have also used that $(E-h_o^{(0)})^3n_\mathrm{B}(E-h_o^{(0)})$ is analytic in $h_o^{(0)}$.\\
In order to first reproduce the gluo-dissociation approach with the BP cross section, let us evaluate the dipole matrix element squared $\left\vert\langle1S\vert r^i\vert \bp\rangle\right\vert^2$ in the absence of the octet potential. This is tantamount to using simple plane waves for $\vert\bp\rangle$, which trivially satisfy Eq.~\eqref{defnormp}. The matrix element squared then reads
\begin{equation}
	\label{peskinmatrix}
	\left\vert\langle1S\vert r^i\vert \bp\rangle\right\vert^2_{\nc\to\infty}=\frac{2^{10}\pi \tilde{a}_0^7\mbp^2}{(1+\tilde{a}_0^2\mbp^2)^6},
\end{equation}
which agrees with BP. Plugging this in Eq.~\eqref{widthinsert} and performing a change of integration variable from $p$ to $q=\mbp^2/m-E_1$, together with the large-$\nc$ limit $E_1\to-\epsilon_0$, we obtain
\begin{equation}
	\label{peskinEFT}
		\Gamma_{1S,\nc\to\infty}=\frac{g^2\cf}{3\pi}\int_{\epsilon_0}^\infty\frac{d^3q}{(2\pi)^3}\frac{2^{10}\pi \tilde{a}_0^7m^{5/2}(\mbq-\epsilon_0)^{3/2}\mbq}{2(1+\tilde{a}_0^2m(\mbq-\epsilon_0))^6}n_\mathrm{B}(\mbq)\,,
\end{equation} 
where $\epsilon_0$ and $\infty$ are the extremes for the $q^2dq$ integration. Comparing this Equation with Eq.~\eqref{gluodiss} and substituting $\cf=4/3$ we can readily identify the cross section, which reads
\begin{equation}
	\label{eftcrosslargenc}
	\sigma_{\nc\to\infty}=16\frac{2^9\pi\als}{9}\frac{\epsilon_0^{5/2}}{m}\frac{(\mbq-\epsilon_0)^{3/2}}{\mbq^5}.
\end{equation}
The discrepancy with the Peskin cross section is easily identified in a multiplicative factor of 16. The Peskin cross section for $g+\Phi(1S)\to(Q\overline{Q})_8$ is averaged over the polarization and colour of the initial gluon, while in our case we naturally sum over all possible colours (8) and polarizations (2). This factor is explicitly included in Eq. (4) of \cite{Park:2007zza} and also the authors of \cite{Zhao:2010nk} multiply the BP cross section by 16 in Eq.~\eqref{gluodiss}.\footnote{Private communications from Miguel Angel Escobedo and Xingbo Zhao are acknowledged.} Multiplicative prefactors aside, we then see how our EFT computation in the large-$\nc$ limit naturally includes the Peskin cross section and reproduces the results of the gluo-dissociation analysis. We now set out to do the calculation with the octet potential, thereby also quantifying the approximation introduced by neglecting it.

\begin{figure}
	\begin{center}
		\includegraphics{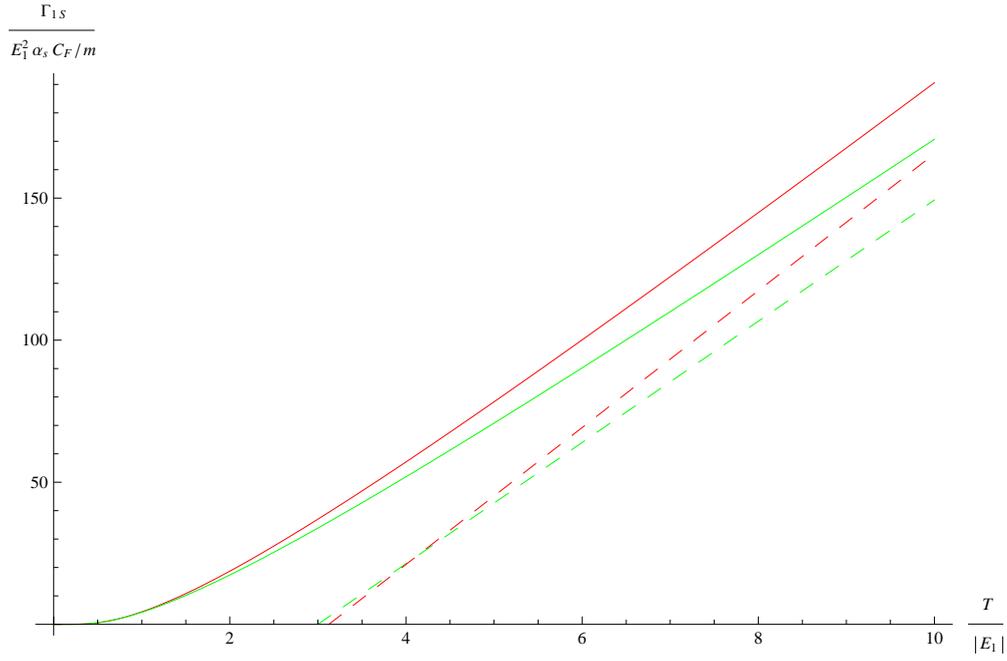}
	\end{center}
	\caption{The width $\Gamma_{1S}$ from a numerical integration of Eq.~\eqref{widthEFT} is shown in red and the corresponding $T\gg \vert E_1\vert$ analytical result, coming from Eq.~\eqref{widthanalytic} with $\rho=1/8$ is plotted in dashed red. Similarly the width $\Gamma_{1S\,\nc\to\infty}$ in the BP approximation from a numerical integration of Eq.~\eqref{peskinEFT} is plotted in green and its analytical counterpart, Eq.~\eqref{widthanalytic} with $\rho=0$, in dashed green.}
	\label{fig_peskin}
\end{figure}

\begin{figure}[ht]
	\begin{center}
		\includegraphics{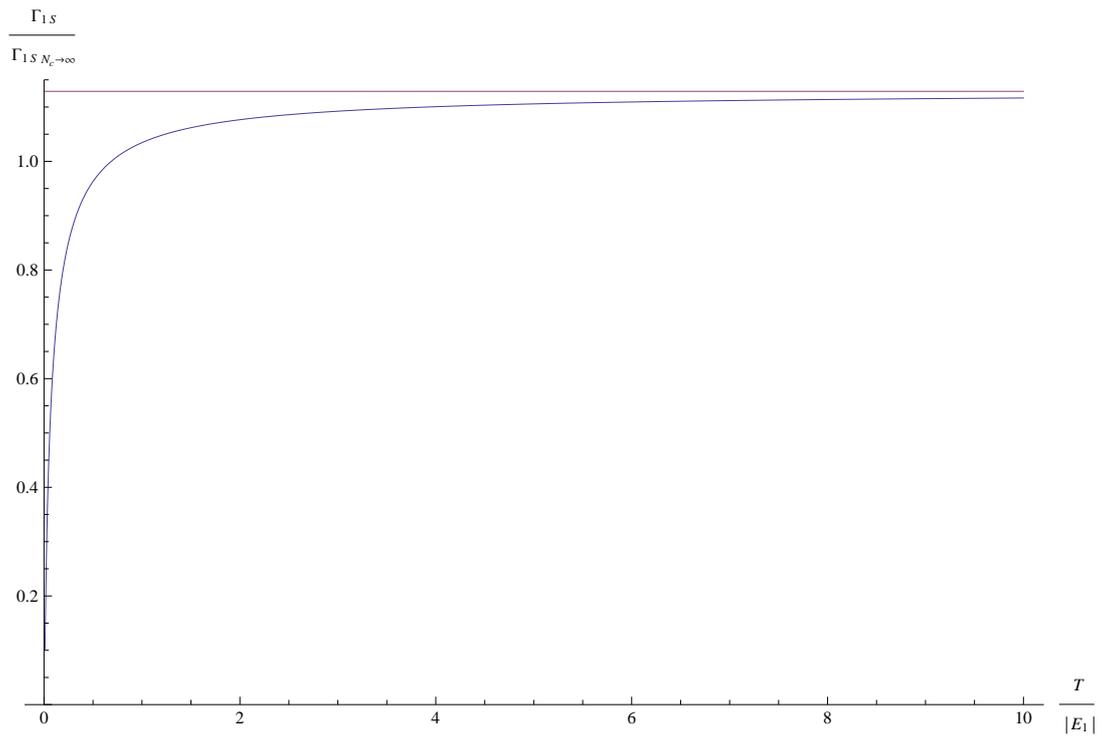}
	\end{center}
	\caption{The ratio $\Gamma_{1S}/\Gamma_{1S\,\nc\to\infty}$ is plotted, as obtained from from numerical integrations of Eqs.~\eqref{widthEFT} and \eqref{peskinEFT}. The horizontal line is the asymptotic limit $289/256$.}
	\label{fig_ratio}
\end{figure}

The calculation of the dipole matrix element squared in this case is more involved, and requires the explicit integration over the continuum octet wavefunction. Coulombic wavefunction in the continuum region can be found in \cite{hydrocont,abramowitz+stegun}. The same integration has been performed in \cite{Kniehl:1999ud,Kniehl:2002br} for the QCD Bethe logarithms and we have used it for the evaluation of Eq.~\eqref{defin}. The dipole introduces a $\Delta l=1$ selection rule, so that only octet $P$ waves contribute and the matrix element squared reads \cite{Kniehl:1999ud,Kniehl:2002br}
\begin{equation}
	\label{penin}
	\left\vert\langle1S\vert r^i\vert \bp\rangle\right\vert^2=\frac{512 \pi ^2 \rho  (\rho +2)^2a_0^6\mbp \left(1+\frac{\rho ^2}{a_0^2 \mbp^2}\right)
	   e^{\frac{4 \rho}{a_0 \mbp}  \tan ^{-1}(a_0 \mbp)}}{
	   \left(e^{\frac{2 \pi  \rho }{a_0 \mbp}}-1\right) \left(1+a_0^2\mbp^2\right)^6},
\end{equation}
where $\rho\equiv1/(2\nc\cf)=1/(\nc^2-1)$. It is easy to see that the $\nc\to\infty$ ($\rho\to0$) limit of this Equation gives back Eq.~\eqref{peskinmatrix}. Plugging this matrix element in Eq.~\eqref{widthinsert} and performing again the change of variables $p\to q=\mbp^2/m-E_1$ yields
\begin{eqnarray}
	\Gamma_{1S}&=&\frac{g^2\cf}{3\pi}\int_{\vert E_1\vert}^\infty \frac{d^3q}{(2\pi)^3}2^8 \pi ^2  \rho  (\rho +2)^2 \frac{E_1^{4}}{m\mbq^5}
	    \left(\frac{\mbq}{\vert E_1\vert}+\rho ^2-1	\right)\left( e^{2 \pi  \rho 
		   \sqrt{\frac{\vert E_1\vert}{\mbq-\vert E_1\vert}}}-1\right)^{-1}\nn\\
	&& \hspace{2.8cm}\times\exp\left(4 \rho  \sqrt{\frac{\vert E_1\vert}{\mbq-\vert E_1\vert}} \tan
	   ^{-1}\left(\sqrt{\frac{\mbq}{\vert E_1\vert}- 1}\right)\right)n_\mathrm{B}(\mbq)\,.
	\label{widthEFT}
\end{eqnarray}
From this expression one can extract the corresponding cross section, which reads
\begin{equation}
	\sigma=\frac{\als\cf}{3} 2^{10} \pi^2  \rho  (\rho +2)^2 \frac{E_1^{4}}{m\mbq^5}
	    \left(\frac{\mbq}{\vert E_1\vert}+\rho ^2-1	\right)\frac{\exp\left(4 \rho  \sqrt{\frac{\vert E_1\vert}{\mbq-\vert E_1\vert}} \tan
	   ^{-1}\left(\sqrt{\frac{\mbq}{\vert E_1\vert}- 1}\right)\right)}{ e^{2 \pi  \rho 
		   \sqrt{\frac{\vert E_1\vert}{\mbq-\vert E_1\vert}}}-1}.
	\label{crossEFT}
\end{equation}
The limit $\rho\to0$ gives back Eq.~\eqref{peskinEFT} as expected. In order to estimate the approximation introduced by ignoring the octet potential, in Fig.~\ref{fig_peskin} we plot the widths $\Gamma_{1S}$ and $\Gamma_{1S\,\nc\to\infty}$ as a function of the temperature as obtained by numerical integrations of Eqs.~\eqref{widthEFT} and \eqref{peskinEFT}. In the latter case we perform the substitutions $\tilde{a}_0\to a_0$ and $\epsilon_0\to\vert E_1\vert$ for a meaningful comparison. $\Gamma_{1S}$ is the continuous red line, whereas $\Gamma_{1S\,\nc\to\infty}$ is the continuous green line. We also plot the analytical result obtained in Sec.~\ref{sub_trans} by the $q\ll T$ expansion of the Bose--Einstein distribution, which can be obtained from the first two lines of Eq.~\eqref{finalwidth} with $n=1$, $l=0$. It reads
\begin{equation}
	\label{widthanalytic}
	\frac{\Gamma_{1S}}{E_1^2\cf\als/m}=\frac{16}{3}\left[(2+\rho)^2\frac{T}{\vert E_1\vert}-(2+\rho)^2(3+\rho)\right]+\order{\frac{\vert E_1\vert}{T}}.
\end{equation}
The value for $\nc=3$ ($\rho=1/8$), corresponding to the inclusion of the octet potential and therefore to the analytical results of Chapter~\ref{chap_rggT} and to the expansion of Eq.~\eqref{widthEFT}, is plotted as a dashed red line, whereas the value for $\rho=0$, corresponding to the BP approximation ($\nc\to\infty$) is plotted in dashed green. Both continuous curves approach the linear regime predicted by our $T\gg \vert E_1\vert$ expansion starting from $T\sim5\vert E_1\vert$, while the ratio $\Gamma_{1S}/\Gamma_{1S\,\nc\to\infty}$ approaches for increasing temperatures the asymptotic value $(2+\rho)^2/4$, which for $\nc=3$ is $289/256\approx1.13$. 
For greater clarity, in Figure~\ref{fig_ratio} we plot the ratio $\Gamma_{1S}/\Gamma_{1S\,\nc\to\infty}$, which shows clearly how the deviation from the asymptotic limit of $289/256$ is very small starting already from $T\sim 2\vert E_1\vert$, where both widths start to become significant. We then see how the error introduced by neglecting the octet potential  is of the order of $10\%$ in this region.

	\chapter{Details on the imaginary-time calculations}
	\label{app_imtime}
	In this Appendix we report some technical details on the calculations that lead to the results of Part~\ref{part_imtime}.
\section{The gluon self energy in the static gauge\label{app_pi00}}
We proceed to the computation of the Matsubara sums in
Eq.~\eqref{defmaster} in order to obtain Eqs.~\eqref{pi00vacuum},
\eqref{pi00matter},~\eqref{pi00zero} and~\eqref{pi00sing}. 
We recall the two basic bosonic Matsubara sums \cite{Kapusta:2006pm}
\begin{eqnarray}
T\sum_{n=-\infty}^{+\infty}\frac{1}{\bp^2+\omega_n^2}&=&\frac{1+2n_\mathrm{B}(\mbp)}{2\mbp},
\label{matsup2}
\\
T\sum_{n=-\infty}^{+\infty}\frac{1}{(\bp^2+\omega_n^2)(\bq^2+\omega_n^2)}&=&\frac{1}{2
  \mbp \mbq}\left(	
\frac{1+n_{\mathrm{B}}(\mbp)+n_{\mathrm{B}}(\mbq)}{\mbp+\mbq}
+\frac{n_{\mathrm{B}}(\mbq)-n_{\mathrm{B}}(\mbp)}{\mbp-\mbq}\right),
\nonumber\\
\label{matsup2q2}
\end{eqnarray}
where $n_\mathrm{B}$ is the Bose--Einstein distribution and the first sum is simply the mixed representation~\eqref{matsumixed} for $\tau=0$. 
Since the sums include also the zero mode, in 
evaluating the master sum integrals defined in
Eqs.~\eqref{defmaster} and~\eqref{shorthand} 
we will have to subtract it. 
Furthermore, we identify the temperature-independent part (the unity) in the numerators
on the r.h.s of Eqs.~\eqref{matsup2} and~\eqref{matsup2q2}
as the vacuum part and the part proportional to the thermal distributions 
as the matter part.

For $I_0$, we have
\begin{equation}
\label{sumi0}
I_0=\int_p^\prime\frac{1}{p^2}=\m2 \int\frac{d^dp}{(2\pi)^d}\left(\frac{1+2n_\mathrm{B}(\mbp)}{2\mbp}-\frac{T}{\bp^2}\right)
=\frac{T^2}{12};
\end{equation}
the subtracted zero mode along with the vacuum part vanish in dimensional regularization. 

For $I_1$, we have ($\bq = \bk -\bp$)
\begin{eqnarray*}
I_1&=&\m2 \int\frac{d^dp}{(2\pi)^d}\left[\frac{\mbp}{2\mbq}\left(	
\frac{1+n_{\mathrm{B}}(\mbp)+n_{\mathrm{B}}(\mbq)}{\mbp+\mbq}
+\frac{n_{\mathrm{B}}(\mbq)-n_{\mathrm{B}}(\mbp)}{\mbp-\mbq}\right)-\frac{T}{\bq^2}\right]
\\
&=&\m2 \int\frac{d^dp}{(2\pi)^d}\left[\frac{\bp^2}{2\mbp\mbq(\mbp+\mbq)}
+\frac{\mbp n_{\mathrm{B}}(\mbp)}{2  \mbq}\left(\frac{-2\mbq}{\bp^2-\bq^2}\right)
\right.
\\
&&
\hspace{5.6cm}
\left. +\frac{|\bq'|n_{\mathrm{B}}(\mbp)}{2 \mbp }\left(\frac{-2|\bq'|}{\bp^2-\bq'^2}\right)-\frac{T}{\bq^2}\right],
\end{eqnarray*}
where we have operated a shift $\bp\to \bq'= \bp + \bk$, $\bq \to -\bp$ in some terms of the matter part. 
The vacuum part can be brought into a more standard form by noting that
\begin{equation}
\label{3to4}
\int_{-\infty}^{+\infty}\frac{dp_0}{2\pi}\frac{1}{(\bp^2+p_0^2)(\bq^2+p_0^2)}=\frac{1}{2\mbp\mbq(\mbp+\mbq)}.
\end{equation}
This allows to write the three-dimensional integral as a standard Euclidean four-dimen\-sional integral,  
which can be computed with the formulas listed in appendix~\ref{sub_oneloop} setting  
$d+1=4-2\epsilon$. We thus have
\begin{equation}
\label{i1vacuum}
(I_1)_\mathrm{vac}=
\m2 \int\frac{d^{d+1}p}{(2\pi)^{d+1}}\frac{p^\mu p^\nu (\delta_{\mu\nu}-\delta_{\mu0}\delta_{\nu0})}{p^2q^2}
=(\delta_{\mu\nu}-\delta_{\mu0}\delta_{\nu0})\m2 L_{d+1}^{\mu\nu}(k,1,1)\vert_{k^0=0}.
\end{equation}
The zero-mode integral vanishes in dimensional regularization, 
whereas the remaining matter part is finite and gives 
\begin{eqnarray}
(I_1)_{\mathrm{mat}}
&=&\frac{1}{2\pi^2}\int_0^\infty d\mbp\,\mbp n_{\mathrm{B}}(\mbp)\left(1+\frac{\mbp}{2\mbk}
\ln\left\vert\frac{\mbk+2\mbp}{\mbk-2\mbp}\right\vert\right).
\label{i1matter}
\end{eqnarray}

Analogously, we have for $I_2$
\begin{equation*}
I_2=\bk^2\m2 \int\frac{d^dp}{(2\pi)^d}\left[\frac{1}{2\mbp\mbq}\left(	
\frac{1+n_{\mathrm{B}}(\mbp)+n_{\mathrm{B}}(\mbq)}{\mbp+\mbq}
+\frac{n_{\mathrm{B}}(\mbq)-n_{\mathrm{B}}(\mbp)}{\mbp-\mbq}\right)-\frac{T}{\bp^2\bq^2}\right].
\end{equation*}
The vacuum part is 
\begin{equation}
(I_2)_\mathrm{vac}=\m2 \int\frac{d^{d+1}p}{(2\pi)^{d+1}}\frac{\bk^2}{p^2q^2}=\bk^2 \m2 L_{d+1}(k,1,1)\vert_{k^0=0},
\label{i2vacuum}
\end{equation}
the matter part is
\begin{equation}
(I_2)_\mathrm{mat}=
\frac{1}{2\pi^2}\left(\int_0^\infty d\mbp\, n_{\mathrm{B}}(\mbp)
\frac{\mbk}{2}\ln\left\vert\frac{\mbk+2\mbp}{\mbk-2\mbp}\right\vert\right), 
\label{i2mat}
\end{equation}
and the subtracted zero-mode part is 
\begin{equation}
(I_2)_\mathrm{zero}= - \m2 \int\frac{d^dp}{(2\pi)^d}\frac{T\bk^2}{\bp^2\bq^2},
\label{i2zero}
\end{equation}
which has been kept in dimensional regularization.

We consider now $I_3$:
\begin{eqnarray}
I_3&=&\bk^2\m2 \int\frac{d^dp}{(2\pi)^d}\left[\frac{1+2n_{\mathrm{B}}(\mbp)}{2\mbp^3}-\frac{T}{\bp^4}\right],
\\
\label{i3vac}
(I_3)_\mathrm{vac}&=&\bk^2\m2 \int\frac{d^dp}{(2\pi)^d}\frac{1}{2\mbp^3}=0,
\\
\label{i3mat}
(I_3)_\mathrm{mat}&=& \frac{1}{2\pi^2}\int_0^\infty d\mbp\,\mbp n_{\mathrm{B}}(\mbp)\frac{\bk^2}{\bp^2}.
\end{eqnarray}
In dimensional regularization the subtracted zero mode vanishes. 
The matter part is infrared divergent.
Since this divergence will cancel against terms from $I_4$ in the sum~\eqref{defp00}, 
we present the result directly in the three-dimensional limit.

$I_4$ is given by
\begin{equation}
I_4
=I_4^a-I_4^b-I_4^c
=\int_p^\prime\frac{\bk^4}{\bp^2\bq^2\omega_n^2}-\int_p^\prime\frac{\bk^4}{p^2q^2\bp^2}-\int_p^\prime\frac{\bk^4}{\bp^2\bq^2q^2}.
\label{i4rewrite}
\end{equation}
$I_4^a$ is
\begin{equation}
\label{i4a}
I_4^a=\frac{2T}{(2\pi T)^2}\frac{\bk^4}{8\mbk}\sum_{n=1}^\infty\frac{1}{n^2}=\frac{\mbk^3}{96T},
\end{equation}
which is a term peculiar to this gauge; it is singular in the $T\to0$ limit and
constitutes $\Pi_{00}^\mathrm{NS}(\bk)_{\mathrm{sing}}$. 
$I_4^b$ is
\begin{equation}
\label{i4b}
I_4^b=\bk^4\m2 \int\frac{d^dp}{(2\pi)^d}\left[\frac{1}{2\mbp^3\mbq}\left(
\frac{1+n_{\mathrm{B}}(\mbp)+n_{\mathrm{B}}(\mbq)}{\mbp+\mbq}+\frac{n_{\mathrm{B}}(\mbq)
-n_{\mathrm{B}}(\mbp)}{\mbp-\mbq}\right)-\frac{T}{\bp^4\bq^2}\right].
\end{equation}
The vacuum part can be brought into a more familiar form by adding and subtracting $1/(2\mbp^3\bq^2)$
\begin{eqnarray}
(I_4^b)_{\mathrm{vac}}&=&
\bk^4\m2 \int\frac{d^dp}{(2\pi)^d}\left[\frac{1}{2\mbp^3\mbq(\mbp+\mbq)}
-\frac{1}{2\mbp^3\mbq^2}\right]+\bk^4\m2 \int\frac{d^dp}{(2\pi)^d}\frac{1}{2\mbp^3\mbq^2}
\nonumber\\
&=&-\bk^2\frac{\ln2}{2\pi^2}+\frac{\bk^4}{2} \m2 L_d(\bk,3/2,1).
\label{i4bvac}
\end{eqnarray}
Although the matter part of $I_4$ is infrared divergent, its infrared divergence cancels against 
the matter part of $I_3$, i.e. Eq.~\eqref{i3mat},  in the sum~\eqref{defp00}.
Hence, we may evaluate it directly in three dimensions. In contrast, we will keep regularized 
the subtracted zero modes. As discussed in the main text, 
these subtracted zero modes behave like $\epsilon \mbk^{1-2\epsilon}$ and are going to contribute 
when evaluating the Fourier transform of $\mbk^{1-2\epsilon}/\mbk^4$ in the Polyakov-loop 
correlator calculation, like in Eq.~\eqref{defselfenergyinsertr}.
Therefore, $(I_4^b)_{\mathrm{mat}}$ and $(I_4^b)_{\mathrm{zero}}$ read
\begin{eqnarray}
(I_4^b)_{\mathrm{mat}}&=&
\frac{1}{2\pi^2}\int_0^\infty d\mbp\,\mbp n_{\mathrm{B}}(\mbp)
\frac{\mbk^3 }{2\mbp^3}\left[\ln\left\vert\frac{\mbk+2\mbp}{\mbk-2\mbp}\right\vert 
+\ln\left\vert\frac{\mbk-\mbp}{\mbk+\mbp}\right\vert\right],
\label{i4bmat}\\
(I_4^b)_{\mathrm{zero}}&=&
-\m2 \int\frac{d^dp}{(2\pi)^d}\frac{T\bk^4}{\bp^4\bq^2}.
\label{i4bzero}
\end{eqnarray}
Similarly $I_4^c$ reads
\begin{equation}
I_4^c=\bk^4\m2 \int\frac{d^dp}{(2\pi)^d}\left[\frac{1+2n_{\mathrm{B}}(\mbq)}{2\bp^2\mbq^3}-\frac{T}{\bp^2\bq^4}\right],
\label{i4c}
\end{equation}
which can be decomposed as
\begin{eqnarray}
(I_4^c)_{\mathrm{vac}}&=&\bk^4\m2 \int\frac{d^dp}{(2\pi)^d}\frac{1}{2\mbp^2\mbq^3}=\frac{\bk^4}{2} \m2 L_d(\bk,1,3/2),
\label{i4cvac}\\
(I_4^c)_{\mathrm{mat}}&=&
\frac{1}{2\pi^2}\int_0^\infty d\mbp\,\mbp n_{\mathrm{B}}(\mbp)\frac{\mbk^3}{2\mbp^3}
\ln\left\vert\frac{\mbk+\mbp}{\mbk-\mbp}\right\vert, 
\label{i4cmat}
\\
(I_4^c)_{\mathrm{zero}}&=& -\m2 \int\frac{d^dp}{(2\pi)^d}\frac{T\bk^4}{\bp^2\bq^4}.
\label{i4zero}
\end{eqnarray}
Notice that, as we anticipated, 
the sum $(I_3)_\mathrm{mat}/2-(I_4^b)_\mathrm{mat}/4-(I_4^c)_\mathrm{mat}/4$, 
which is the combination appearing in $\Pi_{00}^{\mathrm{NS}}(\bk)$, is infrared finite.
It is also worthwhile noticing that the vacuum parts $(I_4^b)_{\mathrm{vac}}$ and 
$(I_4^c)_{\mathrm{vac}}$ are infrared divergent, but that in the sum 
$(I_3)_\mathrm{vac}/2-(I_4^b)_\mathrm{vac}/4-(I_4^c)_\mathrm{vac}/4$,
these infrared divergences are canceled and replaced by an ultraviolet 
divergence eventually removed by renormalization. The canceling infrared divergence and the 
remaining ultraviolet one come from $(I_3)_\mathrm{vac}$, 
which vanishes, like in Eq.~\eqref{i3vac}, if the two are set equal, 
as usually done in dimensional regularization.

Putting all pieces together in Eq.~\eqref{defp00} and using  
\begin{eqnarray*}
\m2 \int\frac{d^dp}{(2\pi)^d}\frac{\bk^4}{\bp^4\bq^2} &=& 
\mbk^{1-2\epsilon}\m2 (4\pi)^{-3/2+\epsilon}
\frac{\Gamma(3/2+\epsilon)\Gamma(1/2-\epsilon)\Gamma(-1/2-\epsilon)}{\Gamma(-2\epsilon)}
\\
&=& \epsilon\frac{\mbk^{1-2\epsilon} \m2 }{4}\left[1 + {\cal O}(\epsilon)\right],
\\
\m2 \int\frac{d^dp}{(2\pi)^d}\frac{\bk^2}{\bp^2\bq^2} &=& 
\mbk^{1-2\epsilon}\m2 (4\pi)^{-3/2+\epsilon}\frac{\Gamma(1/2+\epsilon)\Gamma(1/2-\epsilon)^2}{\Gamma(1-2\epsilon)}
\\
&=& \frac{\mbk^{1-2\epsilon} \m2 }{8}\left[1+\epsilon(-\gamma_E + \ln (16\pi)) + {\cal O}(\epsilon^2)\right],
\end{eqnarray*}
we obtain Eqs.~\eqref{pi00vacuum},~\eqref{pi00matter},~\eqref{pi00zero} and~\eqref{pi00sing}.

\subsection{One-loop integrals\label{sub_oneloop}}
We list here the loop integrals $L_d$, $L_d^{\mu}$ and $L_d^{\mu\nu}$, 
obtained with the Gegenbauer polynomials technique \cite{Pascual:1984zb}:
\begin{eqnarray}
L_d(k,r,s)&=&\int\frac{d^d p}{(2\pi)^d}\frac{1}{(p+k)^{2r}p^{2s}}\nonumber\\
\label{iscalar}
&=&\frac{k^{d-2(r+s)}}{(4\pi)^{d/2}}\frac{\Gamma\left(r+s-{d}/{2}\right)}
{\Gamma(r)\Gamma(s)}\,\frac{\Gamma\left({d}/{2}-s\right)
\Gamma\left({d}/{2}-r\right)}{\Gamma(d-s-r)},
\end{eqnarray}	
\begin{eqnarray}
L_d^{\mu}(k,r,s)&=&\int\frac{d^d p}{(2\pi)^d}\frac{p_{\mu}}{(p+k)^{2r}p^{2s}}
\nonumber\\
\label{ivector}
&=&-k^{\mu}\,\frac{k^{d-2(r+s)}}{(4\pi)^{d/2}}
\frac{\Gamma\left(r+s-{d}/{2}\right)}{\Gamma(r)\Gamma(s)}\,\frac{\Gamma\left({d}/{2}+1-s\right)
\Gamma\left({d}/{2}-r\right)}{\Gamma(d+1-s-r)},
\\
L_d^{\mu\nu}(k,r,s)&=&\int\frac{d^dp}{(2\pi)^d}\frac{p^{\mu}p^{\nu}}{(p+k)^{2r}p^{2s}}
\nonumber\\
&=&\frac{k^{d-2(r+s)}}{(4\pi)^{d/2}}\biggl[\frac{k^2}{2}
\frac{\Gamma\left(r+s-1-{d}/{2}\right)}{\Gamma(r)\Gamma(s)}\,\frac{\Gamma\left({d}/{2}+1-s\right)
\Gamma\left({d}/{2}+1-r\right)}{\Gamma(d+2-s-r)}\enspace\delta^{\mu\nu}
\nonumber\\
\label{itensor}
&&\qquad\qquad+\frac{\Gamma\left(r+s-{d}/{2}\right)}{\Gamma(r)\Gamma(s)}\,
\frac{\Gamma\left({d}/{2}+2-s\right)\Gamma\left({d}/{2}-r\right)}
{\Gamma(d+2-s-r)}\enspace k^{\mu}k^{\nu}
\biggr].
\end{eqnarray}

\section{Expansions\label{app_exp}}
In this appendix, we list the expansions of the gluon self energy for
temperatures much greater or smaller than the momentum $k$.

We start with $T\gg\mbk$. In the non-static sector, 
$I_0$ gives its exact result~\eqref{sumi0} and $I_3$ reads in dimensional regularization
\begin{equation}
I_3= - \frac{2 T \bk^2\Gamma(1-d/2)(2\pi T)^{d-4}\mu^{2\epsilon}}{(4\pi)^{d/2}}\zeta(4-d).
\label{i3exact}
\end{equation}
For the other integrals, we first carry out the integral, then
Taylor expand the result in ${\bk^2}/{\omega_n^2}$ and finally perform the
sums with the zeta function, thus obtaining 
\begin{eqnarray}
I_1&=&
\frac{T^2}{12}-\frac{\Gamma(2-d/2)\mu^{2\epsilon}\left(\sqrt{\pi}T\right)^d}{2\pi^2 T}
\sum_{l=0}^\infty\frac{\Gamma(d/2-1)\Gamma(l+1)}{\Gamma(d/2-1-l)\Gamma(2l+2)}\zeta(2l+2-d)
\left(\frac{k}{2\pi T}\right)^{2l},
\nonumber\\
\\
I_2&=&
\frac{\bk^2\Gamma(2-d/2)\mu^{2\epsilon}\left(\sqrt{\pi}T\right)^{d}}{8\pi^4T^3}
\sum_{l=0}^\infty\frac{\Gamma(d/2-1)\Gamma(l+1)}{\Gamma(d/2-1-l)\Gamma(2l+2)}\zeta(2l+4-d)
\left(\frac{k}{2\pi T}\right)^{2l},\nn\\
\\
I_4&=&
\frac{\bk^4 \Gamma(2-d/2)\mu^{2\epsilon}\left(\sqrt{\pi} T\right)^{d}}{32\pi^6T^5}
\sum_{l=0}^\infty\frac{\Gamma(d/2-1)\Gamma(l+1)}{\Gamma(d/2-1-l)\Gamma(2l+2)}\zeta(2l+6-d)
\left(\frac{k}{2\pi T}\right)^{2l}.\nn\\
\end{eqnarray}
In the fermionic, sector we have
\begin{equation}
\tilde{I}_0=-\frac{T^2}{24},
\label{i0f}
\end{equation}
and we can derive the expansions for $\tilde{I}_1$ and $\tilde{I}_2$ following the 
same procedure used for the bosonic integrals, but ending up with the generalized (Hurwitz) zeta
function as a result of the odd frequency sums. 
Thus we have
\begin{align}
\tilde{I}_1=&\frac{\Gamma(2-d/2)\mu^{2\epsilon}\left(\sqrt{\pi} T\right)^{d}}{2\pi^2T}
\sum_{l=0}^\infty\frac{\Gamma(d/2-1)\Gamma(l+1)}{\Gamma(d/2-1-l)\Gamma(2l+2)}
\zeta(2l+2-d,1/2)\left(\frac{k}{2\pi T}\right)^{2l},
\\
\tilde{I}_2=&\frac{\bk^2\Gamma(2-d/2)\mu^{2\epsilon}\left(\sqrt{\pi} T\right)^{d}}{8\pi^4T^3}
\sum_{l=0}^\infty\frac{\Gamma(d/2-1)\Gamma(l+1)}{\Gamma(d/2-1-l)\Gamma(2l+2)}\zeta(2l+4-d,1/2)
\left(\frac{k}{2\pi T}\right)^{2l}.
\end{align}
Plugging these expressions in Eqs.~\eqref{defp00} and~\eqref{pi00fermion} we
obtain the high-temperature expansion~\eqref{kllt}.

We consider now the low-temperature expansion.
The vacuum part gives the order $\bk^2$ term in the expansion,
whereas, for the matter part, the condition $\mbk\gg T$ translates 
in Eq.~\eqref{pi00matter} into $\mbk\gg \mbp$, since the internal
momentum $\mbp$ is of order $T$. Expanding this expression in
$\mbp/\mbk\ll 1$, as previously done in Sec.~\ref{secvacuumpolkoo}, yields
\begin{eqnarray} 
\Pi_{00}^{\mathrm{NS}}(\mbk\gg T)_{\mathrm{mat}}&=&
- g^2C_A\frac{T^2}{18}
+g^2T^2\mathcal{O}\left(\frac{T^2}{\bk^2}\right).
\label{nonstaticcontrlowt}
\end{eqnarray}
The singular term ($\propto \mbk^3/T$) and the subtracted zero-mode part 
also contribute in this region. 
The sum of Eq.~\eqref{nonstaticcontrlowt} with the vacuum, subtracted zero-mode and
singular parts yields Eq.~\eqref{kggt}.
For what concerns the static modes, the only scales are $\mbk$ and $m_D$, 
thus the condition $\mbk\gg T$ becomes $\mbk\gg m_D$ and we end up with 
Eq.~\eqref{staticcontrhighk}.
Finally, the fermionic contribution is suppressed in this region, i.e. the first 
nonzero term in the expansion of Eq.~\eqref{matterfermion} is of order 
$g^2T^4/{\bk^2}$.

\section{Non-static two-loop sum-integrals\label{app_integrals}}
We set on the evaluation of the two-loop sum-integrals defined by Eq.~\eqref{defji}.
$J_0$ does not contribute in dimensional regularization because the integral over $\bk$ has no scale.
$J_1$ can be rewritten as
\begin{equation}
\label{rewritej1}
J_1=\m2 \int \frac{d^dk}{(2\pi)^d}\int_p^\prime\frac{\bp^2}{\bk^4p^2q^2}
=\m2 \int \frac{d^dk}{(2\pi)^d}\frac{1}{\bk^4}\int_p^\prime\frac{1}{q^2}
-\m2 \int \frac{d^dk}{(2\pi)^d}\int_p^\prime\frac{\omega_n^2}{\bk^4p^2q^2}.
\end{equation}
The first term vanishes in dimensional regularization, whereas the second one
yields\footnote{
A convenient way to proceed is by performing first the momentum integrations,
by means of two Feynman parameters, and then the frequencies sum, which gives
$\zeta(0) = -1/2$.}
\begin{equation}
J_1=-\m2 \int
\frac{d^dk}{(2\pi)^d}\int_p^\prime\frac{\omega_n^2}{\bk^4p^2q^2}
=-\frac{T}{8(4\pi)^2}.
\label{finalj1}
\end{equation}
$J_2$ can be read from \cite{Arnold:1994ps},
\begin{equation}
J_2 = \m2 \int \frac{d^dk}{(2\pi)^d}\int_p^\prime\frac{1}{\bk^2p^2q^2}
=\frac{T}{(4\pi)^2}\left(-\frac{1}{4\epsilon}+\ln\frac{2T}{\mu}
-\frac{1}{2}+\frac{\gamma_E}{2}-\frac{\ln (4\pi)}{2}\right).
\end{equation}
$J_3$ vanishes in dimensional regularization because the $\bk$ integral has no scale and finally $J_4$ yields
\begin{equation}
J_4=\m2 \int\frac{d^dk}{(2\pi)^d}\int_p^\prime\frac{1}{p^2q^2\omega_n^2}
=T\sum_{n\ne0}\frac{1}{\omega_n^2}\left(-\frac{\vert\omega_n\vert}{4\pi}\right)^2
=-\frac{T}{(4\pi)^2}.
\end{equation}

We consider now the fermionic integrals. $\tilde{J}_0$ vanishes because it has
a scaleless $\bk$ integration, whereas $\tilde{J}_1$ can be computed along
the lines of its bosonic counterpart, performing the sum over odd frequencies
by means of the the generalized (Hurwitz) zeta function, 
\begin{equation}
\tilde{J}_1=\m2 \int \frac{d^dk}{(2\pi)^d}T\sum_{n}\m2 
\int\frac{d^dp}{(2\pi)^d}\frac{\tilde{\omega}_n^2}{\bk^4p^2q^2}=-\frac{T\zeta(0,1/2)}{4(4\pi)^2}=0.
\label{finalj1f}
\end{equation}
$\tilde{J}_2$ can be read from \cite{Arnold:1994eb},
\begin{equation}
\tilde{J}_2=
\m2 \int \frac{d^dk}{(2\pi)^d}T\sum_{n}\m2 \int\frac{d^dp}{(2\pi)^d}\frac{1}{\bk^2p^2q^2}=-\frac{T}{(4\pi)^2}\ln2.
\end{equation}

\section{Static-modes contribution to the Polyakov loop\label{app_3d}}
In this appendix, we evaluate the 6-dimensional two-loop
integral entering Eq.~\eqref{mcontribnlos}. 
We will perform the calculation modifying the magnetostatic propagator in Eq.~\eqref{propnonstatic} into
\begin{equation}
\frac{1}{\bk^2}\left(\delta_{ij}-(1-\xi)\frac{k_ik_j}{\bk^2}\right)\delta_{n0}
\to\frac{1}{\bk^2+m_m^2}\left(\delta_{ij}-(1-\xi)\frac{k_ik_j}{\bk^2}\right)\delta_{n0}, 
\label{magnprop}
\end{equation}
where $m_m$ may be interpreted as a small magnetic mass to be put to zero at the end of the calculation.
The magnetic mass modifies the static gluon self-energy expression with resummed gluon propagators 
from Eq.~\eqref{staticvacpol} to 
\begin{equation}
\Pi_{00}^{\mathrm{S}}(\bk)=g^2\ca T\m2 \int\frac{d^dp}{(2\pi)^d}
\left[
\frac{d-(1-\xi)}{\bp^2+m_m^2} 
-\frac{(\bk+\bq)^2-(1-\xi)\displaystyle\frac{((\bk+\bq)\cdot\bp)^2}{\bp^2}}{(\bp^2+m_m^2)(\bq^2+m_D^2)}
\right],
\label{magnmass}
\end{equation}
where $q=k-p$.\footnote{
We use here a different parameterization of the integrand  with respect to  Eq.~\eqref{staticvacpol}.
} 

In  Eq.~\eqref{mcontribnlos}, the integral over the first term in
Eq.~\eqref{magnmass}, i.e. the tadpole contribution, gives 
\begin{equation}
-\frac{d-(1-\xi)}{4\pi}\frac{g^4C_R\ca m_m}{2}\m2 \int \frac{d^dk}{(2\pi)^d}\frac{1}{(\bk^2+m_D^2)^2}
=-\left[d-(1-\xi)\right]\frac{g^4C_R\ca}{4(4\pi)^2}\frac{m_m}{m_D}.
\label{tadpolecontrib}
\end{equation}
For the second term, we start by considering the term proportional to $(\bk+\bq)^2$.
We rewrite 
\begin{equation}
(\bk+\bq)^2=2(\bk^2+m_D^2)+2(\bq^2+m_D^2)-(\bp^2+m_m^2)+(m_m^2-4m_D^2),
\label{algebraicrewrite}
\end{equation}
and consider the contributions given by each of the four terms in brackets. 
The first one gives
\begin{eqnarray}
&& 2\mu^{4\epsilon} \int \frac{d^dp}{(2\pi)^{d}}\int \frac{d^dk}{(2\pi)^{d}}
\frac{1}{(\bk^2+m_D^2)(\bp^2+m_m^2)(\bq^2+m_D^2)}
\nonumber\\
&=&\frac{2}{(4\pi)^2}\left[\frac{1}{4\epsilon}+\ln\frac{\mu}{2m_D+m_m}+\frac{1}{2}
-\frac{\gamma_E}{2}+\frac{\ln (4\pi)}{2}+\mathcal{O}(\epsilon)\right],
\label{staticozero}
\end{eqnarray}	
the second one gives 
\begin{equation}
2\mu^{4\epsilon}\int \frac{d^dp}{(2\pi)^{d}}\int \frac{d^dk}{(2\pi)^{d}}
\frac{1}{(\bk^2+m_D^2)^2(\bp^2+m_m^2)}= -\frac{1}{(4\pi)^2}\frac{m_m}{m_D},
\label{noq}
\end{equation}
the third one gives 
\begin{equation}
-\mu^{4\epsilon}\int \frac{d^dp}{(2\pi)^{d}}\int \frac{d^dk}{(2\pi)^{d}}
\frac{1}{(\bk^2+m_D^2)^2(\bq^2+m_D^2)}=\frac{1}{2(4\pi)^2},
\end{equation}
and the last one 
\begin{eqnarray*} 
&&(m_m^2-4m_D^2)\m2 \int \frac{d^dk}{(2\pi)^d}\m2 \int \frac{d^dp}{(2\pi)^d}
\frac{1}{(\bk^2+m_D^2)^2(\bp^2+m_m^2)(\bq^2+m_D^2)}
\\
=&&\frac{m_m^2-4m_D^2}{-2m}\frac{\partial}{\partial m}\m2 
\left. 
\int \frac{d^dk}{(2\pi)^d}\m2 
\int\frac{d^dp}{(2\pi)^d}\frac{1}{(\bk^2+m^2)(\bp^2+m_m^2)(\bq^2+m_D^2)}\right|_{m=m_D}
\\
=&& \frac{1}{(4\pi)^2}\frac{m_m^2-4m_D^2}{2m_D(2m_D+m_m)}.
\end{eqnarray*}
Finally, we consider the term proportional to ${((\bk+\bq)\cdot\bp)^2}/{\bp^2}$
in Eq.~\eqref{magnmass}. We rewrite the numerator as
\begin{equation}
(1-\xi)\frac{((\bk+\bq)\cdot\bp)^2}{\bp^2}=
\frac{1-\xi}{\bp^2}[(\bk^2+m_D^2)^2+(\bq^2+m_D^2)^2-2(\bk^2+m_D^2)(\bq^2+m_D^2)].
\end{equation}
The first term gives
\begin{equation}
\mu^{4\epsilon} \int \frac{d^dp}{(2\pi)^{d}}\int \frac{d^dk}{(2\pi)^{d}}
\frac{1}{\bp^2(\bp^2+m_m^2)(\bq^2+m_D^2)}=
-\frac{1}{(4\pi)^2}\frac{m_D}{m_m},
\end{equation}
the third term is $-2$ times this one and the second term gives
\begin{eqnarray*}
&&\m2 \int \frac{d^dp}{(2\pi)^{d}}\frac{1}{\bp^2(\bp^2+m_m^2)}\m2 
\int\frac{d^dk}{(2\pi)^{d}}\frac{\bk^2+\bp^2-2\bp\cdot\bk+m_D^2}{(\bk^2+m_D^2)^2}
\\
&=&\m2 \int \frac{d^dp}{(2\pi)^{d}}\frac{1}{\bp^2(\bp^2+m_m^2)}
\left[\frac{\bp^2}{8\pi m_D}-\frac{m_D}{4\pi}\right]
=-\frac{1}{(4\pi)^2}\left[\frac{m_D}{m_m}+\frac{m_m}{2m_D}\right].
\end{eqnarray*}
The static contribution is thus
\begin{eqnarray}
\nonumber\delta\langle L_R\rangle_{\mathrm{S}\,m_D}
&=&\frac{g^4C_AC_R}{2(4\pi^2)}\left[
-\frac{1}{2\epsilon}-\ln\frac{\mu^2}{(2m_D+m_m)^2}+\gamma_E -\ln (4\pi)
+\frac{2-d}{2}\frac{m_m}{m_D}
\right.\\
&&\hspace{2cm} 
\left.-\frac{3}{2}-\frac{m_m^2-4m_D^2}{2m_D(2m_D+m_m)}\right].
\label{finalcontribmmagn}
\end{eqnarray}
The final result is independent of the gauge parameter $\xi$. 
The expression is well behaved for $m_m\to0$ and yields Eq.~\eqref{finalcontribm}.

\section{The Polyakov loop in Feynman gauge\label{app_feynman}}
In this section, we sketch the computation of the vacuum expectation
value of the Polyakov loop in Feynman gauge. We restrict ourselves to  
the fundamental representation ($L\equiv L_F$). 
Since the fermionic contribution, evaluated in
Sec.~\ref{sec_ploop}, is to that order gauge-invariant, 
we do not need to compute it here again.

The perturbative expansion of the Polyakov line through the Baker--Campbell--Haus\-dorff formula is, 
following \cite{Curci:1984rd} and up to order $g^4$, 
\begin{eqnarray}
\nonumber\langle \trt L\rangle &=& 
\frac{1}{\nc}\left\langle{\rm Tr} {\rm P} \exp\left(-ig\int_0^{1/T} d\tau A_0(\bx,\tau)\right)\right\rangle=
\frac{1}{\nc}\left\langle{\rm Tr}\left(1+\frac{g^2}{2}(H_0^2+g^2H_1^2\right.\right.\\
&&\left.\left.+2gH_0H_1+2g^2H_0H_2)+\frac{1}{3!}g^3(H_0^3+3gH_0^2H_1)+\frac{1}{4!}g^4H_0^4\right)\right\rangle+\ldots,
\label{bch}
\end{eqnarray}
where
\begin{eqnarray}
 H_0&=&-i\int_0^{1/T} d\tau A_0(\tau),
\nonumber\\
 H_1&=&-\frac{1}{2}\int_0^{1/T} d\tau_1 \int_0^{\tau_1} d\tau_2 \left[A_0(\tau_2),A_0(\tau_1)\right],
\nonumber \\
H_2 &=&-\frac{1}{6}\left[H_0,H_1\right]
+\frac{i}{6}\int_0^{1/T} d\tau_1 \int_0^{\tau_1} d\tau_2\left[A_0(\tau_2),\left[A_0(\tau_2),A_0(\tau_1)\right]\right]
\nonumber\\
&&+\frac{i}{3}\int_0^{1/T} d\tau_1 \int_0^{\tau_1} d\tau_2\int_0^{\tau_2} d\tau_3
\left[A_0(\tau_3),\left[A_0(\tau_2),A_0(\tau_1)\right]\right],
\label{defh2}
\end{eqnarray}
and $A_0(\tau,\bx)\equiv A_0(\tau)$. 
We recall that 
\begin{eqnarray*}
D_{00}(\tau)&\equiv& 
\theta(\tau)\langle A_0(\tau) A_0(0) \rangle + \theta(-\tau)\langle A_0(0) A_0(\tau) \rangle \\
&=& T\sum_{n}e^{i\omega_n\tau}\m2 \int\frac{d^dk}{(2\pi)^d}D_{00}(\omega_n,\bk),
\end{eqnarray*}
where in Feynman gauge the free temporal-gluon propagator is 
\begin{equation}
D_{00\,F}(\omega_n,\bk)=\frac{1}{\omega_n^2+\bk^2}.
\end{equation}

\begin{figure}[ht]
\begin{center}
\includegraphics{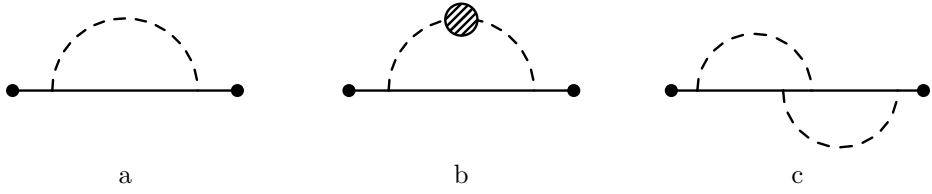}
\end{center}
\caption{Diagrams contributing to the Polyakov loop up to order $g^4$
in Feynman gauge. The blob stands for the one-loop gluon self energy,
the solid line for the Polyakov line and the dots at its beginning/end
represent the points $(0,\bx)$ and $(1/T,\bx)$, which are
compactified by the periodic boundary conditions. 
When integrating over loop momenta of order $m_D$, the dashed lines 
stand for resummed temporal propagators, elsewhere for free ones.}
\label{fig:feynman}
\end{figure}

We can now start working on the different terms in Eq.~\eqref{bch}. 
The first one gives 
\begin{equation}
\frac{1}{\nc}\left\langle{\rm Tr}\frac{g^2}{2}H_0^2\right\rangle
=-\frac{1}{2}\frac{g^2\cf}{T}\m2 \int\frac{d^dk}{(2\pi)^d}D_{00}(0,\bk).
\label{lofeynman}
\end{equation}
Following the same approach as in Sec.~\ref{sec_ploop}, at order $g^4$, 
the relevant diagrams contributing to~\eqref{lofeynman} are shown in 
Fig.~\ref{fig:feynman} a and b. At leading order, the Debye mass is gauge invariant, 
whereas the full one-loop gluon self-energy is not. It is
convenient to separate non-static from static modes. The former
yield \cite{Arnold:1994ps}
\begin{equation}
\Pi_{00}^{\mathrm{NS}}(0,\bk)=-2g^2\ca\left(\frac{d-1}{2}I_0 - (d-1) I_1 + I_2\right),
\label{pi00nsfeynman}
\end{equation}
where the master integrals $I_j$ are those defined in Eq.~\eqref{defmaster}, 
hence Eq.~\eqref{pi00nsfeynman} equals the first three terms
of the static-gauge expression~\eqref{defp00}. The static mode contribution 
to the self energy is common to all gauges that share the same static propagator as the 
static gauge and the Feynman gauge do. Therefore, the static part of the 
self energy in Feynman gauge is just Eq.~\eqref{staticvacpol} with $\xi=1$. 
We then have, separating the contributions coming from the scale $T$ 
from those coming from the scale $m_D$, 
\begin{equation}
\m2 \int\frac{d^dk}{(2\pi)^d}D_{00}(0,\bk)=
\m2 \int\frac{d^dk}{(2\pi)^d}
\left[\frac{1}{\bk^2+m_D^2}-\frac{\Pi_{00}^{\mathrm{NS}}(\mbk\sim T)}{\bk^4}
-\frac{\Pi_{00}^{\mathrm{S}}(\mbk)}{(\bk^2+m_D^2)^2}\right]+\ldots,
\label{feynpropg2}
\end{equation}
where the dots stand for higher orders in the perturbative expansion.  
We have omitted the non-static contribution at the scale $m_D$
(cf. Eq.~\eqref{mcontribnlons}) since it can be shown that also in
Feynman gauge $\Pi_{00}^\mathrm{NS}(\mbk\sim m_D)-m_D^2
=\mathcal{O}\left(g^2\bk^2\right)$, leading to a higher-order
contribution, whereas the contribution of the static modes at the
scale $T$ leads to a scaleless integral. Plugging Eq.~\eqref{feynpropg2}
into Eq.~\eqref{lofeynman} and using the results of appendices
\ref{app_integrals} and~\ref{app_3d} we obtain most of the final,
order $g^4$, result, except for the contribution of $J_4$ in
Eq.~\eqref{tcontribnlonsj}.

We then consider the other terms in the
Baker--Campbell--Hausdorff expansion, starting from $H_1^2$:
\begin{eqnarray}
\frac{1}{\nc}\left\langle{\rm Tr}\,\frac{g^4}{2}H_1^2\right\rangle&=&
\frac{\cf\ca}{8}g^4\int_0^{1/T} d\tau_1 \int_0^{\tau_1} d\tau_2\int_0^{1/T} d\tau_3 
\int_0^{\tau_3} d\tau_4\left[D_{00}(\tau_2-\tau_3)\right.
\nonumber\\
&&\hspace{35mm}
\left.\times D_{00}(\tau_1-\tau_4)-D_{00}(\tau_2-\tau_4)D_{00}(\tau_1-\tau_3)\right]
\nonumber\\
&=&-2\zeta(0)\frac{g^4\cf\ca }{4(4\pi)^2}+\ldots=\frac{\als^2\cf\ca}{4}+\ldots,
\label{nlofeynman}
\end{eqnarray}
where we have used free propagators and the dots stand for higher orders. 
This result corresponds exactly to the contribution of $J_4$ in the static
gauge. The contribution can be traced back to diagram c in Fig.~\ref{fig:feynman} and 
it corresponds to the term called $\mathcal{L}_4$ in Eq. (4) of \cite{Gava:1981qd}.

We now need to show that the sum of the remaining terms yields zero at order $g^4$. 
$\langle{\rm Tr}\, 2gH_0H_1\rangle$ vanishes because it involves 
a three temporal-gluon vertex.
$\langle{\rm Tr} \, 2g^2H_0H_2\rangle$ is a more complicated object,
however one can show that, working with free propagators \cite{Curci:1984rd}, 
\begin{equation}
\frac{1}{\nc}\langle{\rm Tr}\,g^4H_0H_2\rangle=0+\mathcal{O}(g^5,g^4\times(m_D/T)).
\label{hoh2}
\end{equation}
The $H_0^3$ term vanishes, again due to the three temporal-gluon vertex 
and the $H_0^2H_1$ term can be easily shown to be zero after performing 
the colour trace.
The $H_0^4$ term gives
\begin{equation}
\frac{1}{4!\nc}\langle{\rm Tr}\,g^4H_0^4\rangle=\frac{g^4}{4!}
\left(3\cf^2-\frac{\cf\ca}{2} \right)\frac{1}{T^2}
\left(\m2 \int\frac{d^dk}{(2\pi)^d}D_{00}(0,\bk)\right)^2,
\label{h04}
\end{equation}
which is at least of order $g^4\times(m_D/T)^2$. 
This, finally, shows that the Feynman-gauge computation of the Polyakov loop 
agrees with the static-gauge computation that led to Eq.~\eqref{finalg4loop}.

\section{Octet contributions\label{app_octet}}
In this appendix, we want to prove that, up to order $g^6 (rT)^0$, 
$\displaystyle 
\delta_{o,T}^{\, {\cal O}(r^2)\, {\rm NS}} =  \delta_{s,T}^{\, {\cal O}(r^2)\, {\rm NS}}\vert_{V_s\leftrightarrow V_o}$,
$\displaystyle 
\delta_{o,T}^{\, {\cal O}(r^2)\, {\rm S}}  =  - \delta_{s,T}^{\, {\cal O}(r^2)\, {\rm S}}$,
and $\displaystyle \delta_{o,T}^{\,\delta {\cal L}_{\rm pNRQCD}} = -\delta_{s,T}^{\,\delta {\cal L}_{\rm pNRQCD}}$, 
where the left- and right-hand sides of the equalities encode non-zero modes, zero-modes and higher-multipole 
one-loop corrections to the pNRQCD octet and singlet 
propagators respectively induced by interaction vertices of the type 
${\rm S}^\dagger r^{i_1}...r^{i_n}\partial_{i_1}...\partial_{i_{n-1}}E^{i_n} {\rm O} \, +$ Hermitian conjugate   
or ${\rm O}^\dagger r^{i_1}...r^{i_n}\partial_{i_1}...\partial_{i_{n-1}}E^{i_n} {\rm O} \, +$ charge conjugate.

\begin{figure}[ht]
\begin{center}
\includegraphics{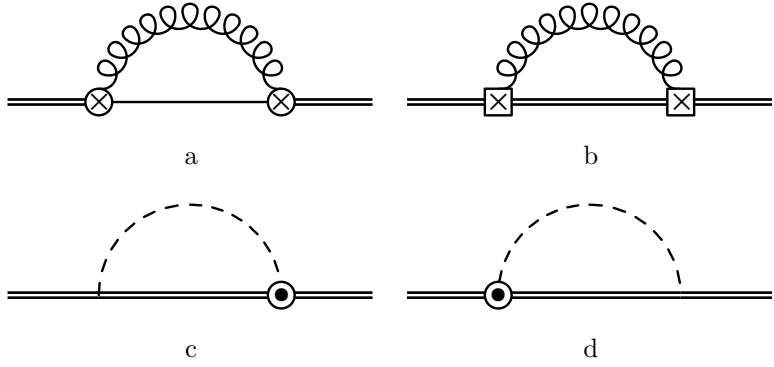}
\put(-189,21.5){$\boldsymbol\bullet$}
\put(-108,21.5){$\boldsymbol\bullet$}
\put(-191,94.2){$\boldsymbol\times$}
\put(-109.5,94.2){$\boldsymbol\times$}
\put(-259,94.2){$\boldsymbol\times$}
\put(-41,94.2){$\boldsymbol\times$}
\end{center}
\caption{The pNRQCD Feynman diagrams giving the leading-order correction to $\delta_o$.
The single continuous line stands for a singlet propagator, 
the double line for an octet propagator, the circle with a cross for the 
chromoelectric dipole vertex proportional to $V_A$ in the Lagrangian~\eqref{pNRQCDeuc}, 
the square  with a cross for the chromoelectric dipole vertex proportional to $V_B$ 
in the Lagrangian~\eqref{pNRQCDeuc},  the circle with a dot for the 
chromoelectric dipole vertex proportional to $V_C$ in the Lagrangian~\eqref{pNRQCDeuc}, 
the curly line for a chromoelectric correlator and the dashed line for a temporal-gluon propagator.
}
\label{fig:octetEEV}
\end{figure}

The general argument goes as follows. Let us first consider contributions coming from the non-zero modes
of the loop integral, Fig.~\ref{fig:hqself}  providing the leading-order contribution to the singlet 
propagator and diagram a in  Fig.~\ref{fig:octetEEV} providing the leading-order contribution 
to the octet propagator. As the leading-order example shows, there is a one to one correspondence between 
diagrams in the singlet and in the octet channel, to each singlet diagram corresponds an octet 
diagram whose contribution is equal to the singlet diagram contribution with $V_s$ replaced by $V_o$ and viceversa.
We note that, since at order $g^4$ these contributions are linear in $V_o-V_s$, they are at that order 
one the opposite of the other.

Let us now consider contributions coming from the zero modes of the loop integral. In order to see how things 
work, we consider, first, the order $r^2$ contribution. In the singlet channel, only one diagram, Fig.~\ref{fig:hqself}, 
contributes; that contribution has been written in Eq.~\eqref{EEVb} and evaluated in Eq.~\eqref{deltasEb}.
In the octet channel, four diagrams contribute, which are shown in  Fig.~\ref{fig:octetEEV}.
Diagram a gives the same contribution as the singlet channel:
\begin{eqnarray}
\delta_{o,T}^{\, a) \, {\rm S}} &=& 
- g^2 \frac{1}{2\nc}\frac{r^ir^j}{2T}
\int \frac{d^dk}{(2\pi)^d} \langle E^{i\,a} U_{ab}E^{j\,b}\rangle(0,\bk)\vert_{|\bk|\sim T}
 + {\cal  O}\left(g^6(rT)\right).
\label{EEVOa}
\end{eqnarray}
Diagram b is like diagram a with the colour factor $1/(2\nc)$ replaced by $d^{abc}d^{abc}/[4(\nc^2-1)]$:
\begin{eqnarray}
\delta_{o,T}^{\, b) \, {\rm S}} &=& 
- g^2\frac{\nc^2-4}{4\nc} \frac{r^ir^j}{2T}
\int \frac{d^dk}{(2\pi)^d} \langle E^{i\,a} U_{ab}E^{j\,b}\rangle(0,\bk)\vert_{|\bk|\sim T}
 + {\cal  O}\left(g^6(rT)\right).
\label{EEVOb}
\end{eqnarray}
Finally, diagrams c and d are like diagram a 
with the colour factor $1/(2\nc)$ replaced by $f^{abc}f^{abc}/[8(\nc^2-1)]$:
\begin{eqnarray}
\delta_{o,T}^{\, c)+d) \, {\rm S}} &=& 
g^2 \frac{\nc}{4} \frac{r^ir^j}{2T}
\int \frac{d^dk}{(2\pi)^d} \langle E^{i\,a} U_{ab}E^{j\,b}\rangle(0,\bk)\vert_{|\bk|\sim T}
 + {\cal  O}\left(g^6(rT)\right),
\label{EEVOc}
\end{eqnarray}
where the positive sign comes from moving a derivative acting on the chromoelectric field in one vertex 
to the temporal gluon in the other one (see also Eq.~\eqref{staticcoulomb}).
Summing Eqs.~\eqref{EEVOa}-\eqref{EEVOc} we obtain the opposite of the singlet contribution in  Eq.~\eqref{EEVb}.

This argument may be easily generalized to any order in the multipole expansion. 
Let us consider diagrams contributing to order $2n$ in the multipole expansion.
The singlet contribution is proportional to 
\begin{equation}
\delta_{s,T}^{\, {\cal O}(r^{2n}) \, {\rm S}} \propto 
r^{2n} \frac{1}{2\nc} \sum_{\ell = 0}^{n-1}\frac{1}{(2\ell+1)!}\frac{1}{(2n-(2\ell+1))!}.
\end{equation}
Again there are three classes of octet contributions that correspond to the three classes discussed 
at order $r^2$. Except for the first class, each one has a different colour factor with respect to the singlet 
contribution, but for the rest they are equal:
\begin{eqnarray}
\delta_{o,T}^{\, a) \, {\rm S}} &\propto& 
- r^{2n} \frac{1}{2\nc} \sum_{\ell = 0}^{n-1}\frac{1}{(2\ell+1)!}\frac{1}{(2n-(2\ell+1))!}, 
\\
\delta_{o,T}^{\, b) \, {\rm S}} &\propto& 
- r^{2n} \frac{\nc^2-4}{4\nc} \sum_{\ell = 0}^{n-1}\frac{1}{(2\ell+1)!}\frac{1}{(2n-(2\ell+1))!}, 
\\
\delta_{o,T}^{\, c)+d) \, {\rm S}} &\propto& 
r^{2n} \frac{\nc}{4} \sum_{\ell = 0}^{n}\frac{1}{(2\ell)!}\frac{1}{(2n-2\ell)!},
\end{eqnarray}
where the positive sign in the last expression comes from moving an odd number of derivatives 
acting on the field in one vertex to the field in the other one. 
Since $\displaystyle \sum_{\ell = 0}^{n}\frac{1}{(2\ell)!}\frac{1}{(2n-2\ell)!} = 
\sum_{\ell = 0}^{n-1}\frac{1}{(2\ell+1)!}\frac{1}{(2n-(2\ell+1))!}$, the sum of 
all octet contributions is just the opposite of the singlet contribution.

	\cleardoublepage
	\phantomsection
	\addcontentsline{toc}{chapter}{{}Bibliography}
	\bibliographystyle{myapsrev}
	\bibliography{biblio}
	\chapter*{Acknowledgments}
		\begin{flushleft}
 	\emph{Turn the light out, say goodnight\\
no thinking for a little while\\
let's not try to figure out everything at once}
\end{flushleft}
So, here we are, finally free for one page from the passive form and the first person plural. I have probably written too much, so let's try to keep this short.\\
First and foremost I would like to thank my advisors Nora and Antonio for their advice and encouragement during these three years of research, through endless hours of discussion and thousands of emails, for giving me an interesting topic to work on and the opportunity to present and discuss it in many interesting locations around the world.

I would also like to thank my collaborators Miguel \'Angel Escobedo, P\'eter Petreczky and Joan Soto for our fruitful work together. I am particularly indebted to P\'eter for getting me interested in the Polyakov-loop correlator and for starting the related project, as well as for the many hours spent discussing the details of the calculation in BNL, Munich and Beijing, and to Miguel for our friendly, ongoing collaboration.

This thesis would not exist without the financial support I have been receiving from the Excellence Cluster Universe. I also acknowledge support from STIBET/DAAD and from the WE Heraeus Stiftung. I would also like to thank Brookhaven National Laboratory, the Kavli Institute for Theoretical Physics in Beijing, the Institute for Nuclear Theory in Seattle and McGill University in Montreal for hospitality while the work that lead to this thesis was being carried out.

A big thanks to my colleagues, past and present, for the nice atmosphere in that narrow corridor. I am in particular indebted to Felix for his countless pieces of advice on German language, proofreading of the Zusammenfassung included, and for his hilarious impressions, to Pablo, for sharing with me the experience of starting anew in a unknown city and to Massi for our long chats and encouraging advice.

Of course a big thanks to my family is in order, for their support from across the alps. This time they were lucky not to have to put up with my thesis mood :) I have also received a lot of transalpine support (Vale even volunteered to organize the PhD party!) from my old friends in Italy (and elsewhere): you are too many to mention, but I do miss you all.
Here in Munich I have made new, wonderful friends and I hope we will stay so in the long run. A special thanks to Laura for putting up with my thesis mood on the u-bahn on occasion.\\
Finally a heartfelt thank-you to Francesca for making these last months of writing lighter and worth remembering.
\end{document}